\documentclass[graybox]{svmult}

\usepackage{type1cm}        
\usepackage{makeidx}         
\usepackage{graphicx}        
\usepackage{multicol}        
\usepackage[bottom]{footmisc}

\usepackage{newtxtext}       %
\usepackage[varvw]{newtxmath}       

\usepackage{cite}

\def\a{\lambda}
\newenvironment{equations}{\equation\aligned}{\endaligned\endequation}
\newcommand{\R}{\mathbb{R}}
\newcommand{\be}{\begin{equation}}
\newcommand{\ee}{\end{equation}}
\newcommand{\fer}[1]{(\ref{#1})}
\def\e{\varepsilon}

\newcommand{\LL}{\Lambda}

\renewcommand{\aa}{a}
\renewcommand{\epsilon}{\varepsilon}
\newcommand{\et}{{\tau}}

\newcommand{\as}{{a}}
\newcommand{\z}{\mathbf{z}}
\newcommand{\fI}{f_I}

\newcommand{\fR}{f_R}
\newcommand{\fS}{f_S}
\newcommand{\fE}{f_E}

\newcommand{\fD}{f_D}
\newcommand{\h}{h}

\newcommand{\RR}{\mathbb{R}}

\def\RR{\mathbb R}

\def\e{\varepsilon}

\def\bbeta{{\bar \beta}}

\renewcommand{\d}{{\rm d}}			

\makeindex             

\begin{document}

\title*{Kinetic modelling of epidemic dynamics: social contacts, control with uncertain data, and multiscale spatial dynamics}

\titlerunning{Kinetic modelling and epidemics}
\author{Giacomo Albi, Giulia Bertaglia, Walter Boscheri, Giacomo Dimarco, Lorenzo Pareschi, Giuseppe Toscani and Mattia Zanella}
\authorrunning{G. Albi, G. Bertaglia, W. Boscheri, G. Dimarco, L. Pareschi, G. Toscani, M. Zanella} 
\institute{Giacomo Albi \at Department of Computer Science, University of Verona, \email{giacomo.albi@univr.it}
\and Giulia Bertaglia, Walter Boscheri, Giacomo Dimarco, Lorenzo Pareschi \at Department of Mathematics and Computer Science, University of Ferrara, Via Machiavelli 30, 44121 Ferrara, Italy \& Center for Modelling Computing and Simulation, University of Ferrara, Via Muratori 9, 44121 Ferrara, Italy. \email{giulia.bertaglia@unife.it, walter.boscheri@unife.it, giacomo.dimarco@unife.it, lorenzo.pareschi@unife.it}
\and Giuseppe Toscani, Mattia Zanella\at Department of Mathematics, University of Pavia, Via Ferrata 1, 44121 Pavia, Italy. \email{giuseppe.toscani@unipv.it, mattia.zanella@unipv.it}
}
%
%
\maketitle

\abstract*{In this survey we report some recent results in the mathematical modeling of epidemic phenomena through the use of kinetic equations. We initially consider models of interaction between agents in which social characteristics play a key role in the spread of an epidemic, such as the age of individuals, the number of social contacts, and their economic wealth. Subsequently, for such models, we discuss the possibility of containing the epidemic through an appropriate optimal control formulation based on the policy maker's perception of the progress of the epidemic. The role of uncertainty in the data is also discussed and addressed. Finally, the kinetic modeling is extended to spatially dependent settings using multiscale transport models that can characterize the impact of movement dynamics on epidemic advancement on both one-dimensional networks and realistic two-dimensional geographic settings.  
}

\abstract{In this survey we report some recent results in the mathematical modeling of epidemic phenomena through the use of kinetic equations. We initially consider models of interaction between agents in which social characteristics play a key role in the spread of an epidemic, such as the age of individuals, the number of social contacts, and their economic wealth. Subsequently, for such models, we discuss the possibility of containing the epidemic through an appropriate optimal control formulation based on the policy maker's perception of the progress of the epidemic. The role of uncertainty in the data is also discussed and addressed. Finally, the kinetic modeling is extended to spatially dependent settings using multiscale transport models that can characterize the impact of movement dynamics on epidemic advancement on both one-dimensional networks and realistic two-dimensional geographic settings. 
}


\section{Introduction}
The recent COVID-19 pandemic has brought mathematical models in epidemiology to unprecedented scientific exposure. Several research groups in all parts of the planet have ventured into the construction and use of mathematical models capable of correctly describing the progress of the epidemic. Such predictive mathematical models are considered critical to understanding the course of the epidemic and planning effective control strategies. 
Most of the proposed models stem from the compartmental approach originally proposed by Kermack and McKendrick^^>\cite{kermack1927,HWH00}, i.e., the classic SIR model based on partitioning the population into susceptible, infected, and recovered.

In compartmental models, the population is divided into groups, where each group or compartment represents a stage of progression in the individual's disease or health. The resulting mathematical model is typically based on a deterministic system of ordinary differential equations (ODEs) that characterizes the transition rates from one compartment to another. The choice of which compartments to include in a model depends on the characteristics of the particular disease being modeled and the purpose of the model.
Recently, numerous extensions of the SIR compartmentalization have been proposed to deal with the specificity of SARS-CoV-2 infection. Such generalizations involve additional compartments to better fit the available experimental data in order to improve the description of disease progression and epidemic characteristics^^>\cite{buonomo2020,Bruno,kantner2020,peirlinck2020,tang2020,lolipiccolomini2020,parolini2021}. 

Most of these models describe the temporal evolution of the epidemic spread only in terms of the average numerical density of individuals in each compartment, thus neglecting other relevant structural properties of individuals, such as their age, wealth status, social contacts, and spatial movements, in favor of an assumption of population and territorial homogeneity. 
Structured epidemic models have been considered by various authors, especially in connection with age dependence, and are a classical topic in mathematical epidemiology^^>\cite{HWH00, Pugliese}. The evolution of these structural quantities, however, is seldom accounted, except for birth and death rates in the context of age-dependent models. 

On the other hand, kinetic models characterized by systems of partial differential equations (PDEs), recently introduced and studied in the social sciences, have shown the ability to accurately describe complex social phenomena such as opinion formation among individuals, the creation of wealth distributions, the emergence of contacts on social networks^^>\cite{APZ3,CMPP10,FPTT19,GCC16,Gup,APZ3}. See for instance the recent monographs and collections^^>\cite{PT13,Naldi}. Similarly, the use of kinetic theory has proven to be very useful in designing feedback controlled models in various fields of social sciences^^>\cite{Albi3,Albi1,DPT} and in modeling the movement dynamics of individuals at different scales^^>\cite{ABFHKPP,aylaj2020,bellomo2019,CPS,SBKT}.  

In addition, available experimental data are often affected by large uncertainty, which must therefore be considered as part of the process of modeling the infectious disease and simulating the potential epidemic scenarios and control strategies^^>\cite{Cetal,Chowell,Rob}. A large amount of research in this direction has been recently carried out in the field of hyperbolic and kinetic equations and it is therefore natural to rely on this scientific background to design new models and numerical methods able to deal efficiently with the presence of uncertain data^^>\cite{bertaglia2020,jin2017,pareschi2020,poette2009,jin2018,Xiu2010}. We also mention some other related research based on modeling the diffusion of COVID-19 using PDEs. Specifically, the multiscale approach in \cite{bellomo2020}, the age- and space-structured model in \cite{colombo2020}, and the space-dependent models in \cite{viguerie2020,viguerie2021,Bere,guglielmi}.

In this survey, we will address these issues in light of a number of recent results in the area of epidemiological modeling using kinetic equations with a focus on applications to the COVID-19 pandemic. More precisely, our presentation is organized into three parts according to the topics covered. In the first part we will focus on the interplay between the evolution of the pandemic and the presence of a social characteristic capable of significantly influencing its behavior and/or that is itself influenced by the pandemic. In the specific case we will consider the influence of the distribution of contacts^^>\cite{DPeTZ,Zanella_mbe,Zanella_m3as} and the distribution of wealth among individuals^^>\cite{DPTZ}.

Next, in the second part our attention will shift to the importance of possible control actions aimed at containing the pandemic \cite{AM,kantner2020}. In this case it is of fundamental importance to take into account in the modeling phase any uncertain data that can significantly change the epidemic scenarios. A particularly significant problem we will address is that of control actions through containment measures based on different age-dependent social interaction functions, e.g., at home, at work or at leisure. Results for various countries affected by the pandemic will illustrate the effectiveness of the proposed methodology^^>\cite{APZ, APZ2, DTZ}. 

The last part of this survey is devoted to the challenging case of evaluating the impact of an infectious disease at a spatial level, including small scale urban dynamics and large scale regional dynamics. The inclusion of the spatial component in epidemiological systems is indeed crucial especially when there is a need to consider spatially heterogeneous interventions, as was and still is the case for the control of the spread of COVID-19^^>\cite{riley2015,pellis2015,dellarossa2020}. These problems will be addressed both in the case of dynamics on networks connecting different cities^^>\cite{bertaglia2021}, and in the case of completely two-dimensional dynamics at  regional level^^>\cite{boscheri2020}. Applications of these models to the early stages of the COVID-19 pandemic in Italy will also be illustrated^^>\cite{bertaglia2021a,bertaglia2021b}. Some open issues and future developments are also discussed at the end of this review along with detailed references to the data sources used in the simulations.

\section{Kinetic modelling of social heterogeneity in epidemic dynamics}
\label{sec:social}

We discuss in this first part an enhancement of the classical compartmental description of epidemic spread that takes into account statistical aspects of the social behavior of individuals ^^>\cite{DPTZ,DPeTZ,Zanella_mbe,Zanella_m3as}. The approach has its roots in the kinetic theory of socio-economic modelling through interacting agents (see the monograph^^>\cite{PT13} for further details). This permits to correlate the social behavior of agents with the dynamics of infection. 

For simplicity we will develop our arguments for the simple SIR model where the entire population is divided into three classes: susceptible (S), infected (I) and recovered (R) individuals. It should be noted, however, that the ideas developed in this section  can be extended to more complex compartmental epidemic models like the ones considered in^^>\cite{BCF,DH,Gatto,buonomo2020,Bruno}. See also Section \ref{sec:control} and \ref{sec:space} of the present survey for some generalizations to more realistic compartmental models including the effects of asymptomatic individuals.

Under a homogeneous mixing assumption the time evolution of the SIR model reads
\be\label{eq:SIRbase}
\begin{split}
\dfrac{d S(t)}{d t} &= -\beta S(t)I(t),\\
\dfrac{d I(t)}{d t} &= \beta S(t)I(t) - \gamma I(t)\\
\dfrac{d R(t)}{d t} &= \gamma I(t),
\end{split}\ee
where $\beta$ is the average number of contacts per person per time, multiplied by the probability of disease transmission in a contact between a susceptible and an infectious person, and $\gamma$ is the transition rate of infected to the recovered compartment. In this situation, it can be shown that the dynamics of the infectious class depends on the ratio $R_0=\beta/\gamma$, the so-called basic reproduction number. 
In this case, it is known  that $I(t)\to 0$, while $S(t)\to S^\infty \in [0,\gamma/\beta]$ solution of
\[
I(0)+S(0)-S^\infty+\frac{\gamma}{\beta}\log\left(\frac{S^\infty}{S(0)}\right)=0.
\]
We refer to^^>\cite{HWH00} for an introduction on compartmental modelling in epidemiology. 

The heterogeneity of the social structure, which impacts the diffusion of the infective disease, is characterized by the variable $w\in \mathbb{R_+}$, characterizing its social state and whose components summarize, for example, the age of the individual, its number of social connections or its economic status^^>\cite{H96,HWH00}. 
For a large system of interacting individuals in a structured population its statistical description is obtained through the introduction of the distribution functions $f_J(w,t)$, $t\ge 0$, denoting the probability of having an individual with the social characteristic $w$ in the class $J$, where $J\in \{S,I,R\}$ and such that 
\be\label{eq:ch}
f_S(w,t) + f_I(w,t) + f_R(w,t) = f(w,t), \qquad  \int_{\mathbb R_+}f(w,t)dw = 1. 
\ee
As a consequence, the quantities 
\begin{equation}\label{mass}
S(t)=\int_{\mathbb{R}^+}f_S(w,t)\,dw,\,\,\, I(t)=\int_{\mathbb{R}^+}f_I(w,t)\,dw,\,\,\,
R(t)=\int_{\mathbb{R}^+}f_R(w,t)\,dw, 
\end{equation}
denote the fractions of susceptible, infected and recovered subjects. 
In the above setting, the time evolution of the functions $f_J(w,t)$, $J \in \{S,I,R\}$, is obtained by supplementing the epidemiological partitioning in \eqref{eq:SIRbase} with the dynamics originating the formation of social heterogeneity by local interactions. Following^^>\cite{PT13,DPTZ} this merging results in the system
\be\label{eq:kinetic_general}
\begin{split}
\dfrac{\partial f_S(w,t)}{\partial t} &= -K(f_S,f_I)(w,t) + {\cal Q}_S(f_S;f_I;f_R)(w,t),\\
\dfrac{\partial f_I(w,t)}{\partial t} &= K(f_S,f_I)(w,t) - \gamma(w) f_I(w,t) + {\cal Q}_I(f_S;f_I;f_R)(w,t)\\
\dfrac{\partial f_R(w,t)}{\partial t} &= \gamma(w) f_I(w,t) + {\cal Q}_R(f_S;f_I;f_R)(w,t)
\end{split}\ee
where 
\be\label{eq:contact_f}
K(f_S,f_I)(w,t) = f_S(w,t) \int_{\mathbb R_+} \beta(v,w) f_I(v,t)dv, 
\ee
represents a nonlinear incidence rate characterized by the number of contacts $\beta(v,w)$ between an infectious individual with social characteristic $v$ and a susceptible individual with social characteristic $w$, while $\gamma(w)$ defines the transition to the recovered compartment of infectious individuals with social feature $w$. In \eqref{eq:kinetic_general} the  operators ${\cal Q}_J(f_S;f_I;f_R)$, $J$ in $\{S,I,R\}$ describe the evolution of social traits by interactions among agents in the various compartments and the formation of the corresponding equilibrium distributions $f_J(w)^\infty$, $J\in \{S,I,R\}$ such that
\be
\label{eq:staz}
{\cal Q}_J(f_S^\infty;f_I^\infty;f_R^\infty)(w) = 0,\qquad J\in \{S,I,R\}.
\ee
Note that, when the epidemic parameters are independent of the social feature, i.e., $\beta(\cdot,\cdot)=\beta$ and $\gamma(\cdot)=\gamma$, thanks to conservation of the number of individuals in each compartment during the evolution of social traits, by direct integration of \eqref{eq:kinetic_general} against $w$ the mass densities \eqref{mass}  satisfy the classical SIR model \eqref{eq:SIRbase}. 

The explicit computation of the equilibrium solutions of \eqref{eq:staz} is extremely difficult in general, as it depends strongly on the evolution dynamics of the specific social feature under consideration. Knowledge of such equilibrium solutions, however, is of paramount importance to gain some understanding of the dynamics and derive simplified reduced-order models. In the following, we will describe in more details the case of social heterogeneity based on the formation of suitable contact distributions^^>\cite{DPeTZ} and the impact of the epidemic on the wealth distribution of individuals^^>\cite{DPTZ}. 

\subsection{Modelling contact heterogeneity}\label{contact}
Let us first consider a kinetic system which suitably describes the spreading of an infectious disease under the dependence of the contagiousness parameters on the  number of social contacts of the agents. 
Aiming to understand social contacts effects on the dynamics, we will not consider in the sequel the role of {other sources of possible} heterogeneity in the disease parameters (such as the personal susceptibility to a given disease), which could be derived from the classical epidemiological models, suitably adjusted to account for new information^^>\cite{Diek, Novo, Van}. 
Therefore, we denote by $f_S(w,t)$, $f_I(w,t)$ and $f_R(w,t)$, the distributions at time $t > 0$ of the number of social contacts of the population of susceptible, infected and recovered individuals. 

For a given constant $\alpha>0$ we denote with $m_{J,\alpha}(t)$, $J \in \{S,I,R\}$ the local moments of order $\alpha$ for the distributions of the number of contacts in each class conveniently divided by the mass of the class
\be\label{means}
m_{J,\alpha}(t)= \frac 1{{J(t)}}\int_{\mathbb{R}^+}w^\alpha f_J(w,t)\,dw, \quad J \in \{S,I,R\}.
\ee
Unambiguously, we will indicate the local mean values, corresponding to $\alpha=1$, by $m_J(t)$, $J \in \{ S,I,R\}$.

In what follows, we assume that the various classes in the model act differently in the social process constituting the contact dynamics. Specifically, we will consider $\gamma(w) \equiv \gamma >0$ and the contact function $\beta(v,w)$ as a nonnegative increasing function with respect to the number of contacts $v$ and $w$ of infected and susceptible individuals, respectively. The choice 
\be
\beta(v,w) = \bbeta v^\alpha w^\alpha, 
\label{eq:betac}
\ee
with constant $\alpha,\bbeta>0$ corresponds to consider an incidence rate dependent on the product of the number of social contacts.

\subsubsection{Kinetic model for contact formation}

To define the dynamics of contacts, we can exploit the results of^^>\cite{DT,GT19,To3,PT13} to obtain a mathematical formulation of the formation of social contacts. In full generality, we assume that individuals in different compartments can have a different mean number of contacts. Then, the microscopic updates of social contacts from $w$ to $w_J'$ of individuals in the class $J\in \{S,I,R\}$ will be taken of the form
 \be\label{coll}
 w_J' = w  - \Phi^\e_\delta (w/m_J) w + \eta_\e w,\qquad J\in \{S,I,R\},
 \ee
where for compactness and simplicity of notation we used the subscript $J$ on the different compartments and kept implicit the dependence on $\e$ in $w_J'$.  
 
In a single update (interaction), the number $w$ of contacts can be modified for two reasons, expressed by two terms, both proportional to the value $w$. In the first one, the function $\Phi^\e_\delta(\cdot)$, which {takes} both positive and negative values, characterizes the typical and predictable variation of the social contacts of agents, namely the personal social behavior of agents. The quantity $\eta_\e$ is  a random variable of zero mean and bounded variance of order $\e>0$, expressed by $\langle \eta_\e \rangle =0$, $\langle \eta_\e^2 \rangle  = \e\sigma^2$, where $\langle \cdot \rangle$ denotes the expectation. Furthermore, we assume that $\eta_\e$ has finite moments up to order three. 
 
The function $\Phi^\e_\delta$ plays the role of the \emph{value function} in the prospect theory of Kahneman and {Tversky}^^>\cite{KT, KT1}. See also^^>\cite{MP1,MP2,PT13,CMPP10} for a related use of the value function in the dynamics of investment propensity. The {main} hypothesis on which {this function} is built is that, in relationship with the mean value $w_J$, $J\in \{S,I,R\}$, it is {considered} normally easier to increase the value of $w$ {(individuals look for larger networks)} than to decrease it {(people maintain as much connections as possible)}.
In terms of the variable $ s = w/m_J$ we consider then as in^^>\cite{DT} the class of value functions {obeying to the above general rule} given by
 \be\label{vd}
 \Phi_\delta^\e(s) = \lambda \frac{e^{\e(s^\delta -1)/\delta}-1}{e^{\e(s^\delta -1)/\delta}+1 } , \quad  s \ge 0,
 \ee
where the value $\lambda$ denotes the maximal amount of variation of $w$ that agents will be able to obtain in a single interaction
\[
  -\lambda \le \Phi_\delta^\e(s) \le \lambda, 
\]
so that the choice $\lambda <1$ implies that, in absence of randomness, the value of $w_J'$ remains positive if $w$ is positive. In \eqref{vd} the parameter $0 < \delta \le 1$ is a suitable constant characterizing the intensity of the individual behavior, while $\e >0$ is related to the intensity of the interaction. We observe that $\e\ll 1$ corresponds to small variations of the expected difference $\langle w_J' -w\rangle$. 

Thus, for a given density $f_J(w,t)$, $J\in \{S,I,R\}$, the operators ${\mathcal Q}_J(f_S;f_I;f_R)(w,t)$ on the right hand side of \eqref{eq:kinetic_general} have a linear structure, depending only on compartment $J$, characterized by the microscopic interaction \eqref{coll}. Denoting by $Q_J^\e(f_J) ={\mathcal Q}_J(f_S;f_I;f_R)$, the interaction terms can be conveniently written in weak form by integration against a smooth function $\varphi(w)$ as^^>\cite{Cer,PT13}
 \begin{equation}
  \label{kin-w}
\int_{\R_+}\varphi(w){Q}_J^\e(f_J)(w,t)\,dw  = 
  \Big \langle \int_{\R_+}B(w) \bigl( \varphi(w_J')-\varphi(w) \bigr) f_J(w,t)
\,dw \Big \rangle.
 \end{equation}
The above operators quantify the variation in density, at a given time, of individuals in the class $J\in \{S,I,R\}$ that modify their value from $w$ to $w_J'$ (r.h.s with negative sign) and agents  that  change their value from  $w_J'$ to $w$  (r.h.s. with positive sign). 
Here, {the} expectation $\langle \cdot \rangle$ takes into account the presence of the random parameter $\eta_\e$ in the microscopic interaction \fer{coll} while the function $B(w)$ measures the interaction frequency. For example, the choice $B(w)=1/w$, which will be used in the sequel, assigns a low probability to interactions where individuals already have a large number of contacts and assigns a high probability to contact transitions when the value of the variable $w$ is small.

\subsubsection{Quasi-invariant scaling and steady states}\label{FP-limit}

Let us focus on the dynamics of social contacts alone in \eqref{eq:kinetic_general}, namely by ignoring the epidemiological terms, and scale time as $t \to t/\epsilon$, in accordance with the parameter $\e$ that measures the intensity of changes in the number of contacts defined by \eqref{coll}. 
Thus, small values of $\e$ correspond to the case in which elementary interactions \fer{coll} produce minimal modification of the number of social contacts and at the same time their frequency increases like $1/\e$. This scaling is usually referred to as quasi-invariant scaling in kinetic socio-economic modelling^^>\cite{PT13}. A general view about this asymptotic passage from  kinetic equations based on general interactions towards Fokker--Planck type equations can be found in \cite{FPTT}. 

Then, as a result of the scaling, the distribution $f_{J}$, $J \in \{S,I,R\}$ is solution of the following problem in weak form
\begin{equation}
    \label{eq:ki}
    \begin{split}
    \dfrac{d}{dt} \int_{\mathbb R_+} \varphi(w) f_{J}(w,t)dw &= \frac1{\e} \int_{\mathbb R_+} \varphi(w) Q_J^\e(f_{J})(w,t)\,dw\\
    & = \dfrac{1}{\e}   \Big \langle \int_{\R_+}B(w) \bigl( \varphi(w_J')-\varphi(w) \bigr) f_{J}(w,t)
    \,dw \Big \rangle.
    \end{split}
\end{equation}
Now, let us concentrate on the analysis of the asymptotic states of the social contact dynamics when $\e\to 0$.
To this aim, note that, {from the definition of $\Phi_\delta^\epsilon$ in \eqref{vd} and the assumptions on the noise term $\eta_\e$ we have}
\be\label{ottimo}
\lim_{\e \to 0} \frac 1\e { \Phi_\delta^\e\left(\frac w{m_J}\right)} = \frac\lambda{2\delta} \left[\left(\frac w{m_J} \right)^\delta -1\right], \qquad \lim_{\e \to 0} \frac 1\e \langle \eta_\e^2\rangle = \sigma^2.
\ee
We can Taylor expand $\phi(w_J')$ in \eqref{eq:ki} as
\[
\varphi(w'_J)-\varphi(w) = (w'_J - w) \varphi'(w) +  \dfrac{1}{2} (w'_J-w)^2  \varphi''(w) + \dfrac{1}{6}(w'_J - w)^3\varphi'''(\hat w_J), 
\]
with $\hat w_J$ a suitable value between $w_J'$ and $w_j$.
Hence, inserting the above expansion in \eqref{eq:ki} and using the microscopic relation \eqref{coll}, as $\e \to 0$ by standard arguments we can prove that the scaled dynamics \eqref{eq:ki} can be approximated by the corresponding Fokker-Planck formulation^^>\cite{DPeTZ,PT13}. More precisely, it can be shown that $f_{J}$, $J \in \{S,I,R\}$ converges to a solution of 
\[
\begin{split}
&\dfrac{d}{dt} \int_{\mathbb R_+}\varphi(w)f_{J}(w,t)dw =\\
&\qquad   \int_{\mathbb{R}_+} \left\{-\varphi'(w) \, \frac{\lambda \,w^{1-\delta}}{2\delta}\left[\left(\frac w{ m_J} \right)^\delta -1\right] +  \frac{\sigma^2}{2}\varphi''(w)\,w^{2-\delta} \right\}f_{J}(w,t)\,dw.
\end{split}\]
Integrating back by parts, the limit equation in strong form coincides with the Fokker-Planck equation 
\[
\dfrac{\partial}{\partial t}f_J(w,t) = \tilde Q_J(f_J)(w,t), \quad J \in \{S,I,R\},
\] 
where
 \begin{equation}\label{q1}
 \begin{split}
 &\tilde Q_J(f_J)(w,t) =\\
 &\qquad\qquad \frac{\lambda}{2\delta}\frac{\partial}{\partial w}\left\{\,w^{1-\delta}\left[\left(\frac w{ m_J} \right)^\delta -1\right]f_{J}(w,t)\right\} +\frac{\sigma^2}{2} \frac{\partial^2}{\partial w^2} (w^{2-\delta} f_J(w,t)),
\end{split}  \end{equation}
complemented with no-flux boundary conditions at $w=0$
\begin{equation}\label{bc}
\frac{\partial}{\partial w} (w^{2-\delta} f_J(w,t))\Big|_{w=0} = 0.
\end{equation}
Following^^>\cite{DPeTZ} we can compute  the explicit equilibrium distribution of the Fokker-Planck model. Let us first observe that equation \eqref{q1} preserves the total number of individuals and the average number of contacts mean values $m_J$, $J \in \{S,I,R\}$, in each compartment. Thus, assuming that the mass of the initial distribution is one and by setting $\mu = \lambda/\sigma^2$, the equilibria can be expressed by the functions
\be\label{equilibrio}
f_J^\infty(w) =  C_J( m_J,\delta,\mu) w^{\mu/\delta +\delta -2}  \exp\left\{ - \frac \mu{\delta^2} \left( \frac w{ m_J} \right)^\delta\right\},\qquad  J \in \{S,I,R\},
\ee 
 where $C_J> 0$ is a normalization constant. 

\begin{tips}{The distribution of contacts}
A particular interesting case, corresponds to the choice $\delta = 1$ for which the steady  states of unit mass are the Gamma densities 
\be\label{gamma-e}
f_J^\infty(w;\theta, \mu) = \left(\frac \mu{ m_J}\right)^\mu \frac 1{\Gamma\left(\mu \right)} w^{\mu -1}\exp\left\{ -\frac\mu{ m_J}\, w\right\}, \quad J\in\{S,I,R\}.
 \ee 
With this particular choice, the mean values and the {energies} of the densities \fer{gamma-e}, $J\in\{S,I,R\}$, are given by
 \be\label{mv}
 \int_{\R^+} w\, f_J^\infty(w;\theta, \mu) \, dw =  m_J, \qquad  \int_{\R^+} w^2 \, f_J^\infty(w;\theta, \mu) \, dw =\frac{\mu +1}\mu m_J^2.
\ee
 It is important to note that the distribution \eqref{gamma-e} is in agreement with that observed experimentally in^^>\cite{Plos}. For this reason, in the rest of the section we will restrict to the case $\delta = 1$ (see^^>\cite{DPeTZ} for a more in-depth discussion). 
 \end{tips}

\subsubsection{{The macroscopic social-SIR dynamics}}\label{splitting}

Referring to Boltzmann's classical legacy concerning the fluid dynamic limits, using the knowledge of the equilibrium states of the kinetic model we can derive the corresponding macroscopic model^^>\cite{Cer}.   
The key assumption is that the dynamics leading to the contact formation is much faster than the epidemic dynamics. This corresponds to introduce the following scaling
\[
t\to t/\tau,\qquad \beta(v,w) \to \et \beta(v,w),\qquad \gamma \to \et\gamma,
\]
being $\et\ll1$ the scaling parameter. 

Hence, considering the linear Fokker-Planck operator \eqref{q1} for $\delta = 1$ as a model for social interactions we can rewrite system \eqref{eq:kinetic_general} as follows
\begin{equations}\label{sir-FP}
& \frac{\partial f_S(w,t)}{\partial t} = -K(f_S,f_I)(w,t) + \frac 1\et\, \tilde Q_S(f_S)(w,t), \\
&\frac{\partial f_I(w,t)}{\partial t} = K(f_S,f_I)(w,t) - \gamma f_I(w,\tau) + \frac 1\et\, \tilde Q_I(f_I)(w,t), \\
& \frac{\partial f_R(w,t)}{\partial t} =   \gamma f_I(w,t)  + \frac1\et\, \tilde Q_R(f_R)(w,t).
\end{equations}
The system \fer{sir-FP} with no-flux boundary conditions at $w=0$ contains all the information on the spreading of the epidemic in terms of the distribution of social contacts. Indeed, the knowledge of the densities $f_J(w,t)$, $J\in\{S,I,R\}$, allows to evaluate by integrations all moments of interest. Due to the incidence rate $K(f_S,f_I)$, as given in \fer{eq:contact_f}, the time evolution of the moments of the distribution functions is not explicitly computable, since the evolution of a moment of a certain order depends on the knowledge of higher order moments, thus producing a  hierarchy of equations, like in classical kinetic theory of rarefied gases^^>\cite{Cer}. However, similarly to the derivation of the fluid dynamic limit we can assume the contact densities to be close to their equilibrium states \eqref{gamma-e}.

Therefore, since for the choice in \eqref{eq:betac} we have
\[
K(f_S,f_I)(w,t) = \bbeta w \,f_S(w,t) m_I(t) \,I(\tau),
\] 
we can compute the time evolution of the number of individuals in each compartment, defined in \fer{mass},
by integrating both sides of the equations in \fer{sir-FP} with respect to $w$. Using the fact that the Fokker-Planck terms preserve the total number of individuals, we obtain the following system of macroscopic equations for the densities 
\begin{equations}\label{sir-mass}
& \frac{d S(t)}{d t} = -\bbeta \, m_S(t) m_I(t) S(t)I(t),  \\
&\frac{d I(t)}{d t} = \bbeta\, m_S(t) m_I(t) S(t)I(t)  - \gamma I(t),  \\
& \frac{d R(t)}{d t} =   \gamma I(t).
\end{equations}
Next, taking the evolution of the first moment in \fer{sir-FP} since the Fokker--Planck operators also preserve momentum, one obtains that the means $m_S(t)S(t)$, $m_I(t)I(t)$ in \eqref{sir-mass} satisfy the differential system
\begin{equations}\label{sir-mass-m2}
& \frac{d }{d t} (m_S(t)S(t)) = -\bbeta \, m_{S,2}(t) m_I(t) S(t)I(t),  \\
&\frac{d }{d t} (m_I(t)I(t)) = \bbeta\, m_{S,2}(t) m_I(t) S(t)I(t)  - \gamma m_I(t) I(t),  
\end{equations}
which depends now on the second order moments. 

The closure of system \fer{sir-mass}-\fer{sir-mass-m2} can be obtained by resorting, at least formally, to the classical equilibrium assumption on the social interaction variable. Indeed, if {$\tau \ll 1$} is sufficiently small, one can easily argue from the exponential convergence of the solution $f_J(w,t)$ of the Fokker-Planck equation towards the equilibrium $f_J^\infty(w;\theta,\nu)$, $J\in\{S,I,R\}$ (see^^>\cite{To4} for example), that the solution remains sufficiently close to the corresponding Gamma density \eqref{gamma-e} for all times. 

The equilibrium distribution $f_J^\infty(w;\theta,\mu)$ can then be inserted into system \fer{sir-mass-m2} and, recalling that for Gamma densities 
\[
 m_{J,2}(t) = \frac{\mu+1}{\mu} m_J^2(t),\qquad J\in\{S,I,R\},
\]
we can derive a closed system that governs the evolution of the local mean values  \begin{equations}\label{sir-mean}
& \frac{d m_S(t)}{d t} = - \frac\bbeta\nu m_{S}(t)^2 m_I(t) I(t),  \\
&\frac{d m_I(t)}{d t} =   \bbeta m_{S}(t) m_I(t)  \left( \frac{1+\mu}{\mu} m_S(t) - m_I(t) \right) S(t),  \\
& \frac{d m_R(t)}{d t} =  \gamma \frac{I(t)}{R(t)}\left( m_I(t) - m_R(t)\right).
\end{equations}
Therefore, the closure of the kinetic system \fer{sir-FP} around a Gamma-type equilibrium of social contacts leads then to the system of six equations \eqref{sir-mass}-\eqref{sir-mean} for the pairs of mass fractions $J(t)$ and local mean values $m_J(t)$, $J\in\{S,I,R\}$. In the following, we refer to the coupled systems \fer{sir-mass} and \fer{sir-mean} as the social SIR model (S-SIR).

It is interesting to remark that system \fer{sir-mean} is explicitly dependent on the positive parameter $\nu =\mu / \lambda$, which measures the heterogeneity of the population in terms of the variance of the statistical distribution of social contacts. More precisely, small values of the constant $\nu$ correspond to high values of the variance, and thus to a larger heterogeneity of the individuals with respect to social contacts. This is an important point which is widely present and studied in the epidemiological literature^^>\cite{AM,BBT, Diek, DH}. 

\begin{tips}{Absence of heterogeneity }
A limiting case of system \fer{sir-mean} is obtained by letting the parameter $\mu \to +\infty$, which corresponds to push the variance to zero (absence of heterogeneity). In this case, if the whole population starts with a common number of daily contacts, say $\bar w$, it is immediate to show that the number of contacts remains fixed in time, thus reducing system \fer{sir-mass} to a classical SIR model with contact rate $\bbeta \bar w^2$. Hence this classical epidemiological model is contained in \fer{sir-mass}-\fer{sir-mean} and corresponds to consider the case of a population that, regardless of the presence of the epidemic, maintains the same fixed number of daily contacts.
\end{tips}

\subsubsection{{A social-SIR model with saturated incidence rate} }

 We consider the case where the average number of social contacts of infected $m_I$ is frozen to $\tilde m_I$ as an effect, for instance, of external interventions aimed at controlling the pandemic spread. In this case, for any $\alpha\ge 1$, one can explicitly solve the equation for the evolution of average contacts of susceptibles
\[
\dfrac{d}{dt}m_S(t) = -\dfrac{\bbeta c_\alpha}{\mu}m_S^{\alpha+1} \tilde m_I^\alpha I(t), \qquad m_S(t) = \dfrac{m_S(0)}{\left(1 + \dfrac{c_\alpha\bbeta\alpha m_S^\alpha(0)}{\mu} \tilde m_I^\alpha \displaystyle\int_0^t I(s)ds \right)^{1/\alpha}},
\]
where $c_\alpha>0$ is such that 
\[
\int_0^{+\infty} w^\alpha f_S^\infty(w)dw = c_\alpha S(t)  w_S^\alpha.
\]
Therefore, approximating the integral $\int_0^t I(s)\, ds\approx t I(t)$ we obtain the closed system for the evolution of mass fractions of the following type
 \begin{equations}\label{sir-mass-closed}
 & \frac{d S(t)}{d t} = -\tilde\beta \, S(t)I(t) H(I(t),t),  \\
 &\frac{d I(t)}{d t} = \tilde\beta\,  H(I(t),t) S(t)I(t)  - \gamma I(t),  \\
 & \frac{d R(t)}{d t} =   \gamma I(t),
\end{equations}
with $\tilde\beta = \bbeta m_S(0)$ and which incorporates the generalized macroscopic incidence function 
\be
H(I(t),t) = \dfrac{1}{\left(1+\phi(t)I(t)\right)^{1/\alpha}}, 
\label{gen_inc_rate}
\ee
with $\phi(t) = c_\alpha\alpha\bbeta m_S^\alpha(0)t/\mu>0$. The system \eqref{sir-mass-closed} corresponds to models with saturated incidence rate, see^^>\cite{Capasso}. We point the interested reader to^^>\cite{DPeTZ} for a detailed discussion. See also Section \ref{sec:control} of this survey for a derivation of the saturated incidence function \eqref{gen_inc_rate} as a feedback control functional. 
 
 \begin{figure}[htb]
    \centering
    \includegraphics[scale = 0.28]{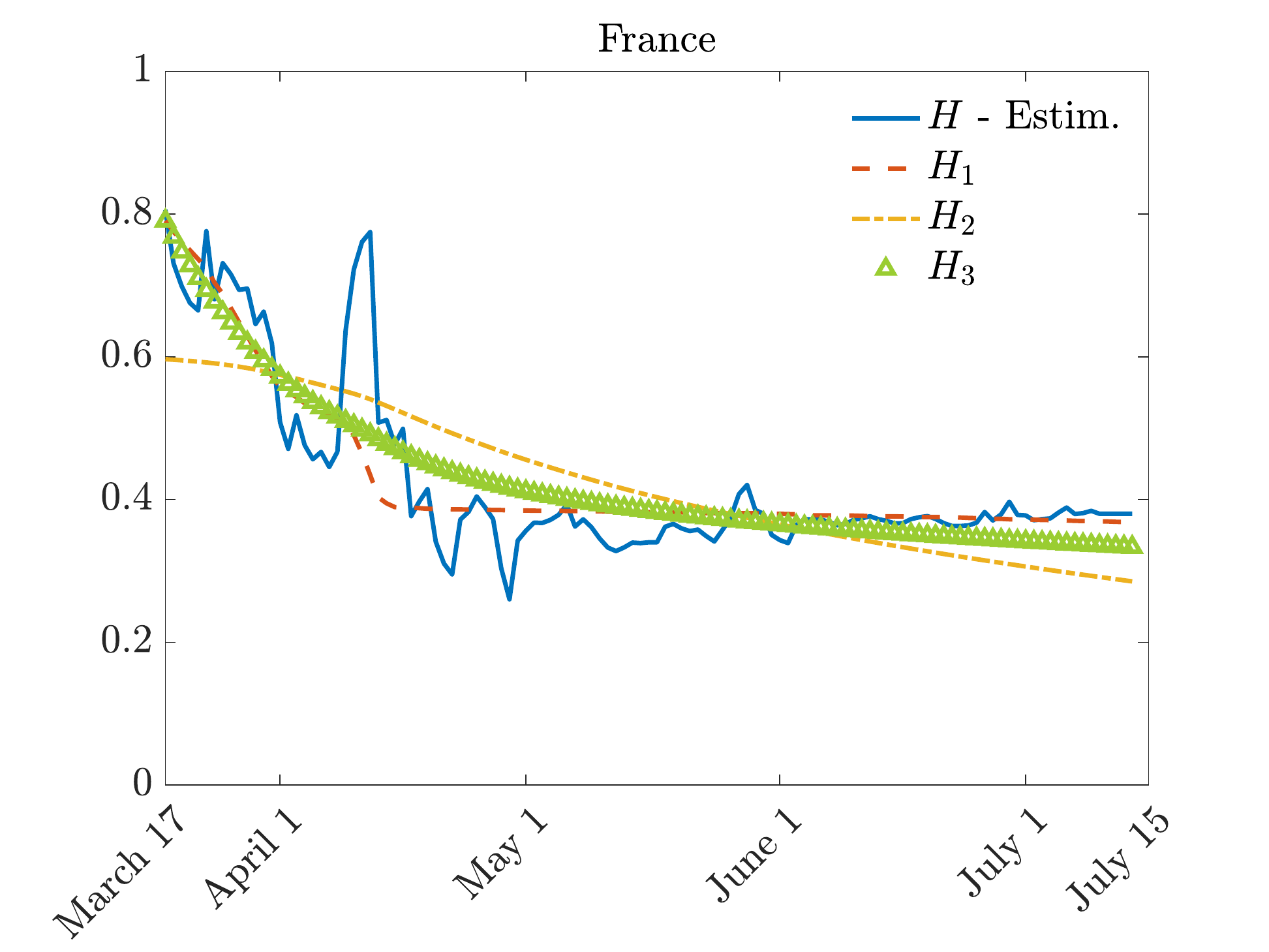}
    \includegraphics[scale = 0.28]{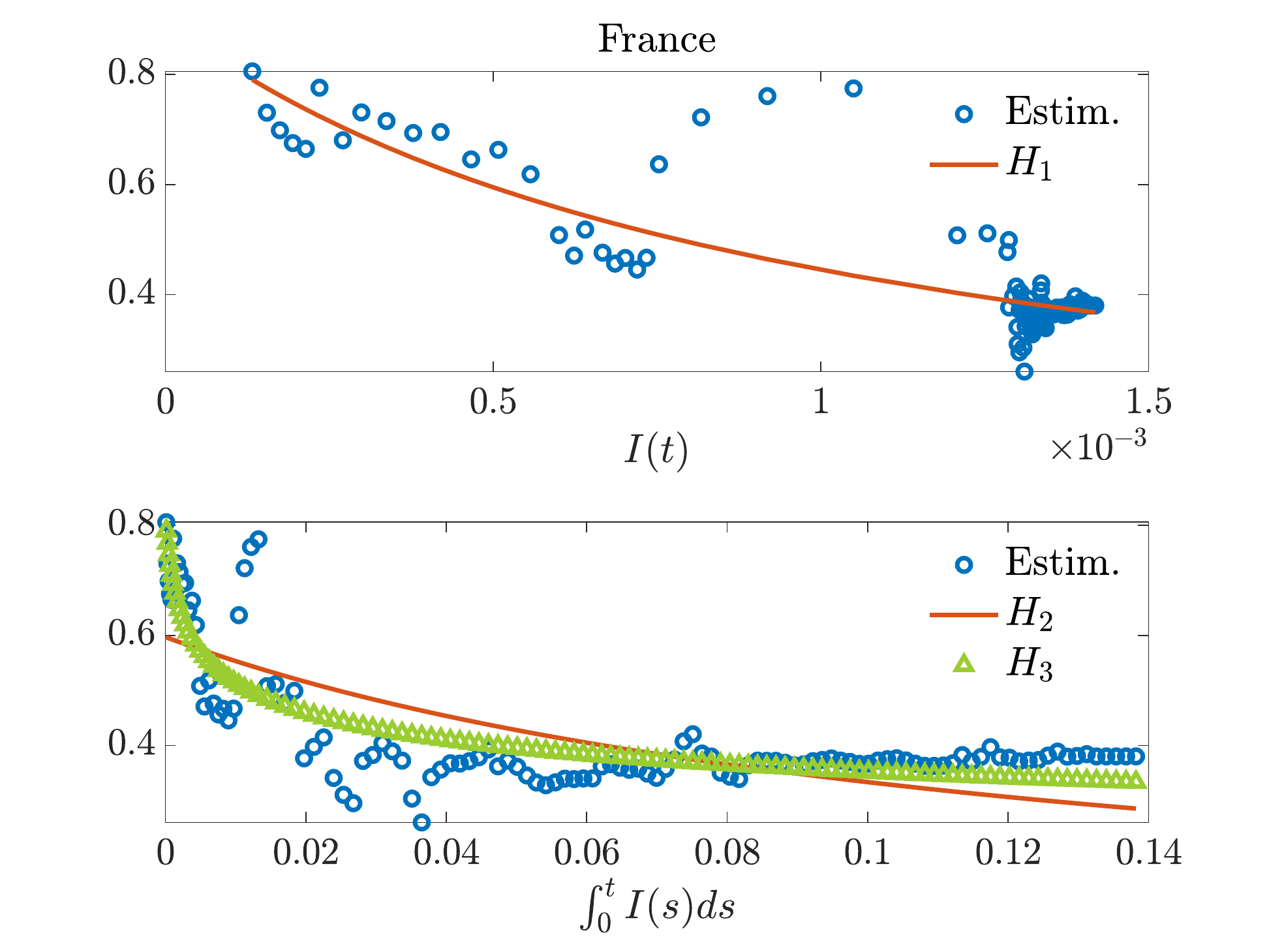}\\
    \includegraphics[scale = 0.28]{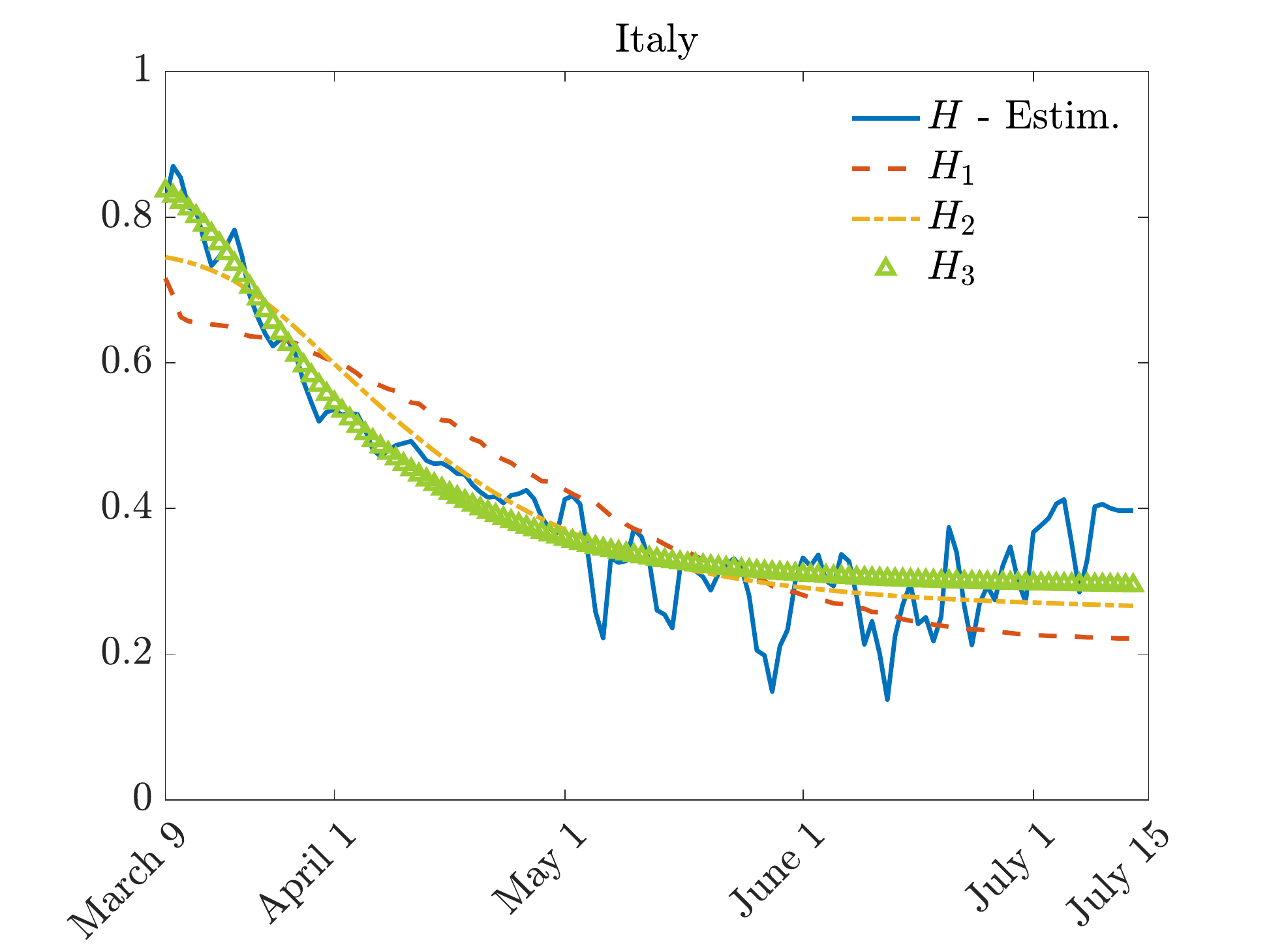}
    \includegraphics[scale = 0.28]{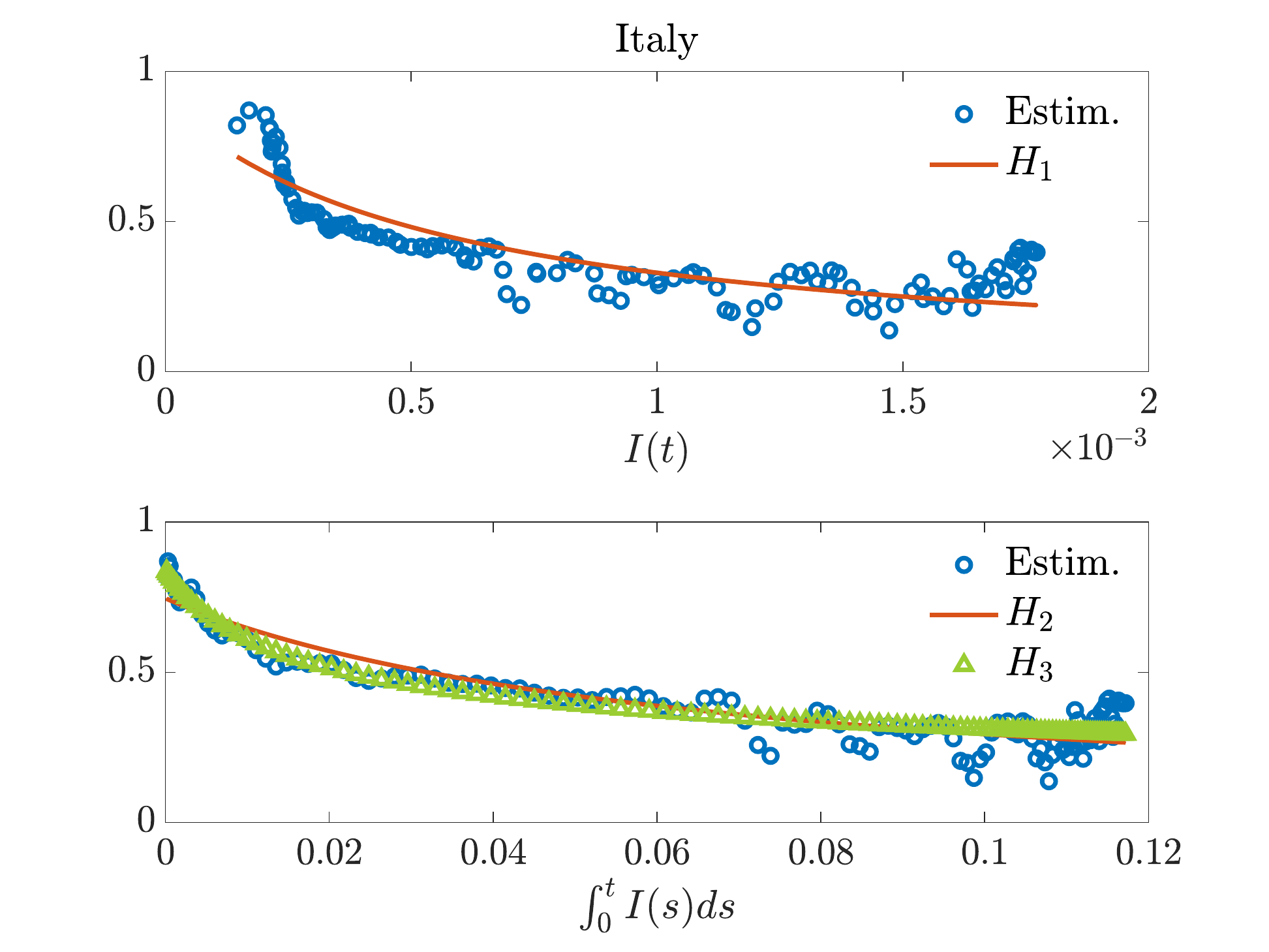}\\
    \includegraphics[scale = 0.28]{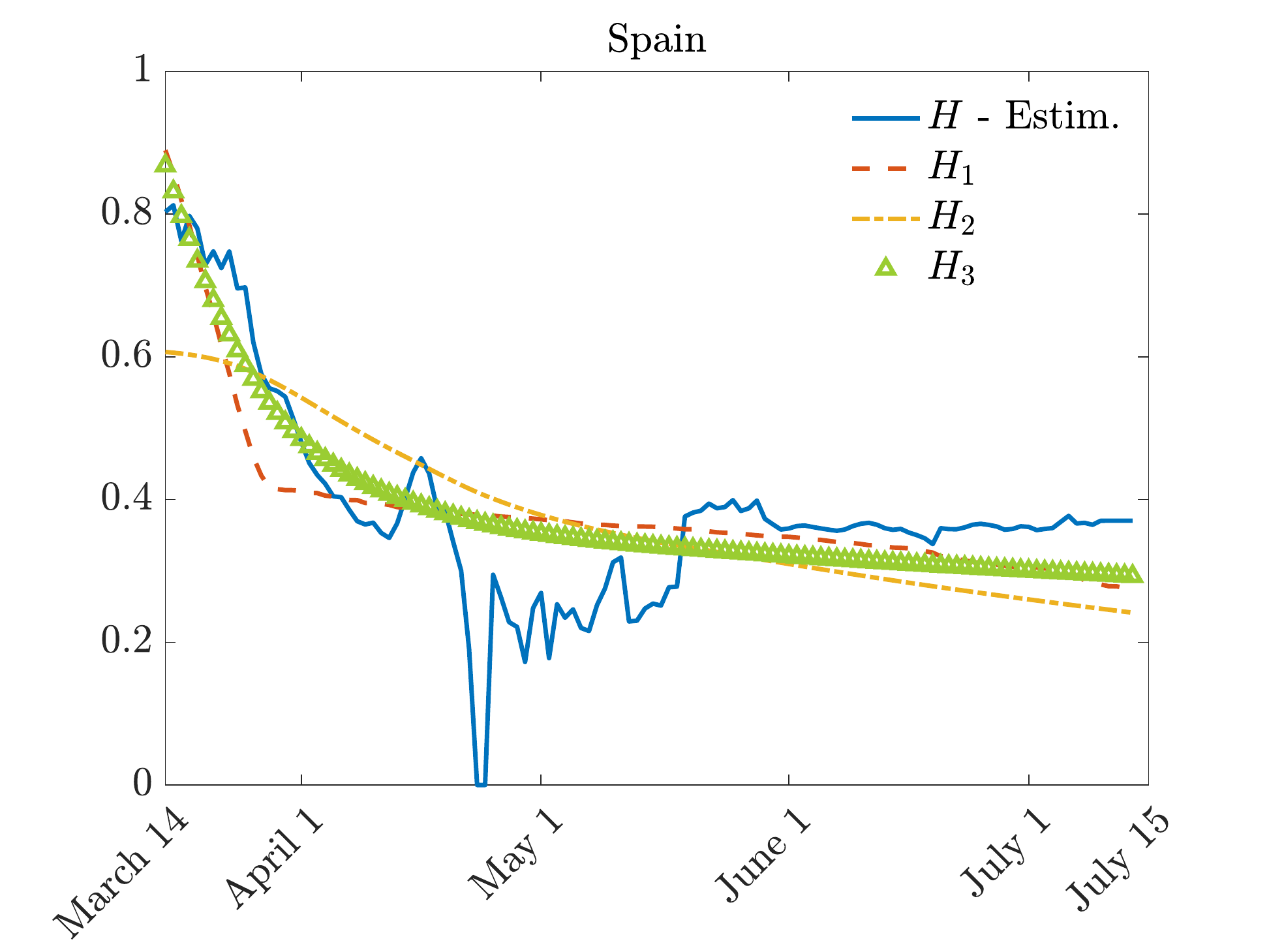}
    \includegraphics[scale = 0.28]{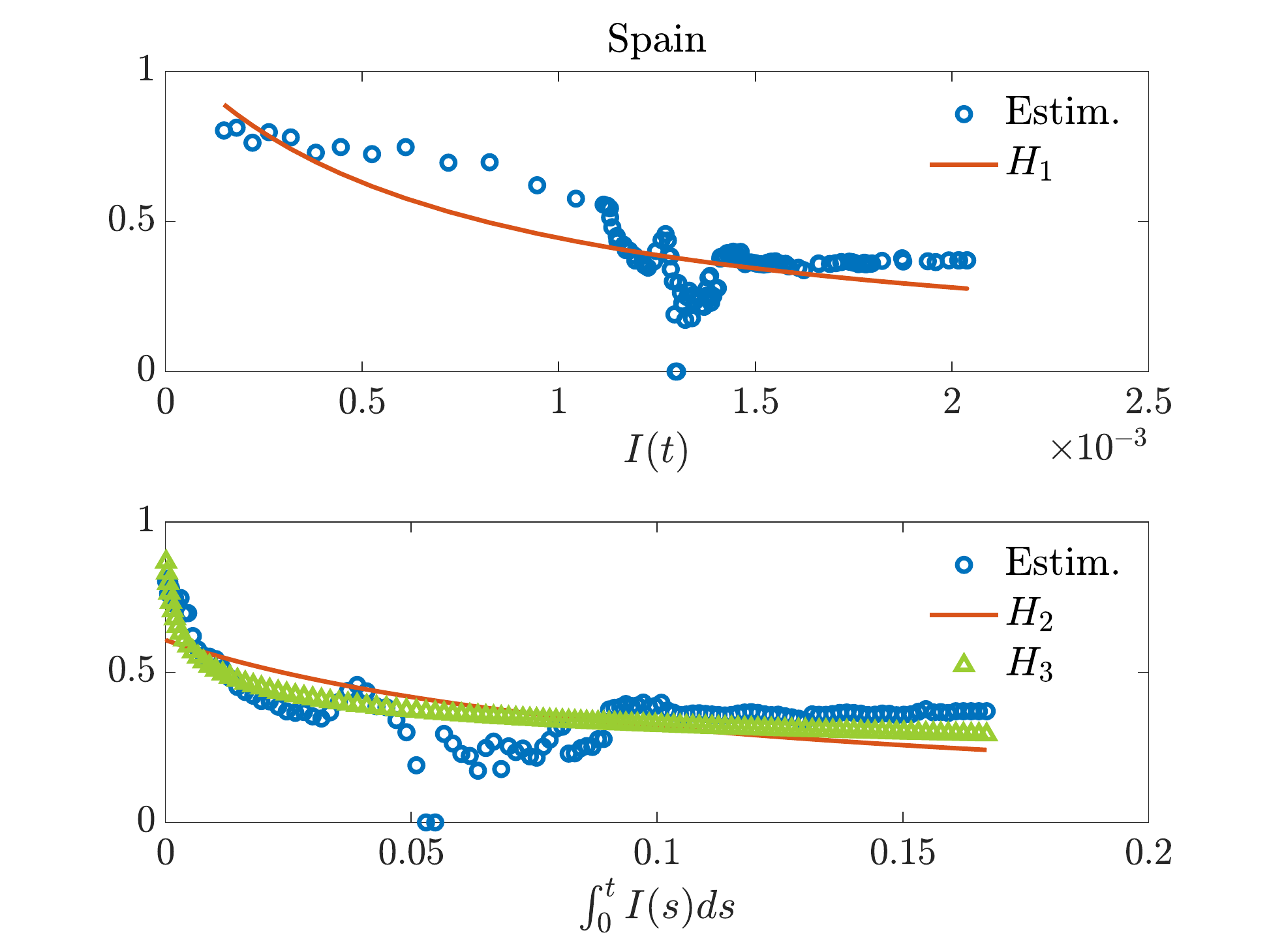}
    \caption{ Estimated shape of the function $H$ in several European countries (left plots) and its dependency on the variables  $I(t)$ and $\int_{0}^{t}I (s) ds $ (right plots). }
    \label{fig:t3_2}
\end{figure}

\subsubsection{Extrapolation of the shape of the incidence rate from data}\label{numericsIII}
In this section, we use the previous model to describe the COVID-19 pandemic in three different European countries: France, Italy and Spain. The data we employ, concerning the actual number of infected, recovered and deaths of COVID-19 are publicly available from the John Hopkins University GitHub repository. For the specific case of Italy, we considered instead the GitHub repository of the Italian Civil Protection Department (see Data Sources in Section \ref{sec:ds}). 
We adopted the fitting procedure described in^^>\cite{DPeTZ,APZ} that is based on a strategy with two optimization horizons (pre-lockdown and lockdown time spans) depending on the different strategies enacted by the governments of the considered European countries. Once the relevant epidemiological parameters have been estimated in the pre-lockdown time span, we successively proceeded with the estimation of the shape of the function $H$ from the data. 

We seek to understand numerically the dependencies of the function $H$ on the number of infected. In particular, we consider the candidate incidence functions $H_1$, $H_2$ and $H_3$ defined as
\[
\begin{split}
H_1(I(t),t) = \dfrac{c}{1 + \displaystyle \phi I(t)}, \quad
 H_2(I(t),t) = \dfrac{c}{1 + \phi \int_0^t I(s)\, ds},
 \end{split}
 \]
 and 
 \[
H_3(I(t),t) = \dfrac{c}{\left(1 + \phi \int_0^t I(s)\, ds\right)^{1/\alpha}},
\]
with $c>0$, accordingly with \eqref{gen_inc_rate} where  $\phi$ and $\alpha$ are free parameters which are determined through a least square minimization approach that best fits the estimated curve with conditions $\phi>0, \alpha\geq 1$. The results of this procedure is presented in Figure \ref{fig:t3_2}. We point the interested reader to^^>\cite{DPeTZ} for a detailed discussion on the estimated parameters. 

We can observe that the optimization gives acceptable results for the different forms of the incidence function especially in the right column of Figure~\ref{fig:t3_2}, where the functions $H_2$ and $H_3$ are clearly able to better explain the estimated values of $H$ especially after the epidemic peak. Note that, the fits of the model with the available data when $H_3$ is used are particularly good. 
This fact may indicate that people are rather fast to apply social distancing, and therefore to reduce their average number of contacts, whereas they tend to restore the pre-pandemic average contact rate more slowly, possibly due to further psychological effects. 

\subsection{The interplay between economy and the pandemic}\label{economic}
The rapid spreading of the COVID-19 epidemic in western countries and the consequent lockdown measures assumed by the governments to control and limit its effects will unequivocally lead to important consequences for their economies. In this section, following^^>\cite{DPTZ}, we introduce a mathematical framework to study the economic impact of the pandemic by integrating epidemiological dynamics with a kinetic model of wealth exchange. 

The description of the evolution of the personal wealth in terms of kinetic-type equations revealed to be successful in the description of emerging wealth distributions, see^^>\cite{BM,CPT05,DPT,PT13}. Clearly, an accurate quantification of the implications due to the pandemic in the distribution of wealth is an extremely difficult problem that requires knowledge of a large number of unknown variables and relationships between them. In an attempt to better understand the mechanisms underlying these dynamics we will consider simplified models that, while based on a few obvious universal characteristics, can be analyzed to provide answers about possible scenarios. 

\subsubsection{Wealth exchanges in epidemic modelling}

The model considered in^^>\cite{DPTZ} has the same structure of the kinetic SIR-type model defined in \eqref{eq:kinetic_general} where now the state of an individual in each class at any instant of time $t\ge 0$ is  completely characterized by the amount of wealth $w\in \R_+$. Therefore, $f_S(w,t)$, $f_I(w,t)$ and $f_R(w,t)$, are the distributions at time $t > 0$ of the amount of wealth of the population of susceptible, infected and recovered individuals, respectively. The distribution of wealth of the whole population is then recovered as in \eqref{eq:ch} and consequently is assumed to be a probability density for all times $t \ge 0$. 

Coherently with the previous notations we denote the relative mean wealths in each compartment as
\begin{equation}\label{me3}
m_J(t)=\frac1{J(t)}\int_{\mathbb{R}^+}wf_J(w,t)\,dw, \quad J \in \{S,I,R\} ,
\end{equation}
and  the total mean wealth as
\[
m(t)=\int_{\mathbb{R}^+}wf(w,t)\,dw.
\]
We emphasize that the above notations differ from those originally used in \cite{DPTZ}, this in order to make them homogeneous with the notations used in Section \ref{contact}. 
 
In equations \eqref{eq:kinetic_general} the choice of a wealth-dependent recovery  rate can be  motivated by considering that  wealth can buy access to better hospitals and better treatments, thus ensuring a higher chance of recovery. Also, a wealth-dependent  contact rate, in the form of a decreasing function of the difference $|w-w_*|$, can be introduced to express  that individuals with different degrees of wealth live in different environments, and this limits contacts in presence of a marked difference. The interaction operators on the r.h.s. of \eqref{eq:kinetic_general} characterize the wealth evolution due to trading between agents of the same class, or between agents of different classes, and are built according to the CPT model \cite{CPT05} with the following structure
\be
{\mathcal Q}_J(f_S;f_I;f_R)=\sum_{H\in \{S,I,R\}} Q_{JH}( f_J, f_H)(w,t),\qquad J\in \{S,I,R\},
\ee
where $Q_{JH}( f_J, f_H)$ describes the changes of wealth in compartment $J$ due to binary interactions among agents in compartments $J$ and $H$, with $J,H\in \{S,I,R\}$.

In details, an interaction between two individuals in compartment $J$ and $H$ with wealth pair $(w,w_*)$ leads to a wealth pair $(w_{JH}',w_{HJ}')$ defined by relations
\begin{equation}
\begin{split}
w_{JH}' &= (1-\a_J) w + \a_H w_* + \eta_{JH} w\\[-.2cm]
&\hskip 6cm J,H\in \{S,I,R\}\\[-.2cm]
w_{HJ}' &= (1-\a_H) w_* + \a_J w + \eta_{HJ} w_*,
\end{split}
\label{eq:bin}
\end{equation}
where $\a_J,\a_H\in (0,1)$ are transaction coefficients, while the market risk variables $\eta_{JH}\ge -\a_J$ and $\eta_{HJ}\ge - \a_H$ are independent and identically distributed random variables with zero mean and the same time-dependent variance $\sigma^2(t)$ (since we assume that the risk in the market does not depend on the particular class of trading agents). 

The trade between agents has been modeled to include the idea that wealth changes hands for a specific reason: one agent intends to invest his wealth in some asset, property etc. in possession of his trade partner. Typically, such investments bear some speculative risk, and either provide the buyer with some additional wealth, or lead to the loss of wealth in a non-deterministic way. Relations \eqref{eq:bin} couple the saving propensity parameter with some risky investment that yields an immediate gain or loss proportional to the current wealth of the investing agent. Hence $0 < \a_J < 1$, $j\in\{S,I,R\}$ are the parameters which identify the saving propensities $1-\lambda_J$, namely the intuitive behavior which prevents the agents to put in a single trade the whole amount of his money. The choice $ \lambda_R > \lambda_S$, for example, reflects the fact that susceptible individuals can be more cautious in the market and tend to save their wealth, since they understand that consuming and working less reduces the probability of infection \cite{macro}. On the other hand, infectious individuals have limited possibilities to act on the market and, as we will see, asymptotically disappear from the wealth dynamics. 
The time-dependence of $\sigma$ has been postulated by assuming that, in the presence of a significant spread of the epidemic, the risk variance tends to increase. This is in agreement, for example, to the market reactions we observed during the COVID-19 spreading at the announcements of the new numbers of infectious people in the various countries^^>\cite{ZHJ}. 

As already observed a convenient way to express the operators $Q_{JH}(f_J,f_H)$ is based on its weak form, namely the way the operator acts on observables^^>\cite{CPT05, PT13}. Let $\phi(v)$ denote a test function and let us define with $\langle \cdot \rangle$ the expected value with respect to the pair $\eta_{JH}$, $\eta_{HJ}$ in the interaction process. 
Thus, for $J,H \in \{S,I,R\}$ we have
\begin{equation}
\begin{split}
& \int_{\mathbb{R}_+}\phi(w)Q_{JH}(f_J,f_H)(w,t)\,dw=\\
&\qquad\qquad\qquad\qquad\qquad \Big\langle\int_{\mathbb{R}^2_+}  (\phi(w_{JH}')-\phi(w))f_J(w,t)f_H(w_*,t)\,dw_*\,dw\Big\rangle,
\end{split}
\label{eq:econQ}
\end{equation}
where $w_{JH}'$ is defined by \eqref{eq:bin}.

\subsubsection{Fokker-Planck scaling and steady states}

To analyze the asymptotic behavior of the model it is useful to resort to the so-called quasi-invariant trading limit which permits to derive the corresponding Fokker-Planck description of the Boltzmann operators \eqref{eq:econQ}. To this aim, in a similar fashion to Section \ref{FP-limit}, following \cite{CPT05, FPTT, PT13,DPTZ}, we scale the binary trades according to
\begin{equation}\label{eq:quasi}
\lambda_J \to \varepsilon \lambda_J,\quad J\in\{S,I,R\}, \quad \sigma \to \sqrt{\varepsilon} \sigma,
\end{equation}
and similarly the functions governing the spread of the disease
\begin{equation}\label{eq:quasi_epi}
\beta(w,w_*) \to \varepsilon \beta(w,w_*),\quad \gamma(w) \to \varepsilon \gamma(w), 
\end{equation}
and denote with $Q_{JH}^{\varepsilon}(\cdot,\cdot)$, $J,H\in\{S,I,R\}$ the scaled interaction terms.

The limit procedure induced by the above scaling corresponds to the situation in which are prevalent the
exchanges of wealth which produce an extremely small modification of wealths, but we are waiting enough time to still see the effects. In fact, rescaling time as $t \to t/\varepsilon$, for small values of $\varepsilon$,  the Boltzmann-type operators converge to Fokker-Planck operator with variable coefficient of diffusion and linear drift. More precisely we have that for small values of $\e \ll 1$ (see \cite{DPTZ} for details)
\[
\frac1{\e}\sum_{H\in\{S,I,R\}} \int_{\RR_+} Q^\e_{JH}(f_J,f_H)(w,t)\phi(w)\,dw \approx \int_{\RR_+} \tilde Q_J (f_J)(w,t)\phi(w)\,dw
\]
where  
\be\label{FPW}
\tilde Q_J (f_J)(w,t) = \frac{\partial}{\partial w}\left[ \frac{\sigma(t)^2}{2} \frac{\partial}{\partial w} (w^2 f_J(w,t)) +
\left(w\lambda_J-\bar{m}(t)\right)f_J(w,t)\right],
\ee
with
\begin{equation}\label{mm2}
\bar{m}(t)=\lambda_S m_S(t)S(t)+\lambda_I m_I(t)I(t)+\lambda_R m_R(t)R(t).
\end{equation}
This gives the system 
\begin{equations}\label{sir-wealth}
\frac{\partial f_S(w,t)}{\partial t} &= -K(f_S,f_I)(w,t) +   \tilde Q_S(f_S)(w,t)
\\
\frac{\partial f_I(w,t)}{\partial t} &= K(f_S,f_I)(w,t)  - \gamma f_I(w,t) +  \tilde Q_I(f_I)(w,t)
\\
\frac{\partial f_R(w,t)}{\partial t} &= \gamma f_I(w,t) + \tilde Q_R(f_R)(w,t). 
\end{equations}
It is immediate to verify that the above Fokker-Planck-type operators are mass and momentum preserving. 
Similarly to Section \ref{splitting}, one can analyze the equilibrium densities associated to the differential system 
 \[
\frac{\sigma^2}{2} \frac{\partial (w^2 f_J^\infty(x))}{\partial w} +   \left( w\lambda_J -\bar m \right)f_J^\infty(w)= 0, \quad J \in \{S,I,R\},
\]
to derive reduced order models for the evolution of the densities of susceptible, infectious and recovered individuals. As we will see in the next section these equilibrium states have the shape of inverse Gamma distributions^^>\cite{Sta,Lie}.

 \subsubsection{The formation of bimodal wealth distributions} 
 We verify in a simplified case, that the Fokker--Planck system \eqref{sir-wealth} possesses as stationary solutions inverse Gamma distributions that may generate a bimodal form of wealth distribution. Bimodal shapes are typical of situations of high stress in economy, and are investigated starting from the Argentinian crisis of the first year of the new century \cite{Gup,GCC16}. This example also shows that a similar behavior can be expected in reason of the epidemic spreading.

\begin{figure}[t]
\centering
\includegraphics[scale = 0.45]{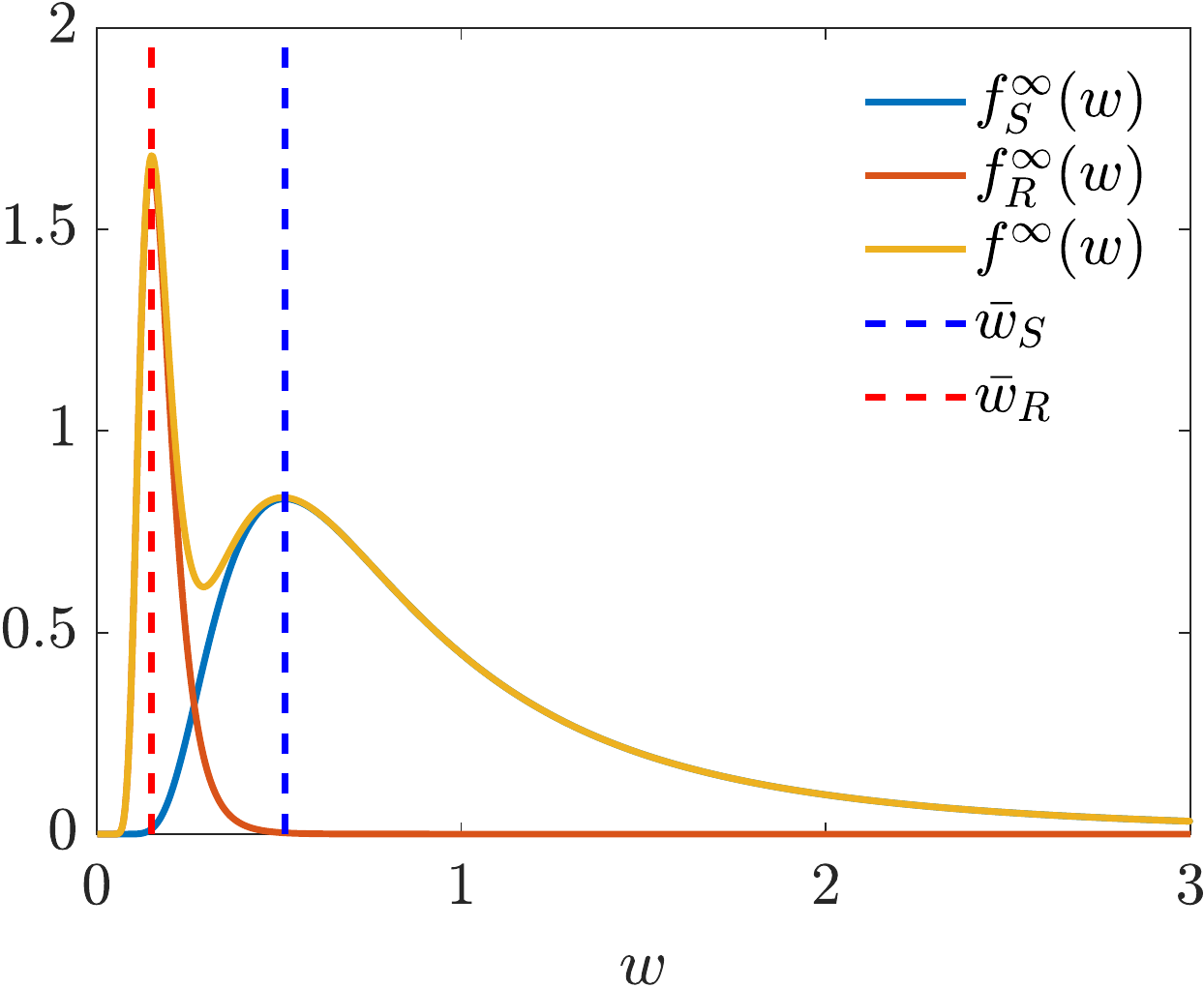}
\includegraphics[scale = 0.45]{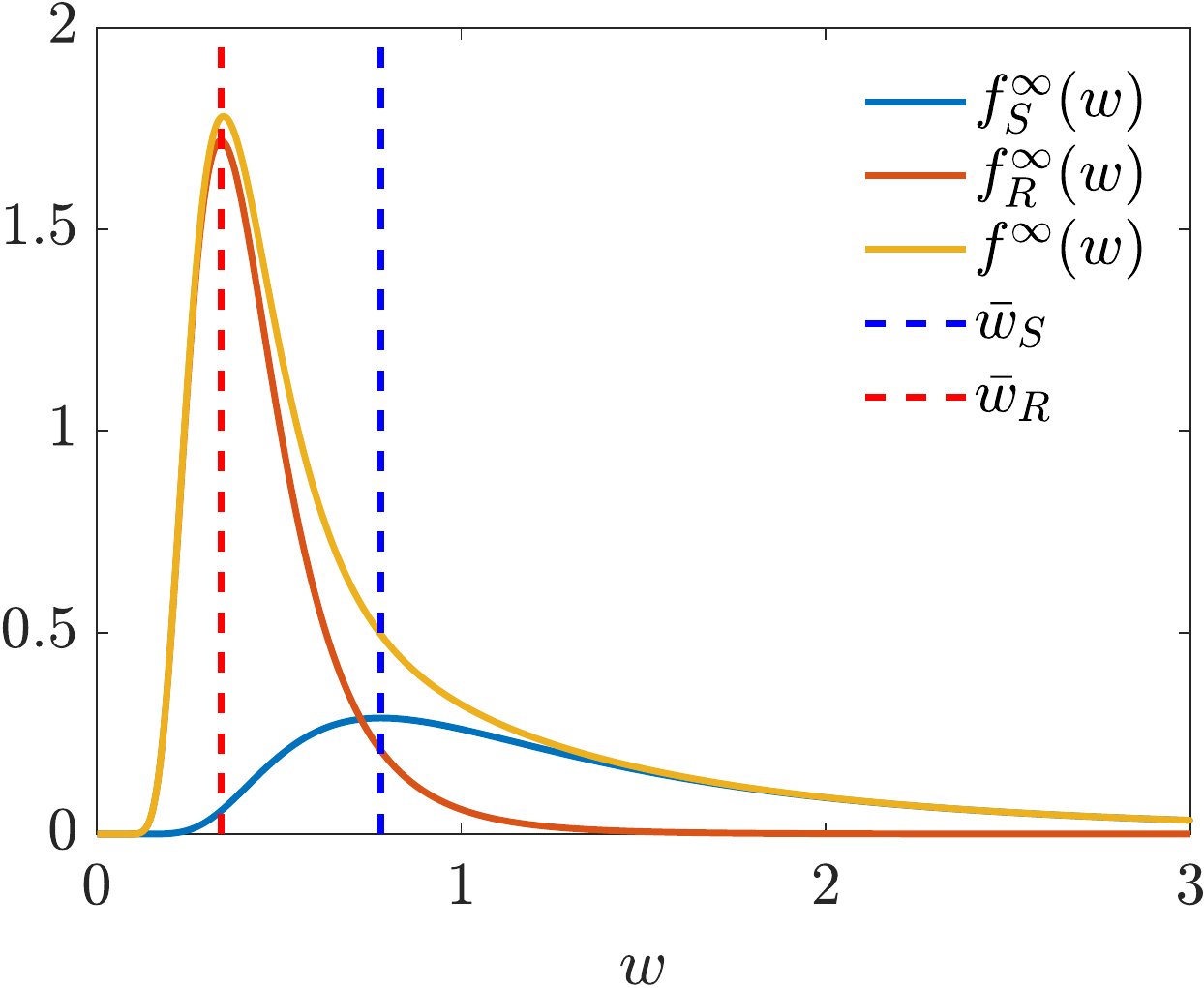}
\caption{Exact solutions for wealth distributions at the end of the epidemic \eqref{sir-wealth} in the Fokker-Planck approximation for $\mu_S =2.5$, $\mu_R =7.0$, $S^\infty =0.4$, and $R^\infty =0.6$ (top) and $\mu_S =2.5$, $\mu_R = 11.0$, $S^\infty = 0.8$, and $R\infty = 0.2$ (bottom).}
\label{fig:bimo}
\end{figure}

Suppose that $\beta(w,w_*)=\beta$, $\gamma(w)=\gamma$ and $\sigma(t) =\sigma$ are constant. Then, integrating with respect to the wealth variable, thanks to conservation of the total wealth, we obtain that the relative mass densities satisfy the classical SIR model \eqref{eq:SIRbase}. In this case, it is known that $I(t)\to 0$, while $S(t)\to S^\infty \in [0,\gamma/\beta]$. Likewise, the system for the mean values reads
\begin{eqnarray}
\frac{d (m_S(t) S(t))}{d t} &=& -\beta I(t) m_S(t)\,S(t)+ (\bar{m}(t) - \lambda_S m_S(t)) S(t)
\label{eq:SIRm1}\\
\frac{d (m_I(t) I(t))}{d t} &=& \beta I(t) m_S(t)\,S(t) - \gamma m_I(t)\,I(t)+ (\bar{m}(t) - \lambda_I m_I(t)) I(t)
\label{eq:SIRm2}\\
\frac{d (m_R(t) R(t))}{d t} &=& \gamma m_I(t)\,I(t)+(\bar{m}(t) - \lambda_R m_R(t)) R(t). 
\label{eq:SIRm3}
\end{eqnarray}
Since, as $t\to+\infty$ we have $I(t)\to 0$, $m_S(t)\to m_S^\infty$ and $m_R(t)\to m_R^\infty$, the asymptotic values of the means satisfy
\[
\lambda_R m^\infty_R = \lambda_S m^\infty_S, 
\]
together with the constraint $m^\infty_R\,R^\infty+m^\infty_S\,S^\infty=m$ by conservation of the total mean wealth. This gives the asymptotic values
\begin{equation}
m^\infty_S = \frac{\lambda_R}{\lambda_R S^\infty+\lambda_S R^\infty} m,\qquad m^\infty_R = \frac{\lambda_S}{\lambda_R S^\infty+\lambda_S R^\infty} m.
\label{eq:mean}
\end{equation}
Thus, formally as $t \to \infty$ in the Fokker-Planck system \eqref{sir-wealth} we get that the stationary states $f^\infty_S(w)$ and $f^\infty_R(w)$ are given by two inverse Gamma densities 
\begin{equation}
f^\infty_S(w)=S^\infty \frac{\kappa^{\mu_S}}{\Gamma(\mu_S)}\frac{e^{-\frac{\kappa}{w}}}{w^{1+\mu_S}}, \qquad f^\infty_R(w)=R^\infty\frac{\kappa^{\mu_R}}{\Gamma(\mu_R)}\frac{e^{-\frac{\kappa}{w}}}{w^{1+\mu_R}}
\label{eq:fRS}
\end{equation}
with
\begin{equation}
\begin{split}
\mu_S=1+2\frac{\lambda_S}{\sigma},\qquad \mu_R=1+2\frac{\lambda_R}{\sigma},\quad
\kappa = {(\mu_S-1)}{m_S^\infty} = {(\mu_R-1)}{m_R^\infty}.
\label{eq:mu}
 \end{split}
\end{equation}
The  details of the trading activity at the basis of the kinetic description allow to characterize the tails of the distributions from \eqref{eq:mu}. Hence, a low value of the Pareto index is obtained in presence of small values of the parameter $\lambda_S$, $\lambda_R$  (small saving propensity of agents), or to high values of the parameter $\sigma$ (highly risky market).  
Therefore, the asymptotic wealth distribution is the mixture of two inverse Gamma densities of mass $S^\infty$ and $R^\infty$ respectively
\begin{equation}\label{eq:finf}
f^\infty(w)=f^\infty_S(w)+f^\infty_R(w),
\end{equation}  
with asymptotic means \eqref{eq:mean} and variances given by 
\[
{\rm Var}^\infty_S = \frac{\kappa^2}{(\mu_S-1)(\mu_S-2)},\qquad {\rm Var}^\infty_R = \frac{\kappa^2}{(\mu_R-1)(\mu_S-2)},\qquad \mu_R,\mu_S > 2.
\]
As a consequence, the wealth distribution has a bimodal structure, since the maximum of $f^\infty_S(w)$ and $f^\infty_R(w)$ are achieved, respectively, at the points 
\begin{equation}\label{eq:wbar_SR}
\begin{split}
\bar{w}_S = \frac{\kappa}{\mu_S+1},\qquad 
\bar{w}_R = \frac{\kappa}{\mu_R+1}.
\end{split}
\end{equation}  
We report in Figure \ref{fig:bimo} the resulting profiles for various choices of $\mu_S<\mu_R$, and $S^\infty$, $R^\infty$. Note that the mixture of the two inverse Gamma densities \fer{eq:fRS}  does not always result in an evident bimodal shape. Indeed, while the profile on the right of Figure \ref{fig:bimo} is clearly bimodal, a different choice of parameters on the left produces a unimodal steady profile.  

\subsubsection{The increase of wealth inequalities}

Next, we compare the evolution of the wealth distribution of the system under more realistic hypotheses about the dependence of the risk coefficient $\sigma$ on the epidemic spread. We consider the kinetic model \eqref{sir-wealth} in the case of the following two infectious-dependent market risk coefficients
\begin{equation}
\label{eq:sigma_12}
\sigma_1(t)=\sigma_0 (1+\alpha I(t)), \qquad
 \sigma_2(t) = \sigma_0\left(1+\alpha \int_0^t I(\tau)d\tau\right),
\end{equation}
where $\alpha>0$, $\sigma_0>0$. In details, $\sigma_1(t)$ characterizes the instantaneous influence of the epidemic based on the observed number of infected, whereas $\sigma_2(t)$ takes into account possible long time memory effects on the market based on the epidemic impact.

We consider, as initial distribution, an inverse Gamma distribution 
\be
f(w) = \dfrac{(\mu-1)^\mu}{\Gamma(\mu)} \dfrac{\textrm{exp} \left( -\frac{\mu-1}{w}\right)}{w^{1+\mu}}
\ee
with $\mu = 3$, representing an initial economic equilibrium state. 

To get a more detailed view of the emerging equilibria, we resort to the Gini index calculation, see^^>\cite{DPT}. This value should be understood as a measure of a country's wealth inequality and varies in $[0,1]$, where $0$ indicates perfect equality and $1$ maximum inequality. 

\begin{figure}[t]
\centering
\includegraphics[scale = 0.4]{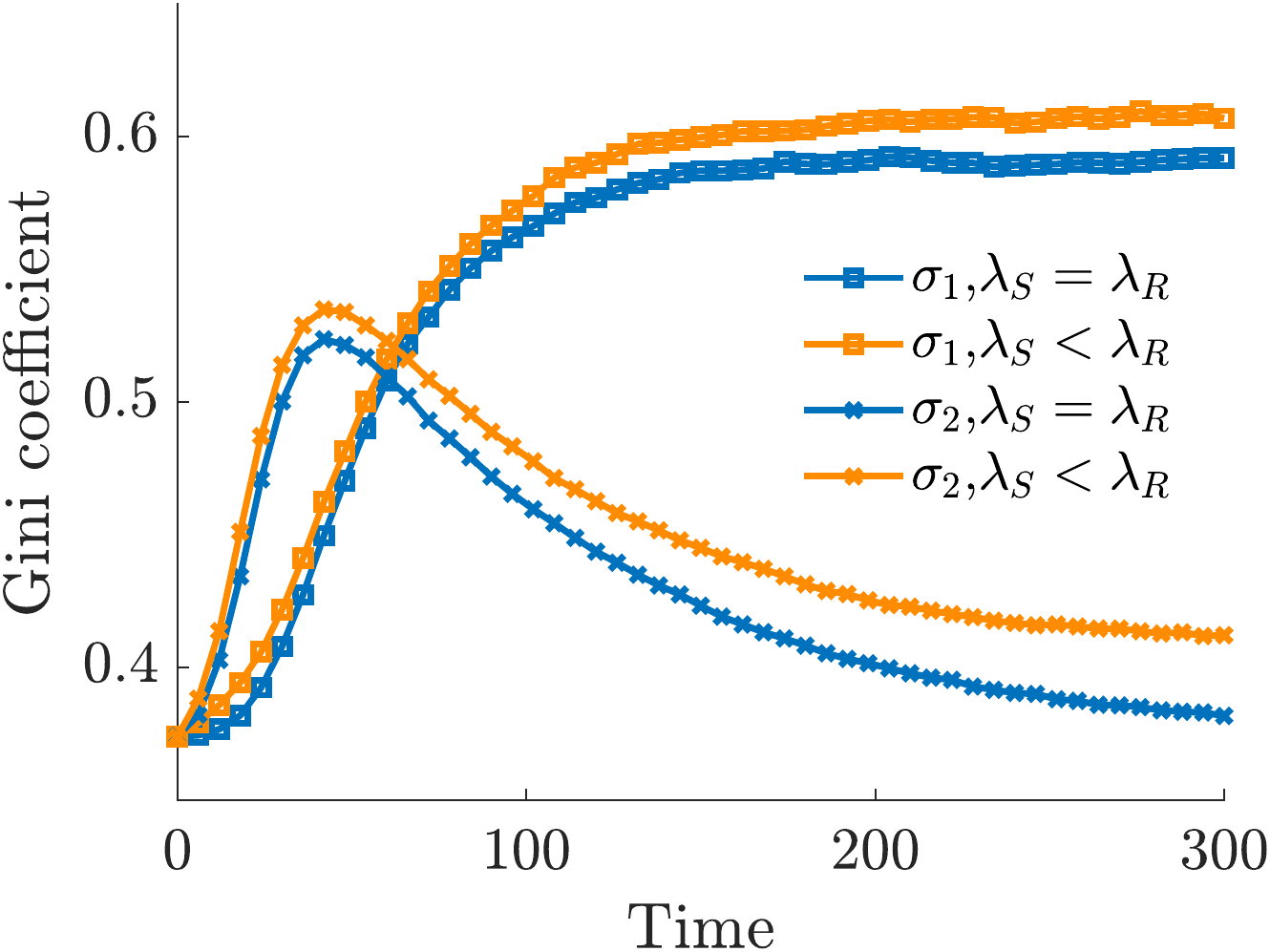}
\includegraphics[scale = 0.4]{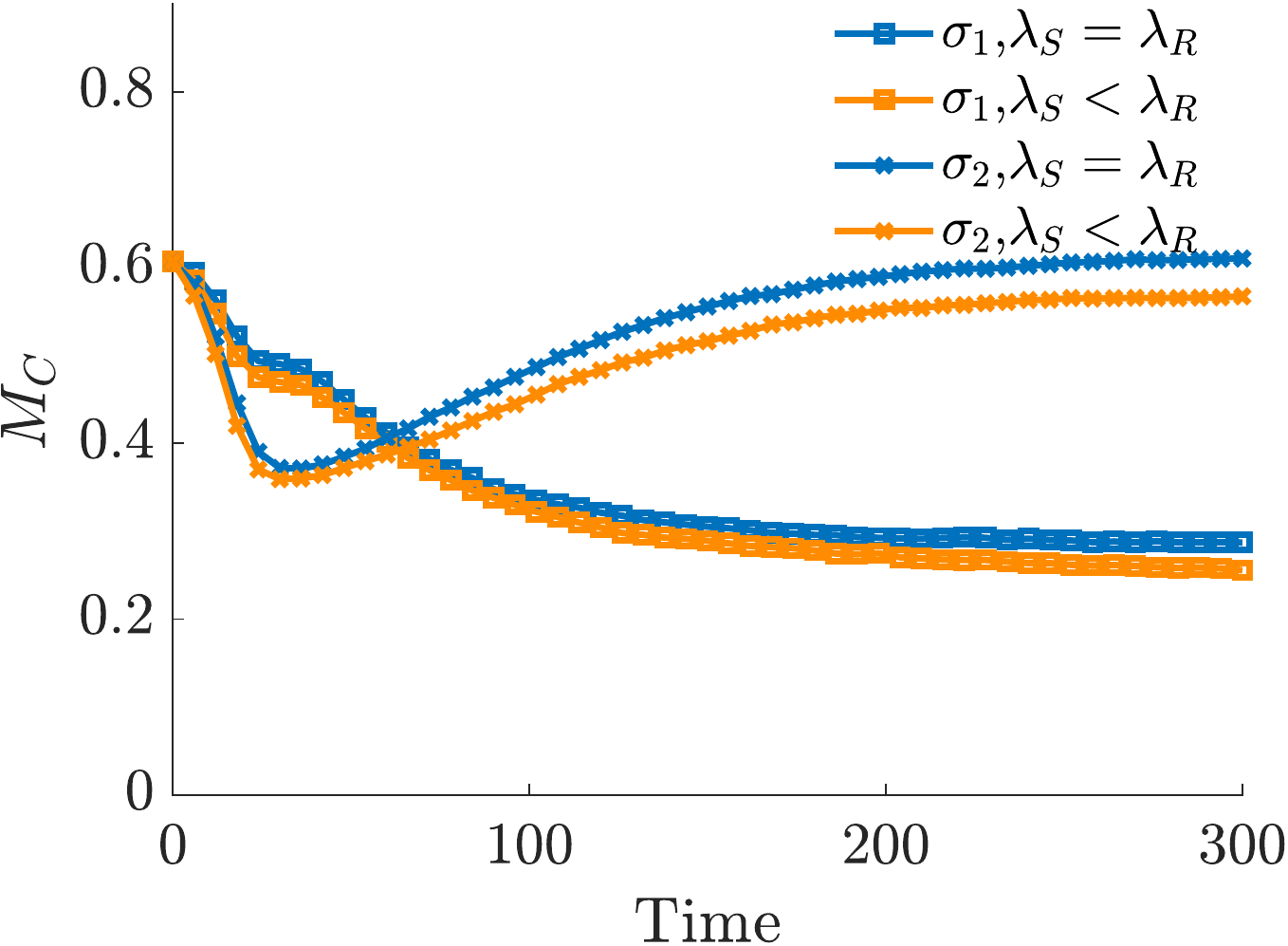}
\caption{\textbf{Test 2}. Behavior of the Gini index (left) and of the middle class fraction (right) defined in \eqref{eq:MC} during the outbreak of the epidemic for the different risk measures in \eqref{eq:sigma_12} with $\alpha =5$, $\sigma_0 = 0.1$. }
\label{fig:3}
\end{figure}

In Figure \ref{fig:3} we represent the evolution of the Gini index. We clearly observe an inequality of wealth that grows with the epidemiological dynamics. Moreover, even in the case of $\sigma_1$ with $\lambda_S = \lambda_R$, where these effects are absorbed in the long-lasting trends, the recovery of the economy occurs at a much lower rate than the worsening rate. 

Epidemiological dynamics may translate into additional wealth inequalities, in particular we can measure the evolution of the total number of individuals belonging to the middle class. Although there are several ways to give a technical definition of the middle class, it is often more of an idea or estimate than a fixed number. Generally speaking, the middle class is loosely defined as those who fall into the middle group of workers compared to the bottom $20\%$ or top $20\%$. We can define it using an interval $[w_L,w_R]$ such that
\be\label{eq:classes} 
\int_{0}^{w_L} f(w,0) \approx 0.2,\qquad \int_{w_R}^{\infty} f(w,0) \approx 0.2,
\ee 
and computing the time evolution of
\be\label{eq:MC}
M_C(t) = \int_{w_L}^{w_R} f(w,t)\,dw,  
\ee
gives us an estimate of the percentage of people living in middle-income households. In Figure \ref{fig:3} (right plot), we represent the evolution of $M_C(t)$ corresponding to the considered $\sigma_1(t)$, $\sigma_2(t)$. We can clearly see how the emerging inequalities mainly affect the middle class, which is constantly decreasing in the case of $\sigma_2$ and undergoes a transitory decrease for $\sigma_1$. In particular, in this last scenario and in the $\lambda_S<\lambda_R$ regime, at the end of the epidemic dynamics only a partial recovery to the original pre-epidemic level is observed. 

\section{Social control and data uncertainty}
\label{sec:control}
The adoption of containment measures to reduce the amplitude of the epidemic peak is a key aspect in tackling the rapid spread of an epidemic. Classical compartmental models need to be modified and studied to correctly describe the effects of forced external actions to reduce the impact of the disease. The importance of the social structure, such as age dependence, which was shown to be essential in the recent COVID-19 pandemic, must be considered. In addition, available data are often incomplete and heterogeneous, so a high degree of uncertainty must be incorporated into the model. In this section we deal with both these aspects following^^>\cite{APZ,APZ2}.

\subsection{Control of socially structured models}
The heterogeneity of the social structure, which impacts the diffusion of the infective disease has been already discussed in Section \ref{sec:social}. Among the social characteristics of fundamental importance in the context of the COVID-19 pandemic certainly the age of individuals is among the most significant given the heterogeneity of the contagiousness of the virus and the related health risks. We assume that the rapid spread of the disease and the low mortality rate allow to ignore changes in the social structure, such as the aging process, births and deaths. This is equivalent to assuming in \eqref{eq:kinetic_general} that the interaction operators $\mathcal{Q}_J(f_S;f_I;f_R)$, $J\in\{S,I,R\}$ vanish. 

In order to keep to a standard notation in age-structured models^^>\cite{HWH00,Pugliese} in the following we will use the variable $a$ to denote the social feature instead of $w$ used in the general formulation \eqref{eq:kinetic_general}. Thus, we will denote by $\fS(\aa,t)$, $\fI(\aa,t)$ and $\fR(\aa,t)$, the  distributions at time $t > 0$ of susceptible, infected and recovered individuals with a given age $a \in \LL \subset \RR^+$.
In this situation the nonlinear incidence rate \eqref{eq:contact_f} reads
\be
K(f_S,f_I)(a,t)=\fS(\aa,t)\int_{\LL} \beta_{\textrm{social}}(\aa,\aa_*){\fI(\aa_*,t)}\ d\aa_*,
\ee
where the function $\beta_{\textrm{social}}(\aa,\aa_*) \geq 0$ represents the interaction rate among individuals with different ages. 

Typically, in socially structured models the interaction rate between people is assumed to be separable, and proportionate to the activity level of the social feature^^>\cite{H96,HWH00},  alternative approaches are based on preferential mixing^^>\cite{GFMDC12,CHALL89}. In what follows we will assume an age-dependent social interaction function such that
\begin{equation}
\beta_{\textrm{social}} (\aa,\aa_*) = \sum_{j\in\mathcal A}\beta_j(\aa,\aa),
\end{equation}
where the set $\mathcal A$ indicates the social activies, such as family, work, school.

In the following, although we will derive our feedback-controlled formulation for an age-structured SIR model, the extension to more realistic compartmental models^^>\cite{Gatto,Dutta2020} and other social characteristics, such as the total number of social contacts and the wealth of individuals discussed in Section \ref{sec:social}, can be done in a similar fashion.

\subsubsection{Optimal control formulation}
We consider an optimal control framework to define the strategy of a policy maker in introducing non-pharmaceutical interventions such as social distancing and other containment measures linked to the social structure. In what follows we assume that policy maker aims to minimize the impact of the epidemic through an age dependent control action depending both on time and pairwise interactions among individuals with different ages.  Thus, we introduce the optimal control problem
\begin{equation}\label{eq:func}
\begin{split}
\min_{{\mathbf u}\in \mathcal U}J({\mathbf u}):=& \int_0^T\psi(S(t),I(t))  dt \\
&+\sum_{j\in\mathcal A}\dfrac{1}{2}\int_0^T\int_{\LL\times\LL} {\nu_j(\as,\as_*,t)}|u_j(\as,\as_*,t)|^2\ d\as d\as_*  dt,
\end{split}
\end{equation}
subject to
\begin{equation}\label{eq:SIRc}
\begin{split}
\dfrac{\partial\fS(\as,t)}{\partial t}&= - \fS(\as,t)\sum_{j\in\mathcal{A}}\int_{\LL} (\beta_j(\as,\as_*) -u_j(\as,\as_*,t)){\fI(\as_*,t)}\ d\as_* \\
\dfrac{\partial\fI(\as,t)}{\partial t} &=  \fS(\as,t)\sum_{j\in\mathcal{A}}\int_{\LL} (\beta_j(\as,\as_*) -u_j(\as,\as_*,t)){\fI(\as_*,t)}\ d\as_*\\&\qquad\qquad\qquad - \gamma (\as) \fI(\as,t) \\
\dfrac{\partial\fR(\as,t)}{\partial t}&= \gamma (\as) \fI(\as,t),
\end{split}
\end{equation}
with initial condition $\fS(\as,0) = \fS^0(\as)$, $\fI(\as,0) = \fI^0(\as)$, and $\fR(\as,0) = \fR^0(\as)$. The function $\psi(S,I)$ accounts for the total number of the infected population $I(t)$ and susceptibles $S(t)$, such that $\psi(\cdot,\cdot)$ is positive and $\partial_I \psi(S,I)\geq0$. This function models the policy maker's perception of the impact of the epidemic by the number of people currently infected and susceptibles {and in the sequel will be referred to as \emph{perception function}}. Each component of the control ${\mathbf u}=(u_1,\ldots,u_{L})$ acts selectively on the interaction between individuals of ages $a$ and $a_*$  for a specific activity in $\mathcal{A}$, with $L$ the total number of activities. We consider a quadratic penalization of the control, weigthed by a specific function $\nu_j(a,t) > 0$ associated to each activity.

In \eqref{eq:func} the set $\mathcal U \subseteq \mathbb{R}^L$ is the space of admissible controls $u_j$, $j\in \mathcal A$ defined as 
\[
\mathcal U=\left\{{\mathbf u}\in\mathbb{R}^L\,|\, 0 \leq \mathcal{I}(u_j)(\as,t) \leq\min\{M,\mathcal{I}(\beta_j)(\as,t)\},\,\, \forall\, (\as,t),\, M>0\right\},
\] 
where $\mathcal I $ corresponds to the integral operator
\begin{equation}\label{eq:intfeed}
\begin{split}
\mathcal{I}(\varphi)(\as,t)=\frac1{I(t)}{\int_{\LL} \varphi(\as,\as_*,t)\fI(\as_*,t)\,da_*},\quad \end{split}
\end{equation}
which ensures the admissibility of the solution for \eqref{eq:SIRc}. The above restriction on admissible controls can be relaxed if we consider controls that violate the previous condition locally but preserve the inequality in integral form after integration against $i(a_*,t)$.
 
Solving the above optimization problem, however, is generally quite complicated and computationally demanding when there are uncertainties as it involves solving simultaneously the forward problem \eqref{eq:func}- \eqref{eq:SIRc} and the backward problem derived from the optimality conditions^^>\cite{APZ}. Furthermore, the assumption that the policy maker follows an optimal strategy over a long time horizon seems rather unrealistic in the case of a rapidly spreading disease such as the COVID-19 epidemic. 

\begin{tips}{Examples of perception function}
	
We report two relevant examples of the perception function $\psi(\cdot)$, given by a convex function underestimating the number of infected 
\begin{equation} 
\label{eq:psi1}
\psi(S,I)(t)=C\frac{I^q(t)}{q},\qquad q \geq 1,
\end{equation}
and a concave function overestimating such number
\begin{equation} 
\label{eq:psi2}
\psi(S,I)(t)=C\frac{\ln(1+\tau I(t))}{\tau S(t)},\quad \tau > 0,
\end{equation}
with $C>0$ a suitable renormalization constant. The function in \eqref{eq:psi1} has been introduced^^>\cite{APZ}, whereas the function in \eqref{eq:psi2} is related to well-known epidemic models with saturated incidence rates^^>\cite{Capasso,franco2020,KM}.

{Let us emphasize that extending the above optimal control formulation to more complex compartmental models designed specifically for COVID-19, like SEPIAR or SIDHARTE^^>\cite{Bruno,Gatto}, can be done by generalizing the perception function in \eqref{eq:func} to include, for example, the hospitalized compartment, or other specific indicators that can be measured from the data.} 	
\end{tips}
\subsubsection{Feedback controlled compartmental models}\label{sect:ins}
In this section we consider short time horizon strategies which permit to derive suitable feedback controlled models. These strategies are suboptimal with respect to the original problem \eqref{eq:func}-\eqref{eq:SIRc} but they have proved to be very successful in several social modeling problems^^>\cite{Albi1,Albi2,Albi3,Albi4}. To this aim, we consider a short time horizon of length $h>0$ and formulate a time discretize optimal control problem through the functional $J_h(u)$ restricted to the interval $[t,t+\h]$, as follows
\begin{equation}\label{eq:func_I}
\min_{{\bf u}\in\mathcal U} J_h({\bf u}) := \psi(S(t),I(t+\h))+\sum_{j\in\mathcal A}\frac{1}{2}\int_{\Lambda\times\Lambda}{\nu_j(\as,\as_*,t)}|u_j(\as,\as_*,t)|^2d\as d\as_*
\end{equation}
subject to dynamics \eqref{eq:SIRc}. By recalling that the macroscopic information on the infected is
\begin{align*}
I(t+\h)&= I(t)+ \h \int_{\LL} \Big[ {\fS(\as,t)}{}\sum_{j\in\mathcal A}\int_{\LL} \left(\beta_j(\as,\as_*) - u_j(\as,\as_*,t)\right)\fI(\as_*,t) d\as_*
\cr
&\qquad\qquad\qquad\qquad- \gamma(\as) \fI(\as,t)\Big]\ d\as,
\end{align*}
we can derive the minimizer of $J_h$ computing $\nabla_{{\mathbf u}} J_h({\mathbf u})\equiv 0$.
Using \eqref{eq:func_I} and the macroscopic information on $I(t+h)$ and introducing the scaling $\nu_j(\as,\as_*,t) = \h \kappa_j(\as,\as_*,t)$ we retrieve the instantaneous control
\begin{equation}\label{eq:ic0}
u_j (\as,\as_*,t)=\frac{ 1}{\kappa_j(\as,\as_*)}\fS(\as,t) i(\cdot,\as_*,t)\partial_I\psi(S(t),I(t+\h))].
\end{equation}
Passing to the limit for $h\to 0$ and embedding into \eqref{eq:SIRc} the control $u_j$ we obtain an instantaneous feedback controlled dynamics.

\begin{tips}{Explicit form of incidence rates}

To understand the action of the feedback control \eqref{eq:ic0} let us consider the simplest case of a standard SIR model without age dependence (homogeneous mixing), and specific social interactions. In this simplified setting the model has the structure of SIR model with the modified transmission rate
\begin{equation}
\begin{split}
\frac{d}{dt} S(t)&= - \beta_\kappa(t)S(t)I(t) \\
\frac{d}{dt} I(t) &= \beta_\kappa(t)S(t)I(t)-\gamma I(t),
\end{split}
\label{eq:homo}
\end{equation}
where the transmission rate is 
\be
\label{eq:beta1}
\beta_\kappa(t)=\beta-\frac{S(t)I(t)\partial_I \psi(S(t),I(t))}{\kappa}.
\ee
Introducing the explicit expressions of the control term for the perception function \eqref{eq:psi1} and \eqref{eq:psi2}, we obtain in the convex case \eqref{eq:psi1}
\be
\label{eq:beta2}
\beta_\kappa(t)=\beta-\frac{C S(t) I(t)^q}{\kappa} = \beta\left(1-\frac{S(t) I(t)^q}{\kappa}\right),
\ee
whereas in the logarithmic case \eqref{eq:psi2} and assuming $C=\beta$ and $\tau=1/\kappa$, we have
\be
\label{eq:beta3}
\beta_\kappa(t) = \beta-\frac{C I(t)}{\kappa(1+\tau I(t))} = \frac{\beta}{1+\tau I(t)}.
\ee
 { Interestingly enough, the resulting nonlinear incidence rates \eqref{eq:beta2}-\eqref{eq:beta3} embedding the action of feedback controls correspond to the ones considered in^^>\cite{APZ2, XLiu} and^^>\cite{franco2020, Capasso}, respectively.} See also Section \ref{contact} of the present survey for a derivation of saturated incidence rates like \eqref{eq:beta3}. Other nonlinear incidence  rates may be obtained similarly by considering different perception functions, see^^>\cite{XLiu} and the references therein.
 \end{tips}See also Section \ref{contact} of the present survey.

\begin{tips}{Extensions to SEIRD models}
We can extend previous computation to the socially structured compartmental model including additional compartments such as exposed, and dead individuals. The resulting feedback controlled SEIRD model reads  	
\begin{equation}
\label{eq:SEIRD_u}
\begin{split}
\dfrac{\partial\fS(\as,t)}{\partial t} &=\fS(\as,t)\sum_{j\in \mathcal A} \!\int_{\LL}\!\! \left(\beta_j(\as,\as_*) - u_j(\as,\as_*,t)\right)\fI(\as_*,t) d\as_* \\
\dfrac{\partial\fE(\as,t)}{\partial t} &= \fS(\as,t)\sum_{j\in \mathcal A}\!\int_{\LL}\!\! \left(\beta_j(\as,\as_*) - u_j(\as,\as_*,t)\right)\fI(\as_*,t) d\as_*  \\
&\qquad\qquad\qquad- \sigma(\as) \fE(\as,t),\\
\dfrac{\partial\fI(\as,t)}{\partial t} &= \sigma(\as) \fE(\as,t+h)-(\gamma(\as)+\alpha(\as)) \fI(\as,t).\\
\dfrac{\partial\fR(\as,t)}{\partial t}&= \gamma (\as) \fI(\as,t)\\
\dfrac{\partial\fD(\as,t)}{\partial t} &= \alpha (\as) \fI(\as,t),
\end{split}
\end{equation}
with initial condition $\fS(\as,0) = \fS^0(\as)$, $\fE(\as,0) = \fE^0(\as)$, $\fI(\as,0) = \fI^0(\as)$,  $\fR(\as,0) = \fR^0(\as)$ and $\fD(\as,0) = \fD^0(\as)$. Compared to \eqref{eq:SIRc} we introduced the age dependent parameters: $\sigma (\as) \geq 0$, the transition rate of exposed individuals to the infected class, and $\alpha (\as) \geq 0$, the disease-induced death rate of infectious individuals.
The feedback control $u_j$ in this case is defined as follows
\begin{equation}\label{eq:ic2_SEIRD}
u_j (\as,\as_*,t)=\frac{\sigma(\as)}{\kappa_j(\as,\as_*)}\fS(\as,t) \fI(\as_*,t)\partial_I\psi(S(t),I(t)),
\end{equation}
where the main difference with respect to \eqref{eq:ic0} is the additional scaling parameter $\sigma(a)$. 
We refer to ^^>\cite{APZ} for the derivation of the control form \eqref{eq:ic2_SEIRD}, and to^^>\cite{APZ2} for further extension to SEPIAR model and an extensive study on $\psi(S,I) = I^q/q$.

	\end{tips}

\subsubsection{Containment in homogeneous social mixing dynamics}
To illustrate the effects of introduced controls that mimic containment procedures, let us first consider the case where the social structure is not present. Hence we consider model \eqref{eq:homo} with initial small number of infected and recovered $I(0) = 3.68 \times 10^{-6}$, $R(0) = 8.33\times 10^{-8}$. These normalized fractions refer specifically to the first reported values in the case of the Italian outbreak of COVID-19, even if in this simple test case we will not try to match the data in a quantitative setting but simply to illustrate the behavior of the feedback controlled model.

Based on recent studies^^>\cite{Zhang_etal,Liu}, the initial infection rate of COVID-19 $R_0 = \beta/\gamma$ has been estimated between $2$ and $6.5$. Here, to exemplify the possible evolution of the pandemic we consider a value close to the lower bound, taking $\beta =0.25$ and $\gamma = 0.10$, {namely a recovery rate of $10$ days}, so that $R_0 = 2.5$. 

\begin{figure}
	\centering
	\includegraphics[scale = 0.27]{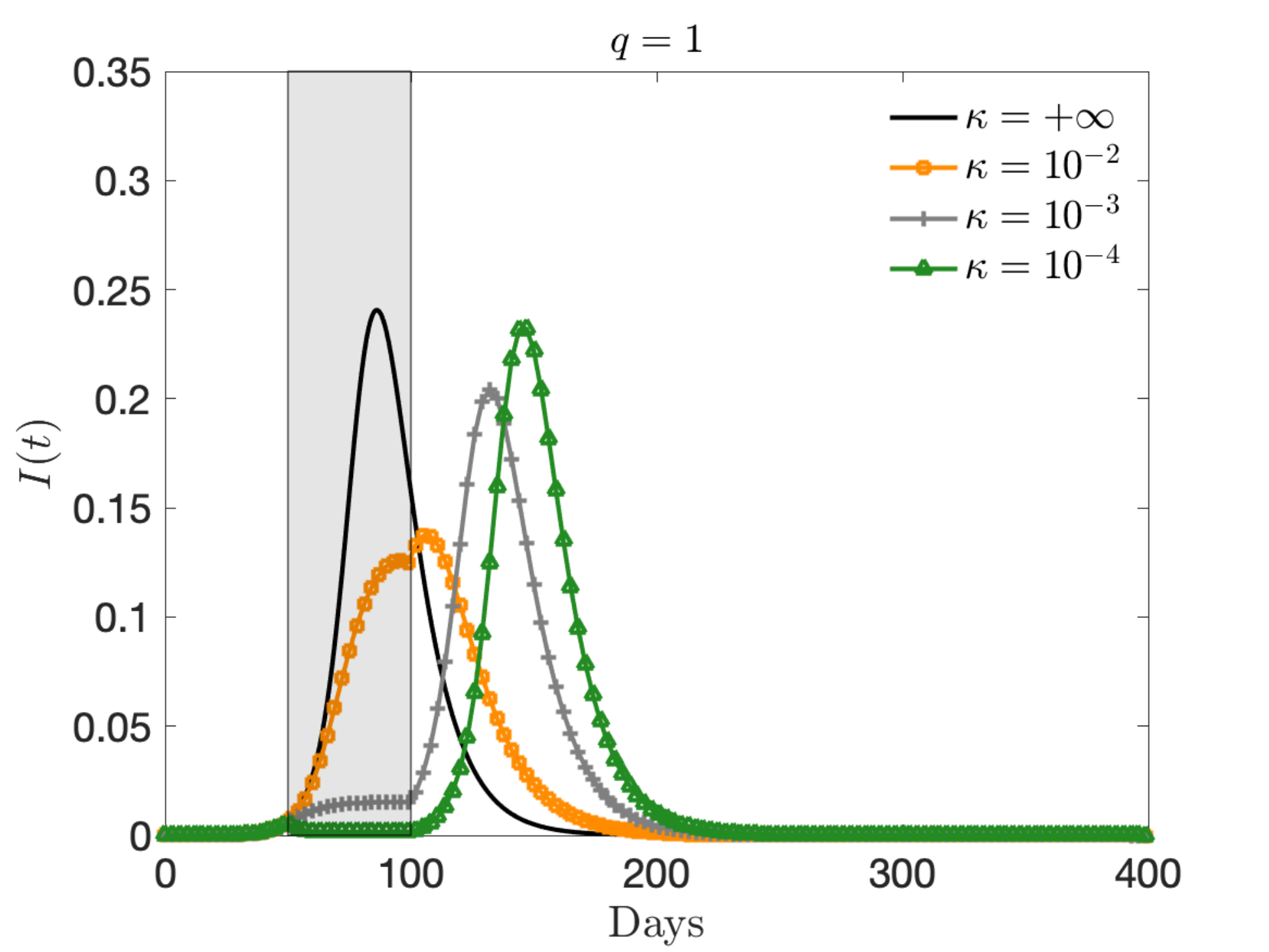}
	\includegraphics[scale = 0.27]{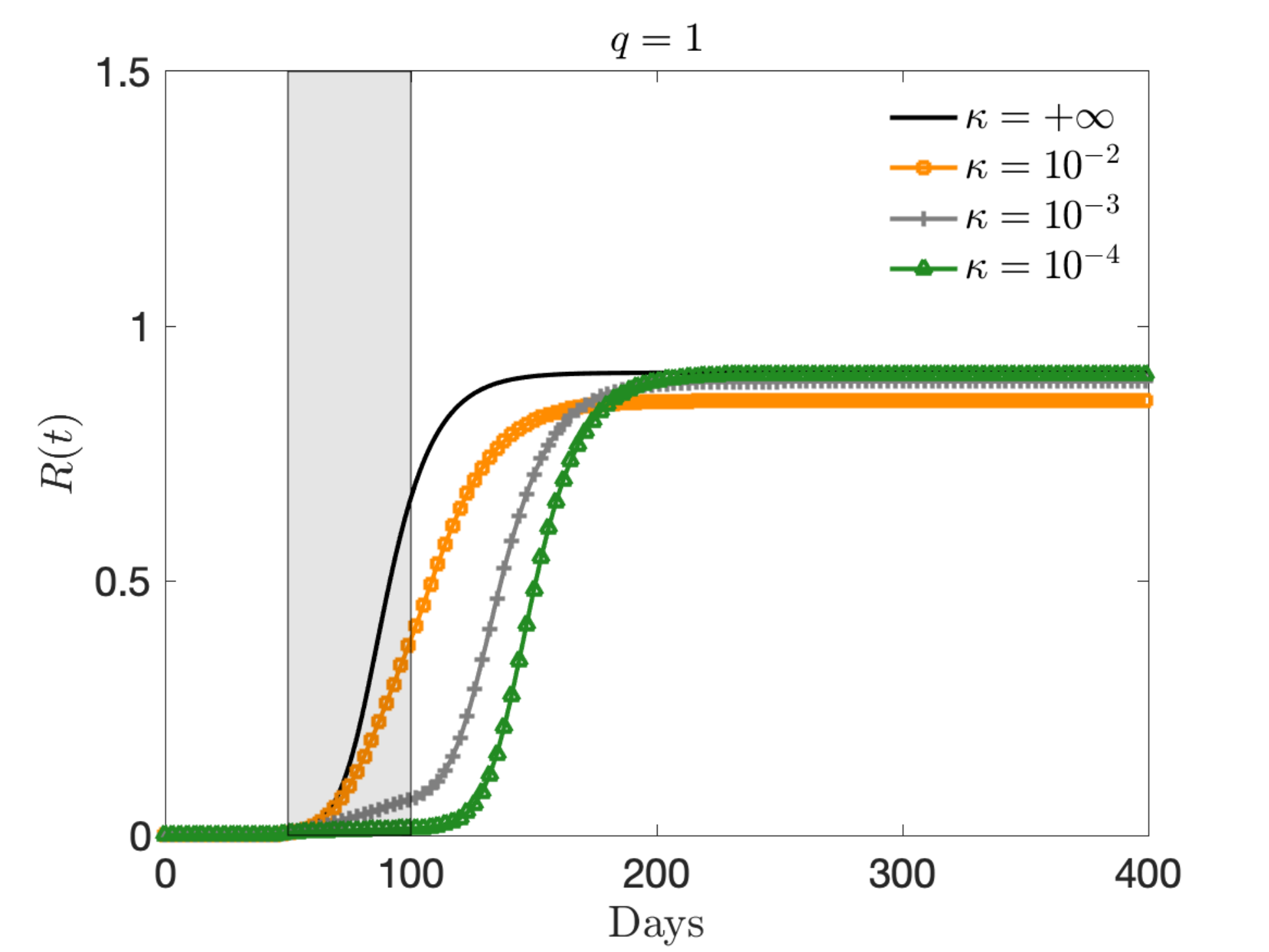} \\
	\includegraphics[scale = 0.27]{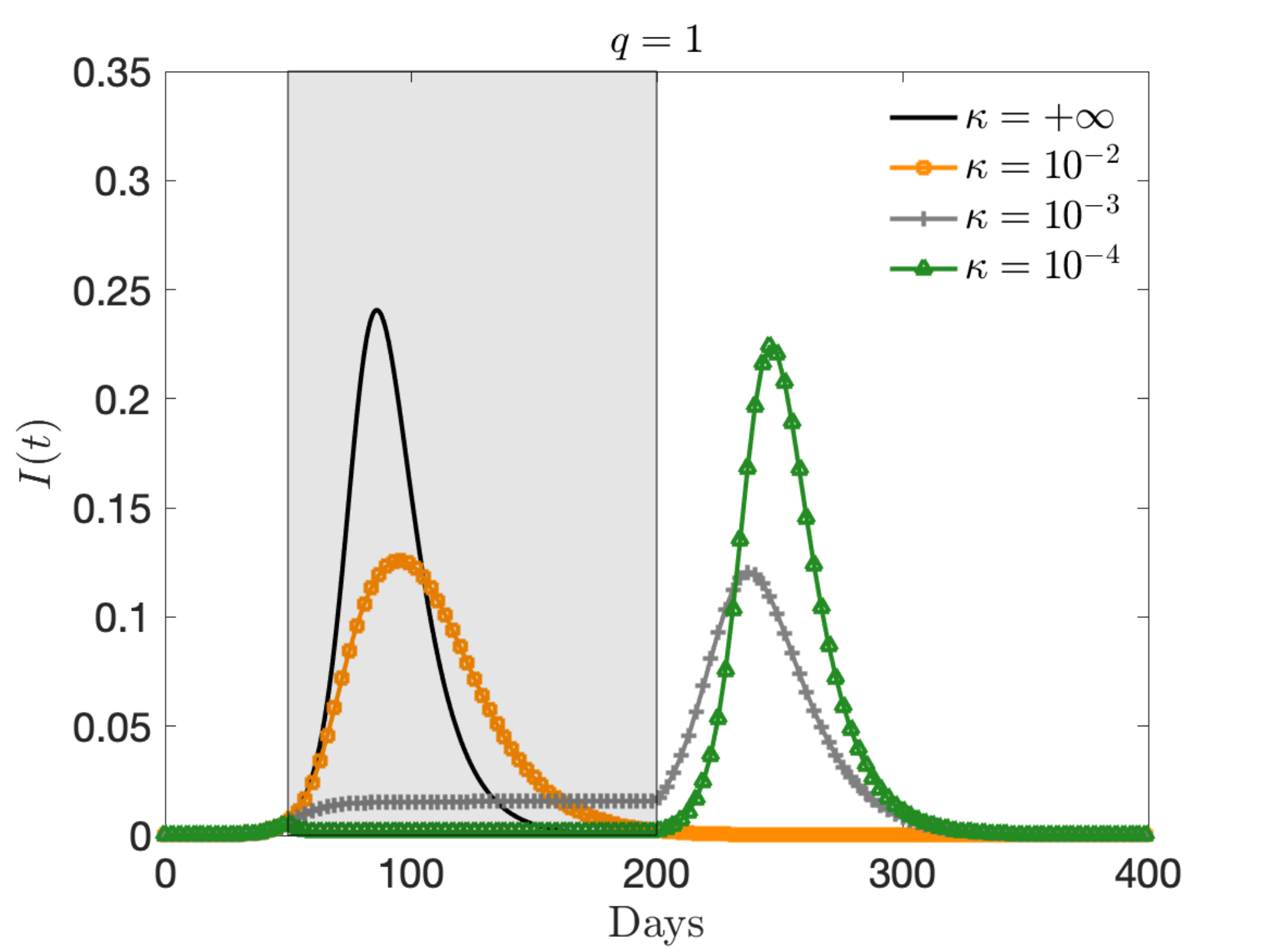}
	\includegraphics[scale = 0.27]{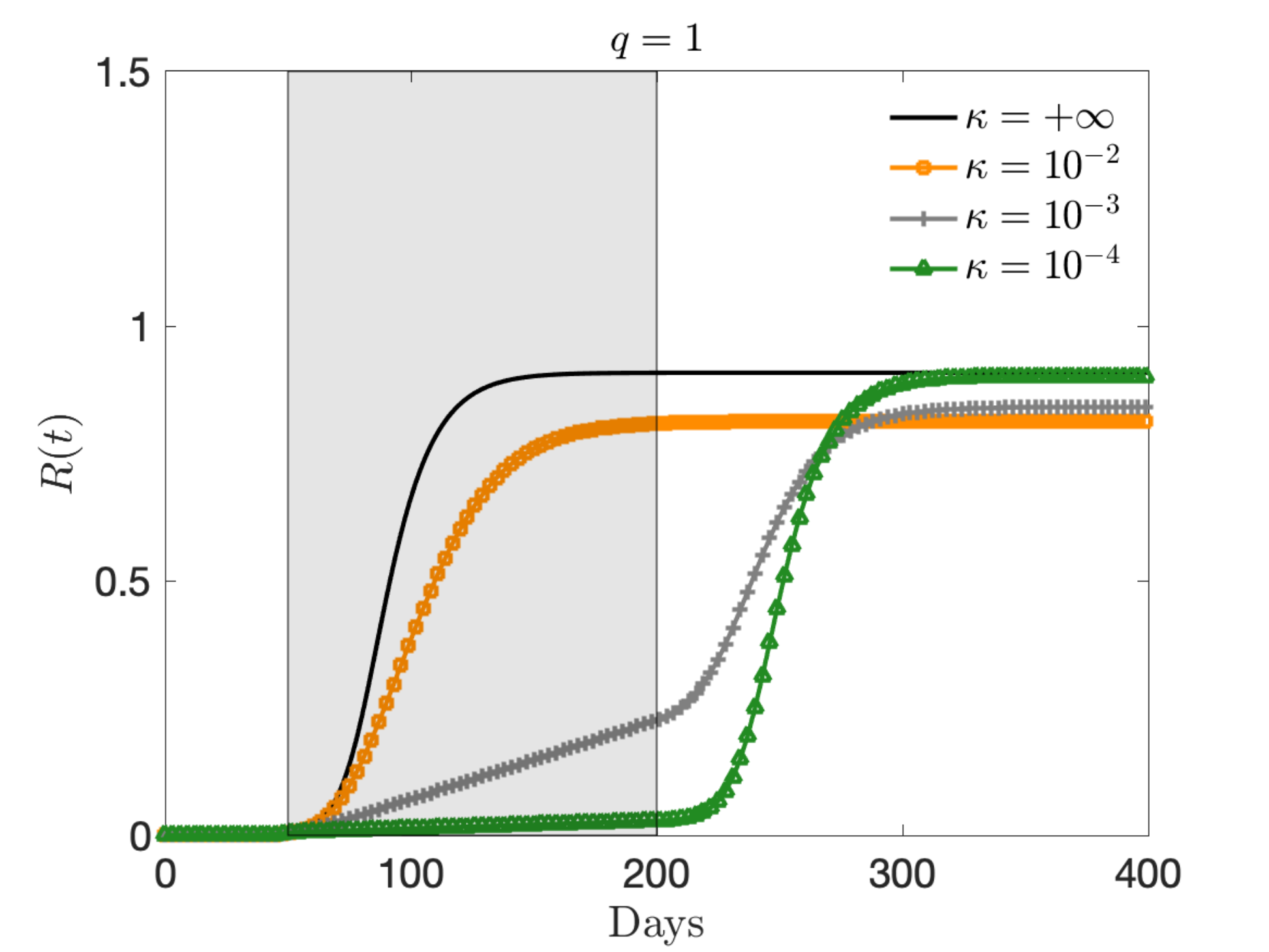}
	\caption{\textbf{Test 1}. Evolution of the fraction of infected (left) and recovered (right) based on {the SIR feedback constrained model \eqref{eq:homo} with perception function} $\psi(I)=I$ and several penalizations $\kappa = 10^{-2}, 10^{-3}, 10^{-4}$. The choice $\kappa = +\infty$ corresponds to the unconstrained case. 
		First row, the control is applied for $t\in [50,100]$; Second row, the control has a longer action in $[50,200]$.}
	\label{fig:test1a}
\end{figure}

In Figures \ref{fig:test1a} we report the infected and recovered dynamics based on the activation of the control in two different time frames. Top images show the case in which the activation time is bounded $t\in [50,100]$, which means that after $100$ days we suppose that all containment restrictions are cancelled. In bottom row we consider a larger activation time frame $t\in[50,200]$. With the choice of the perception function $\psi(I)=I$ we can observe how the control term is able to flatten the curve. 

\subsection{Dealing with data uncertainty}
Early in the outbreak of new infectious diseases, the actual number of people infected and recovered is typically underestimated, causing fatal delays in implementing public health policies in the face of spreading epidemic fronts. This is the case of the spreading of COVID-19 worldwide, often mistakenly underestimated due to deficiencies in surveillance and diagnostic capacity^^>\cite{RR,MKZC,Zhang_etal}. 
Among the common sources of uncertainties for dynamical systems modeling epidemic outbreaks we may consider: noisy and incomplete available data, and structural uncertainty due to the possible inadequacy of the mathematical model used to describe the  phenomena under consideration.
In the following we consider the effects on the dynamics of uncertain data, such as the initial conditions on the number of infected people or the interaction and recovery rates. On the numerical level we consider techniques based on stochastic Galerkin methods, for which spectral convergence on random variables is obtained under appropriate regularity assumptions^^>\cite{Xiu2010}.

\subsubsection{Feedback controlled and socially structured models with uncertain inputs}
\label{uncertain_data}
We introduce the random vector $\z = (z_1,\dots,z_{d_z})$ whose components are assumed to be independent real valued random variables
$z_k:(\Omega,F) \rightarrow (\mathbb R,\mathcal B_{\mathbb R})$, $k = 1,\dots,d_z$
with $\mathcal B_{\mathbb R}$ the Borel set. We assume to know the probability density $p(\z): \mathbb R^{d_z} \rightarrow \mathbb R^{d_z}_+$ characterizing the distribution of $\z$. 
Here, $\z\in\mathbb{R}^{d_z}$ is a random vector taking into account various possible sources of uncertainty in the model.

In presence of uncertainties we generalize the initial modeling by introducing the quantities $\fS(\z,\as,t)$, $\fI(\z,\as,t)$ and $\fR(\z,\as,t)$ representing the distributions at time $t\ge 0$ of susceptible, infectious and recovered individuals. The total size of the population is a deterministic conserved quantity in time, i.e.
\[
\fS(\z,\as,t)  + \fI(\z,\as,t) + \fR(\z,\as,t) = p(\as), \qquad \int_{\LL} p(\as)d\as = 1, 
\]
and the uncertain fractions of the population that are susceptible, infected and recovered are defined as follows 
\[
S(\z,t)=\int_{\LL}\fS (\z,\as,t)\,d\as,\quad I(\z,t)=\int_{\LL}\fI (\z,\as,t)\,d\as,\quad R(\z,t)=\int_{\LL}\fR (\z,\as,t)\,d\as.
\]
Hence, the controlled system \eqref{eq:SIRc} in presence of uncertainty reads
\begin{equation}
\label{eq:SIR_z}
\begin{split}
\partial_t\fS(\z,\as,t) &=  - {\fS(\z,\as,t)}\sum_{j\in \mathcal A} \!\int_{\LL}\!\! \left(\beta_j(\z,\as,\as_*) - u_j(\as,\as_*,t)\right)\fI(\z,\as_*,t) d\as_* \\
\partial_t\fI(\z,\as,t) &= {\fS(\z,\as,t)} \sum_{j\in \mathcal A}\!\int_{\LL}\!\! \left(\beta_j(\z,\as,\as_*) - u_j(\as,\as_*,t)\right)\fI(\z,\as_*,t) d\as_*\\
&\quad  - \gamma(\z,\as)) \fI(\z,\as,t)\\
\partial_t\fR(\z,\as,t) &= \gamma(\z,\as) \fI(\z,\as,t),
\end{split}
\end{equation} 
where the controls terms are assumed to be deterministic and defined as
\begin{equation*}\label{eq:ic00}
u_j (\as,\as_*,t)=\frac{ 1}{\kappa_j(\as,\as_*)}{\mathcal R}[\fS(\cdot,\as,t) \fI(\cdot,\as_*,t)\partial_I\psi(S(\cdot,t),I(\cdot,t))],
\end{equation*}
being $\mathcal R[\psi(S(\cdot,t),I(\cdot,t))]$ a suitable operator taking into account the presence of the uncertainties $\z$. Examples of such operator that are of interest in epidemic modelling rely on the expectated value
\begin{equation}
\label{eq:R1}
\mathcal R [\psi(S,I)(\cdot,t)]=\mathbb E[\psi(S,I)(\cdot,t)]= \int_{\mathbb R^{d_z}} \psi(S,I)(\z,t) \; p(\z)d\z,
\end{equation} 
or on deterministic data which underestimate the number of infected
\begin{equation} 
\label{eq:R2}
\mathcal R [\psi(S,I)(\cdot,t)]=\psi(S,I)(\z_0,t),
\end{equation}
where $\z_0$ is a given value such that $I(\z_0,t) \leq I(\z,t)$, for all $ \z\in \mathbb{R}^{d_z}$ and $t>0$. We refer to^^>\cite{APZ,APZ2} for further details on the derivation, and further extensions.

\begin{tips}{A solvable example}
	We consider a simplified version of model \eqref{eq:SIR_z} in absence of control, with homogeneous mixing $\beta_{\textrm{social}}(z)$ and recovery rate $\gamma(z)$, $z\in \mathbb{R}$ distributed as $p(z)$. Integrating against $a$ we obtain the following SIR model with uncertainty 
	\begin{equation}
	\begin{split}
	\frac{d}{dt} S(z,t)&= - \beta_{\textrm{social}}(z)S(z,t)I(z,t) \\
	\frac{d}{dt} I(z,t) &= \beta_{\textrm{social}}(z)S(z,t)I(z,t)-\gamma(z) I(z,t),
	\end{split}
	\label{eq:homo2}
	\end{equation}
	with deterministic initial values $I(z,0)=I_0$ and $S(z,0)=S_0$.
	Following ^^>\cite{Rob}, we assume a linear source of uncertainty $\beta_{\textrm{social}}(z)=\beta+\alpha z$, $\alpha > 0$, and constant recovery rate $\gamma(z) = \gamma>0$. The solution for the proportion of infected during the initial exponential phase is
	{\[
		I(z,t) = I_0 e^{(\beta+\alpha z)S_0 t - \gamma t},
		\]}
	and its expectation
	{
		\begin{equation}
		{\mathbb E}[I(\cdot,t)]= I_0 e^{\beta S_0 t - \gamma t} \int_{\mathbb{R}} e^{\alpha z S_0 t} p(z)\,dz=I_0 e^{\beta S_0 t - \gamma t}W(t),
		\end{equation}}
	where $W(t)$ represents the statistical correction factor to the standard deterministic exponential phase of the disease {$I_0 e^{\beta S_0 t - \gamma t}$}. If $z$ is uniformly distributed in $[-1,1]$ we can explicitly compute 
	{
		\[
		W(t)=\frac{\sinh \left({\alpha S_0 t}\right)}{\alpha S_0 t} > 1,\quad t>0. 
		\]}
	More in general, if $z$ has zero mean then by Jensen's inequality we have $W(t)>1$ for $t > 0$, so that the expected exponential phase is amplified by the uncertainty.
	
	In a similar way, keeping $\beta_{\textrm{social}} (z)= \beta$ constant, but introducing a source of uncertainty in the initial data $I(z,0)=I_0+\mu z$, $\mu > 0$ and $z\in \mathbb{R}$ distributed as $p(z)$ the solution in the exponential phase reads
	{
		\[
		I(z,t) = (I_0+\mu z) e^{\beta S_0 t - \gamma t},
		\]}
	and then its expectation 
	{
		\begin{equation}
		\begin{split}
		{\mathbb E}[I(\cdot,t)]&=\int_{\mathbb{R}}(I_0+\mu z) e^{\beta S_0 t - \gamma t}p(z)\,dz= (I_0 + \mu\bar{z}) e^{\beta S_0 t - \gamma t},
		\end{split}
		\end{equation}}
	where $\bar{z}$ is the mean of the variable $z$. Therefore, the expected initial exponential growth behaves as the one with deterministic initial data $I_0+\mu \bar{z}$. Of course, if both sources of uncertainty are present the two effects just described sum up in the dynamics.
\end{tips}

The presence of a large number of undetected infected is at the basis of the construction of numerous epidemiological models with an increasingly complex compartmental structure in which the original compartment of the infected is subdivided into further compartments with different roles in the propagation of the disease^^>\cite{Bruno,Gatto,Flax}. The following remark clarifies the relationships to other deterministic compartmental models.

\begin{tips}{Connection to other compartmental models}
	Let us consider model \eqref{eq:homo2} with a one-dimensional random input $z \in \mathbb{R}$ distributed as $p(z)$. Furthermore, for a function $F(z,t)$ we will denote its expected value as $\bar F(t) = \mathbb{E}[F(\cdot,t)]$. Now, starting from a discrete probability density function
	\[
	p_k=P\left\{Z=z_k\right\},\qquad \sum_{k=1}^n p_k = 1,
	\]
	we have $\bar F(t) = \sum_{k=1}^n p_k F_k$, with $F_k=F(z_k)$. Taking the expectation in \eqref{eq:homo2}, we can write
	\begin{equation}
	\begin{split}
	\frac{d}{dt} \bar S(t)&= - \bar S(t) \sum_{k=1}^n \tilde \beta_k  p_k I_k(t) \\
	\frac{d}{dt} \bar I(t) &= \bar S(t) \sum_{k=1}^n \tilde \beta_k  p_k I_k(t)-\sum_{k=1}^n \gamma_k p_k I_k(t),\\
	\frac{d}{dt} \bar R(t) &= \sum_{k=1}^n \gamma_k p_k I_k(t),
	\end{split}
	\label{eq:homopar}
	\end{equation}
	with $\tilde \beta_k = S_k \beta_k/\bar S $, $k=1,\ldots,n$. For example, in the case $n=2$, by identifying $I_d=p_1 I_1$ and $I_u=p_2 I_2$ with the compartments of detected and undetected infectious individuals, we have the same structure of a SIAR compartmental model including the undetected (or the asymptomatic) class.

	The additional dependence of the epidemiological parameters on the random variable allows us to take into account changes in the corresponding dynamics of disease transmission and recovery. 
	\end{tips}

\subsubsection{Application to the COVID-19 outbreak}\label{sec:appcovid}
In this section, we first present the impact of social structure in feedback-controlled models with uncertain data, which account for the presence of symptomatic and asymptomatic unreported cases, at the first wave of the COVID-19 pandemic.
In particular we will focus on different scenarios of possible containment measures for different countries.

\paragraph{\bf Model calibration and estimating actual infection trends}

Estimating epidemiological parameters is a very difficult problem that can be addressed with different approaches^^>\cite{Cetal, Chowell, Rob}. In the case of COVID-19 due to the limited number of data and their great heterogeneity this becomes an even bigger problem that can easily lead to unrealistic results. 

Similarly to Section^^>\ref{numericsIII} we calibrate the model using data publicly available from the John Hopkins University GitHub repository, and GitHub repository of the Italian Civil Protection Department for the Italian case (see Data Sources in Section \ref{sec:ds}).
We adopted the fitting procedure described in^^>\cite{APZ2} that is based on a strategy with two optimization horizons (pre-lockdown and lockdown time spans) depending on the different strategies enacted by the governments of the considered European countries. Once the relevant epidemiological parameters have been estimated in the pre-lockdown time span, i.e.  $\beta_e>0$ and $\gamma_e>0$, we successively proceeded with the estimation of the control penalty parameter $\kappa_e=\kappa(t)>0$. These two calibration steps were analyzed under the assumption of homogeneous mixing.

\begin{figure}
	\centering
	\includegraphics[scale =.195]{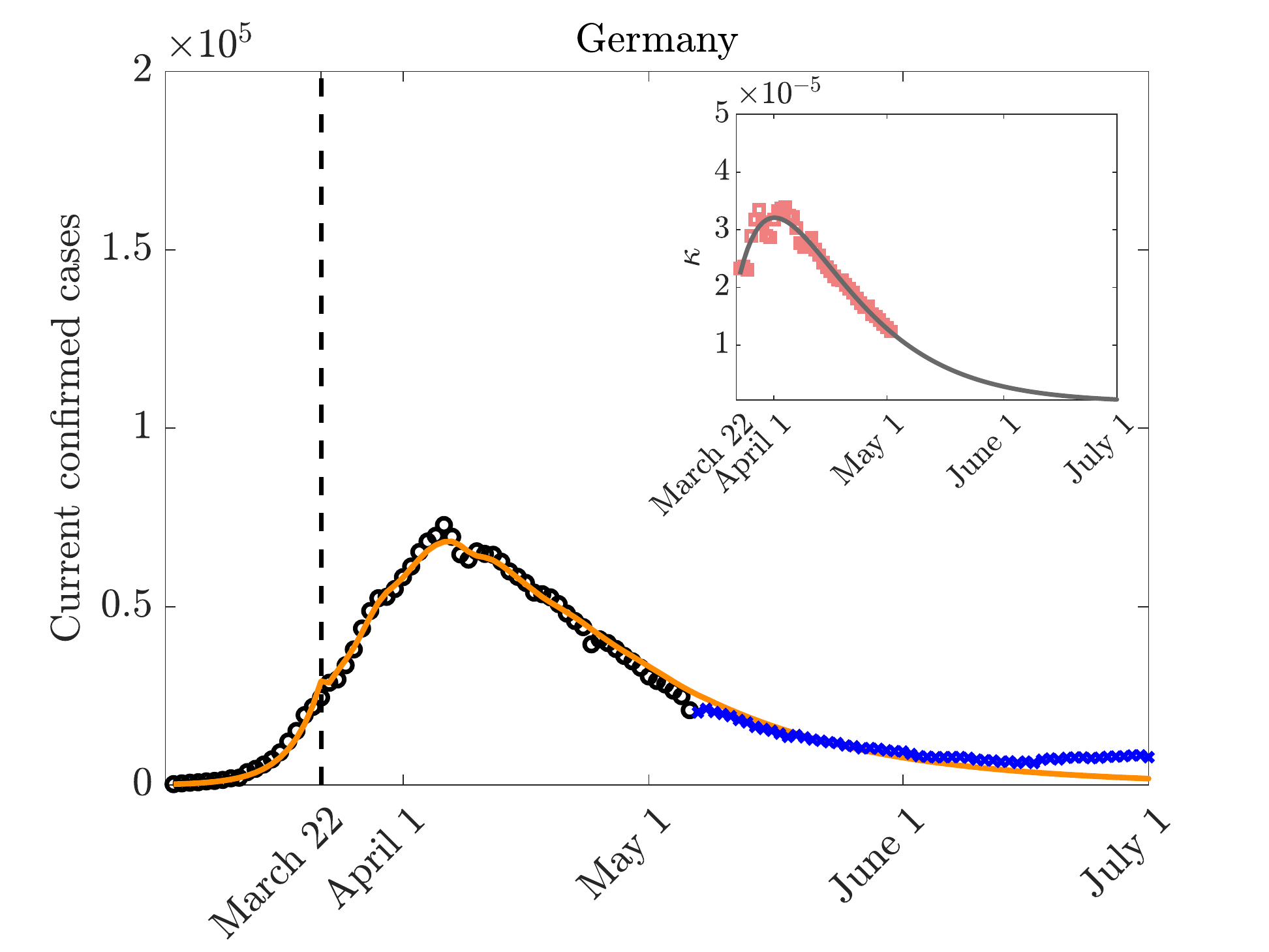}
	\includegraphics[scale =.195]{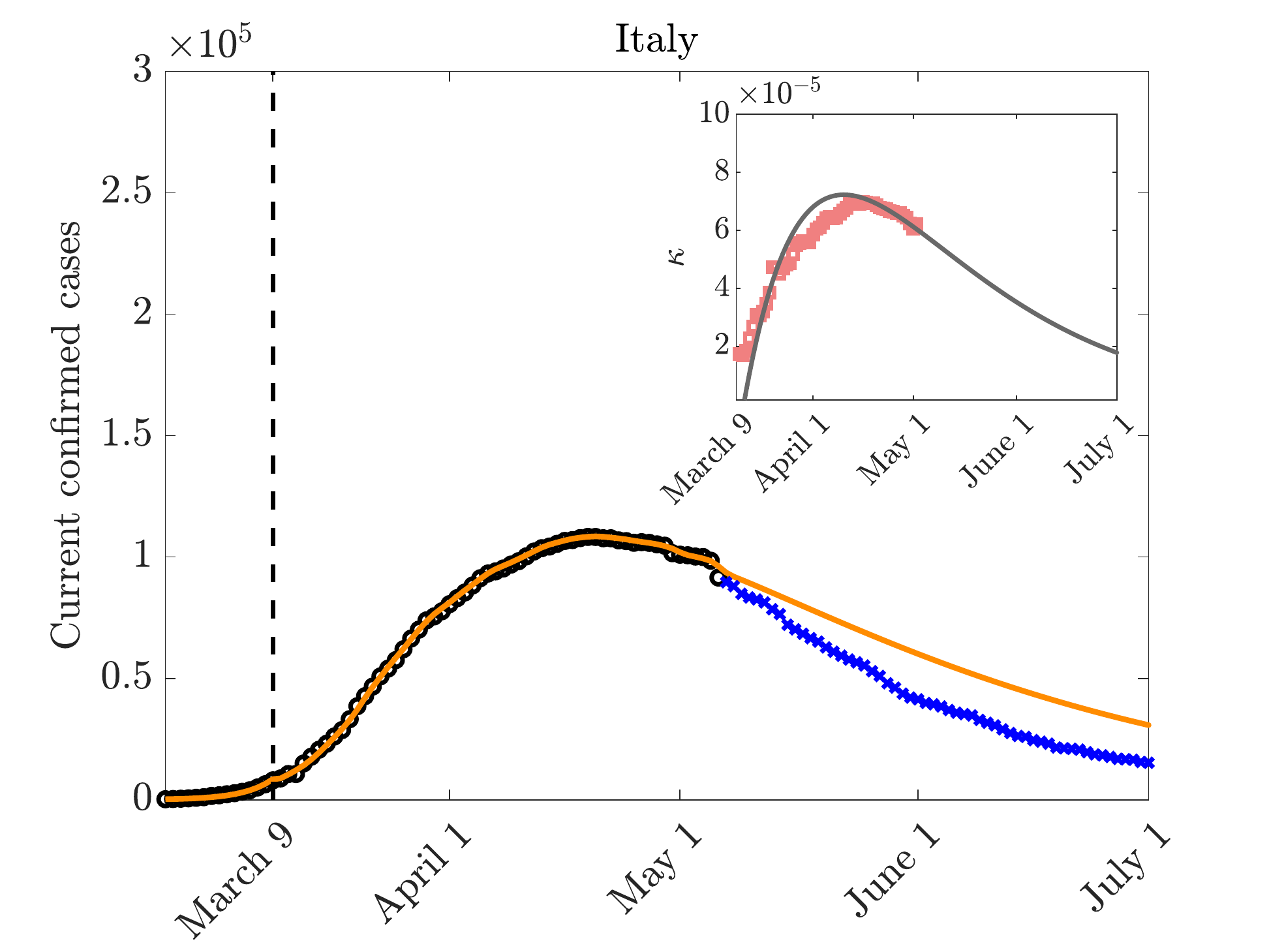} 
	\includegraphics[scale =.195]{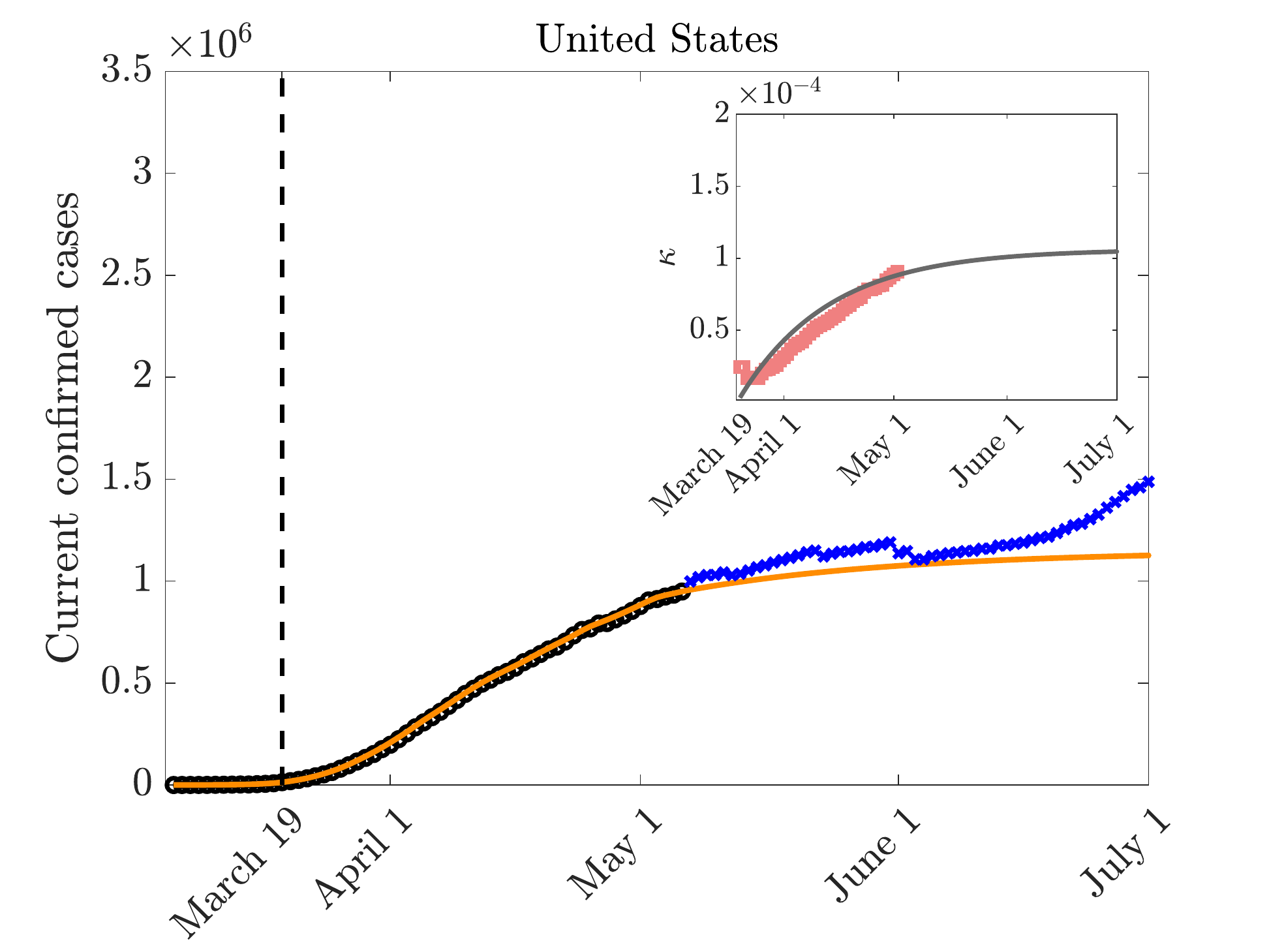}
	\caption{Model behavior with fitting parameters and actual trends in the number of reported infectious using the estimated control penalization terms after lockdown over time in the various countries.}
	\label{fig:datak}
\end{figure}

We report in Figure \ref{fig:datak} the corresponding time dependent values for the control parameter $k(t)$, as well as results of the model fitting with the actual trends of infected individuals. 

Next we focus on the influence of uncertain quantities on the controlled system with homogeneous mixing. According to recent results on the diffusion of COVID-19 in many countries the number of infected, and therefore recovered, is largely underestimated on the official reports, see e.g.^^>\cite{JRGL,MKZC}.    As discussed in^^>\cite{APZ2} a parameter estimation based on the previous fitting has some limitations and in particular overestimates the reproduction rate in the early phase of the pandemic. For this purpose, to have an insight on global impact of uncertain parameters we consider a two-dimensional uncertainty $\z = (z_1,z_2)$ with independent components such that 
\be\label{eq:ir_z1}
I(\z,0) = I_0(1+\mu z_1),\qquad R(\z,0) = R_0(1+\mu z_1), \qquad \mu>0
\ee
and
\be\label{eq:beta_z2}
\beta(\z) = \beta_e-\alpha_\beta z_2,\qquad \gamma(\z)=\gamma_e + \alpha_\gamma z_2,\qquad \alpha_\beta,\alpha_\gamma>0 
\ee
where $z_1$, $z_2$ are chosen to be distributed as symmetric Beta distributions in $[0,1]$, $\fI^0$ and $\fR^0$ are the initial number of reported cases and recovered taken from^^>\cite{Zhang_etal}. The parameter $\mu=2(c-1)$ is common for all countries such that $\mathbb E[I(\z,0)]=c I(0)$, $\mathbb E[R(\z,0)]=c R(0)$ where $c=8.56$, corresponding to average disagreement in the total number of cases based on an estimated infection fatality rate (IFR) of $1.3\%$ in the range $0.9\%-2.0\%$. The feedback controlled model has been computed using an estimation of the total number of susceptible and infected reported, namely we have the control term 
\begin{equation}
u(t)=-\frac1{k(t)} S_r(t)I_r(t),
\label{eq:r0e}
\end{equation}
where $S_r(t)$ and $I_r(t)$ are the model solution obtained from the registered data, and thus $I_r(t)$ represents a lower bound for the uncertain solution $I(\z,t)$. 

\begin{figure}
	\centering
	\hspace{-0.5cm}
	\includegraphics[scale =.195]{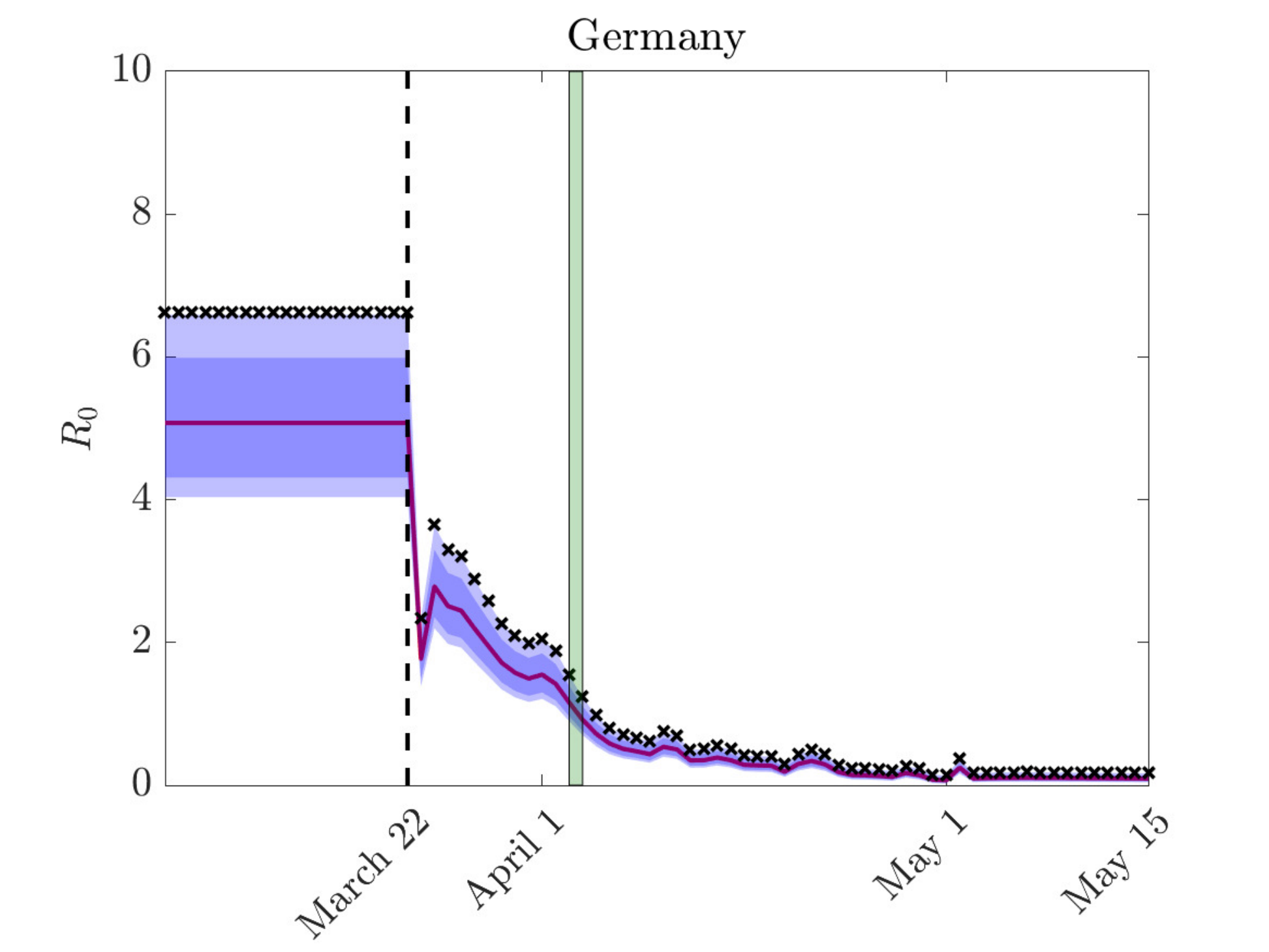}\hspace{-0.25cm}
	\includegraphics[scale =.195]{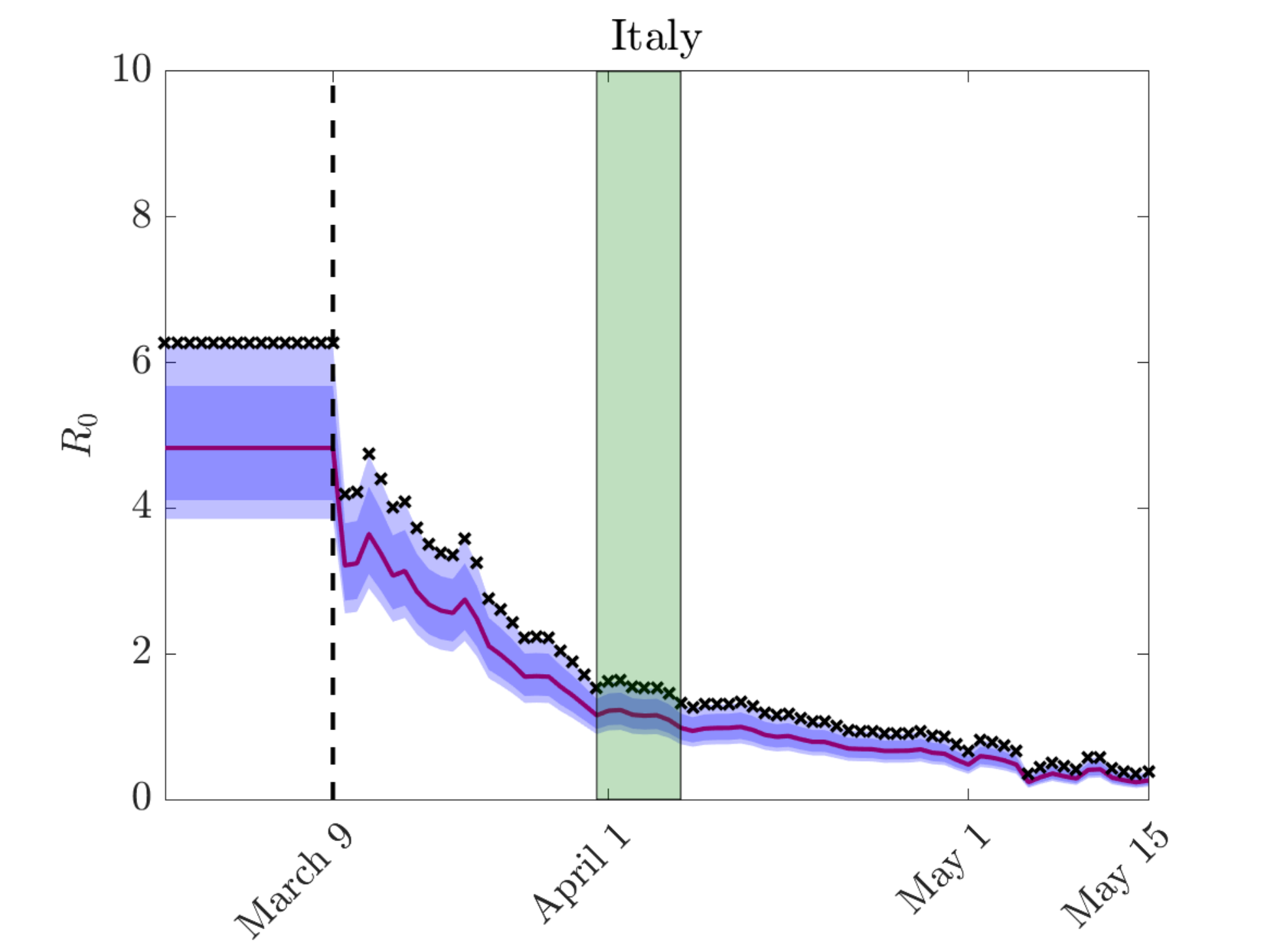} \hspace{-0.25cm}
	\includegraphics[scale =.195]{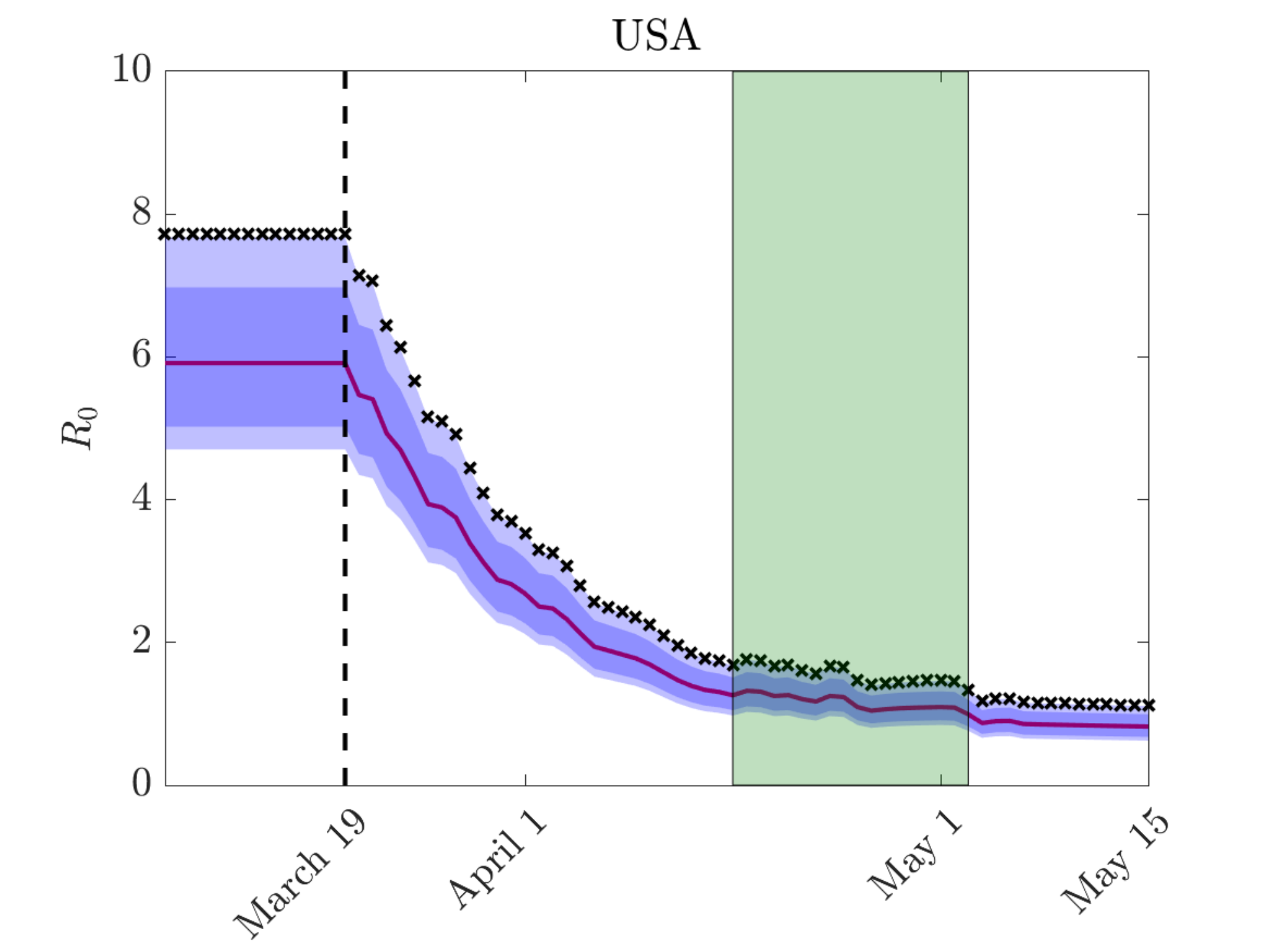}
	\caption{Evolution of estimated reproduction number $R_0$ and its confidence bands for uncertain data in as in \eqref{eq:beta_z2}. The $95\%$ and $50\%$ confidence levels  are represented as shaded and darker shaded areas respectively. The green zones denote the interval between the first day  the $50\%$ confidence band and the expected value fall below $1$. }
	\label{fig:R0}
\end{figure}

In Figure \ref{fig:R0} we report the evolution of reproduction number $R_0$ for the considered countries under the uncertainties in \eqref{eq:beta_z2} obtained with $\alpha_\beta=0.03$, $\alpha_\gamma=0.05$ and $z_2\sim B(2,2)$. The reproduction number is estimated from 
\[
R_0(z_2,t) = \dfrac{\beta(z_2) - u(t) \chi(t>\bar t)}{\gamma(z_2)}, 
\]
being the control $u(t)$ defined in \eqref{eq:r0e} and $\bar t$ is the country-dependent lockdown time. The estimated reproduction number relative to data  is reported with {\tt x}-marked symbols and represents an upper bound for $R_0(z_2,t)$.

\paragraph{{\bf Effect of social contacts in the population.}}

We first analyze the effects of the inclusion of age dependence and social interactions in the above dynamics with uncertainty in the case of COVID-19 outbreak for the Italian case.
	 {The age dependent social interaction rate $\beta(a,a_*)$ is defined as follows,
	\be\label{eq:beta_aa}
	\beta(a,a_*) = (1-\xi)\beta_e + \xi \sum_{j\in\mathcal{A}} \beta_j(a,a_*),  
	\ee
	where $0\leq\xi\leq 1$, thus for $\xi=0$ we recover the {homogeneuos mixing}, whereas for $\xi = 1$ we have a full {social mixing} behavior.}

	The social interaction function, $\beta_{\rm social}(a,a_*)$, accounts for the interactions due to specific activities $\mathcal A=\{\textrm{Family,\ Education,\, Profession}\}$. 
    This function is normalized using the estimated parameters $\beta_e$ in accordance with
	\be\label{eq:bg}
	\beta_e =C_\beta\int_{\LL\times \LL} \beta(a,a_*)f(a)f(a_*)\,da\,da_*,\quad \gamma_e = C_\gamma\int_{\LL} \gamma(a)f(a)\,da,
	\ee
	where $f(a)$ is the age distribution with $\Lambda = [0,a_{\rm max}]$, $a_{\max}=100$, and $C_\beta,C_\gamma$ normalization constants.
	We refer to the Appendix of^^>\cite{APZ} for specific definition of the social interaction $\beta_j(a,a_*)$.

\begin{figure}
	\centering
	\includegraphics[width=3.85cm]{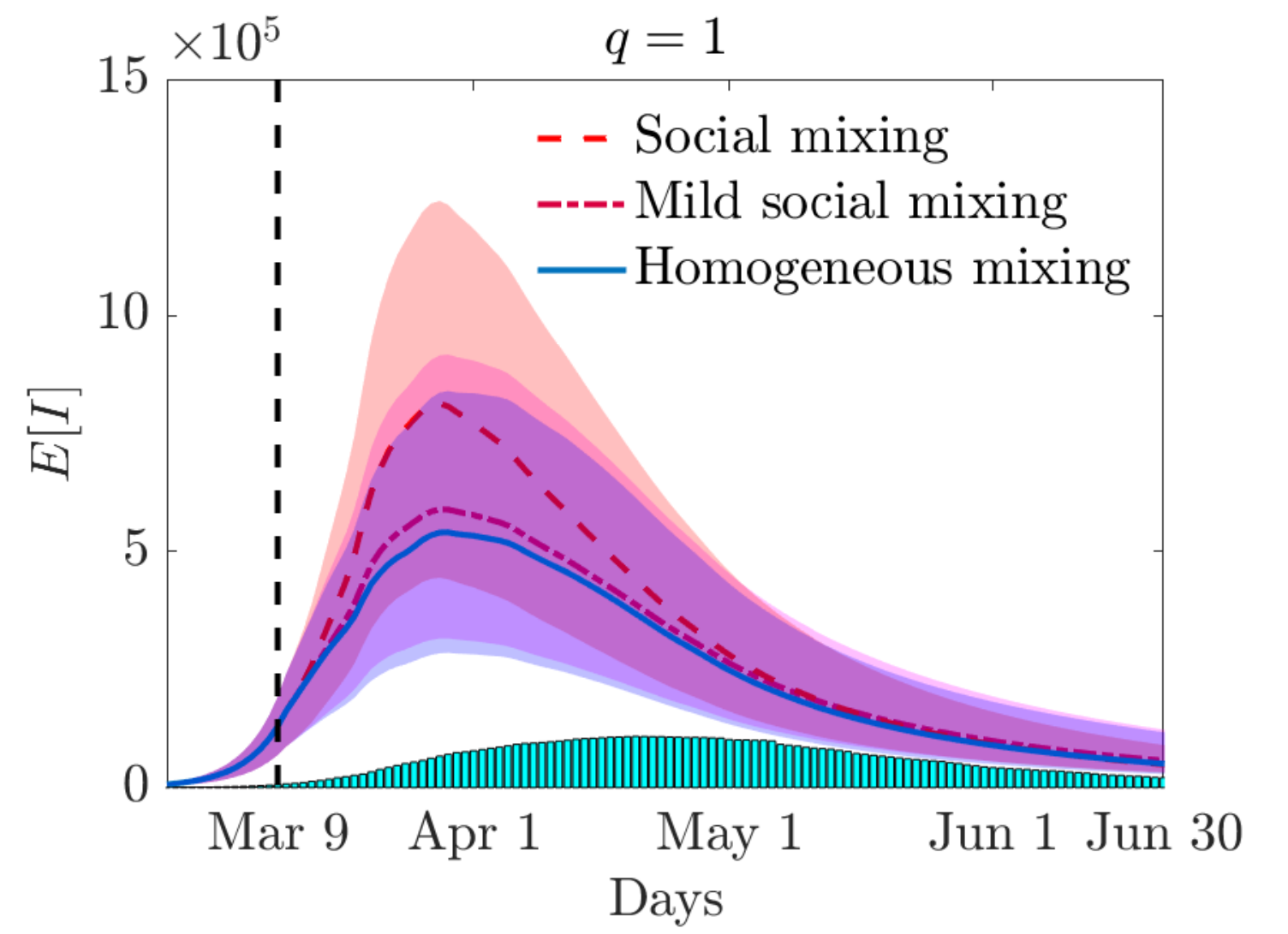}
	\includegraphics[width=3.85cm]{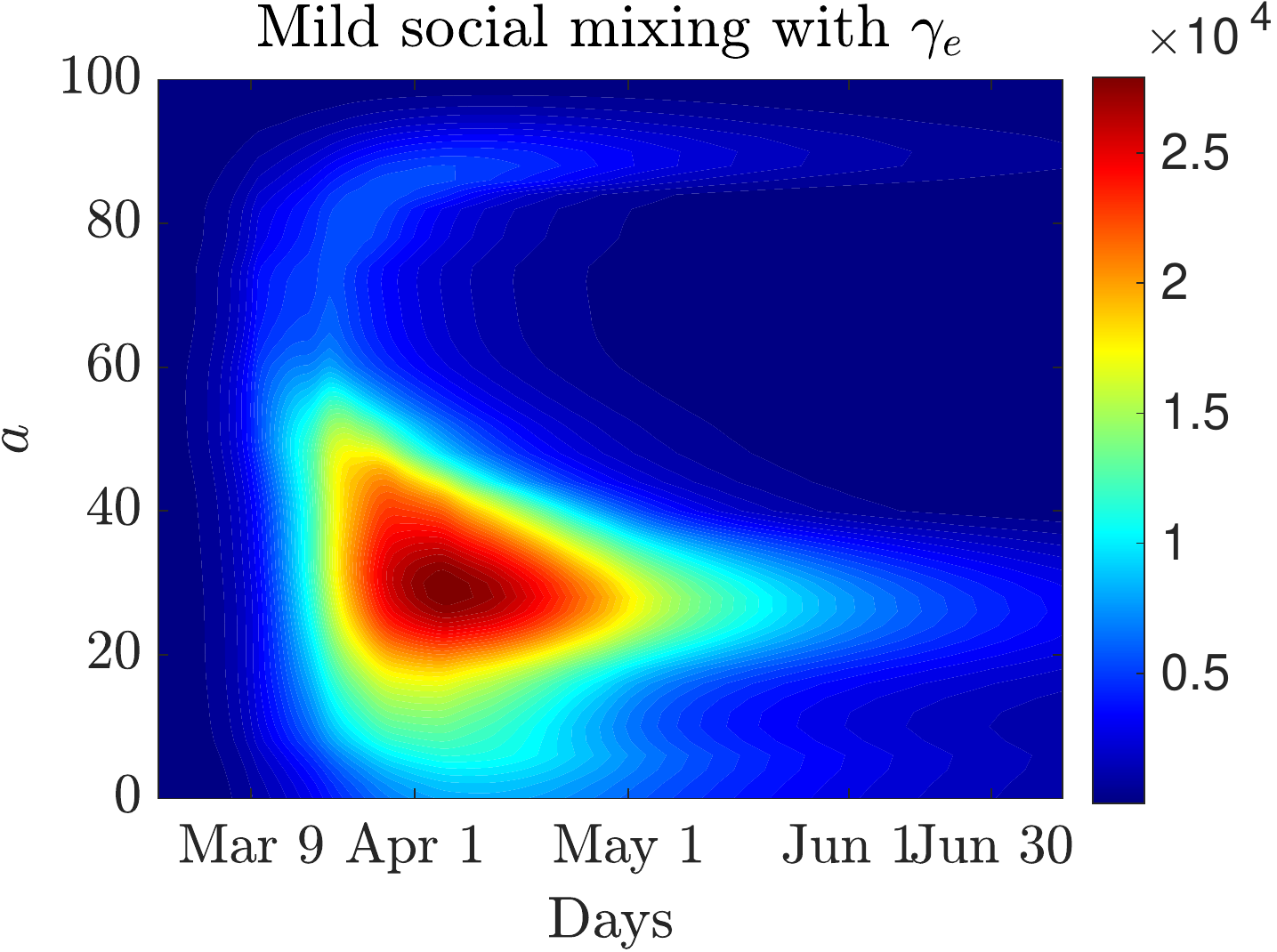}
	\includegraphics[width=3.85cm]{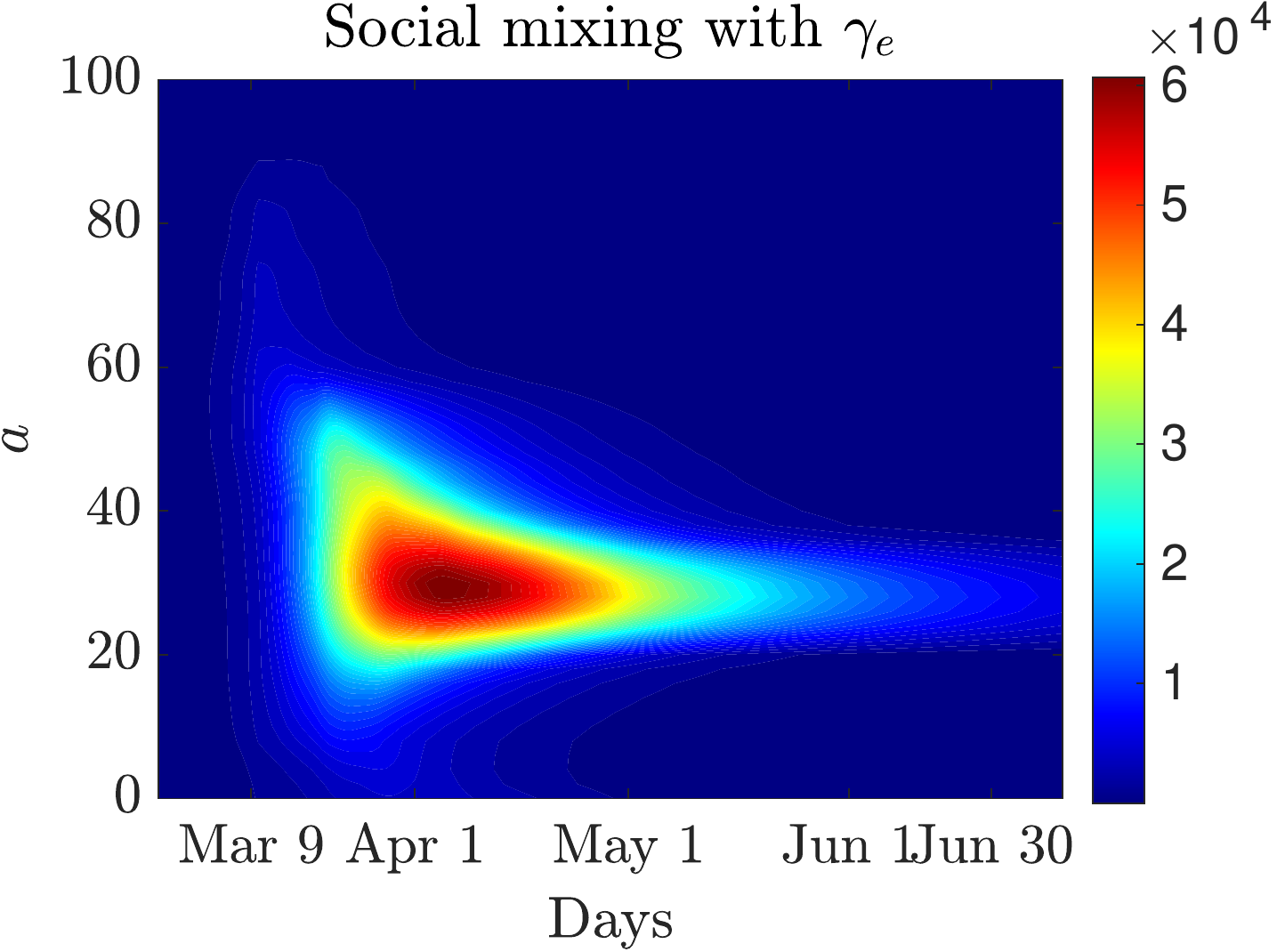}
	\caption{Left: Expected number of infected in time for {the perception function $\psi(I)=I$, and a constant recovery rate $\gamma_e$ together with the confidence bands for homogeneous mixing ($\xi=0$), mild social mixing $(\xi = 0.75)$ and full social mixing ($\xi = 1$)}. Middle and right: Expected age distribution of infectious individuals with constant $\gamma_e$.}
	\label{fig:predict}
\end{figure}

In Figure \ref{fig:predict} we report the results of the expected number of infected with the related confidence bands in case of homogeneous mixing and different levels of social mixing ($\xi = 0.75,~ \xi =1$) for the constant recovery rate $\gamma_e$. Middle and right figures report the corresponding expected density of infected individuals $\fI(a,t) = \mathbb{E}(\fI(z,a,t))$ for mild and full social mixing. 
Uncontrolled homogeneous mixing model is used in the pre-lockdown phase (before 9 March), whereas the feedback controlled age dependent model \eqref{eq:SIR_z} is used in the lockdown phase.

\paragraph{{\bf Relaxing control on the various social activities.}} 

\begin{figure}
	\centering
	\includegraphics[scale = 0.25]{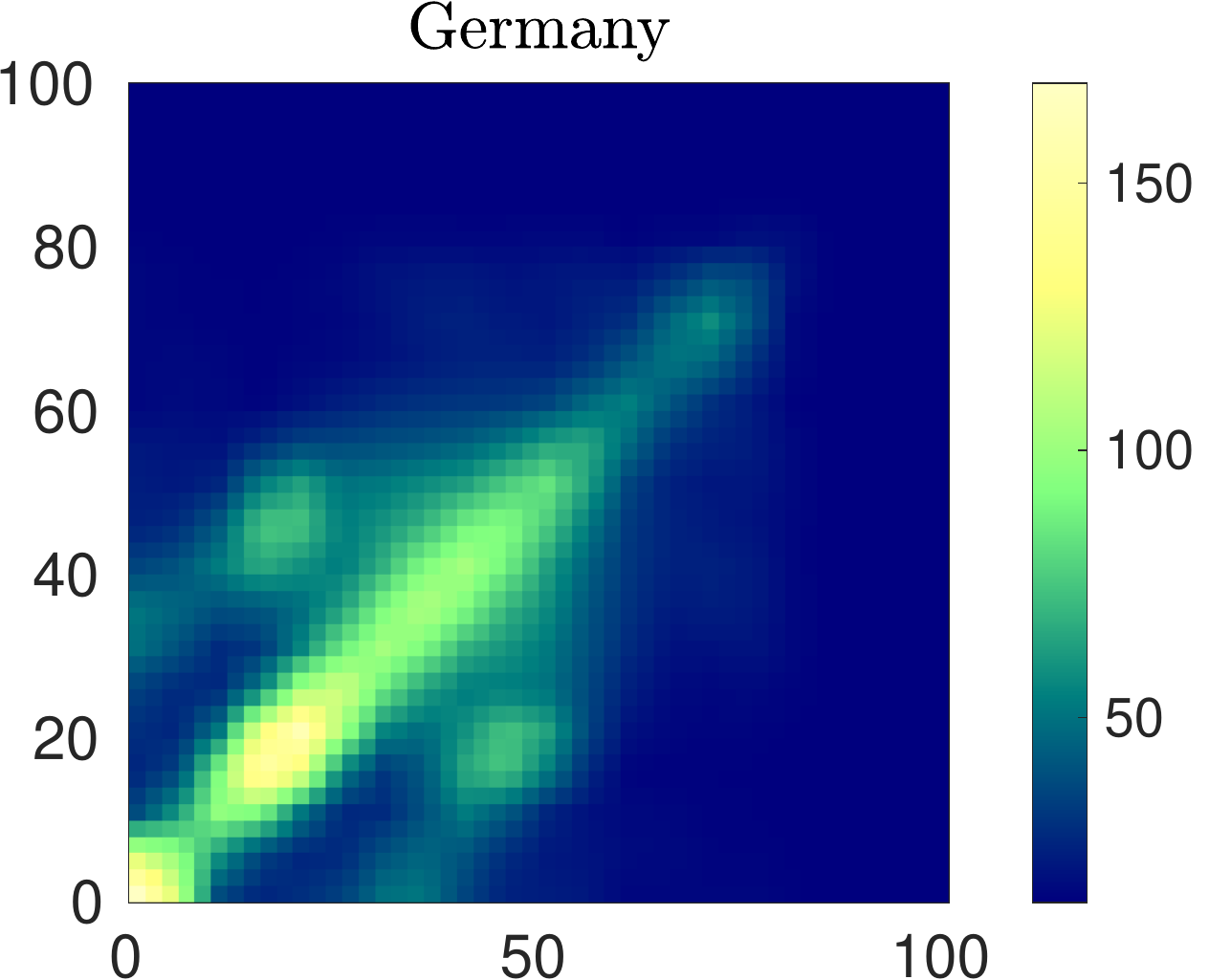}
	\includegraphics[scale = 0.25]{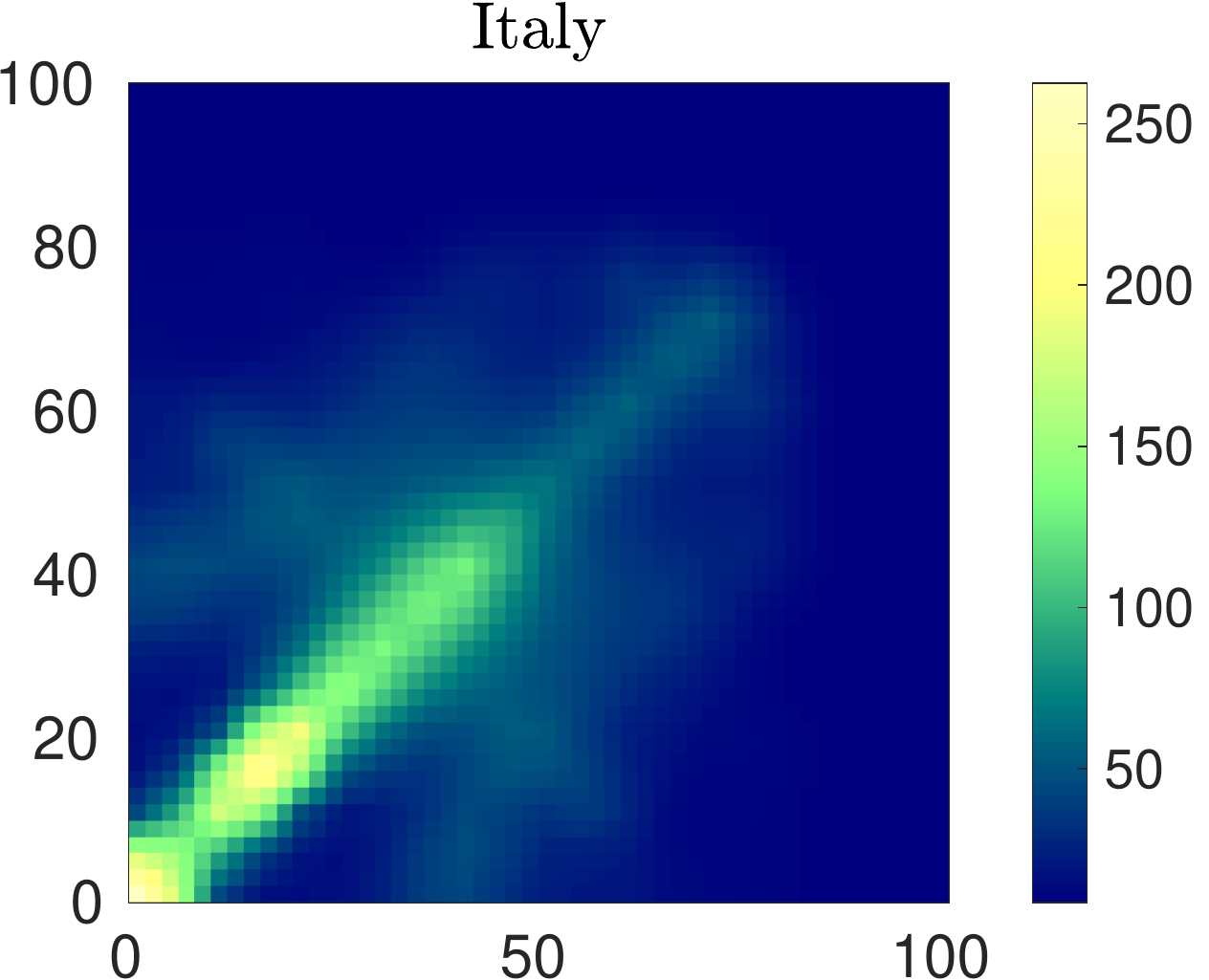}
	\includegraphics[scale = 0.25]{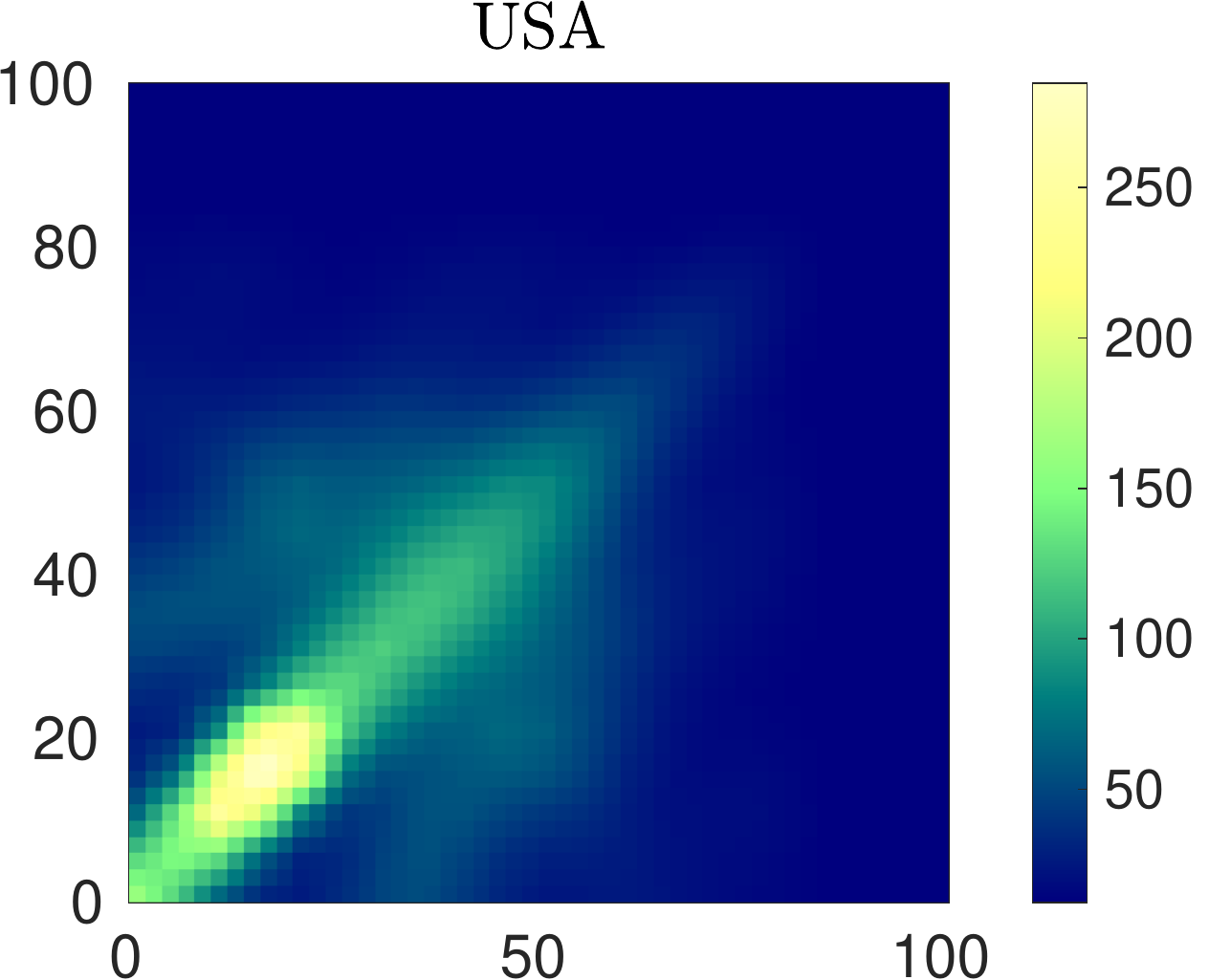}
	\vskip .2cm
	\caption{The total contact interaction function $\beta=\beta_F+\beta_E+\beta_P+\beta_O$ taking into account the contact rates of people with different ages. Family and home contacts are characterized by $\beta_F$, education and school contacts by $\beta_E$, professional and work contacts by $\beta_P$, and other contacts by $\beta_O$.}
	\label{fg:matrices}
\end{figure}

We consider the social interaction functions corresponding to the contact matrices in^^>\cite{PCJ} for the various countries. As a result we have four interaction functions characterized by $\mathcal A = \{F,E,P,O\}$, where we identify family and home contacts with $\beta_F$, education and school contacts with $\beta_E$,  professional and work contacts with $\beta_P$, and other contacts with $\beta_O$. We report in Figure \ref{fg:matrices}, as an example, the total social interaction functions for the various countries. The functions share a similar structure but with different scalings according to the country specific features.

An age-related recovery rate $\gamma(a)$ is selected according to^^>\cite{APZ2} as a decreasing function of the age, 
\be\label{eq:gamma_a}
\gamma(a) =  \gamma_e + C e^{-ra},  
\ee
with $r=5$ and $C\in\mathbb R$, in accordance with^^>\cite{GammaAge,Zanella_mbe}, and such that the normalization \eqref{eq:bg} is satisfied. 

To match the single control applied in the extrapolation of the penalization term $\kappa(t)$ to age dependent penalization factors $\kappa_j(a,t)$ we redistribute the values of the penalization parameters as
\[
\kappa_j(a,t)^{-1}=\frac{w_j(t)\int_{\LL} \beta_j(a,a_*)\,da_*}{\sum_{j\in\mathcal{A}}w_j(t)\int_{\LL\times \LL} \beta_j(a,a_*)\,da\,da_*}\kappa(t)^{-1},\quad j\in \mathcal{A}
\]
where $w_j(t)\geq 0$, are weight factors denoting the relative amount of control on a specific activity. According to^^>\cite{PCJ}, we assume $w_E=1.5$, $w_H=0.2$, $w_P=0.5$, $w_O=0.6$, namely the largest effort of the control is due to the school closure which as a consequence implies more interactions at home. Work and other activities are equally impacted by the lockdown. 

\begin{figure}
	\centering
	\includegraphics[scale =.25]{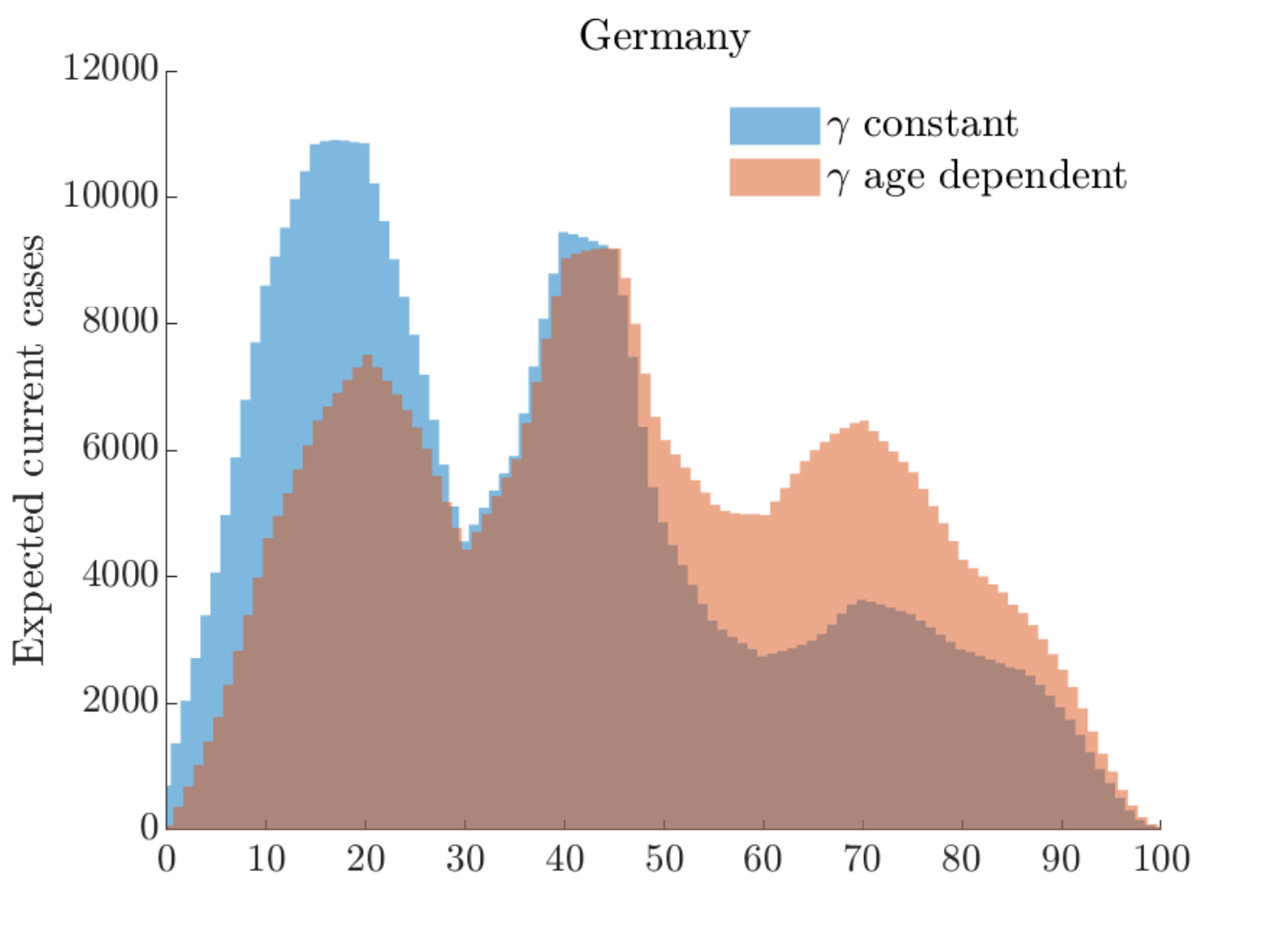}
	\includegraphics[scale =.25]{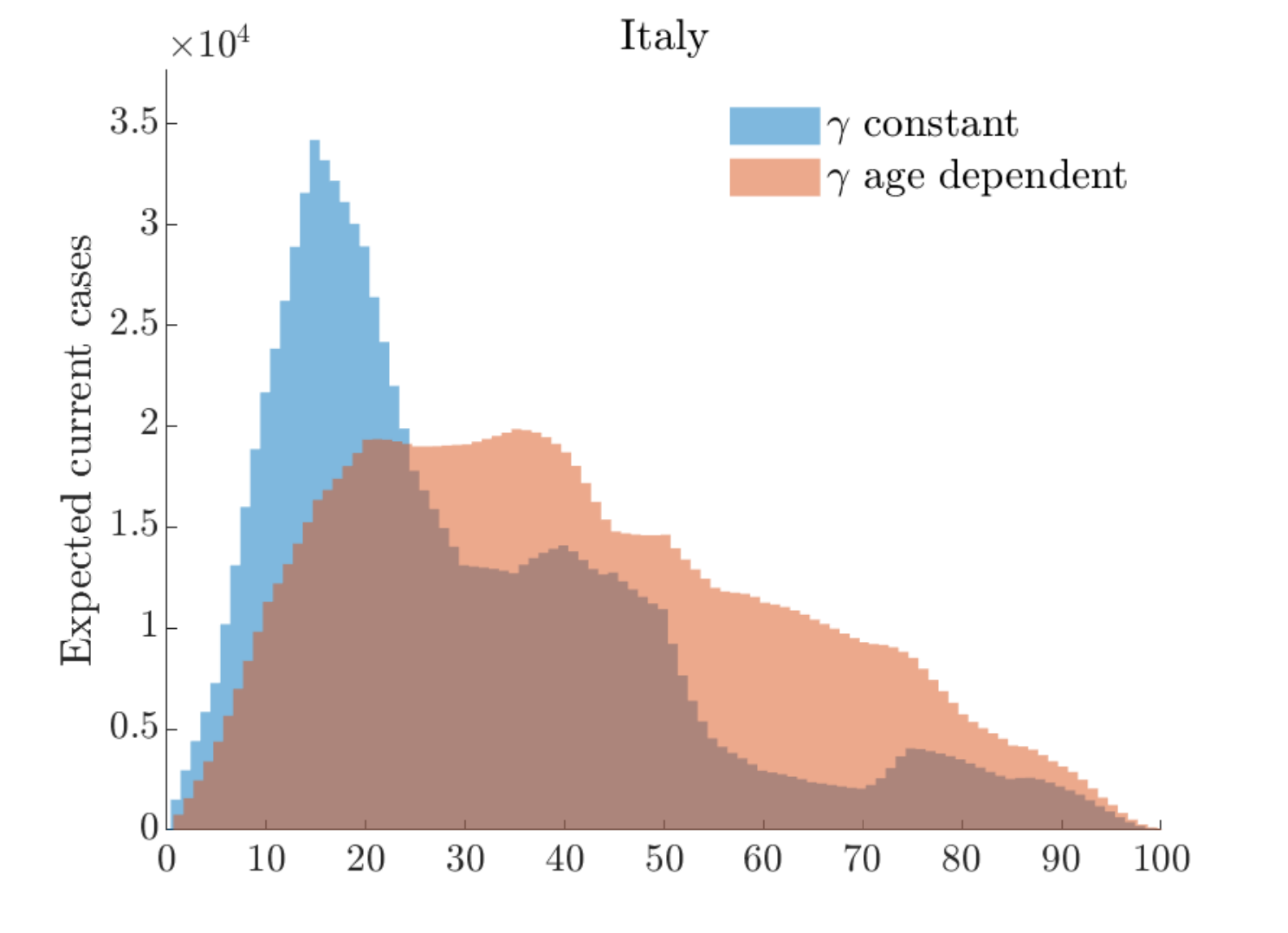} 
	\includegraphics[scale =.25]{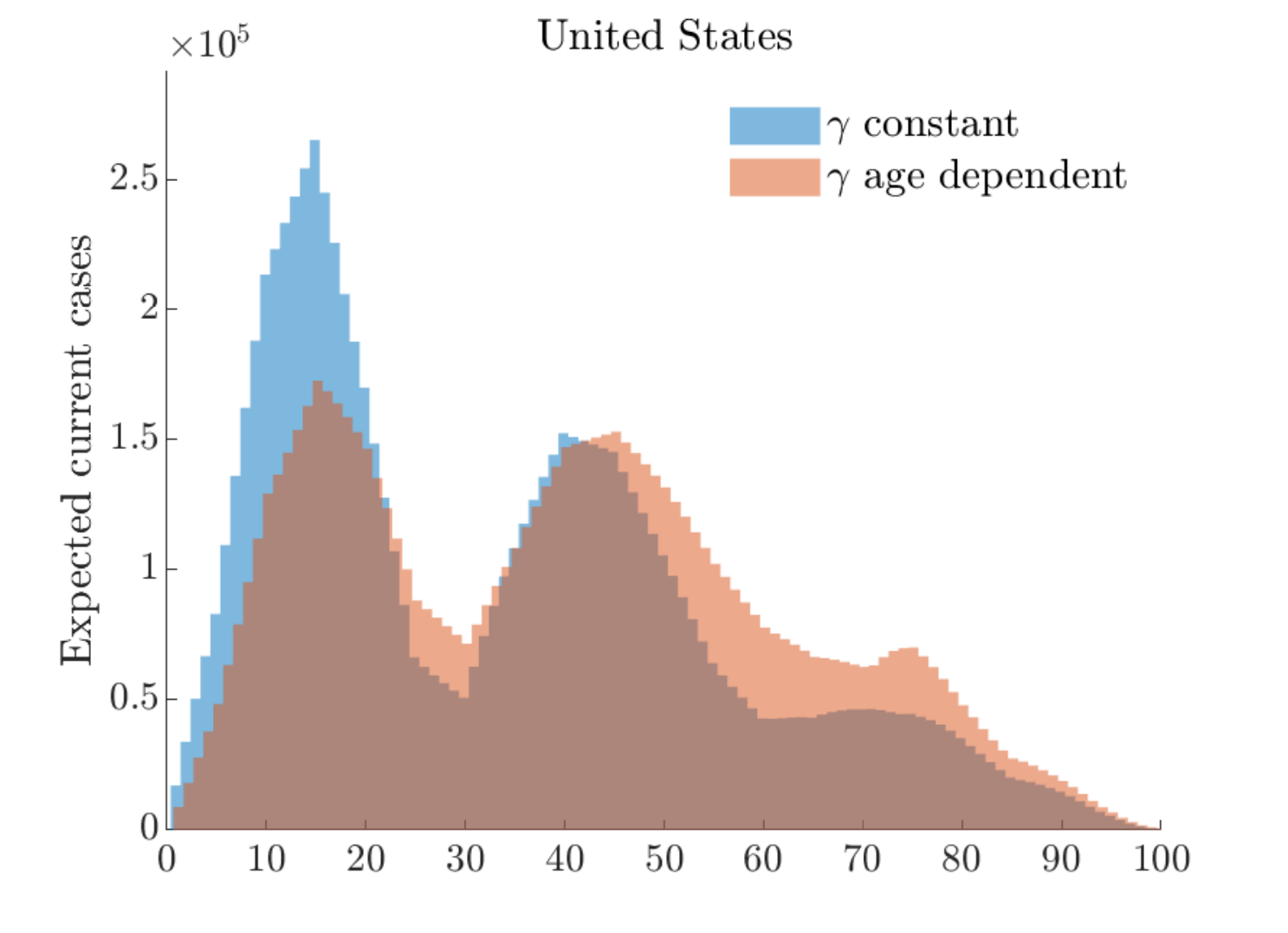}
	\caption{Age distribution of infected using constant and age dependent recovery rates as in \eqref{eq:gamma_a} at the end of the lockdown period in different countries.}
	\label{fig:age2}
\end{figure}

In Figure \ref{fig:age2} we report the age distribution of infected computed for each country at the end of the lockdown period using an age dependent recovery and a constant recovery. The differences in the resulting age distributions are evident. In subsequent simulations, to avoid an unrealistic peak of infection among young people, we decided to adopt an age-dependent recovery^^>\cite{GammaAge}.

\begin{figure}
	\centering
	\includegraphics[scale =.25]{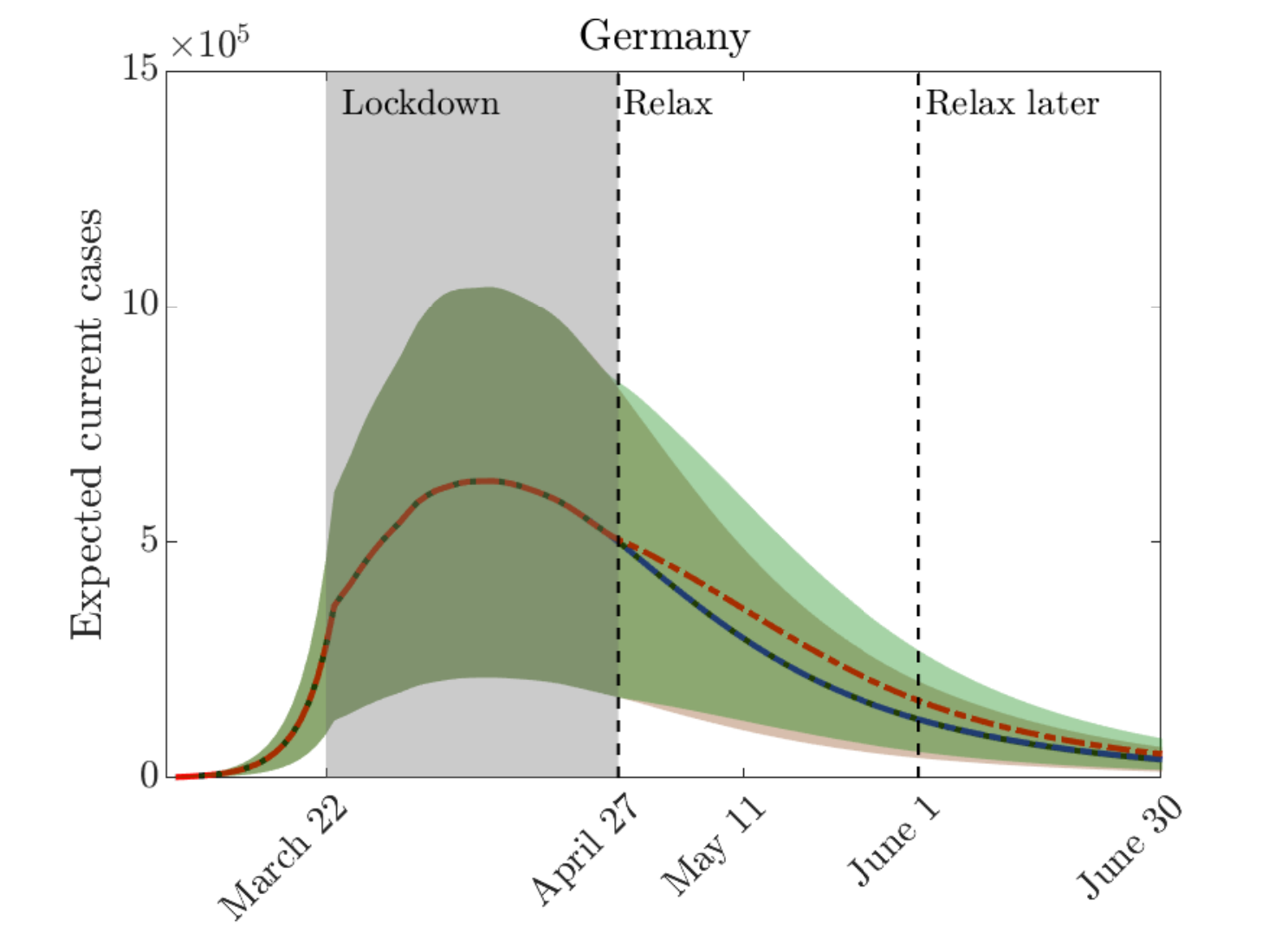}
	\includegraphics[scale =.25]{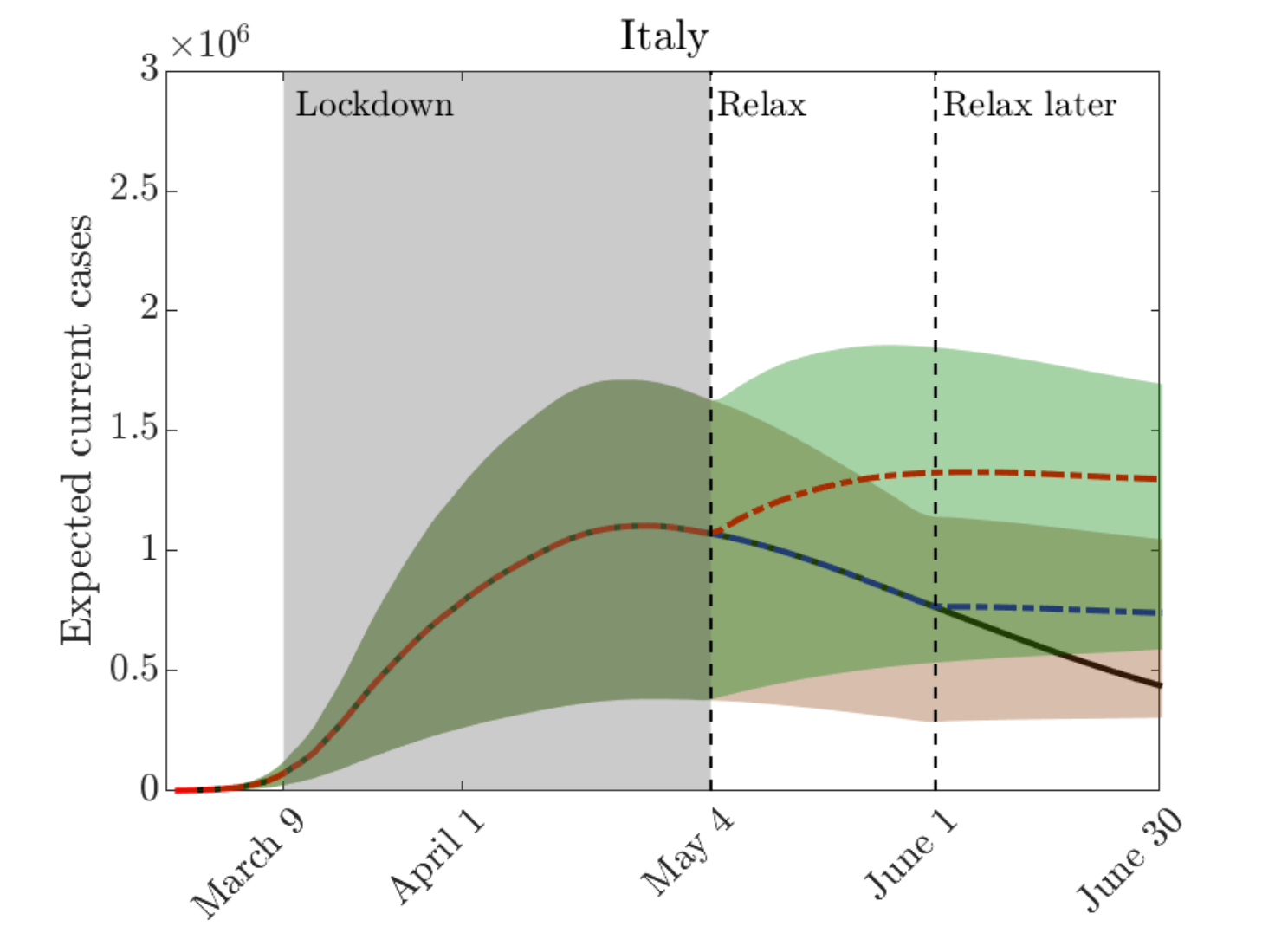} 
	\includegraphics[scale =.25]{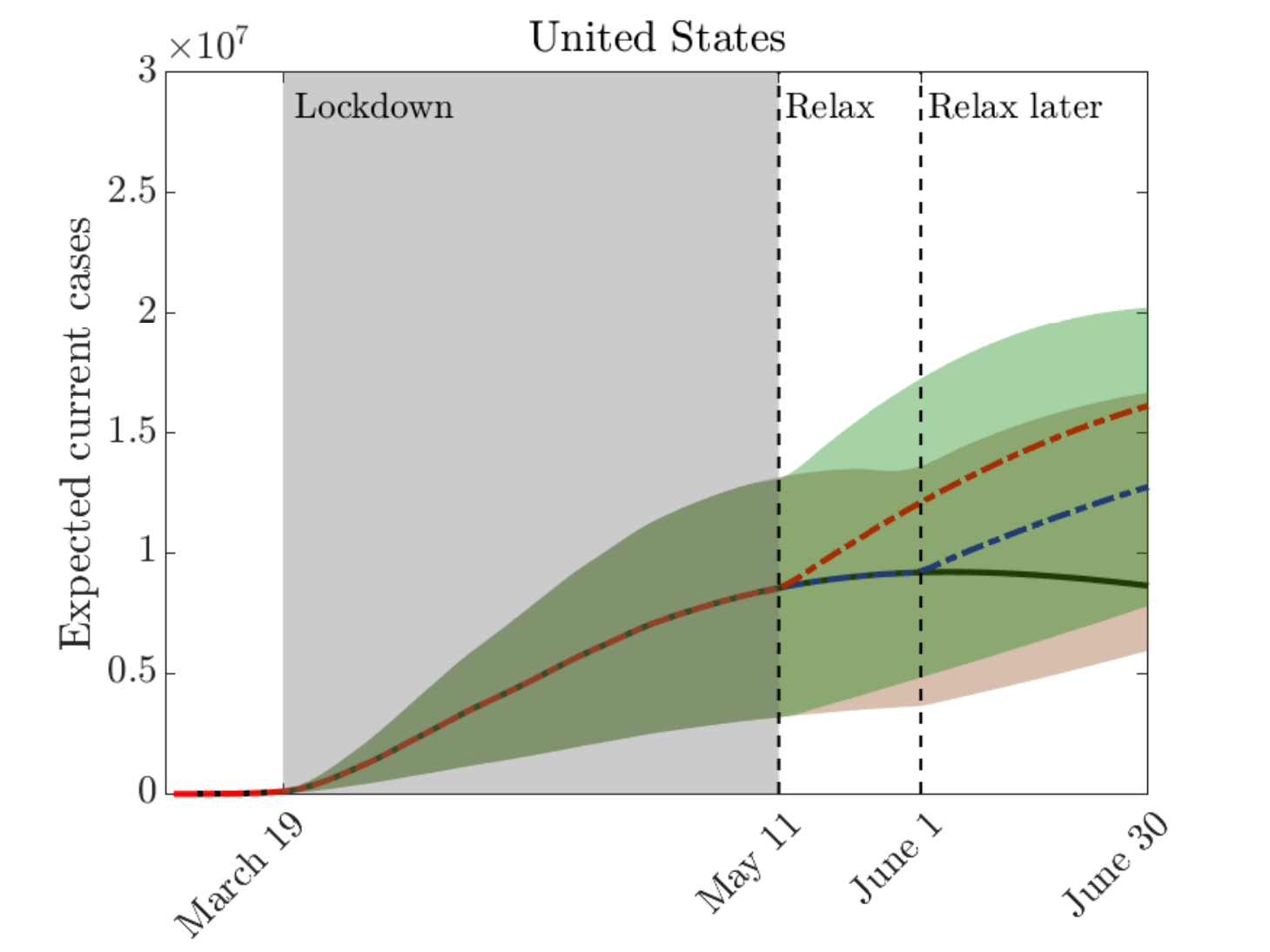}
	\caption{{\bf Scenario 1:} Effect on releasing containment measures in various countries at two different times. In all countries after lockdown we assumed a reduction of individual controls on the different activities by $20\%$ on family activities, $35\%$ on work activities and $30\%$ on other activities by keeping the lockdown over the school.}
	\label{fig:release1}
\end{figure}

We analyze the effects on each country of the same relaxation of the lockdown measures at two different times. The first date is country specific according to current available informations, the second is June 1st for all countries.
For all countries we assumed a reduction of individual controls on the different activities by $20\%$ on family activities, $35\%$ on work activities and $30\%$ on other activities without changing the control over the school. The behaviors of the curves of infected people together with the relative $95\%$ confidence bands are reported in Figure \ref{fig:release1}. 

The results show well the substantial differences between the different countries, with a situation in US which highlight that the relaxation of lockdown measures could lead to a resurgence of the infection. On the contrary, Germany was in the most favorable situation to ease the lockdown without risking a new start of the infection. 

\section{Multiscale transport models}
\label{sec:space}

In this section, we introduce multiscale hyperbolic transport models designed to study the propagation of an epidemic phenomenon described by the diffusive behavior of the non-commuting part of the population, acting only over an urban scale, and the spatial movement and interaction of commuters, moving also on an extra-urban scale. This makes it possible to describe more realistically the typical dynamic of commuters, which affects only a small fraction of individuals, and to distinguish it from the epidemic process which, instead, involves the entire population, including non-commuters. The presence of a group of non-commuting population, indeed, prevents the entire population in a compartment from moving indiscriminately through space generating an unrealistic mass migration effect.
In the following, we will consider a spatial domain either structured as a network, whose nodes identify cities of interest and arcs represent common mobility paths, as discussed in Section^^>\ref{Sect.network}, or representing realistic 2D geographical regions, as further detailed in Section^^>\ref{Sect.realistic}.

\subsection{Spatial dynamics on networks}
\label{Sect.network}

\subsubsection{1D hyperbolic compartmental model}
\label{sect:1D}
To simplify the presentation, the epidemiological starting point of the model is given by a compartmental structure with a simple SIR partitioning^^>\cite{kermack1927,HWH00}. We assume to have a population with individuals having no prior immunity and the vital dynamics represented by births and deaths is neglected because of the time scale considered. 
To account for the spatial movement of the population, individuals of each compartment are subdivided in three classes, $S_{\pm,0}$, $I_{\pm,0}$, $R_{\pm,0}$, traveling in a 1D bounded space domain $\Omega\subseteq \RR$ with characteristic speeds $+\lambda_i,-\lambda_i$ and $0$ respectively, with $i \in \{S,I,R\}$. Therefore, we consider a stationary part of the population, of non-commuters, characterized by a null characteristic speed. The total compartmental densities are defined as the sum of all the components of the subgroups
\begin{equation}
S_T=S_++S_-+S_0,\qquad I_T=I_++I_-+I_0,\qquad R_T=R_++R_-+R_0\,.
\label{eq:sirt}
\end{equation}
The discrete-velocity system of the SIR epidemic transport model for commuters, associated to relaxation times $\tau_i$, then reads
\begin{equation}
\begin{aligned}
\frac{\partial S_{\pm}}{\partial t} \pm \lambda_S \frac{\partial S_{\pm}}{\partial x} &= -F_I(S_{\pm}, I_T) + \frac{1}{2\tau_S}\left(S_\mp - S_\pm\right)\,,\\
\frac{\partial I_{\pm}}{\partial t} \pm \lambda_I \frac{\partial I_{\pm}}{\partial x} &= F_I(S_{\pm}, I_T) -\gamma_I I_{\pm} + \frac1{2\tau_I}\left(I_\mp - I_\pm\right)\,,\\
\frac{\partial R_{\pm}}{\partial t} \pm \lambda_R \frac{\partial R_{\pm}}{\partial x} &= \gamma_I I_{\pm} + \frac1{2\tau_R}\left(R_\mp - R_\pm\right)\, .
\end{aligned}
\label{eq.SIRkinetic}
\end{equation}
This system is coupled with a classical ODE SIR model, which describes the evolution of the stationary population of non-commuters:
\begin{equation}
\frac{\d S_{0}}{\d t}  = -F_I(S_{0}, I_T) \,, \quad
\frac{\d I_{0}}{\d t}  = F_I(S_{0}, I_T) -\gamma_I I_{0}\,, \quad
\frac{\d R_{0}}{\d t}  = \gamma_I I_0 \, .
\label{eq.SIRkinetic_noncommuters}
\end{equation}
Let us observe that, under no inflow/outflow boundary conditions, summing up the equations in \eqref{eq.SIRkinetic}-\eqref{eq.SIRkinetic_noncommuters} and integrating in $\Omega$ yields the conservation of the total population.

All the epidemic densities and, eventually, the epidemic parameters depend on $(\mathbf{z},x,t)$, where $(x,t)$ are the physical variables of space $x \in \Omega \subseteq \mathbb{R}$ and time $t>0$, while $\mathbf{z} = (z_1,\ldots,z_{d})^T \in \mathbb{R}^{d}$ is a random vector characterizing the possible sources of uncertainty as introduced in Section^^>\ref{uncertain_data}. The same applies for the incidence function $F_I$, defined with respect to the infectious compartment $I$ as
\begin{equation}
F_I(g,I_T)=\beta_I \frac{g I_T^p}{1+\kappa_I I_T^p}, \qquad p\geq 1,
\label{eq:incf}
\end{equation}
where $\beta_I(\mathbf{z},x,t)$ is the transmission rate, accounting for both number of contacts and probability of transmission, hence it may vary based on the effects of government control actions, such as mandatory wearing of masks, shutdown of specific work/school activities, or full lockdowns^^>\cite{HWH00,Bruno,APZ}. The parameter $\kappa_I(\mathbf{z},x,t)$ acts as incidence damping coefficients based on the self-protective behavior of the individual that arises from awareness of the risk associated with the epidemic^^>\cite{Capasso,bertaglia2021,franco2020}. We refer also to Section \ref{contact} and \ref{sec:control} for the derivation of saturated incidence functions of the form \eqref{eq:incf}.
Note that, the classic bilinear case corresponds to $p = 1$ and $\kappa_I = 0$. Finally, the parameter $\gamma_I(\mathbf{z},x,t)$ is the recovery rate of infected (inverse of the infectious period).

\subsubsection{Macroscopic formulation and diffusion limit}
\label{Sect.diff}
Introducing now the macroscopic variables $S_c,I_c,R_c$ for the commuters, with
$S_c = S_++S_-$, $I_c = I_++I_-$, $R_c = R_++R_-$,
and defining the fluxes
\begin{equation}
J_S = \lambda_S(S_+-S_-),\quad J_I = \lambda_I(I_+-I_-),\quad J_R = \lambda_R(R_+-R_-),
\label{eq:flux}
\end{equation}
a hyperbolic model underlying the macroscopic formulation of the spatial propagation of an epidemic at finite speeds, equivalent to the mesoscopic one^^>\cite{aylaj2020}, presented in system \eqref{eq.SIRkinetic}, is obtained^^>\cite{bertaglia2021}:
\begin{equation}
\begin{aligned}
\frac{\partial S_{c}}{\partial t} + \frac{\partial J_S}{\partial x} &= -F_I(S_c, I_T)  \,,\\
\frac{\partial I_{c}}{\partial t} + \frac{\partial J_I}{\partial x} &= F_I(S_c, I_T)  - \gamma_I I_{c}\,,\\
\frac{\partial R_{c}}{\partial t} + \frac{\partial J_R}{\partial x} &= \gamma_I I_{c} \,,\\
\frac{\partial J_S}{\partial t} + \lambda_S^2 \frac{\partial S_c}{\partial x} &= -F_I(J_S, I_T) - \frac1{\tau_S} J_S\,,\\
\frac{\partial J_I}{\partial t} + \lambda_I^2 \frac{\partial I_c}{\partial x} &=  \frac{\lambda_I}{\lambda_S}F_I(J_S, I_T) - \gamma_I J_I-\frac1{\tau_I} J_I\,,\\
\frac{\partial J_R}{\partial t} + \lambda_R^2 \frac{\partial R_c}{\partial x} &= \frac{\lambda_R}{\lambda_I} \gamma_I J_I -\frac1{\tau_R} J_R \,.
\end{aligned}
\label{eq.SIRmacro}
\end{equation}
Note that here the above system is coupled with the equations for the non-commuting population \eqref{eq.SIRkinetic_noncommuters} through identities \eqref{eq:sirt}.
It is easy to verify that system \eqref{eq.SIRmacro} is symmetric hyperbolic in the sense of Friedrichs-Lax^^>\cite{friedrichs1971}.

From a formal viewpoint, it can be shown that the proposed model recovers the parabolic behavior expected from standard space-dependent epidemic models in the diffusion limit^^>\cite{barbera2013,bertaglia2021}.
Introducing the diffusion coefficients
$
D_i=\lambda_i^2 \tau_i, i\in\{S,I,R\}
$
that characterize the diffusive transport mechanism of $S,I,R$ respectively, and letting $\tau_i \to 0$, while keeping the diffusion coefficients finite^^>\cite{Lions1997}, from the last three equations of system \eqref{eq.SIRmacro} we recover Fick’s laws^^>\cite{bertaglia2021}, which, inserted in the rest of the equations of system \eqref{eq.SIRmacro}, yield the following parabolic reaction-diffusion system for the commuters^^>\cite{berestycki2021,murray2003}
\begin{equation}
\begin{split}
\frac{\partial S_{c}}{\partial t}  &= \frac{\partial}{\partial x}\left(D_S \frac{\partial }{\partial x}S_c \right) - F_I(S_{c}, I_T) \,,\\
\frac{\partial I_{c}}{\partial t}  &= \frac{\partial}{\partial x}\left(D_I \frac{\partial }{\partial x}I_c \right)+ F_I(S_{c}, I_T) -\gamma_I I_{c}\,,\\
\frac{\partial R_{c}}{\partial t}  &= \frac{\partial}{\partial x}\left(D_R \frac{\partial }{\partial x}R_c \right)+\gamma_I I_{c} \,.
\end{split}
\label{eq.SIRdiffusive}
\end{equation}
The relaxation times can modify the nature of the behavior of the solution^^>\cite{bertaglia2021,barbera2013}, which can result either hyperbolic or parabolic (when considering small relaxation times and large speeds). This feature of the model makes it particularly suitable for the description of the dynamics of human populations, which are characterized by movement at different spatial scales^^>\cite{boscheri2020}. It is therefore natural to assume $\tau_i=\tau_i(x)$, since in geographic areas densely populated we can assume a diffusive dynamics while along the main arteries of communication a hyperbolic description will be more appropriate avoiding propagation of information at infinite speed.  

\begin{tips}{Reproduction number in space dependent dynamics}
The standard threshold of epidemic models is the well-known basic reproduction number $R_0$. Its definition in the case of spatially dependent dynamics, as already noted in^^>\cite{viguerie2020,viguerie2021}, is not straightforward particularly when considering its spatial dependence. 

Assuming no inflow/outflow boundary conditions in $\Omega$, summing up the evolutionary equations for the infectious compartment $I$ in \eqref{eq.SIRkinetic}-\eqref{eq.SIRkinetic_noncommuters} and integrating over space we have
\[
\frac{\partial}{\partial t} \int_{\Omega} I_T(\mathbf{z},x,t)\,dx =  \int_{\Omega} F_I(S,I_T)\,dx-\int_{\Omega} \gamma_I(\mathbf{z},x,t) I_T(\mathbf{z},x,t)\,dx \geq 0
\]
when
\begin{equation}
R_0(\mathbf{z},t)=\frac{\int_{\Omega}F_I(S,I_T)\,dx}{\int_{\Omega} \gamma_I(\mathbf{z},x,t) I_T(\mathbf{z},x,t)\,dx} \geq 1.
\label{eq:R0_1}
\end{equation}
If no spatial dependence is assigned to variables and parameters, as well as no uncertainty, and no social distancing effects are taken into account, i.e. $\kappa_I= 0$, we recover the conventional SIR ODE model and the reproduction number results in accordance with its standard definition^^>\cite{HWH00}:
\begin{equation*}
R_0 (t) = \frac{\beta_I S_T}{\gamma_I} \, .
\end{equation*}
\end{tips}

\subsubsection{Extension to multi-compartmental modelling}
\label{SEIAR_modelling}
To account for more complex compartmental models capable of better analyzing the evolution of specific infectious diseases, we consider extending the simple SIR compartmentalization by taking into account two additional population compartments, $E$ and $A$, resulting in a SEIAR model^^>\cite{bertaglia2021a,bertaglia2021b}. Subjects in the $E$ compartment are the exposed, hence infected but not yet infectious, being in the latent period. Moreover, among the infectious subjects, we distinguish the population between a group of individuals $I$ who will develop severe symptoms and a group of individuals $A$ who will never develop symptoms or, if they do, these will be very mild. In fact, as discussed in Section \ref{sec:control}, the presence of undetected asymptomatic individuals turns out to be essential to correctly analyze the evolution of COVID-19^^>\cite{Gatto,peirlinck2020}. 

Note that, the presence of uncertainty in the data, included from the beginning in the modeling process, could allow the compartmentalization of asymptomatic individuals to be eliminated by implicitly including them in the uncertainty about the number of infected individuals, as described in Section^^>\ref{uncertain_data}. In this context, however, in order to highlight the link with similar models used in the literature^^>\cite{tang2020,tang2020a,peirlinck2020}, we keep the asymptomatic compartment separated to the symptomatic one, with the former being affected by the highest level of uncertainty. 

\begin{figure}
\centering
\includegraphics[width=0.65\textwidth]{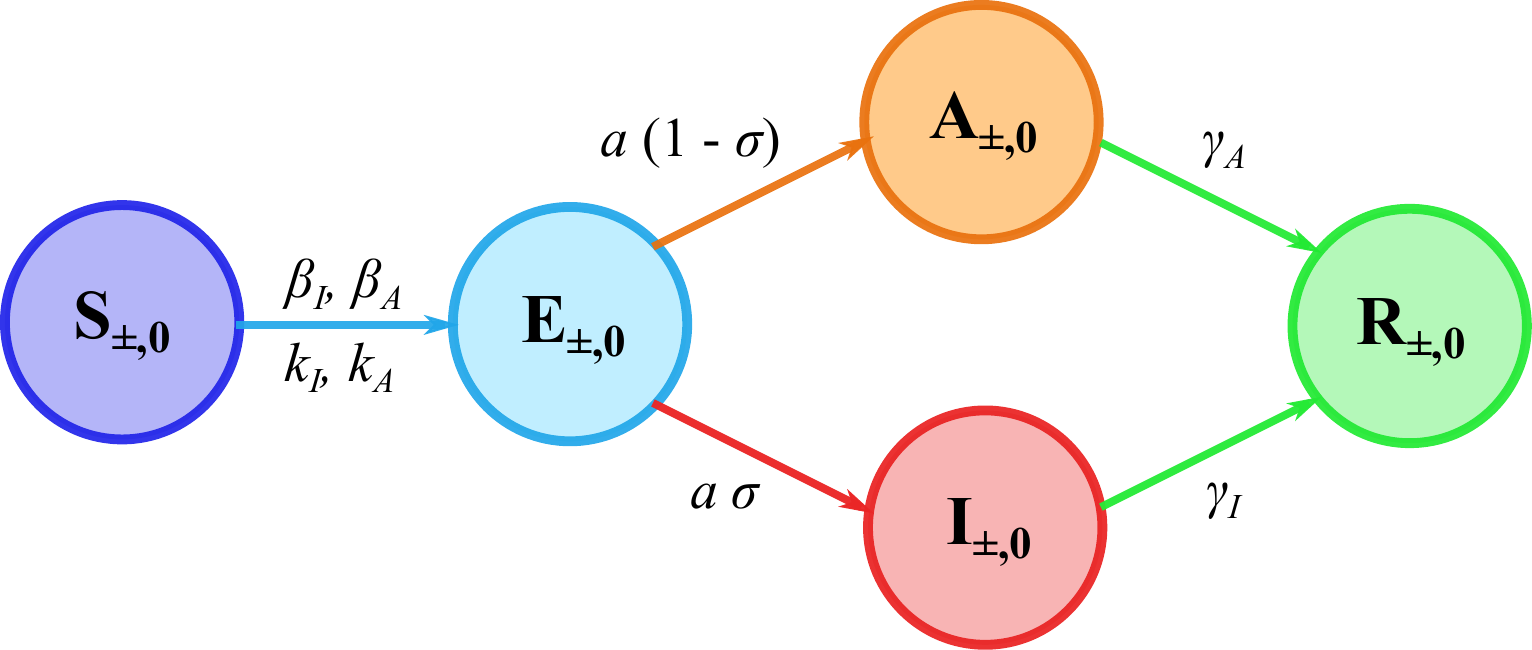}
\caption{Flow chart of the multi-population SEIAR dynamics based on five compartments: susceptible (S), exposed (E), severe symptomatic infectious (I), mildly symptomatic/asymptomatic infectious (A), and removed –healed or deceased– population (R), each one subdivided in three classes of individuals traveling in the domain with characteristic speeds $+\lambda_i$, $-\lambda_i$ and $0$, with $i \in \{S, E, I, A, R\}$}
\label{fig.SEIAR}
\end{figure}

Defining the total density of the additional compartments, $E_T~=~E_+~+~E_-~+~E_0$, $A_T~=~A_+~+~A_-~+~A_0$,
the resulting discrete-velocity system of the SEIAR epidemic transport model for commuters reads
\begin{equation}
\begin{aligned}
\frac{\partial S_{\pm}}{\partial t} \pm \lambda_S \frac{\partial S_{\pm}}{\partial x} &= -F_I(S_{\pm}, I_T) -F_A(S_{\pm}, A_T) + \frac{1}{2\tau_S}\left(S_\mp - S_\pm\right)\,,\\
\frac{\partial E_{\pm}}{\partial t} \pm \lambda_E \frac{\partial E_{\pm}}{\partial x} &= F_I(S_{\pm}, I_T) +F_A(S_{\pm},A_T) -a E_{\pm} + \frac{1}{2\tau_E}\left(E_\mp - E_\pm\right)\,,\\
\frac{\partial I_{\pm}}{\partial t} \pm \lambda_I \frac{\partial I_{\pm}}{\partial x} &= a \sigma E_{\pm} -\gamma_I I_{\pm} + \frac1{2\tau_I}\left(I_\mp - I_\pm\right)\,,\\
\frac{\partial A_{\pm}}{\partial t} \pm \lambda_A \frac{\partial A_{\pm}}{\partial x} &= a(1-\sigma) E_{\pm} -\gamma_A A_{\pm} + \frac{1}{2\tau_A}\left(A_\mp - A_\pm\right)\,,\\
\frac{\partial R_{\pm}}{\partial t} \pm \lambda_R \frac{\partial R_{\pm}}{\partial x} &= \gamma_I I_{\pm} + \gamma_A A_{\pm} + \frac1{2\tau_R}\left(R_\mp - R_\pm\right)\, ,
\end{aligned}
\label{eq.SEIARkinetic}
\end{equation}
which is coupled with the following SEIAR model describing the evolution of non-commuting individuals
\begin{equation}
\begin{aligned}
\frac{\d S_{0}}{\d t}  &= -F_I(S_{0}, I_T) +F_A(S_{0}, A_T) \,,\\
\frac{\d E_{0}}{\d t}  &= F_I(S_{0}, I_T) +F_A(S_{0}, A_T) -a E_0 \,,\\
\frac{\d I_{0}}{\d t}  &= a \sigma E_0 -\gamma_I I_{0}\,,\\
\frac{\d A_{0}}{\d t}  &= a(1-\sigma) E_0 - \gamma_A A_0 \,,\\
\frac{\d R_{0}}{\d t}  &= \gamma_I I_0 + \gamma_A A_0\, .
\end{aligned}
\label{eq.SEIARkinetic_noncommuters}
\end{equation}
The quantity $\gamma_A(\mathbf{z},x,t)$ is the recovery rate of asymptomatic/mildly symptomatic infected, which is distinguished from the recovery rate of highly symptomatic infected previously introduced $\gamma_I(\mathbf{z},x,t)$; while $a(\mathbf{z},x,t)$ represents the inverse of the latency period and $\sigma(\mathbf{z},x,t)$ is the probability rate of developing severe symptoms^^>\cite{tang2020,Gatto,buonomo2020}.
In this model, the transmission of the infection is governed by two different incidence functions, $F_I(\cdot,I_T)$ and $F_A(\cdot,A_T)$, simply to distinguish between the behavior of $I$ and $A$ individuals. Analogously to \eqref{eq:incf},
\begin{equation}
F_A(g,A_T)=\beta_A \frac{g A_T^p}{1+\kappa_A A_T},
\label{eq:incf-SEIAR}
\end{equation}
where a different contact rate, $\beta_A$, and coefficient $\kappa_A$ are taken into account for mildly/no symptomatic people. The flow chart of the multiscale SEIAR model is shown in Fig.~\ref{fig.SEIAR}.

Let us observe that, following the same procedure presented in Section^^>\ref{Sect.diff} for the SIR-type spatial model, introducing the same definition \eqref{eq:flux} of flux for the additional compartments, $J_E$ and $J_A$, it is possible to obtain an analogous macroscopic formulation also for the SEIAR-type spatial model^^>\cite{bertaglia2021a}.
Furthermore, defining also $D_E=\lambda_E^2\tau_E$ and $D_A=\lambda_A^2\tau_A$, we recover the diffusion limit of the SEIAR-type system. The reader can refer to^^>\cite{bertaglia2021a} for details on this derivation.

\begin{tips}{Reproduction number in space dependent SEIAR models}
For the SEIAR-type spatial model, the reproduction number (which is again not straightforward to be determined) can be computed following the \textit{Next-Generation Matrix} (NGM) approach^^>\cite{Diek} considering no flux boundary conditions, which yields the following definition for the average value of $R_0$ in the domain $\Omega$ for $t>0$, given the uncertain input vector $\mathbf{z}$:
\begin{equation}
\begin{aligned}
R_0 (\mathbf{z},t) &= \frac{\int_\Omega F_I(S_T,I_T) \,dx}{\int_\Omega \gamma_I(\mathbf{z},x,t) I_T(\mathbf{z},x,t) \,dx} \cdot \frac{\int_\Omega a(\mathbf{z},x,t)\sigma (\mathbf{z},x,t) E_T(\mathbf{z},x,t) \,dx}{\int_\Omega a(\mathbf{z},x,t) E_T(\mathbf{z},x,t) \,dx} \\&+ \frac{\int_\Omega F_A(S_T,A_T) \,dx}{\int_\Omega \gamma_A(\mathbf{z},x,t) A_T(\mathbf{z},x,t)\, dx} \cdot \frac{\int_\Omega a(\mathbf{z},x,t)(1-\sigma(\mathbf{z},x,t)) E_T(\mathbf{z},x,t) \,dx}{\int_\Omega a(\mathbf{z},x,t) E_T(\mathbf{z},x,t)\, dx} \,.
\end{aligned}
\label{eq.R0_2}
\end{equation}
We refer to^^>\cite{bertaglia2021a} for the details of the derivation of the above expression.
\end{tips}

\subsubsection{Network modelling}
\label{network_modelling}
The hyperbolic transport models here proposed, similarly to other fields of application, like traffic flow models, chemotaxis and cardiovascular modeling, can be embedded into a network of cities following^^>\cite{bretti2014,piccoli2006}. Note that, the approach differs from the classical network modeling in epidemiology based on coupled systems of ODEs^^>\cite{Barth,dellarossa2020,Gatto,LGPWS}.

\begin{figure}[htb]
\centering
\includegraphics[width=0.45\textwidth]{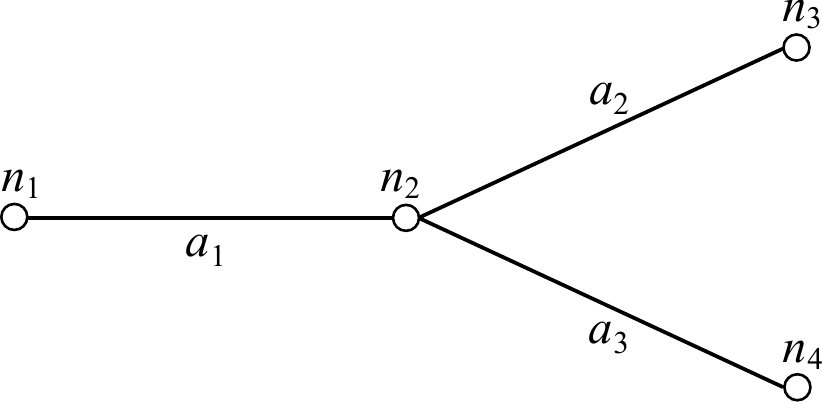}
\caption{Schematic representation of a network composed by 4 nodes ($n_1$, $n_2$, $n_3$, $n_4$) and 3 arcs ($a_1$, $a_2$, $a_3$) in a Y-shape configuration.}
\label{fig.network}
\end{figure}

A network or a connected graph $\mathcal{G = (N,A)}$ is composed of a finite set of $N$ nodes (or vertices) $\mathcal{N}$ and a finite set of $A$ bidirectional arcs (or edges) $\mathcal{A}$, such that an arc connects a pair of nodes^^>\cite{piccoli2006}. An example of network is presented in Fig.~\ref{fig.network}.

Following^^>\cite{bertaglia2021,bertaglia2021b}, it is possible to structure a 1D network considering that the nodes of the network identify locations of interest such as municipalities, provinces or, in a wider scale, regions or nations, while the arcs, enclosing the 1D spatial dynamics, represent the paths linking each location to the others. In this configuration, nodes are \textit{active} since the epidemic state of each one evolves in time influenced by the mobility of the commuting individuals, moving from the other locations included in the network, always considering a part of the population composed by non-commuting individuals which remain at the origin node.

In order to prescribe the proper coupling between nodes and arcs, it is necessary to impose appropriate transmission conditions at each arc-node interface, which ensure the conservation of total density (population) in the network and of fluxes at the interface and further solving the Riemann problem at each interface employing Riemann Invariants.
The complete description of the implementation of transmission conditions at nodes is presented in^^>\cite{bertaglia2021} for a SIR-type spatial model and in^^>\cite{bertaglia2021b} for a SEIAR-type transport model.

\subsubsection{Effect of spatially heterogeneous environments in hyperbolic and parabolic configuration}
\begin{figure}
\centering
\includegraphics[width=0.35\textwidth]{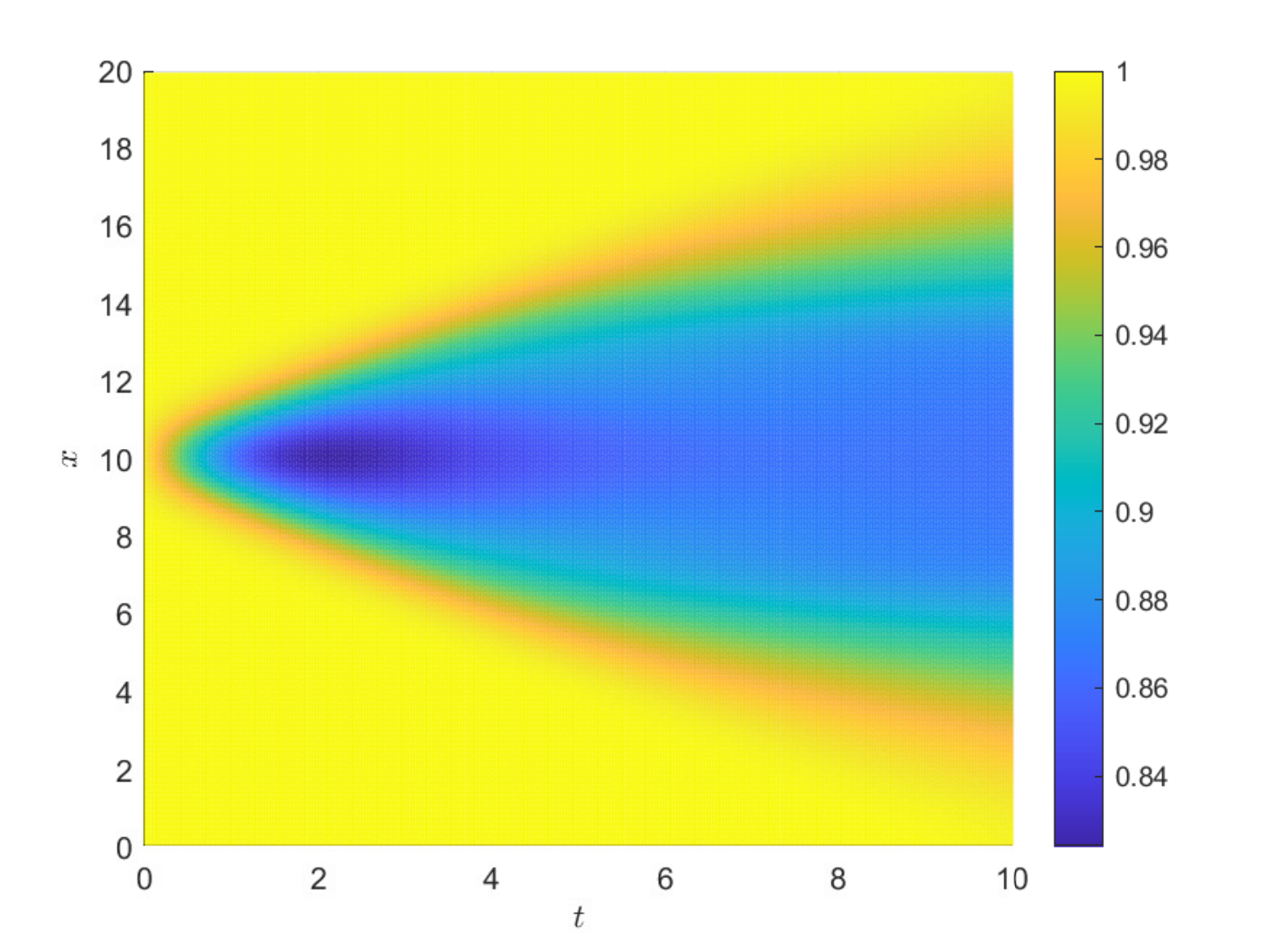}
\includegraphics[width=0.35\textwidth]{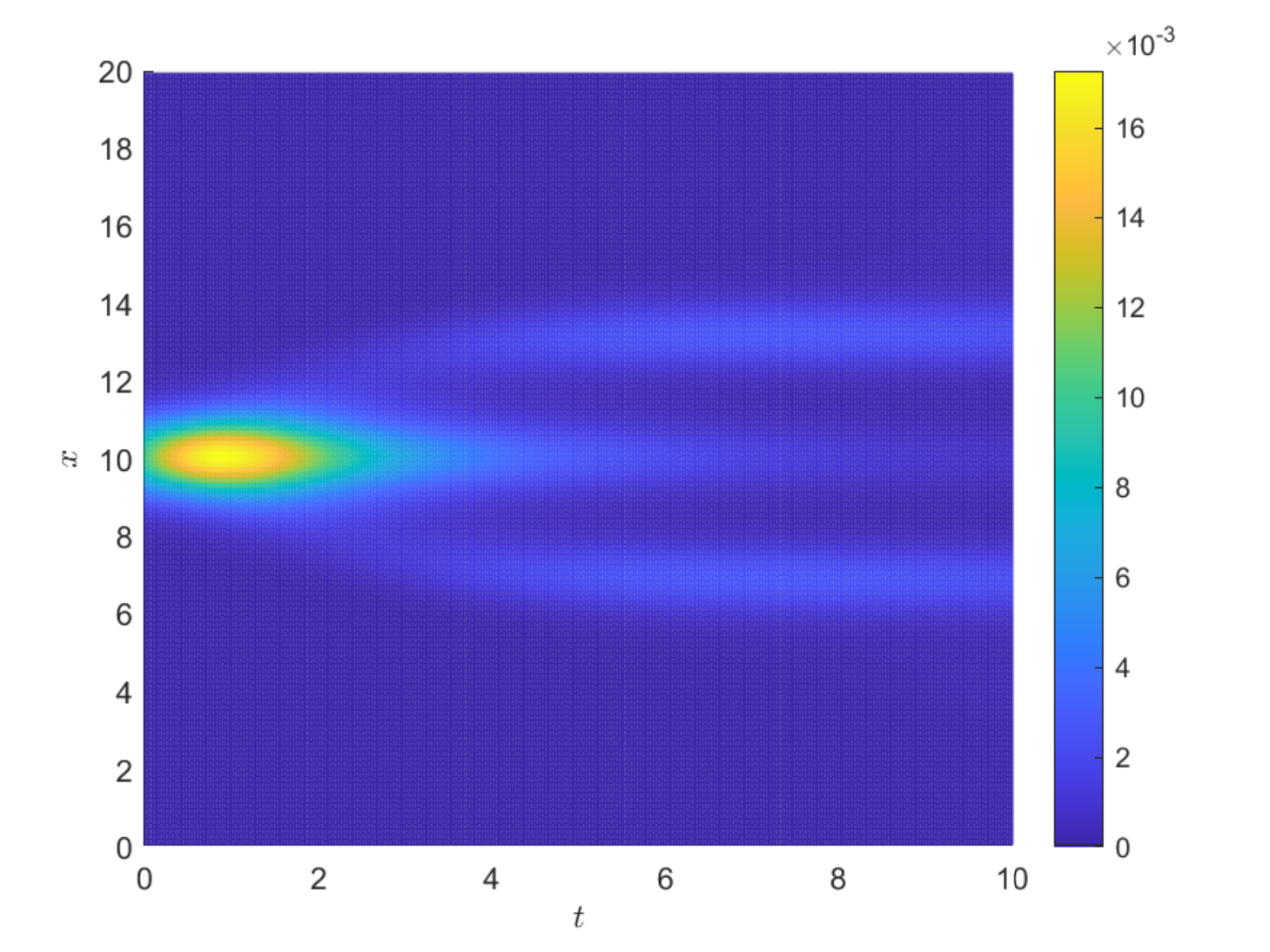}
\includegraphics[width=0.35\textwidth]{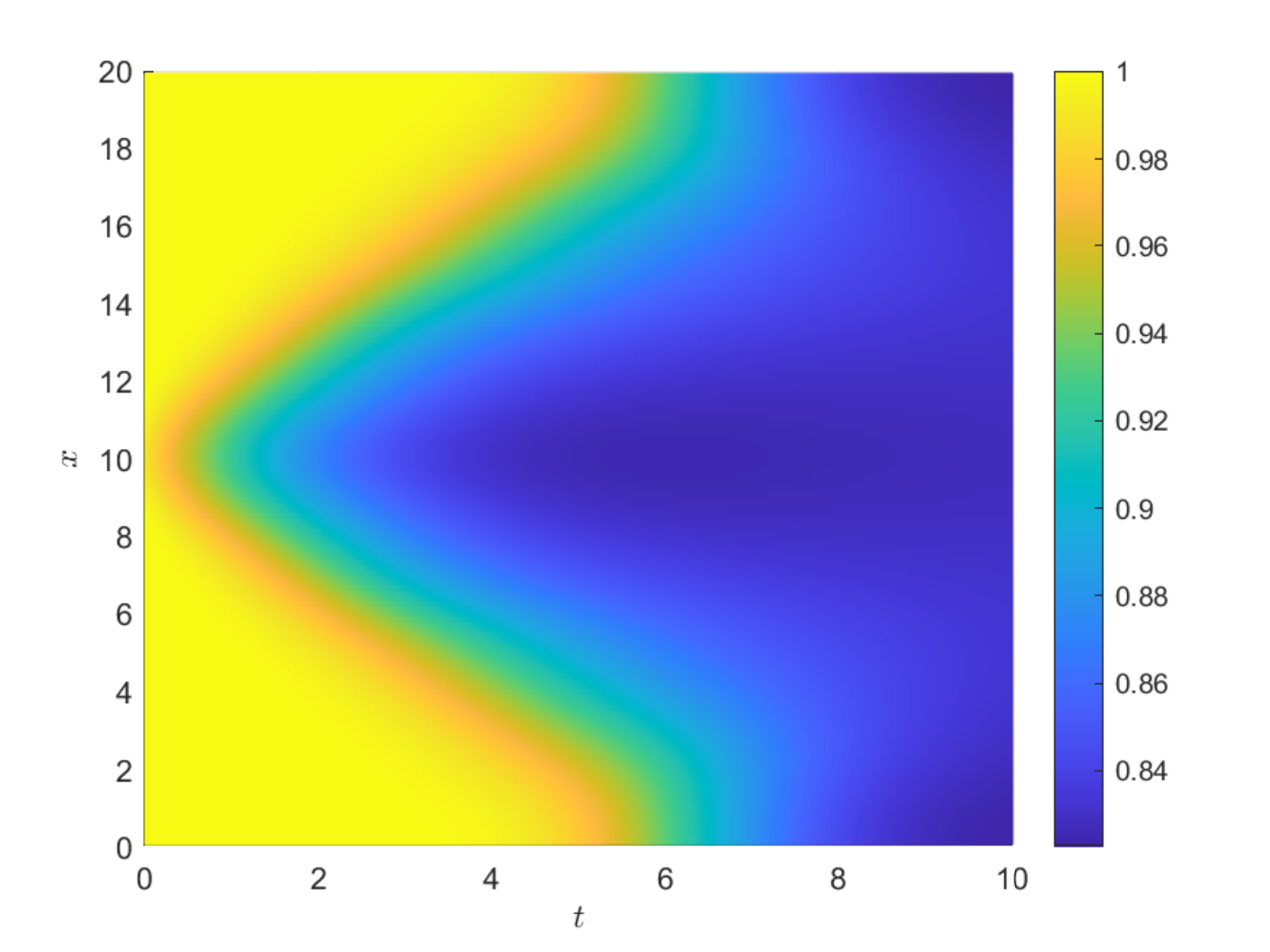}
\includegraphics[width=0.35\textwidth]{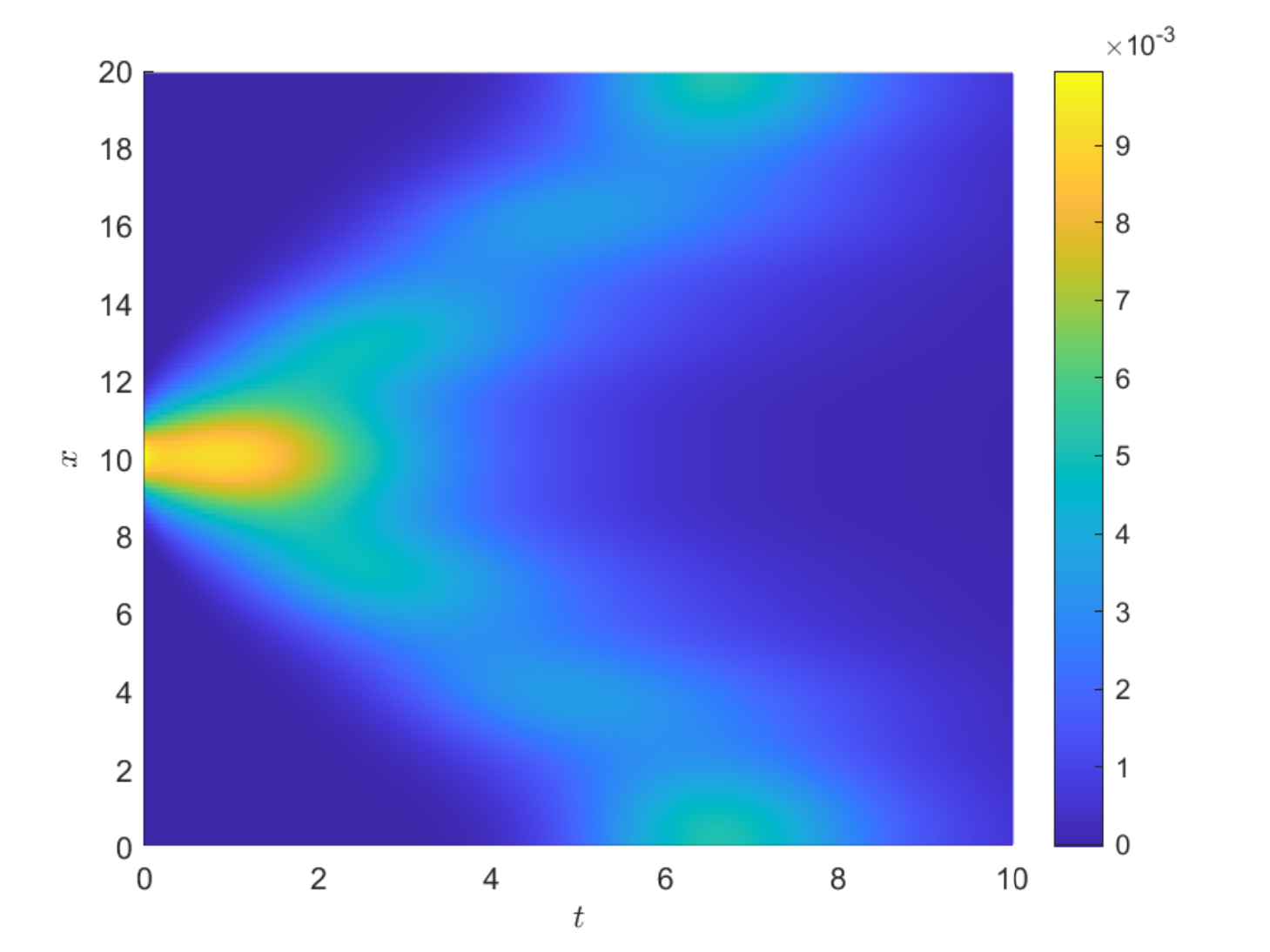}
\caption{Numerical results of the spatially heterogeneous SIR case with hyperbolic configuration of relaxation times and characteristic velocities (first row), with $\tau = 1.0, \lambda^2 = 1.0$, or parabolic configuration (second row), with $\tau = 10^{-5}, \lambda^2 = 10^5$. Time and spatial evolution of $S$ (first column) and $I$ (second column).}
\label{fig.TC4.1}
\end{figure}

Following^^>\cite{wang2020}, we analyze the behavior of the SIR-type model \eqref{eq.SIRkinetic} with a commuter-only population ($i_T = i_c, i \in \{S,I,R\}$) in a single 1D domain concerning spatially heterogeneous environments, taking into account a spatially variable contact rate
\begin{equation*}
\beta_I(x) = \hat \beta_I \left(1 + 0.05 \, \mathrm{sin} \frac{13 \pi x}{20} \right) .
\end{equation*}
Initial conditions are imposed assuming, in this setting, no uncertainty in the input data, with
\begin{equation*}
S_T(x,0) = 1 - I_T(x,0), \qquad I_T(x,0) = 0.01\,e^{-(x-10)^2}, \qquad R_T(x,0) = 0.0 ,
\end{equation*}
fluxes $J_S(x,0) = J_I(x,0) = J_R(x,0) = 0.0$ and zero-flux boundary conditions. The initial reproduction number results $R_0 = 1.111 > 1$, given by the choice $\hat \beta_I = 11.0$, $\kappa_I=0$ and $\gamma_I = 10.0$. Two different scenarios are considered, to concern both the hyperbolic and the parabolic limit of the system of equations. In the hyperbolic configuration, the relaxation times of all the compartments of individuals are $\tau = 1.0$, with the square of the characteristic velocities $\lambda^2 = 1.0$; while in the parabolic configuration $\tau = 10^{-5}$ and $\lambda^2 = 10^5$. The problem is solved applying an asymptotic-preserving (AP) Implicit-Explicit (IMEX) Runge-Kutta Finite Volume method, which permits to consistently simulate the diffusive (and stiff) regime of the system without loosing the expected 2$^{nd}$ order accuracy^^>\cite{bertaglia2021,boscarino2017}.
In Fig.~\ref{fig.TC4.1}, numerical results for both the scenarios are reported. A temporary persistence of the infectious can be noticed, with oscillations that reflect the sinusoidal form of the spatially variable contact rate. Differences of the dynamics of the epidemics in the two configurations of the relaxation times are evident. In particular, observing the evolution of susceptible individuals, it can be seen that in the purely diffusive case the amount of susceptible tends to a much lower equilibrium value than in the hyperbolic case, with almost all the individuals of the system infected by the disease.

\subsubsection{Application to the emergence of COVID-19 in Italy}
\label{res_lombardy_network}
To analyze the effectiveness of the proposed approach in a realistic epidemic scenario, we design a numerical test reproducing the evolution of the first outbreak of COVID-19 in the Lombardy Region of Italy, from February 27, 2020 to March 27, 2020, with respect to uncertainties underlying the initial conditions and chosen epidemic parameters, considering the SEIAR-type multiscale transport SEIAR \eqref{eq.SEIARkinetic}-\eqref{eq.SEIARkinetic_noncommuters} in a network configuration, as described in Section^^>\ref{network_modelling}. The system of equations is solved using a stochastic AP (sAP) IMEX Runge-Kutta Finite Volume Collocation method^^>\cite{bertaglia2021a,bertaglia2020,xiu2005}. This numerical scheme permits to reach spectral accuracy in the stochastic space, if the solution is sufficiently smooth in that space, and to switch from a stochastic Collocation method for the advection problem to a stochastic Collocation method for the diffusive problem in a uniform way with respect to the involved parameters without loosing accuracy, i.e. sAP property^^>\cite{jin2015,jin2018}. For further details regarding the numerical method and its convergence analysis the reader can refer to^^>\cite{bertaglia2021a}.

\begin{figure}[b]
\centering
\includegraphics[width=0.45\linewidth]{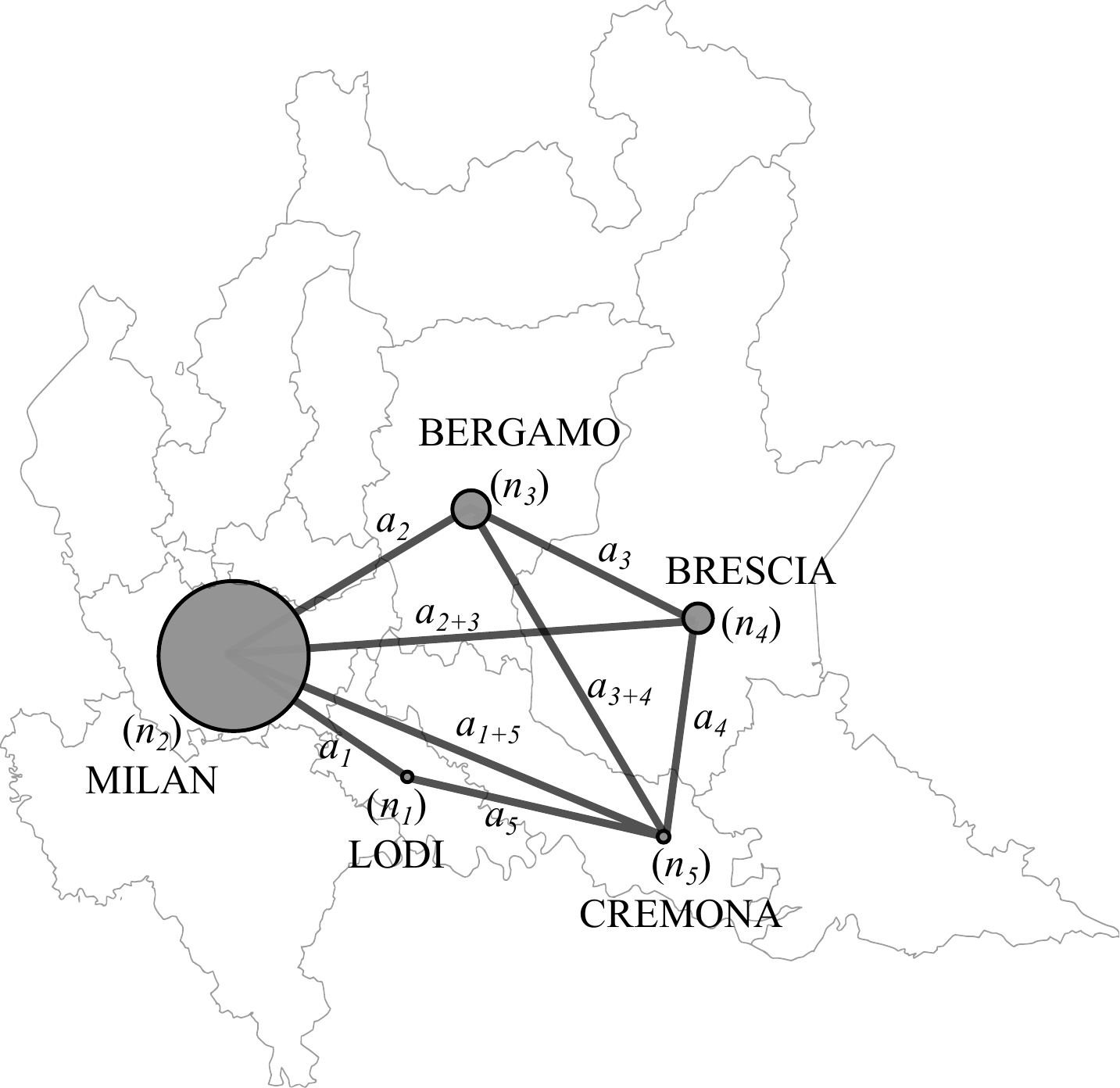}
\caption{Representation of the network of the Lombardy test case, composed of 5 nodes, corresponding to the provinces of interest and 5 arcs, connecting each city to the others, considering all the main paths of commuters. The dimension of the node is proportional to the dimension of the urbanized area of the province.}
\label{fig.network_lombardia}
\end{figure}

A five-node network is considered, whose nodes represent the 5 main provinces interested by the epidemic outbreak in the first months of 2020: Lodi ($n_1$), Milan ($n_2$), Bergamo ($n_3$), Brescia ($n_4$) and Cremona ($n_5$). The arcs $a_{j}$ connecting each node to the others identify the main set of routes and railways viable by commuters each day.  
A schematic representation of this network is shown in Fig. \ref{fig.network_lombardia}. 

The transmission coefficients at each arc-node interface, as well as the percentage of commuters belonging to each province, are imposed using the official national mobility flow assessment. In particular, the matrix of commuters used reflects mobility data provided by Lombardy Region for the regional fluxes of year 2020 (see Data Sources in Section \ref{sec:ds}).

The characteristic speed associated to each arc is fixed to permit a full round trip in each origin-destination section within a day. The characteristic speed of compartment $I$ is fixed to zero in all the nodes of the network. In the arcs, the relaxation time is assigned so that the model recovers a hyperbolic regime, while a parabolic setting is prescribed in the cities for commuters to simulate the diffusive behavior of the disease spread which typically occurs in highly urbanized zones. 

Concerning initial conditions and epidemic parameters of the test, at the beginning of the pandemic tracking of positive individuals cannot be considered reliable, but an information affected by uncertainty. To this aim, we introduce a single source of uncertainty $z$ having uniform distribution, $z \sim \mathcal{U}(0, 1)$, and the initial conditions for compartment $I$, at each node, are prescribed as
\begin{equation}
I_T(x,0,z) = I_T^0(1+z)\,,
\label{IT_stochastic}
\end{equation}
with $I_T^0$ density of infectious people on February 27, 2020, as given by data recorded by the Civil Protection Department of Italy. The amount of total inhabitants of each province is given by 2019 data of the Italian National Institute of Statistics
(see Data Sources in Section \ref{sec:ds}).

Due to the adopted screening policy, we chose to associate all infected individuals detected to the $I$ compartment. 
Furthermore, also $\beta_I$ is considered a random parameter:
\begin{equation*}
\beta_I(0,z) = \beta_{I,0}(1+\mu z)\, .
\end{equation*}
Assuming that highly infectious subjects are mostly detected in the most optimistic scenario, being subsequently quarantined or hospitalized, we set the minimum value $\beta_{I,0} = 0.03\,\beta_A$, as in^^>\cite{Gatto,buonomo2020} and $\mu = 0.06^{-1}$.
The initial value of $\beta_A$ is calibrated as the result of a least square problem, namely the L2 norm of the difference between the observed cumulative number of infected $I(t)$ and the numerical evolution of the same compartment, through a deterministic SEIAR ODE model set up for the whole Lombardy Region, with the result $\beta_A = 0.545$. 
In the above fitting, we also estimated $E_T^0 \approx 10\, I_T^0$ and $A_T^0 \approx 9\, I_T^0$. Consequently, also initial conditions for compartments $E$, $A$ and $S$ are stochastic, depending on the initial amount of severe infectious at each location, while $R_T^0 = 0$ everywhere.
Finally, we fix $\gamma_I$, $\gamma_A$ and $a$ according to^^>\cite{Gatto,buonomo2020}, considering these clinical parameters deterministic and $\sigma$ as in^^>\cite{buonomo2020,kantner2020}, setting then $\kappa_I=\kappa_A=30$.
With the above setup, we obtain an initial expected value of the basic reproduction number in the whole network $\mathbb{E}[R_0] = 3.6$, which is in agreement with estimations reported in^^>\cite{Gatto,buonomo2020,vollmer2020}.

\begin{figure}[t!]
\centering
\includegraphics[width=0.32\textwidth]{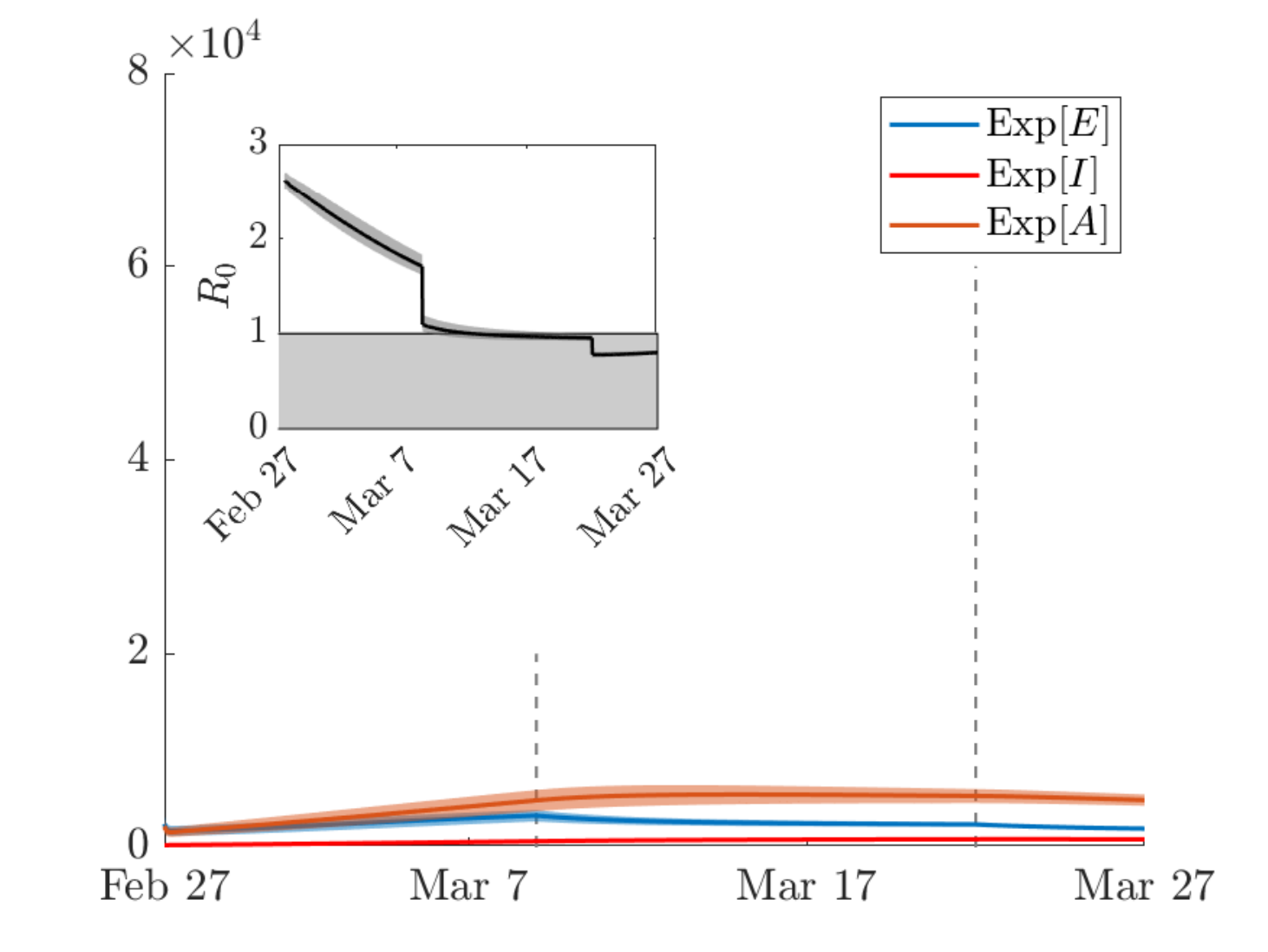}
\includegraphics[width=0.32\textwidth]{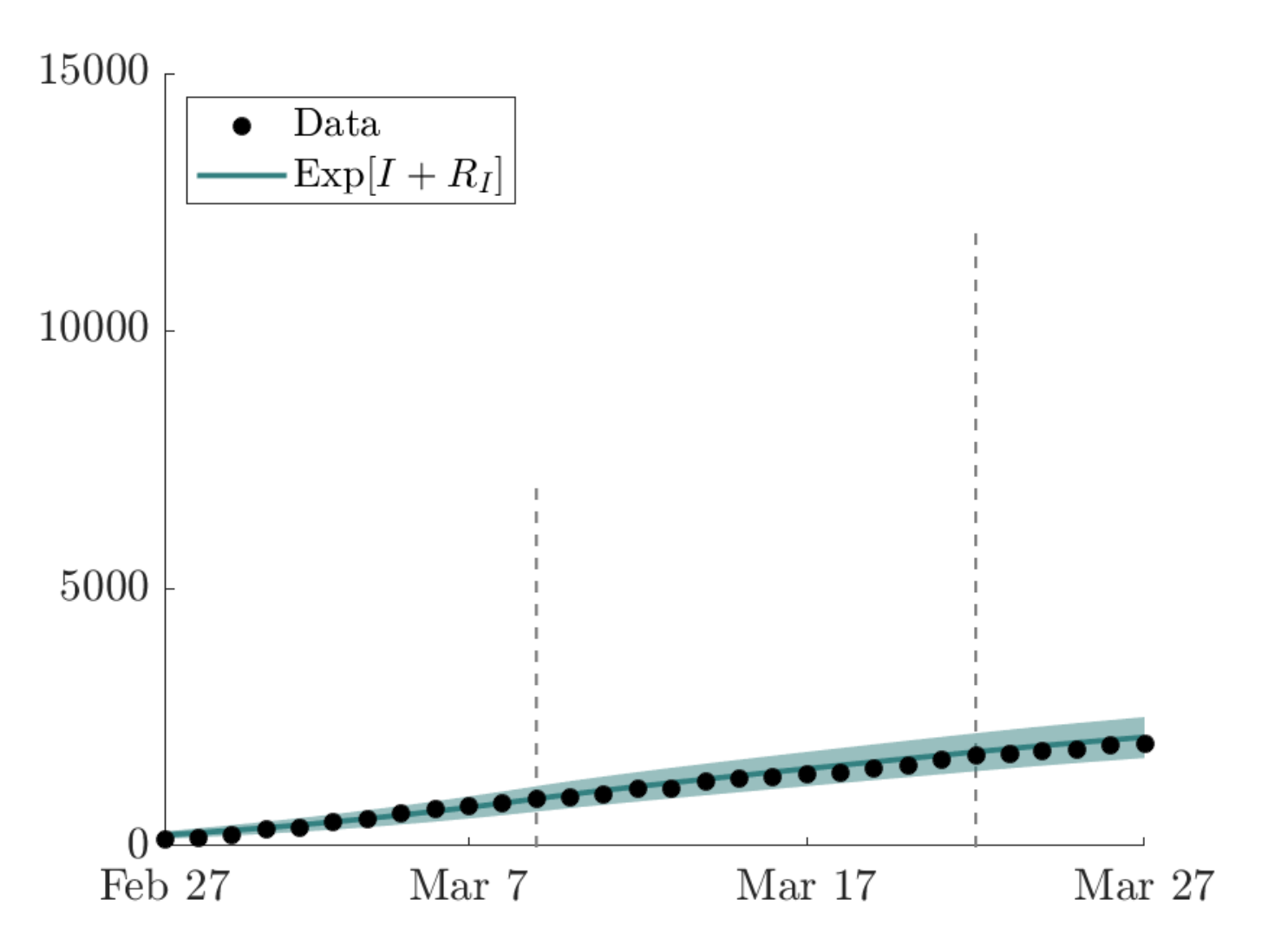}
\includegraphics[width=0.32\textwidth]{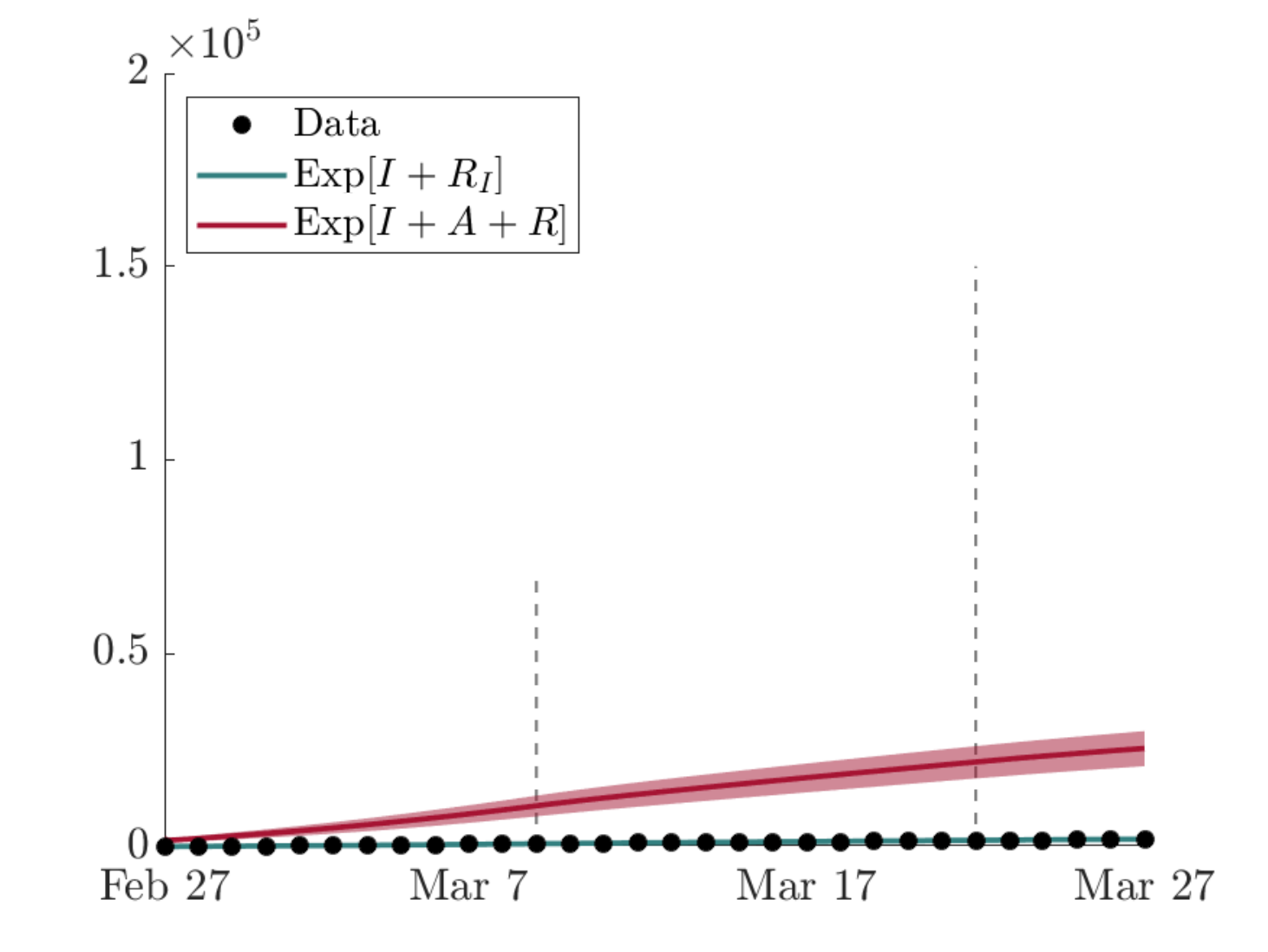}
\includegraphics[width=0.32\textwidth]{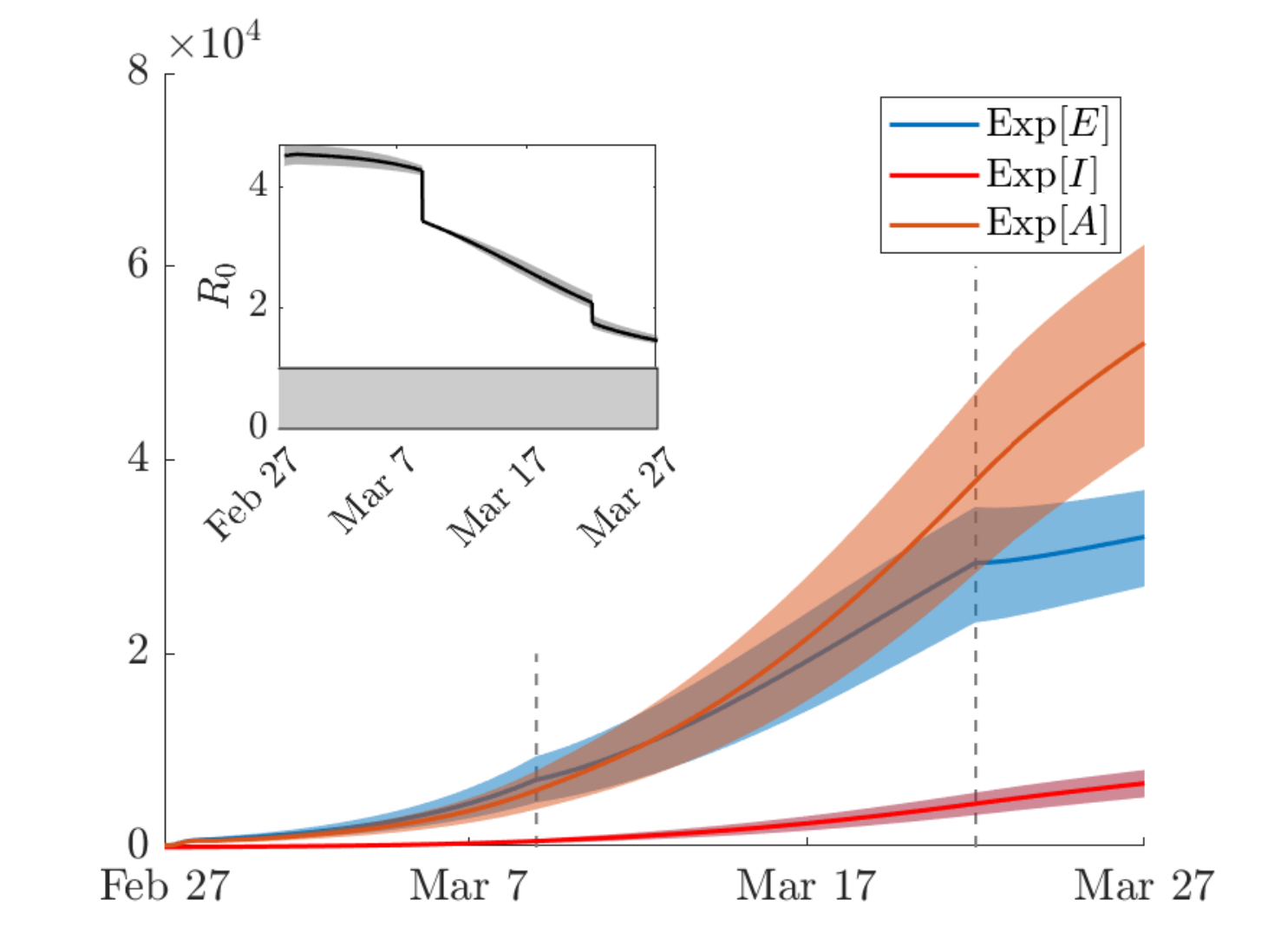}
\includegraphics[width=0.32\textwidth]{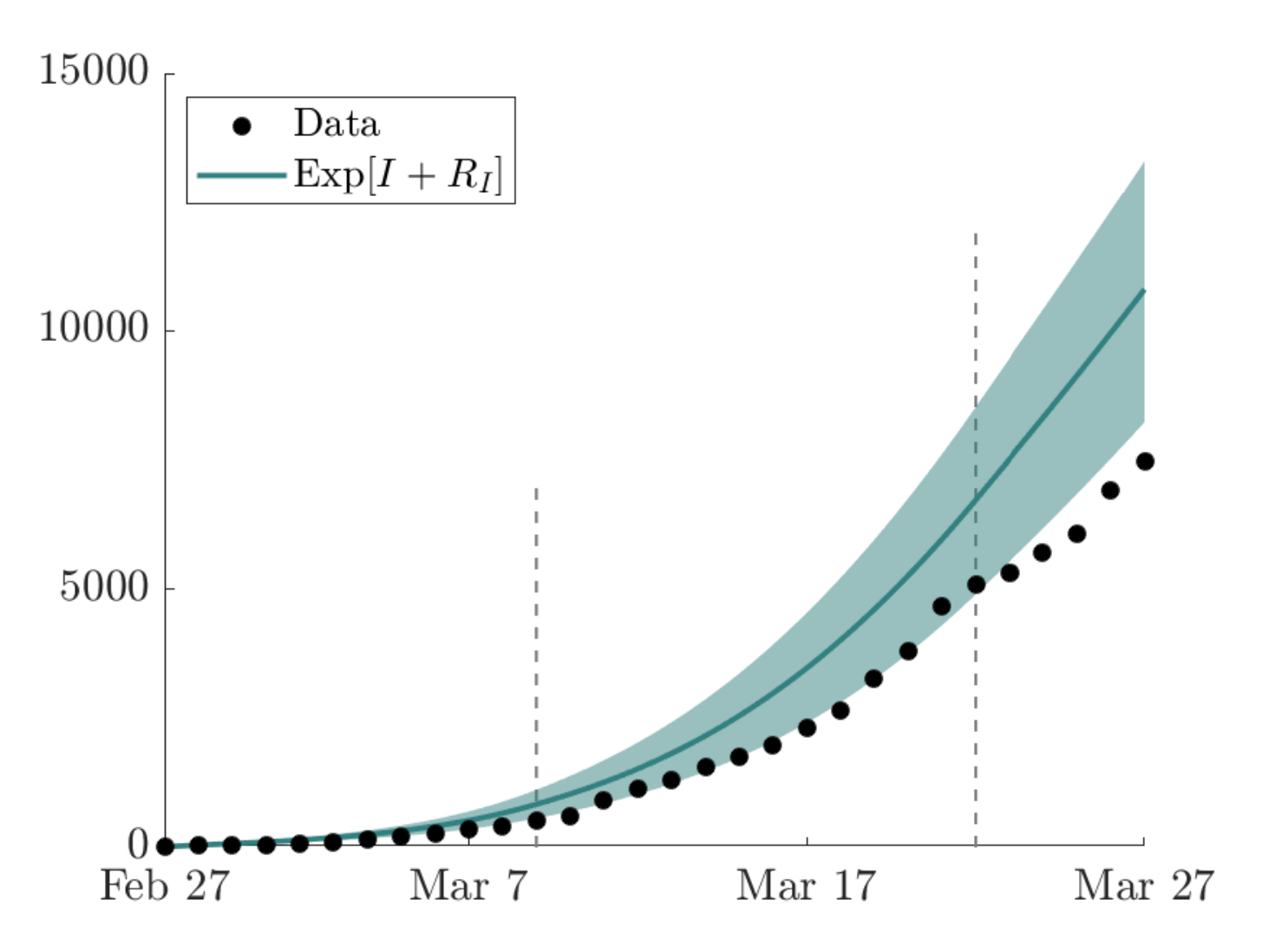}
\includegraphics[width=0.32\textwidth]{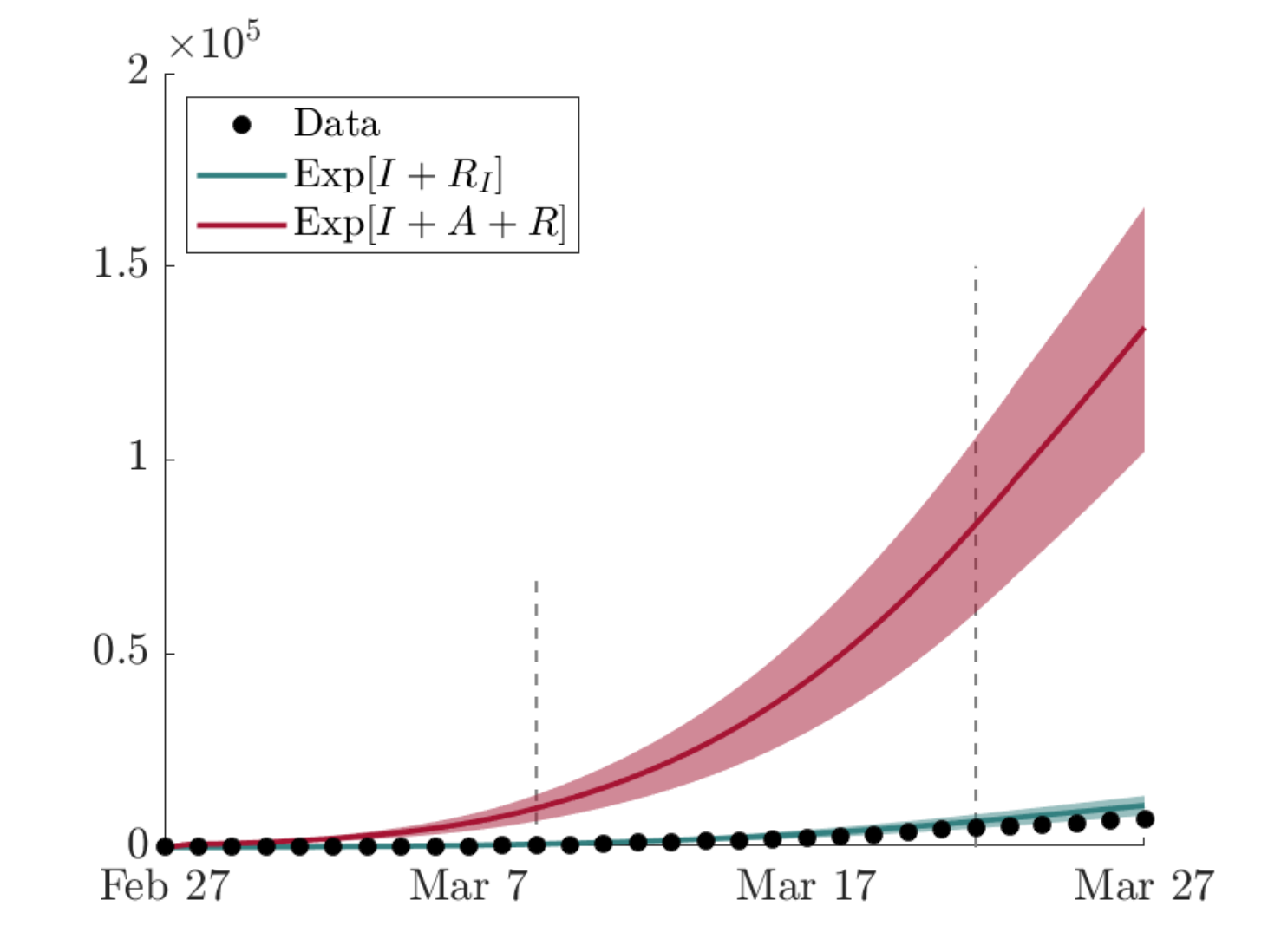}
\includegraphics[width=0.32\textwidth]{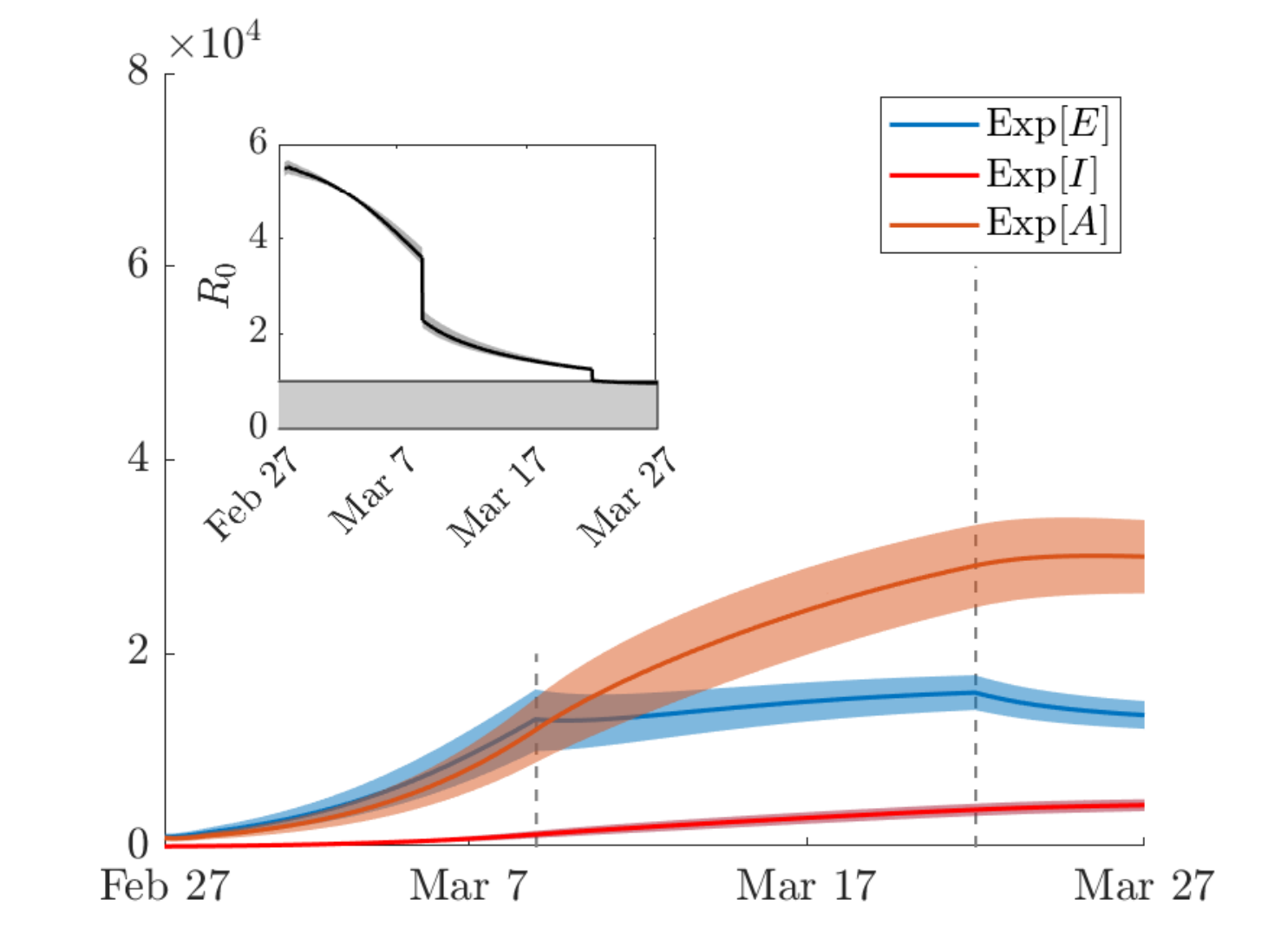}
\includegraphics[width=0.32\textwidth]{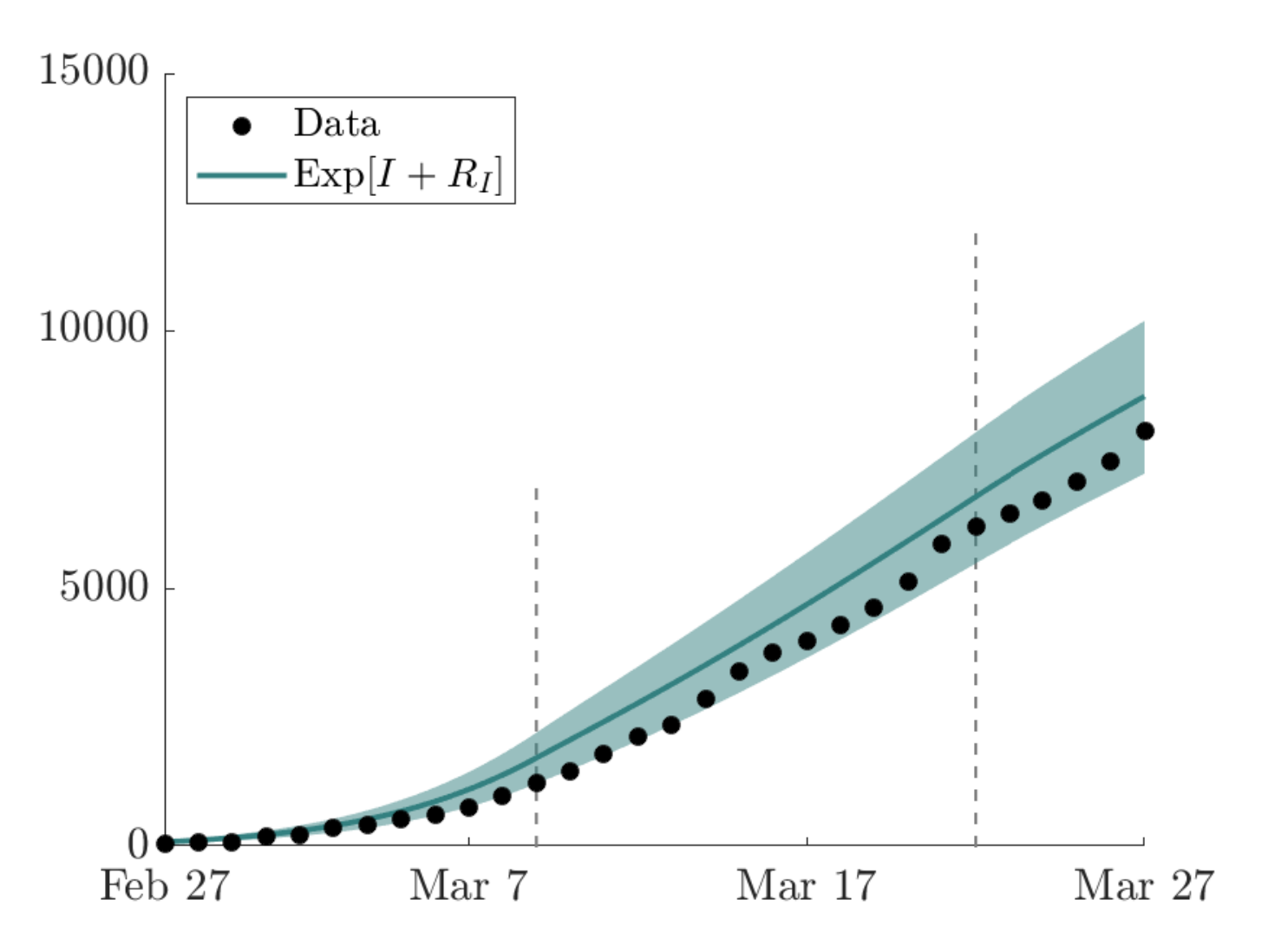}
\includegraphics[width=0.32\textwidth]{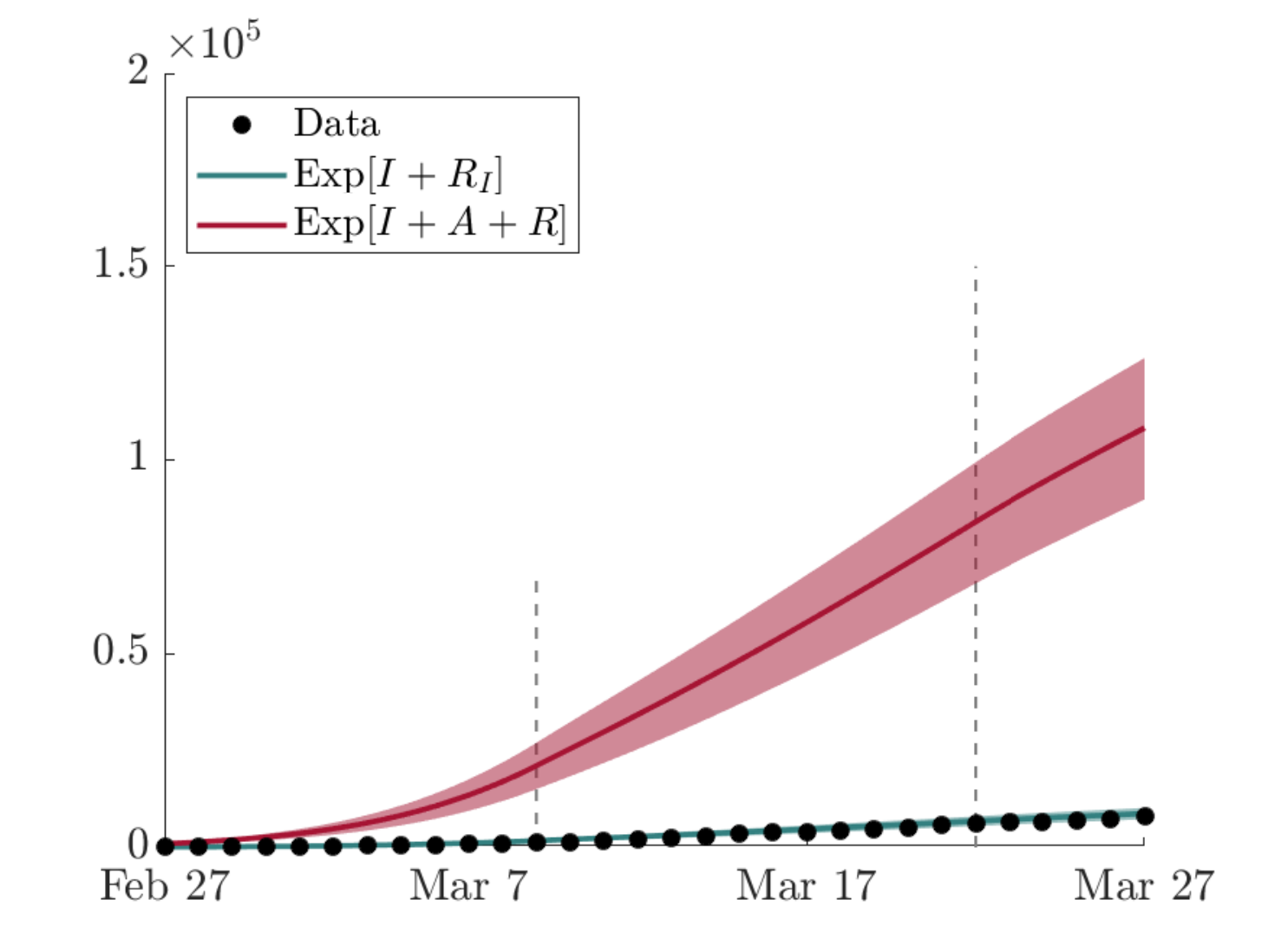}
\includegraphics[width=0.32\textwidth]{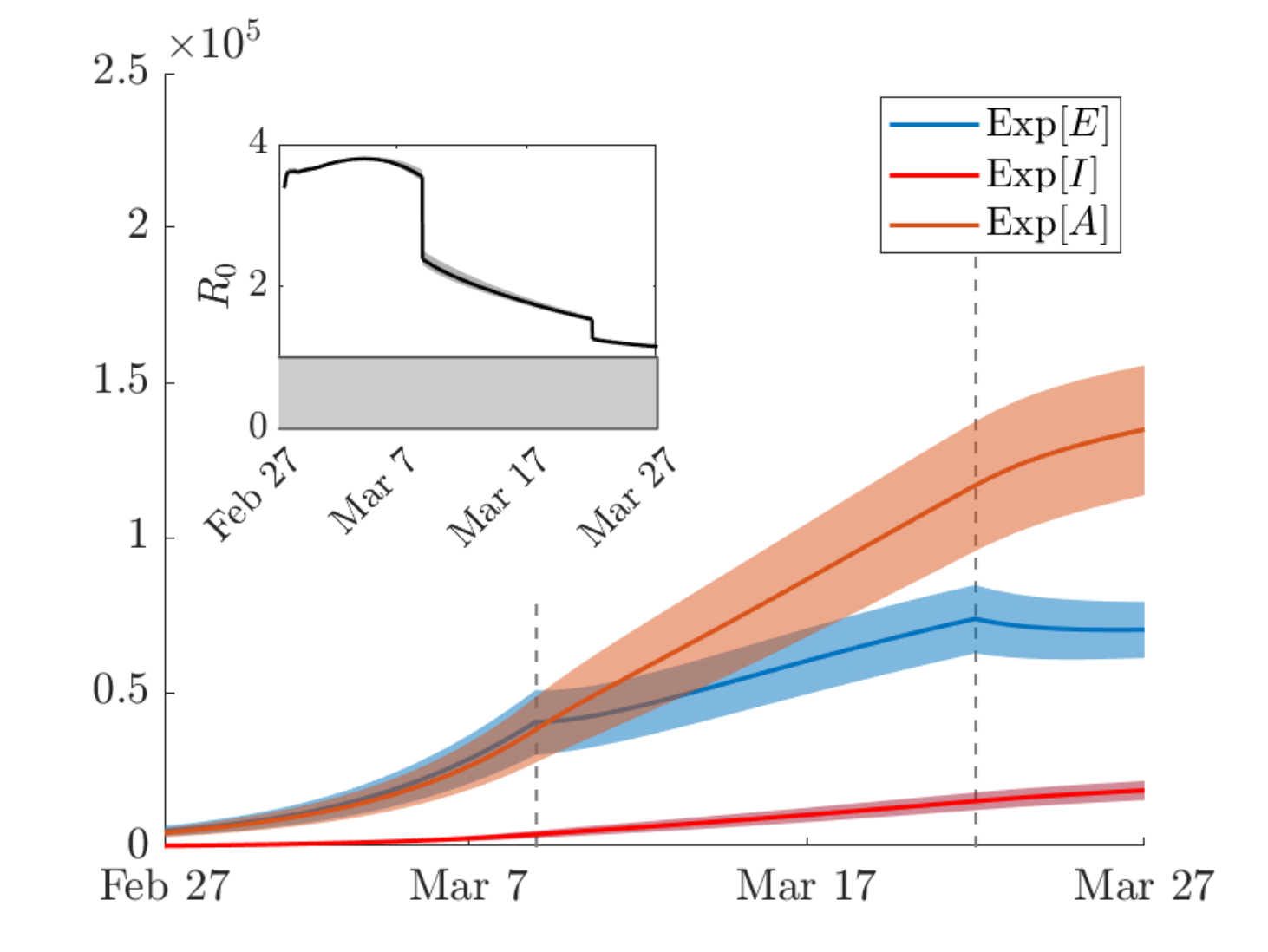}
\includegraphics[width=0.32\textwidth]{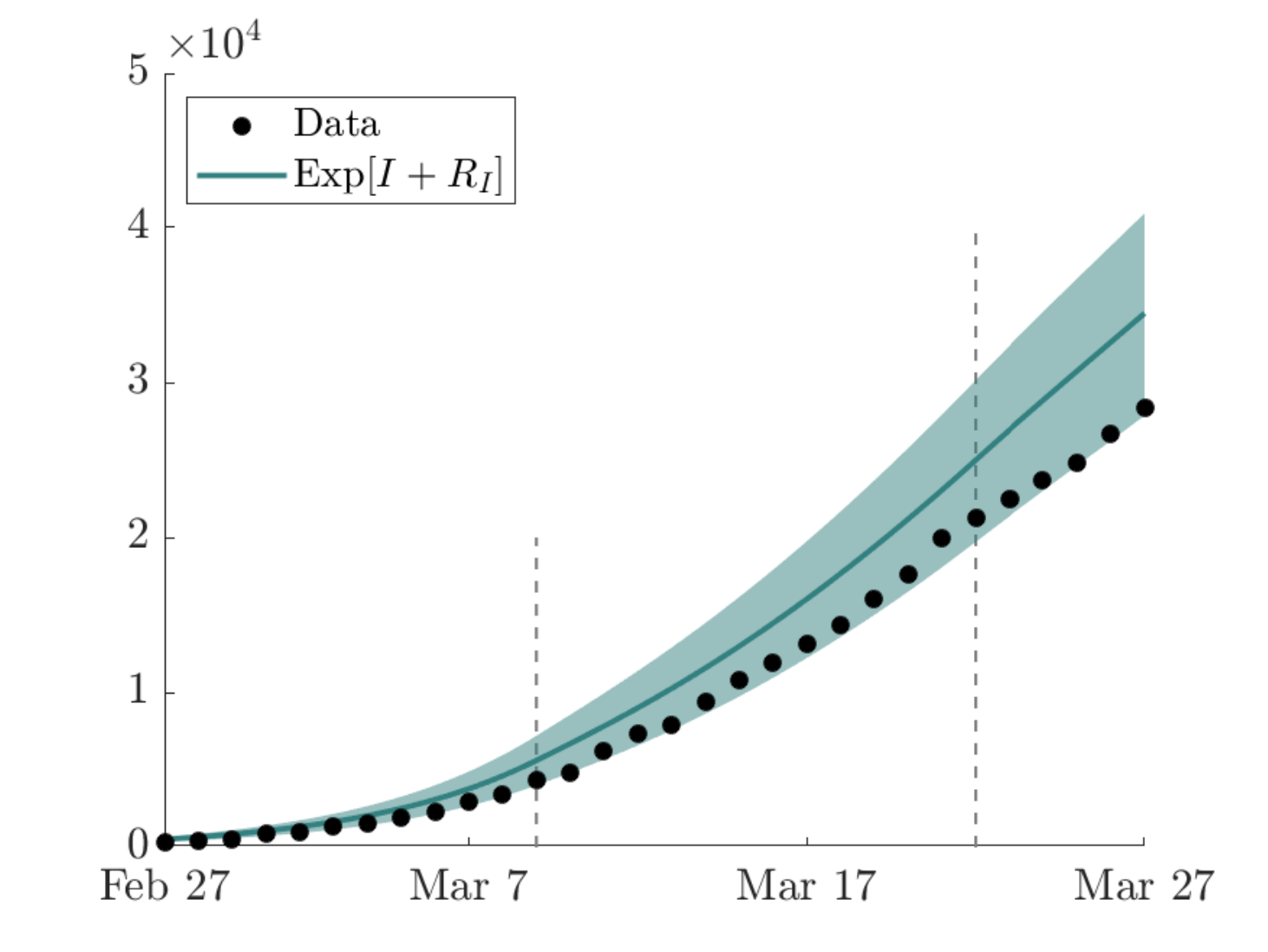}
\includegraphics[width=0.32\textwidth]{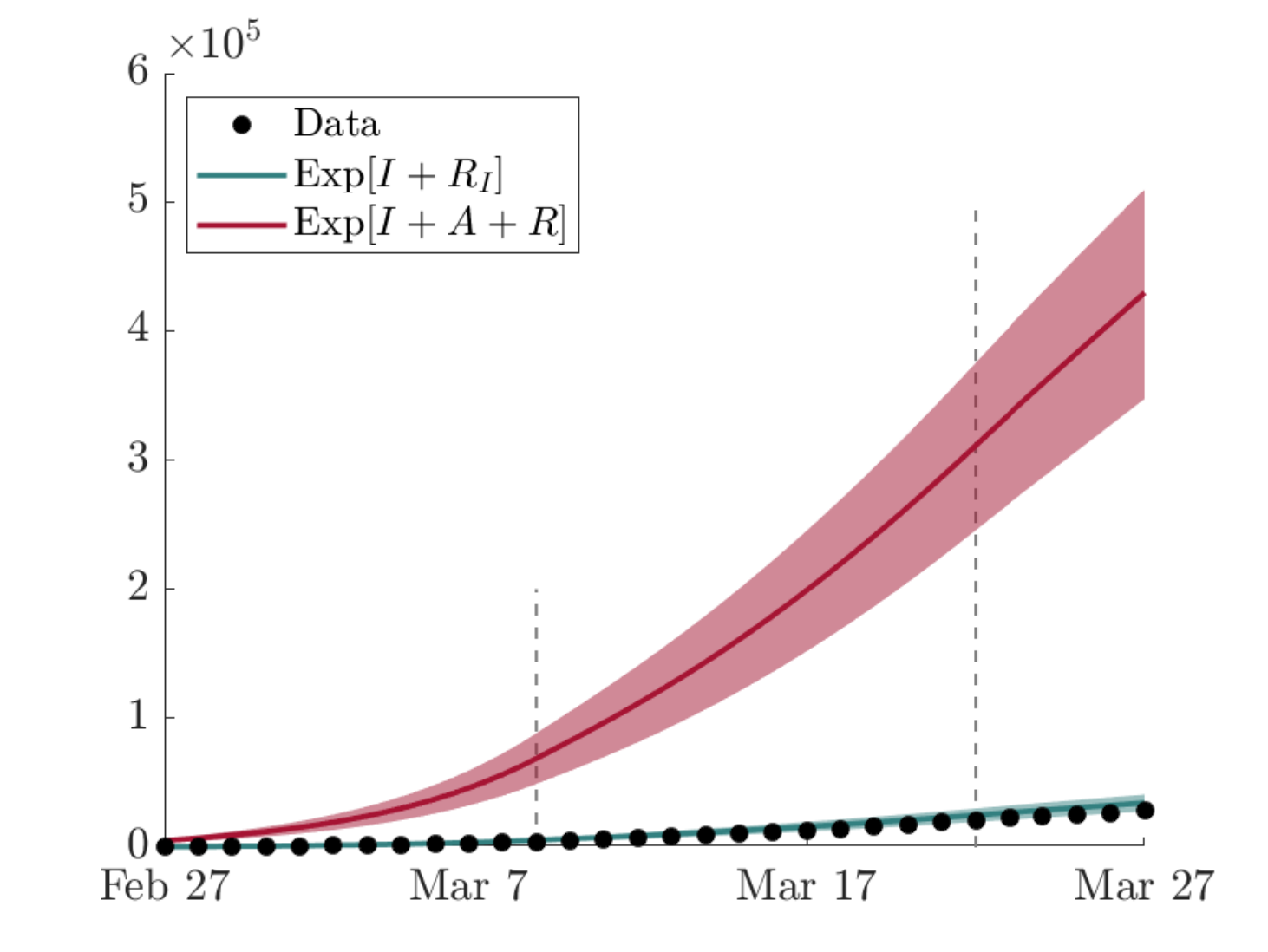}
\caption{Expected evolution (Exp) in time, with 95\% confidence intervals, for chosen representative cities, Lodi (first row), Milan (second row), Bergamo (third row), and the whole Lombardy network (fourth row), of: compartments $E$, $A$, $I$, together with the basic reproduction number $R_0$ (left); cumulative amount of severe infectious ($I+R_I$) compared with data of cumulative infectious taken from the COVID-19 repository of the Civil Protection Department of Italy (middle); cumulative amount of severe infectious ($I+R_I$) with respect to the effective cumulative amount of total infectious people, including asymptomatic and mildly symptomatic individuals ($I+A+R$) (right).
Vertical dashed lines identify the onset of governmental lockdown restrictions.}
\label{fig.NetworkLombardia}
\end{figure}

We model the escalation of lockdown restrictions, starting from March 9, 2020, initial day of the northern Italy lockdown, reducing the transmission rates, increasing $\kappa$ coefficients, due to the public being increasingly aware of the epidemic risks and reducing the percentage of commuting individuals according to mobility data tracked through mobile phones and made available by Google^^>\cite{aktay2020,vollmer2020}. 

Numerical results are reported in Fig.~\ref{fig.NetworkLombardia} for 3 representative cities, namely Lodi, Milan, Bergamo, and the whole Lombardy network. 
In Figs.~\ref{fig.NetworkLombardia} (first column) the expected evolution in time of the infected individuals, together with 95\% confidence intervals, including $E$, $I$ and $A$. Each plot is also associated with the temporal evolution of the reproduction number $R_0(\mathbf{z},t)$.
One can see the capacity of the model to reproduce a very heterogeneous epidemic trend in the network analyzed, which is also reflected in the different ranges and patterns shown for the $R_0$ of each province. It can also be observed the agreement between the evolution of the reproduction number and the epidemic spread. In particular, it is confirmed the decline of the daily number of infected as $R_0$ reaches values below 1, as shown in the plots for Lodi and Bergamo. On the other hand, the persistence of the virus in the complete network, and especially in Milan, is noticed until March 27, 2020 (last day of the simulation), where the reproduction number remains greater than 1.

As visible from Fig. \ref{fig.NetworkLombardia} (second column), the lower bound of the confidence band of the cumulative amount in time of $I$ is comparable with the observed data of the Civil Protection Department of Italy. As expected, the mean value of infected people is higher, especially in Milan, the province most affected by the virus, due to the uncertainty of available data, which certainly underestimate the real amount of infected people.

The comparison between the expected evolution in time of the cumulative amount of $I$ with respect to the effective cumulative amount of total infectious people, including also compartment $A$, is shown in Fig. \ref{fig.NetworkLombardia} (third column). Here, it can be noticed how much of the spread of COVID-19 has actually been lost from the data of the first outbreak in Lombardy and the impact that the presence of asymptomatic or undetected subjects has had on the epidemic evolution. 

\subsection{Realistic geographical settings}
\label{Sect.realistic}
\subsubsection{2D kinetic transport model}
\label{Sect.2D}
Let us now define $ \Omega\subset \RR^2$ a two-dimensional geographical area of interest, still assuming that individuals have been separated into commuting and non-commuting population, with the former at position $x\in\Omega$ moving with velocity directions $v \in \mathbb{S}^1$. Considering initially a simple SIR-dynamics, we denote by $f_S=f_S(\mathbf{z},x,v,t)$, $f_I=f_I(\mathbf{z},x,v,t)$ and $f_R=f_R(\mathbf{z},x,v,t)$, the respective kinetic densities of susceptible, infected and removed individuals.
The kinetic distribution of commuters is then given by
\[
f(\mathbf{z},x,v,t)=f_S(\mathbf{z},x,v,t)+f_I(\mathbf{z},x,v,t)+f_R(\mathbf{z},x,v,t),
\]
and their total density is obtained by integration over the velocity space 
\[
\rho(\mathbf{z},x,t)=\frac1{2\pi}\int_{\mathbb{S}^1} f(\mathbf{z},x,v_*,t)\,dv_*.
\]
As a consequence, the number of commuting susceptible, infectious and removed individuals can be recovered irrespective of their direction of displacement by integration over the velocity space. This gives
\[
S_c(\mathbf{z},x,t)=\frac1{2\pi}\int_{\mathbb{S}^1}  f_S(\mathbf{z},x,v,t)\,dv\,, \quad I_c(\mathbf{z},x,t)=\frac1{2\pi}\int_{\mathbb{S}^1}  f_I(\mathbf{z},x,v,t)\,dv\,, 
\]
\[
R_c(\mathbf{z},x,t)=\frac1{2\pi}\int_{\mathbb{S}^1}  f_R(\mathbf{z},x,v,t)\,dv\,.
\]
In this setting, the densities of the commuters satisfy the kinetic transport equations^^>\cite{boscheri2020}
\begin{equation}
\begin{split}
\frac{\partial f_S}{\partial t} + v_S \cdot \nabla_x f_S &= -F_I(f_S, I_T)+\frac1{\tau_S}\left(S_c-f_S\right)\\
\frac{\partial f_I}{\partial t} + v_I \cdot\nabla_x f_I &= F_I(f_S, I_T) -\gamma_I f_I+\frac1{\tau_I}\left(I_c-f_I\right)\\
\frac{\partial f_R}{\partial t} + v_R \cdot\nabla_x f_R &= \gamma_I f_I+\frac1{\tau_R}\left(R_c-f_R\right)\,,
\end{split}
\label{eq:kineticc}
\end{equation}
where the total densities are still defined by the sum of commuting and non-commuting part
$S_T=S_c+S_0$, $I_T=I_c+I_0$, $R_T=R_c+R_0$.
Densities of non-commuters, who act only at a local scale, satisfy the following diffusion dynamics
\begin{equation}
\begin{split}
\frac{\partial S_0}{\partial t} &= -F_I(S_0, I_T) + \nabla_x\cdot (D_S\nabla_x S_0) \\
\frac{\partial I_0}{\partial t}  &=  F_I(S_0, I_T)-\gamma_I I_0+\nabla_x\cdot (D_I\nabla_x I_0)\\
\frac{\partial R_0}{\partial t}  &= \gamma_I I_0+\nabla_x\cdot (D_R\nabla_x R_0).
\end{split}
\label{eq:diffuse}
\end{equation}
In the resulting multiscale kinetic SIR model \eqref{eq:kineticc}-\eqref{eq:diffuse}, which couples the commuting and non-commuting dynamics, the velocities $v_i=\lambda_i v$  in \eqref{eq:kineticc}, as well as the diffusion coefficients $D_i$ in \eqref{eq:diffuse}, with $i\in\{S,I,R\}$, are designed to take into account the heterogeneity of geographical areas, and are thus chosen dependent on the spatial location. The same stands also for the relaxation times, in analogy with system \eqref{eq.SIRkinetic}-\eqref{eq.SIRkinetic_noncommuters}. We refer to \eqref{eq:incf} for the definition of the incidence function $F_I(\cdot,I_T)$, and to Section^^>\ref{sect:1D} in general for the definition of the epidemic parameters involved. Furthermore, for the definition of the reproduction number of the above system we refer to \eqref{eq:R0_1} and the relative discussion.

\subsubsection{Macroscopic formulation and diffusion limit}
\label{Sect.2D_diff}
Let us introduce the flux functions
\begin{equation}
\begin{split}
J_S=\frac{\lambda_S}{2\pi} \int_{\mathbb{S}^1} & v f_S(x,v,t)\,dv,\quad J_I=\frac{\lambda_I}{2\pi}\int_{\mathbb{S}^1}  v f_I(x,v,t)\,dv,\quad J_R=\frac{\lambda_R}{2\pi}\int_{\mathbb{S}^1}  v f_R(x,v,t)\,dv.
\end{split}
\label{eq:flux-2D}
\end{equation}
Then, integrating system \eqref{eq:kineticc} in $v$, we get the following set of equations for the macroscopic formulation of densities
\begin{equation}
\begin{split}
\frac{\partial S_c}{\partial t} + \nabla_x\cdot J_S &= -F_I(S_c, I_T)\\
\frac{\partial I_c}{\partial t} + \nabla_x\cdot J_I &= F_I(S_c, I_T) -\gamma_I I_c\\
\frac{\partial R_c}{\partial t} + \nabla_x\cdot J_R &= \gamma_I I_c,
\end{split}
\label{eq:density}
\end{equation}
and fluxes
\begin{equation}
\begin{split}
\frac{\partial J_S}{\partial t} +  \frac{\lambda_S^2}{2\pi} \int_{\mathbb{S}^1}  (v\cdot \nabla_x f_S)v\,dv &= -F_I(J_S, I_T)-\frac1{\tau_S} J_S\\
\frac{\partial J_I}{\partial t} +  \frac{\lambda_I^2}{2\pi} \int_{\mathbb{S}^1}  (v\cdot \nabla_x f_I)v\,dv &= \frac{\lambda_I}{\lambda_S}F_I(J_S, I_T) - \gamma_I J_I-\frac1{\tau_I} J_I\\
\frac{\partial J_R}{\partial t} +  \frac{\lambda_R^2}{2\pi} \int_{\mathbb{S}^1}  (v\cdot \nabla_x f_R)v\,dv &= \frac{\lambda_R}{\lambda_I} \gamma_I J_I-\frac1{\tau_R} J_R.
\end{split}
\label{eq:fluxm}
\end{equation}
Note that the above system is not closed because the evolution of the fluxes in \eqref{eq:density}-\eqref{eq:fluxm} involves higher order moments of the kinetic densities. 

The diffusion limit can be formally recovered by introducing the space dependent diffusion coefficients 
$D_i=\frac12\lambda_i^2\tau_i$, with $i\in\{S,I,R\}$,
and letting $\tau_{i} \to 0$. We get, from the r.h.s. in \eqref{eq:kineticc}, $f_S=S_c, \,f_I=I_c, \,f_R=R_c$,
and, consequently, from the last three equations in \eqref{eq:density}-\eqref{eq:fluxm} we recover Fick's laws, which inserted into the first three equations in \eqref{eq:density}-\eqref{eq:fluxm} lead to the diffusion system for the population of commuters^^>\cite{MWW, Sun, Webb}
\begin{equation}
\begin{split}
\frac{\partial S_c}{\partial t} &= -F_I(S_c, I_T) + \nabla_x\cdot ({D_S}\nabla_x S_c) \\
\frac{\partial I_c}{\partial t}  &=  F_I(S_c, I_T)-\gamma_I I_c+\nabla_x\cdot ({D_I}\nabla_x I_c)\\
\frac{\partial R_c}{\partial t}  &= \gamma_I I_c+\nabla_x\cdot ({D_R}\nabla_x R_c)\,.
\end{split}
\label{eq:diff}
\end{equation}
System \eqref{eq:diff} is coupled with \eqref{eq:diffuse} for the non-commuting counterpart. 

Similarly to the one-dimensional case, the capability of the model to account for different regimes, hyperbolic or parabolic, according to the space dependent relaxation times $\tau_i$, $i\in\{S,I,R\}$, makes it suitable for describing the dynamics of human beings. Indeed, it is reasonable to avoid describing the details of movements within an urban area and model this through a diffusion operator. On the other hand, commuters when moving from one city to another follow well-established connections for which a description via transport operators is more appropriate.  

\subsubsection{Extension to multi-compartmental modelling}
As previously presented for the 1D model in Section^^>\ref{SEIAR_modelling}, it is possible to extend the modelling considering more general compartmental subdivisions. For example, more realistic models for COVID-19 should take into account the exposed population as well as the asymptomatic fraction of infected. As an example, we describe the extension of the multiscale kinetic transport modeling presented in the previous sections to a more general compartmental structure, even if still sufficiently simple, where the exposed population is included and infected people are distinguished between highly symptomatic and mildly/no symptomatic (see Fig.~\ref{fig.SEIAR}). We denote the commuter individuals which belong to the newly introduced compartment of exposed by $f_E(\mathbf{z},x,v,t)$ and of asymptomatic (or mildly symptomatic) by $f_A(\mathbf{z},x,v,t)$, resulting
\[
f(\mathbf{z},x,v,t)=f_S(\mathbf{z},x,v,t)+f_E(\mathbf{z},x,v,t)+f_I(\mathbf{z},x,v,t)+f_A(\mathbf{z},x,v,t)+f_R(\mathbf{z},x,v,t)\,.
\]
The kinetic SEIAR-type dynamics of the commuters then reads^^>\cite{bertaglia2021b}
\begin{equation}
\begin{split}
\frac{\partial f_S}{\partial t} + v_S \cdot \nabla_x f_S &= -F_I(f_S, I_T)-F_A(f_S, A_T) +\frac1{\tau_S}\left(S_c-f_S\right)\\
\frac{\partial f_E}{\partial t} + v_E \cdot\nabla_x f_E &=  F_I(f_S, I_T)+F_A(f_S, A_T)-a f_E+\frac1{\tau_E}\left(E_c-f_E\right)\\
\frac{\partial f_I}{\partial t} + v_I \cdot\nabla_x f_I &= a\sigma f_E -\gamma_I f_I+\frac1{\tau_I}\left(I_c-f_I\right)\\
\frac{\partial f_A}{\partial t} + v_A \cdot\nabla_x f_A &= a(1-\sigma) f_E -\gamma_A f_A+\frac1{\tau_A}\left(A_c-f_A\right)\\
\frac{\partial f_R}{\partial t} + v_R \cdot\nabla_x f_R &= \gamma_I f_I+\gamma_A f_A+\frac1{\tau_R}\left(R_c-f_R\right),
\end{split}
\label{eq:kineticSEIAR}
\end{equation}
with $E_T=E_c+E_0$, $A_T=A_c+A_0$, and
\begin{equation*}
E_c(\mathbf{z},x,t)=\frac1{2\pi}\int_{\mathbb{S}^1}  f_E(\mathbf{z},x,v,t)\,dv\,, \quad 
A_c(\mathbf{z},x,t)=\frac1{2\pi}\int_{\mathbb{S}^1}  f_A(\mathbf{z},x,v,t)\,dv\, .
\end{equation*}
Indeed, this system is coupled with the following one describing the dynamics of non-commuters, who act only at the urban scale:
\begin{equation}
\begin{split}
\frac{\partial S_0}{\partial t} &= -F_I(S_0, I_T) -F_A(S_0, A_T) + \nabla_x\cdot ({D_S}\nabla_x S_0) \\
\frac{\partial E_0}{\partial t} &= F_I(S_0, I_T) + F_A(S_0, A_T) -a E_0 + \nabla_x \cdot({D_E}\nabla_x E_0) \\
\frac{\partial I_0}{\partial t}  &=  a\sigma E_0-\gamma_I I_0+\nabla_x\cdot ({D_I}\nabla_x I_0)\\
\frac{\partial A_0}{\partial t}  &=  a(1-\sigma)E_0-\gamma_A A_0+\nabla_x\cdot ({D_A}\nabla_x A_0)\\
\frac{\partial R_0}{\partial t}  &= \gamma_I I_0+\gamma_A A_0+\nabla_x\cdot ({D_R}\nabla_x R_0).
\end{split}
\label{eq:diffuseSEIAR}
\end{equation}
For the definition of the incidence function regarding asymptomatic people $F_A(\cdot,A_T)$, we consider \eqref{eq:incf-SEIAR}.

When introducing the same definition of flux \eqref{eq:flux-2D} for the additional compartments, $J_E$ and $J_A$, integrating system \eqref{eq:kineticSEIAR} in $v$, we get the set of equations for the macroscopic densities^^>\cite{bertaglia2021b}.
Moreover, defining also $D_E=\frac12\lambda_E^2\tau_E$ and $D_A=\frac12\lambda_A^2\tau_A$ and considering the same procedure discussed in Section^^>\ref{Sect.2D_diff}, we recover SEIAR system in the diffusive regime for the commuting individuals^^>\cite{bertaglia2021b} coupled with \eqref{eq:diffuseSEIAR} for the non-commuting counterpart.

To define the reproduction number also for this multiscale SEIAR-type kinetic transport model, we recall the NGM approach^^>\cite{Diek} considering no flux boundary conditions, which yields the same definition \eqref{eq.R0_2}. Details of this derivation are reported in^^>\cite{bertaglia2021b}.

\subsubsection{Application to the spatial spread of COVID-19 in Italy in Emilia-Romagna and Lombardy Region}
Let us underline that the discretization of the resulting multiscale systems of PDEs is not trivial and therefore requires the construction of a specific numerical method able to correctly describe the transition from a convective to a diffusive regime in realistic geometries. For this purpose, we adopt an asymptotic-preserving IMEX Runge-Kutta method on unstructured grids coupled with a stochastic Collocation method which ensures spectral accuracy in the stochastic space^^>\cite{jin2017,jin2018,pareschi2020}. At each collocation node, the numerical scheme combines a discrete ordinate method in velocity with the even and odd parity formulation^^>\cite{DP,JPT} and achieves asymptotic preservation in time using suitable IMEX Runge-Kutta schemes^^>\cite{boscarino2017}, namely, to obtain a scheme which consistently captures the diffusion limit and for which the choice of the time discretization step is not related to the smallness of the scaling parameters $\tau$.
All the details concerning the numerical scheme and its validation in terms of accuracy are reported in^^>\cite{boscheri2020,bertaglia2021b}.

To validate the proposed methodology in realistic geographical and epidemiological scenarios, two numerical tests reproducing, respectively, the epidemic outbreak of COVID-19 in the Emilia-Romagna Region of Italy, from March 1, 2020 to March 10, 2020, and in the Lombardy Region of Italy, from February 27, 2020 to March 22, 2020, are designed. In the former, we solve a multiscale SEIR-type system of PDEs in a deterministic setting (for further details on the chosen SEIR model the reader can refer to^^>\cite{boscheri2020}). In the latter, we also take into account the uncertainty underlying initial conditions of infected individuals, solving the multiscale SEIAR-type system of PDEs \eqref{eq:kineticc}-\eqref{eq:diffuse}.

\begin{figure}
\centering
\includegraphics[width=0.45\textwidth]{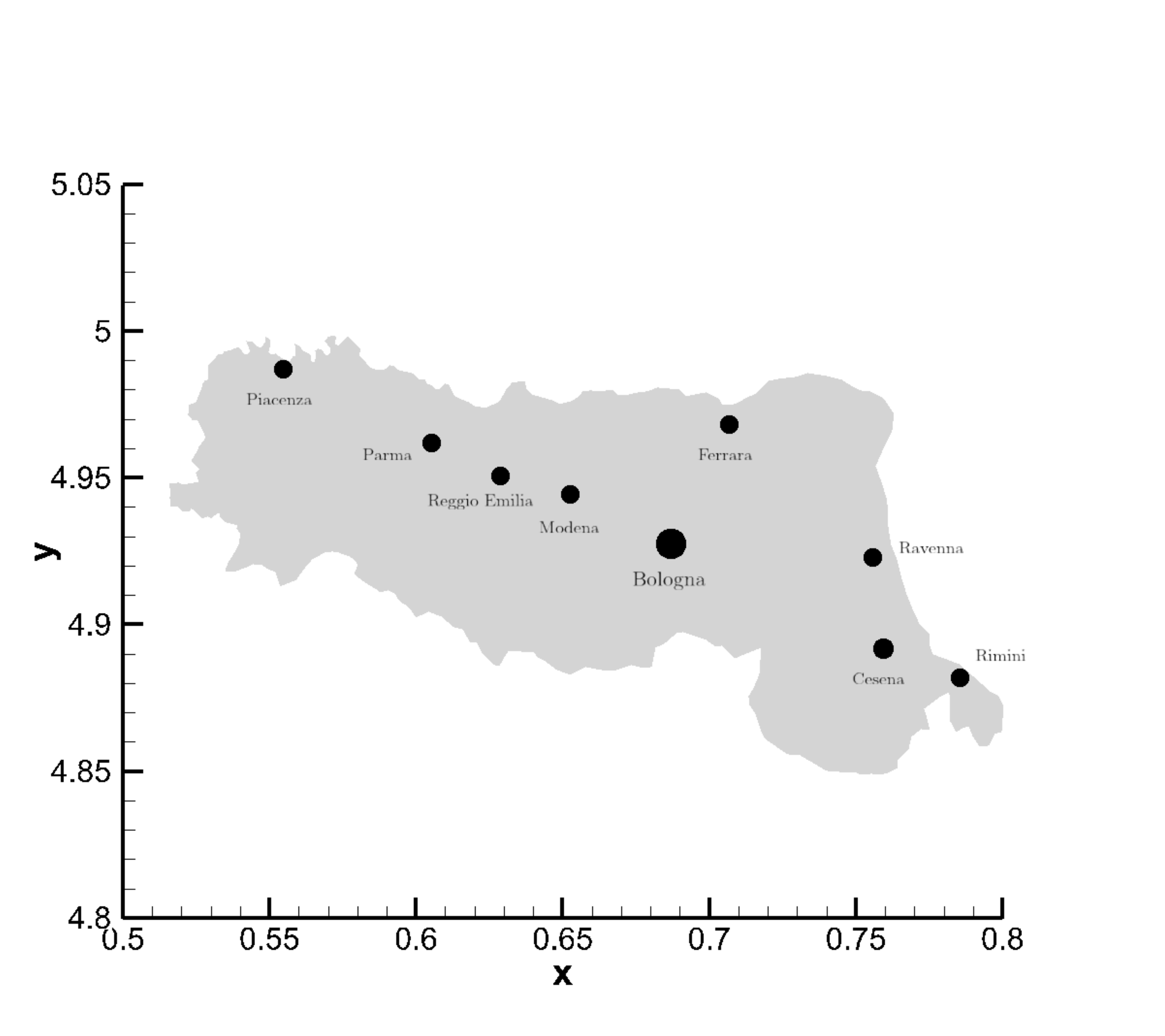}
\includegraphics[width=0.45\textwidth]{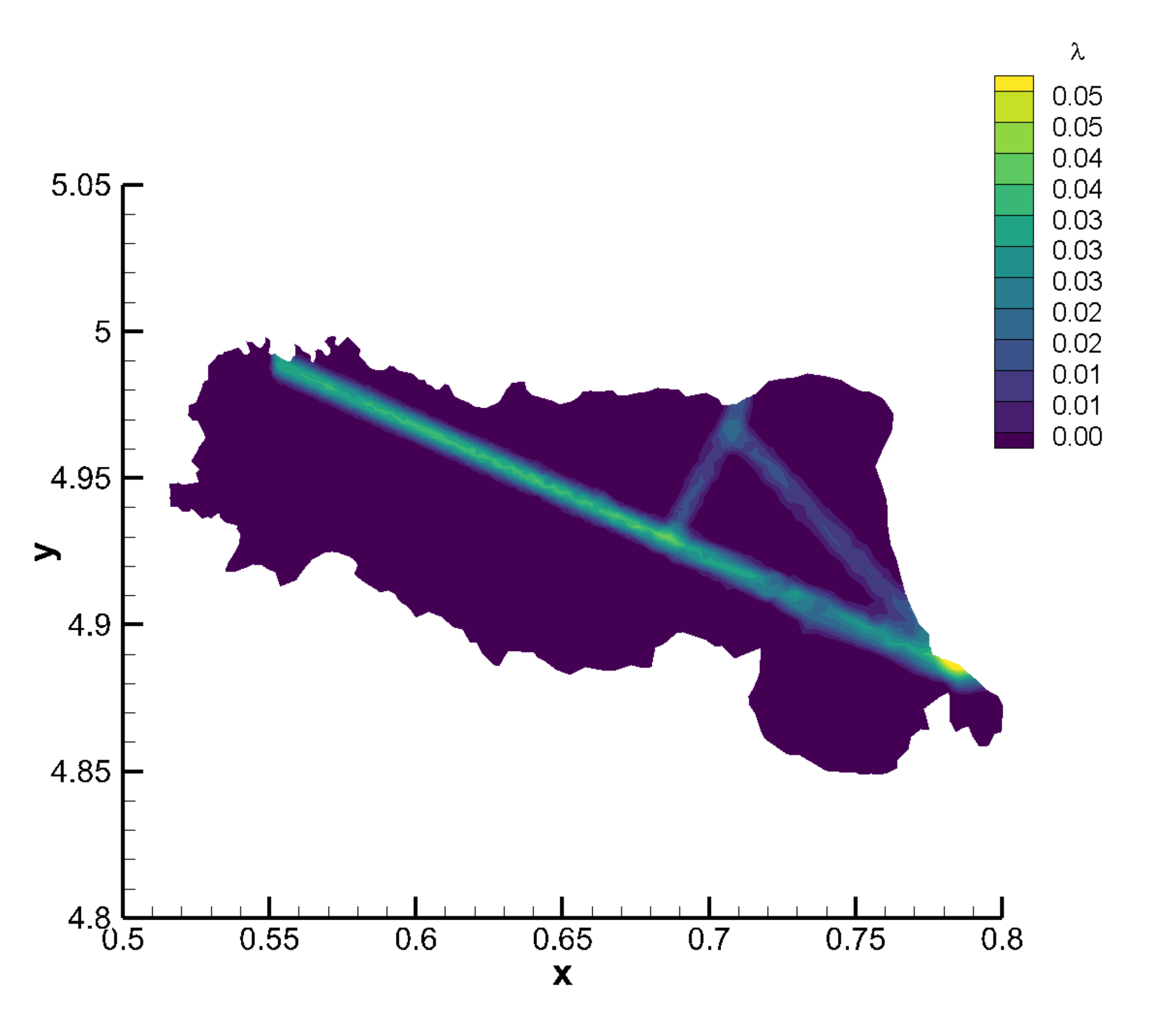}\\
\includegraphics[width=0.45\textwidth]{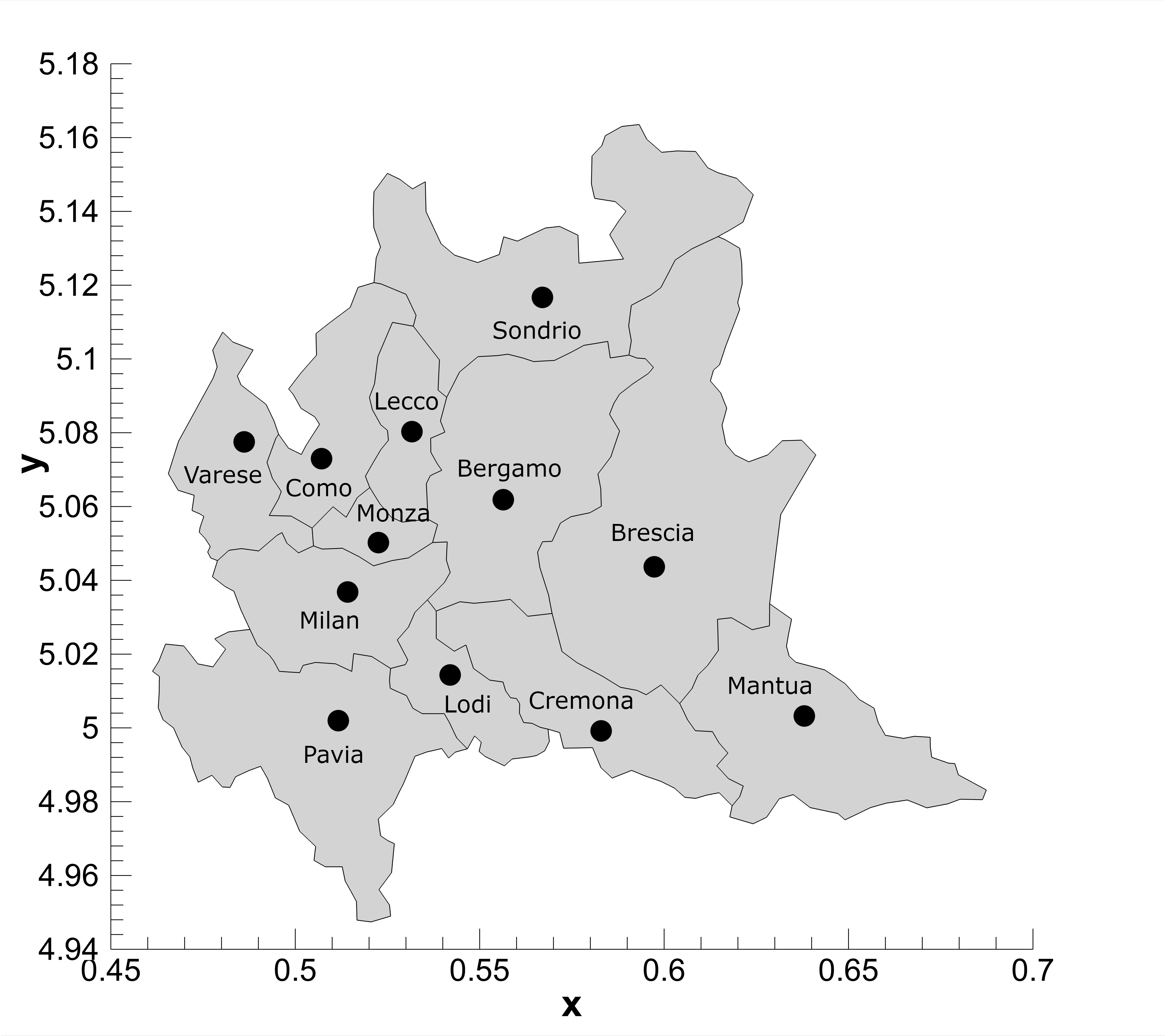}
\includegraphics[width=0.45\textwidth]{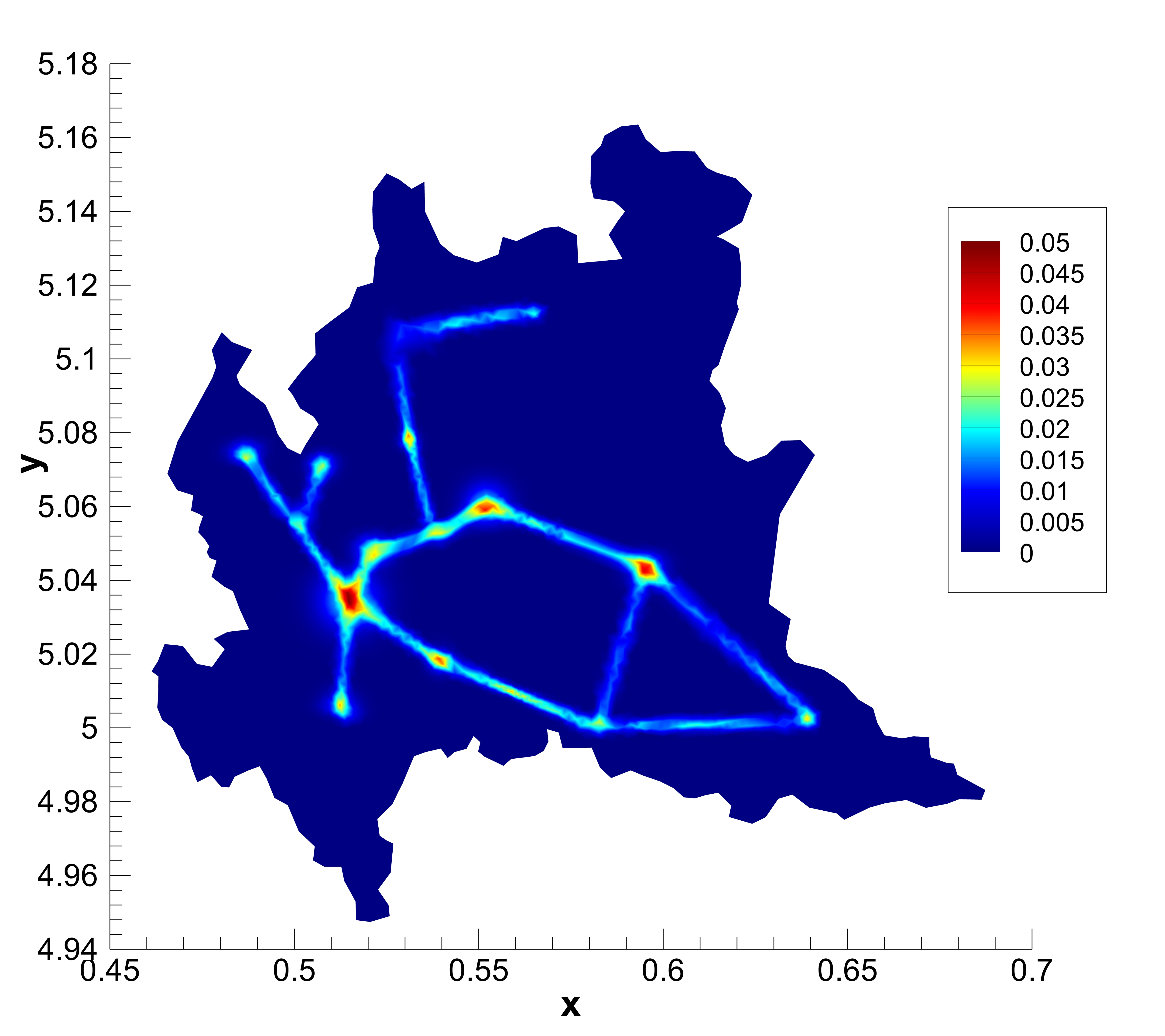}
\caption{Top: identification of the Emilia-Romagna provinces (left) and initial condition imposed for characteristic speeds $\lambda_i$, $i\in\{S,E,R\}$ (right). Bottom: identification of the Lombardy provinces (left), initial condition imposed for characteristic speeds $\lambda_i$, $i\in\{S,E,A,R\}$ (right).}
\label{fig.computational_domain}
\end{figure}

The computational domain is defined in terms of the boundary that circumscribes the Regions as a list of georeferenced points in the ED50/UTM Zone 32N reference coordinate system from Istituto Nazionale di Statistica (see Data Sources in Section \ref{sec:ds}). No-flux boundary conditions are imposed in the whole boundary of the domain, assuming that the population is not moving from/to the adjacent Regions. The domain is then subdivided in the provinces of the specific Region. The identification of these cities is shown in Fig. \ref{fig.computational_domain} (top left) for Emilia-Romagna and in Fig. \ref{fig.computational_domain} (bottom left) for Lombardy.
To avoid the mobility of the population in the entire territory and to simulate a more realistic geographical scenarios in which individuals travel along the main traffic paths of the Region, different values of propagation speeds are assigned in the domain which reflect, as close as possible, the real characteristics of the territory.
The resulting distribution of the characteristic speeds is visible from Fig. \ref{fig.computational_domain} (top right) for Emilia-Romagna and Fig. \ref{fig.computational_domain} (bottom right) for Lombardy case.

The space-dependent relaxation time is assigned so that the model recovers a hyperbolic regime in the entire region, apart from the main cities, where a parabolic setting is prescribed to correctly capture the diffusive behavior of the disease spreading which typically occurs in highly urbanized zones. Considering $p_c$ the number of citizens of a generic city (province) denoted with subscript $c=1, \ldots, \mathcal{N}_c$, the initial spatial distribution of the generic population $f(x,y)$ is assigned, for each province and each epidemiological compartment, as a multivariate Gaussian function with the variance being the radius of the urban area $r_c$:
\begin{equation*}
f(x,y) = \frac{1}{2\pi r_c} e^{-\frac{(x-x_c)^2 +(y-y_c)^2}{2r_c^2}} p_c \,,
\end{equation*}
with $(x_c, y_c)$ representing the coordinates of a generic city center. The initial population setting, for each province, is taken from 2019 data of the Italian National Institute of Statistics. 

For the Emilia-Romagna Region, we estimate the initial number of exposed individuals, including asymptomatic, as $E_{T}^0 = 4\, I_{T}^0$; while, for the Lombardy Region, $E_{T}^0 = 10\, I_{T}^0$ and $A_{T}^0= 9\, I_{T}^0$ in each location, with $I_{T}^0$ given by data recorded by the Civil Protection Department of Italy in the first day simulated. 
As previously discussed in Section^^>\ref{res_lombardy_network}, for the Lombardy test, we introduce a single source of uncertainty $z$ having uniform distribution, $z \sim \mathcal{U}(0, 1)$ so that the initial conditions for compartment $I$, in each control volume, are prescribed as in \eqref{IT_stochastic}. 
Moreover, we refer to regional mobility data to properly subdivide the population in commuters and non-commuters (see Data Sources in Section \ref{sec:ds}).
Concerning the calibration and the choice of clinical epidemic parameters, as well as for the modeling of the governmental restrictions, the reader can refer to^^>\cite{boscheri2020} for the Emilia-Romagna case and to^^>\cite{bertaglia2021a} for the Lombardy case. 

With the chosen parametric setups, we obtain initial reproduction number for Emilia-Romagna $R_0 = 2.3$ and an initial expected value of the basic reproduction number for the Lombardy Region $\mathbb{E}[R_0(0)] = 3.2$, which are in accordance with available literature^^>\cite{Gatto,buonomo2020,vollmer2020}. 
\begin{figure}[b]
\centering
\includegraphics[width=0.45\textwidth]{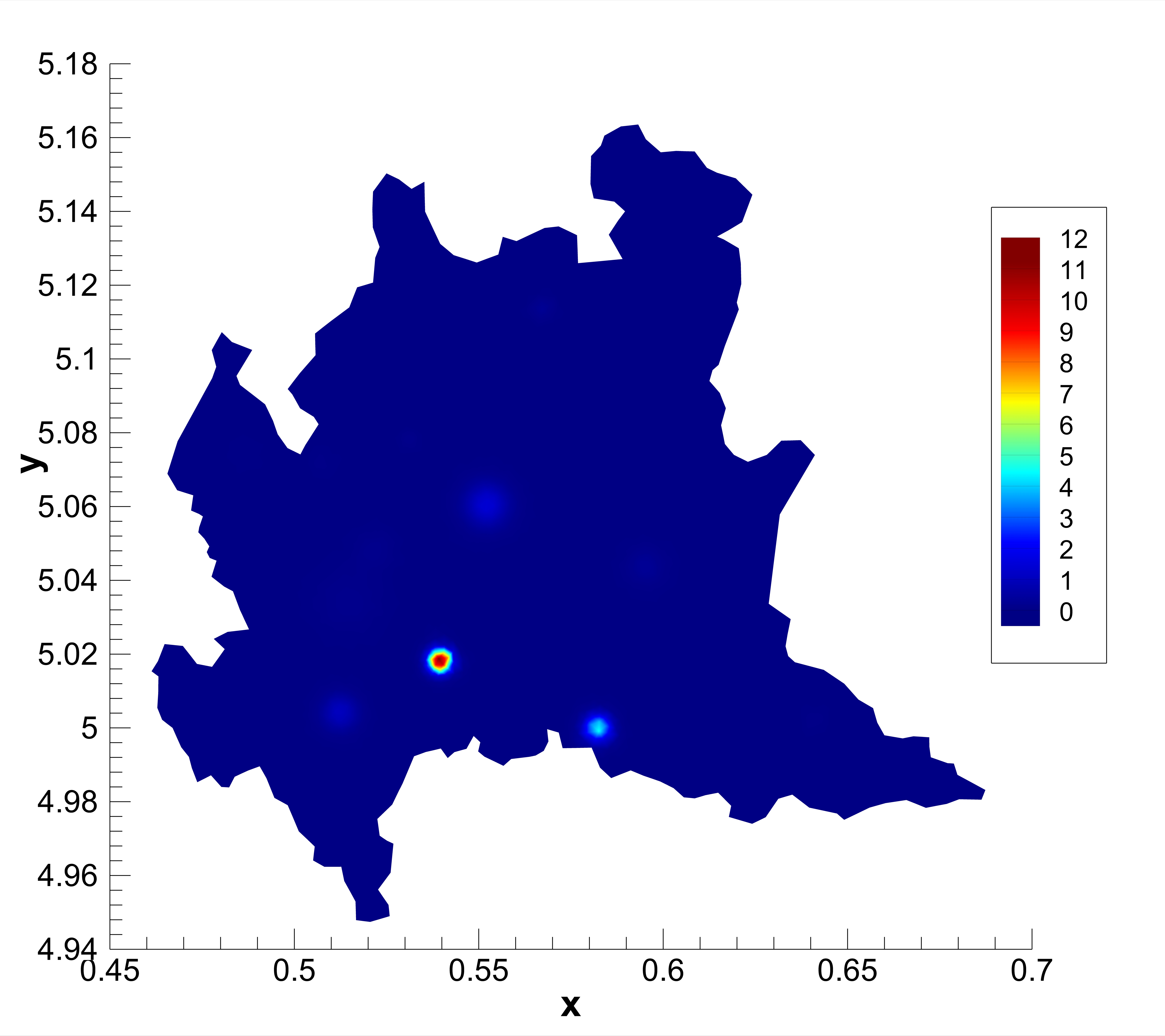}
\includegraphics[width=0.45\textwidth]{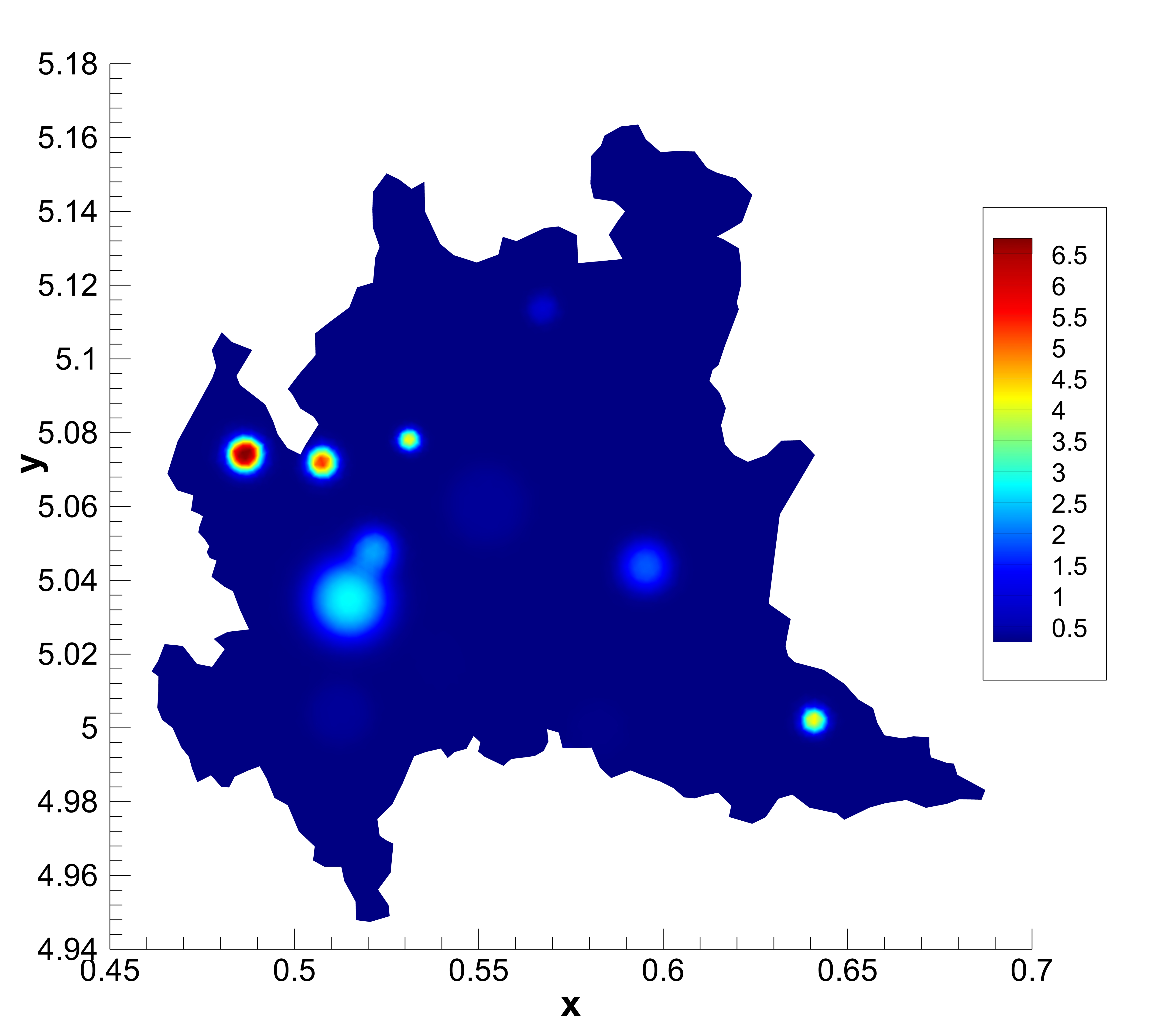}
\caption{Initial distribution (on February 27, 2020) of the infected population $E_{T}^0+I_{T}^0+A_{T}^0$ (left) and of the reproduction number $R_0(0)$ (right) in the Lombardy Region.}
\label{fig.computational_domain_LB}
\end{figure}
Nevertheless, with the proposed methodology it is possible to present the heterogeneity underlying the basic reproduction number at the local scale, as shown for the Lombardy case in Fig. \ref{fig.computational_domain_LB} (bottom), together with the initial global amount of infected people $E_{T}^0(x,y)+I_{T}^0(x,y)+A_{T}^0(x,y)$ present in the domain.

\begin{figure}[t]
\centering
\includegraphics[width=0.45\textwidth]{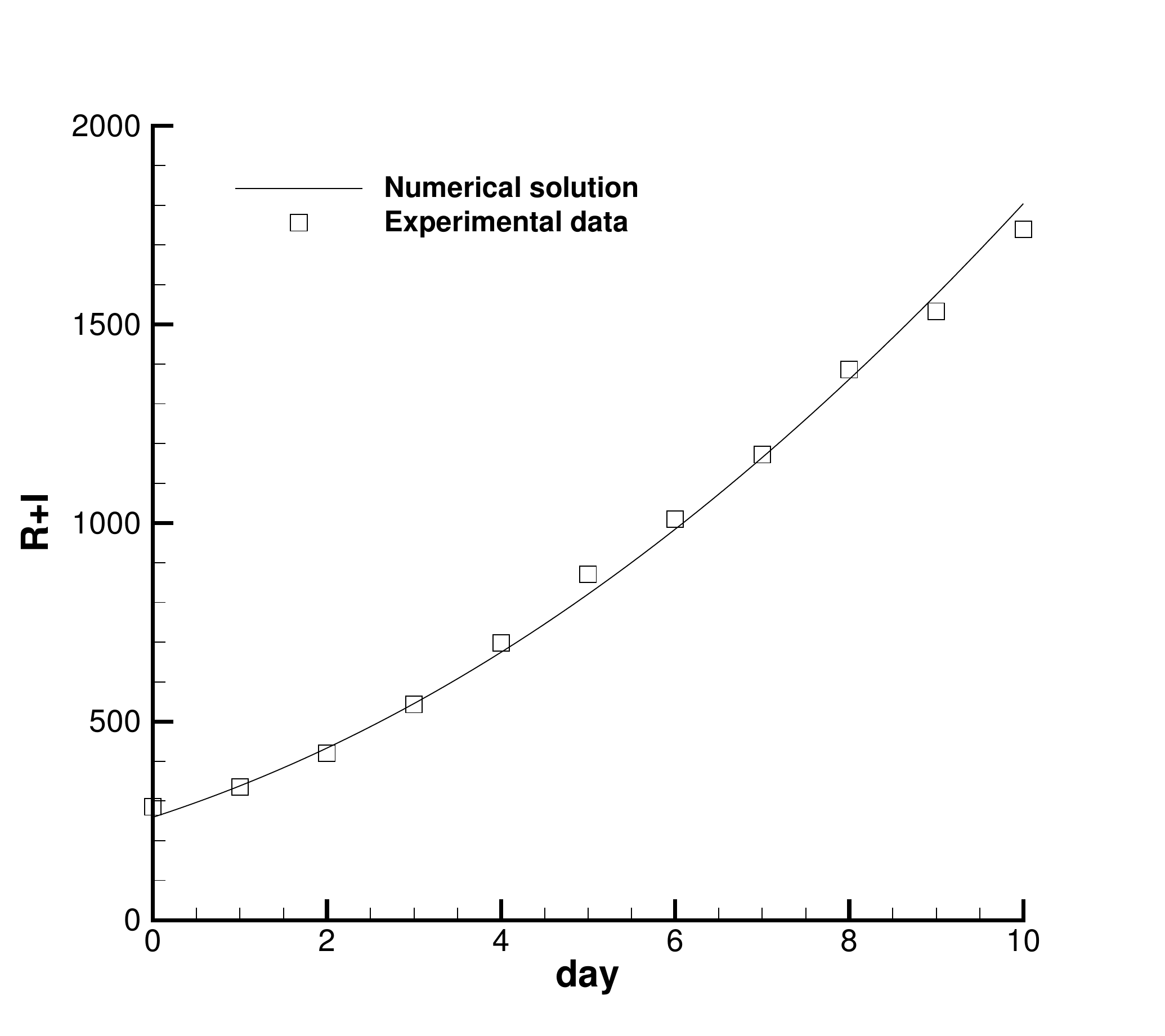}
\includegraphics[width=0.45\textwidth]{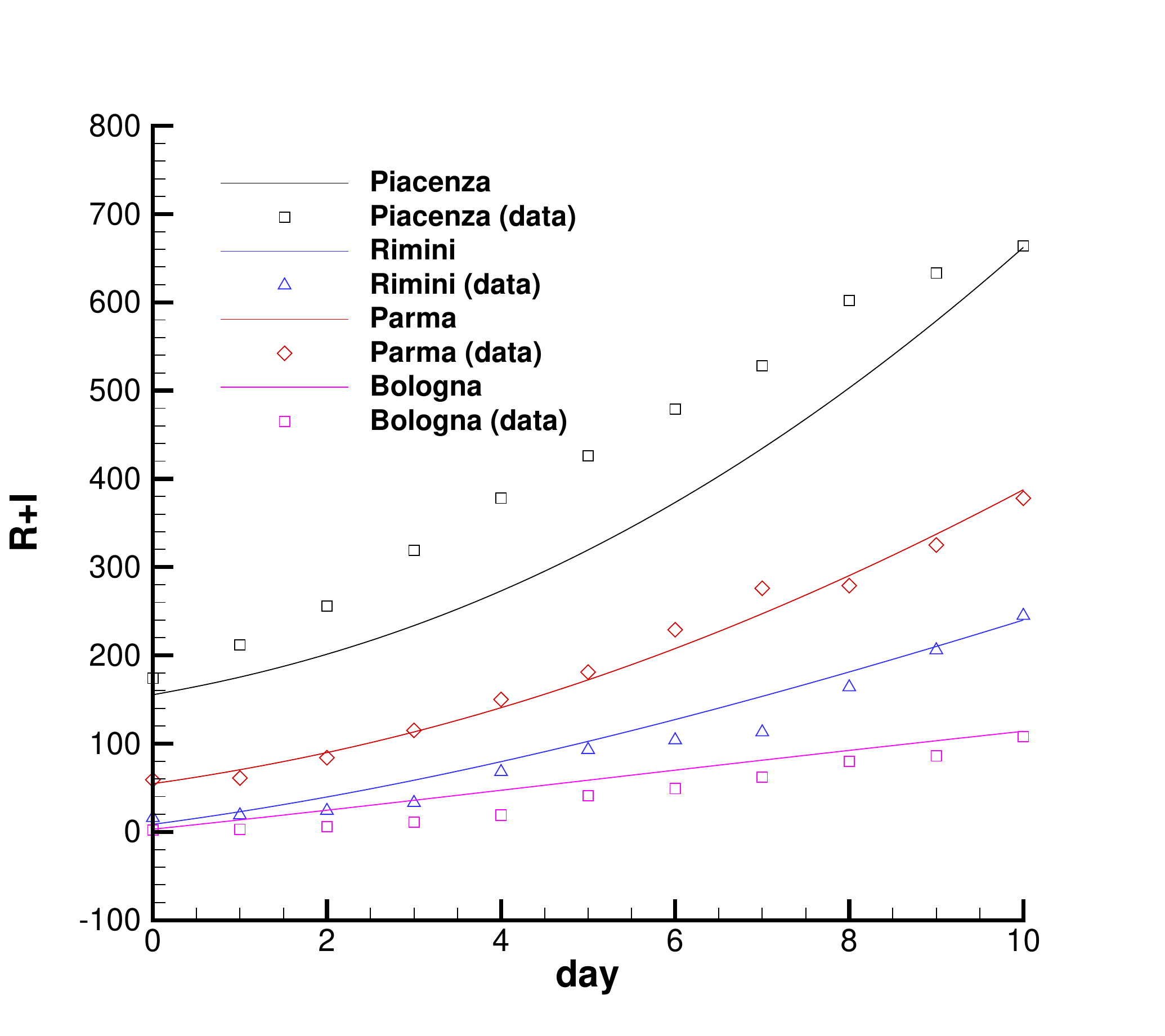}
\caption{Left: time evolution of total infected and recovered population $(R+ I)$ compared against experimental data for the Emilia-Romagna Region. Right: time evolution of total infected and recovered population $(R+I)$ compared against experimental data for the province of Piacenza (black), Parma (red), Bologna (purple) and Rimini (blue).}
\label{fig.results_ER}
\end{figure}
\begin{figure}[t]
\centering
\includegraphics[width=0.32\textwidth]{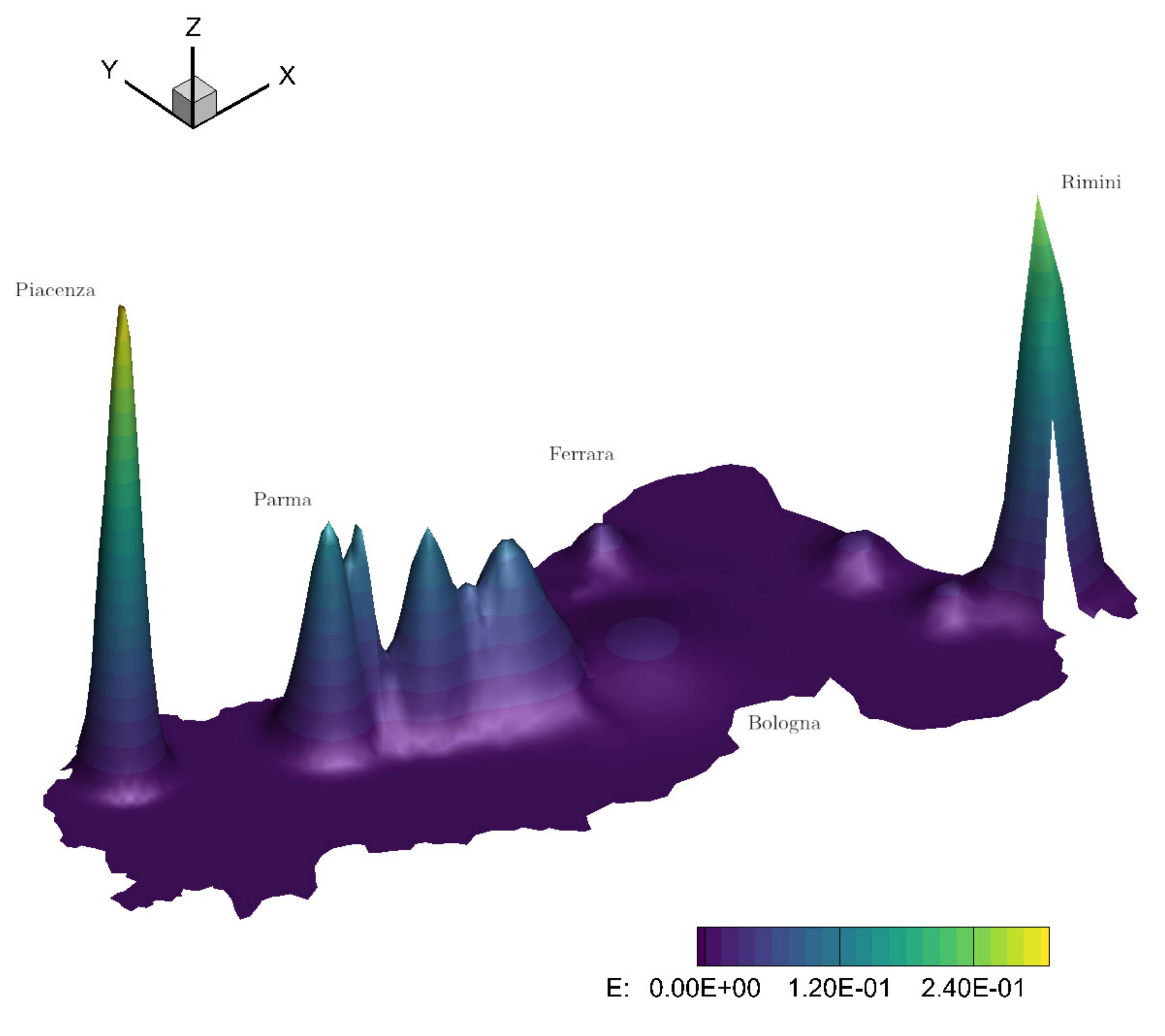}
\includegraphics[width=0.32\textwidth]{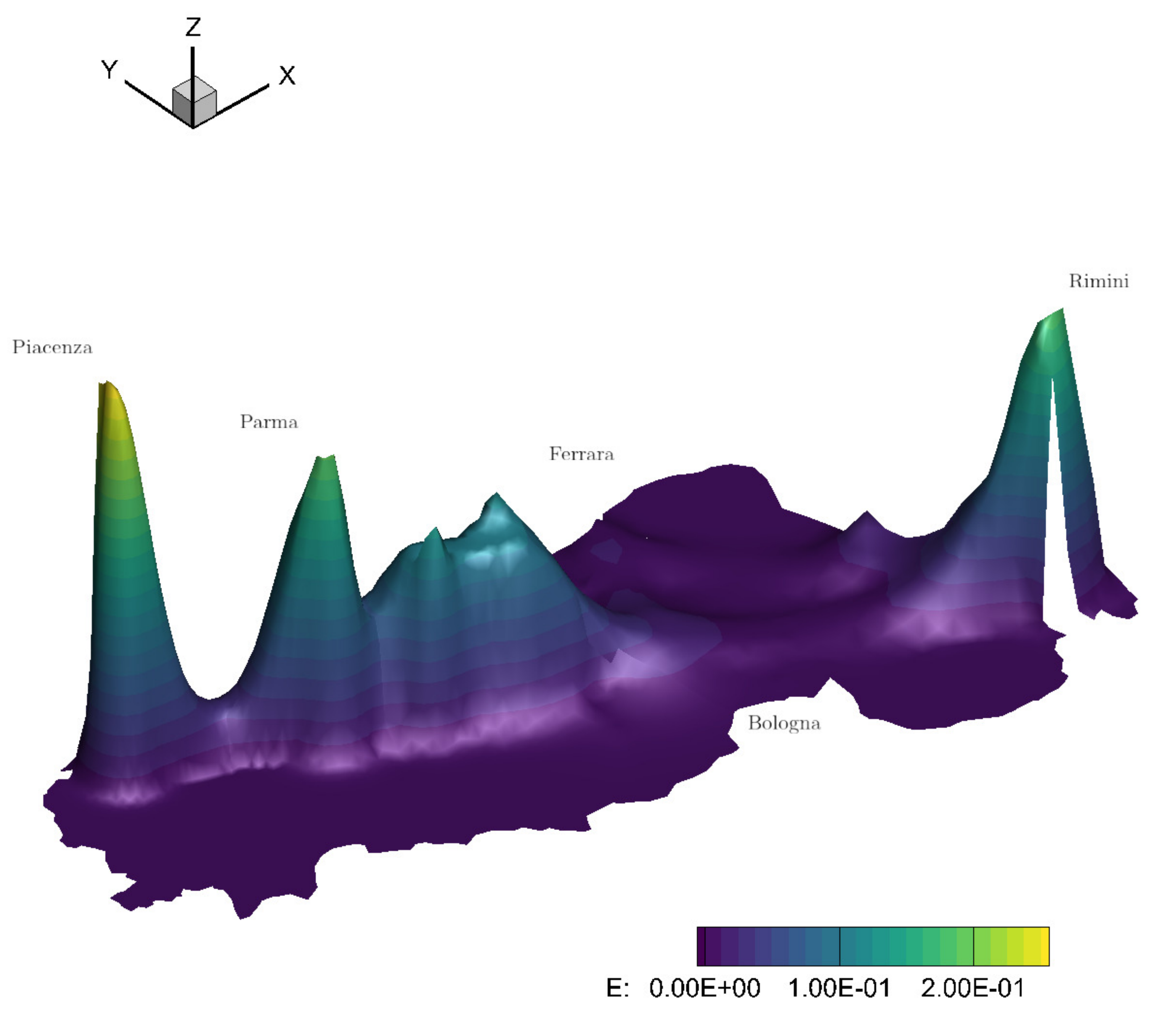}
\includegraphics[width=0.32\textwidth]{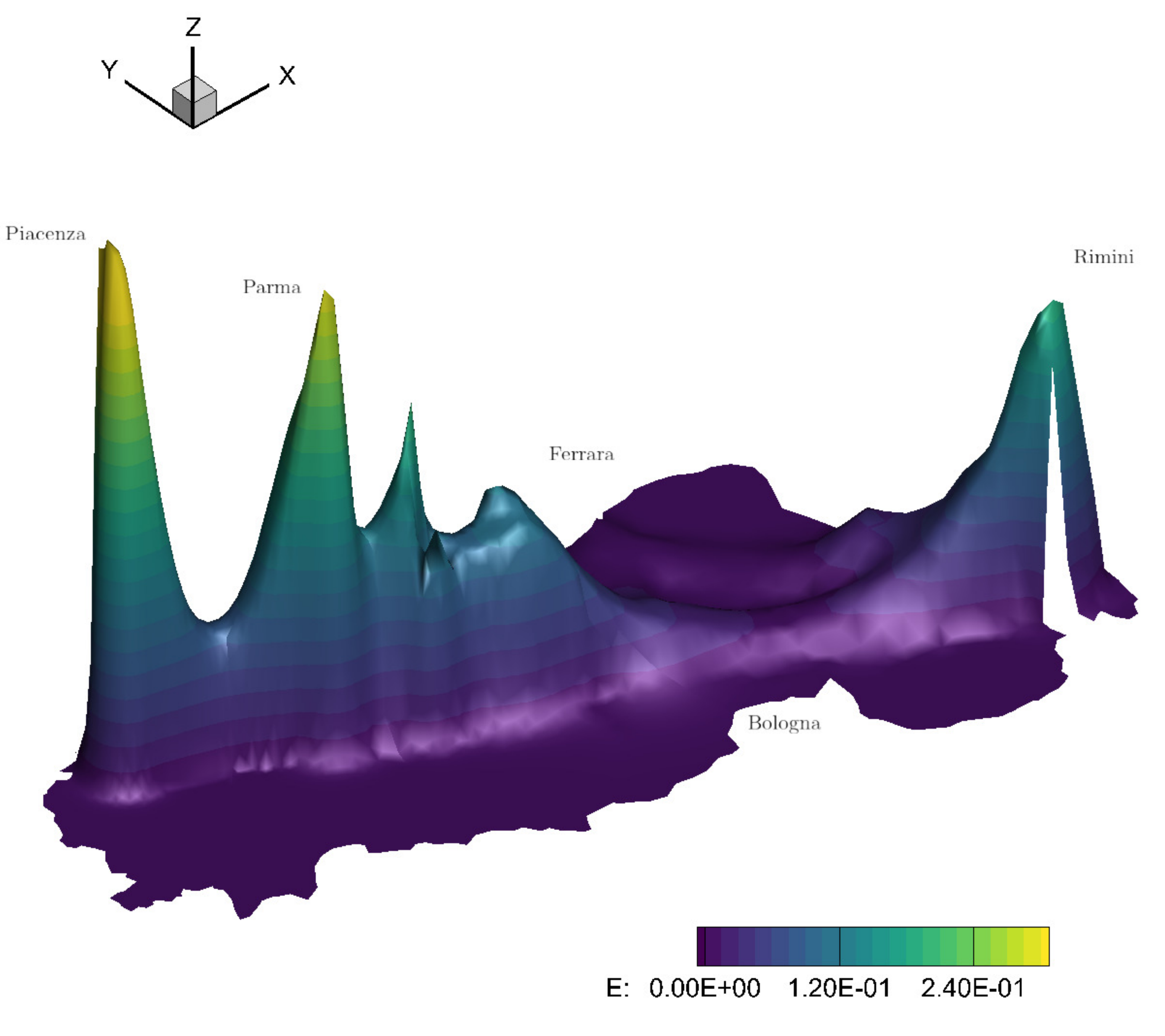}
\caption{Distribution of exposed population $E$, including asymptomatic, on March 1, 4 and 10, 2020 (from left to right) in Emilia-Romagna.}
\label{fig.results_ER_2D}
\end{figure}

Fig.~\ref{fig.results_ER} plots a comparison against the measured data at the Regional level, reported by the Civil Protection Department of Italy, and the same comparison for the province of Piacenza, Parma, Bologna and Rimini, depicting an overall very good agreement.
Figure.~\ref{fig.results_ER_2D} shows the time evolution of the exposed population $E$, including asymptomatic, which is moving from both Piacenza and Rimini towards the center of the region and the city of Bologna, then spreading northern in the direction of Ferrara. The wave of the exposed population is clearly visible, highlighting the hyperbolic regime of the model.

\begin{figure}[t!]
\centering
\includegraphics[width=0.32\textwidth]{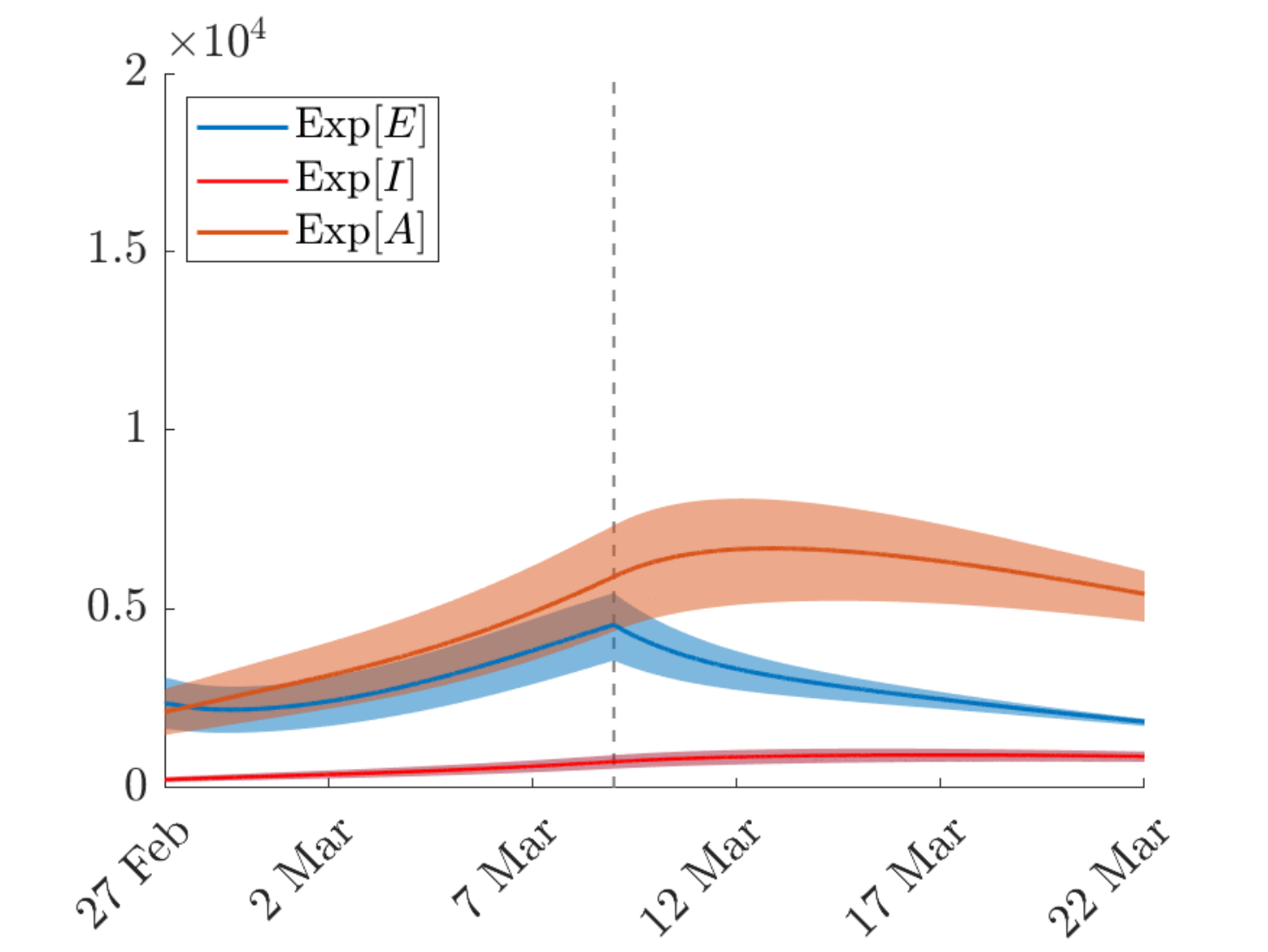}
\includegraphics[width=0.32\textwidth]{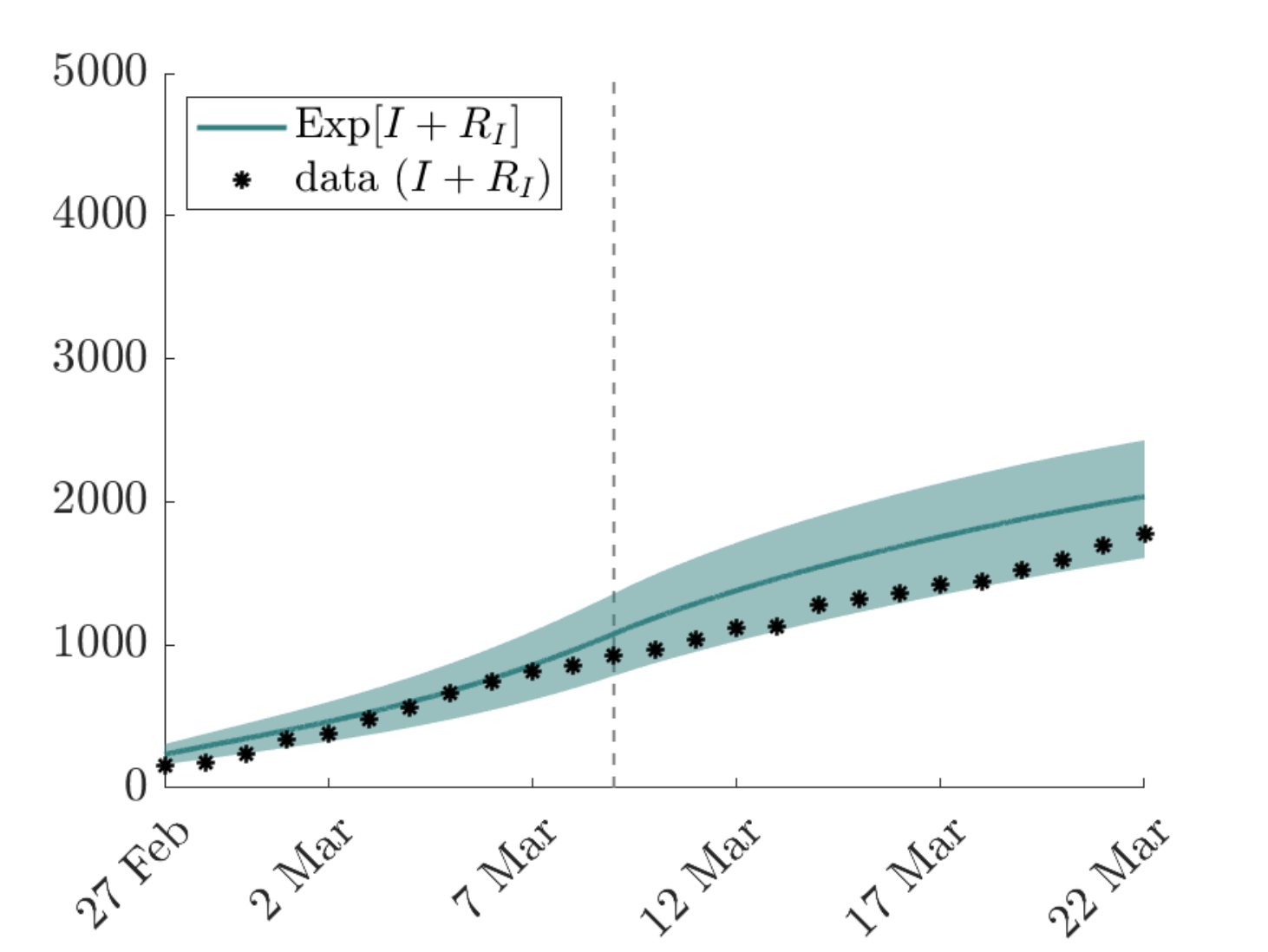}
\includegraphics[width=0.32\textwidth]{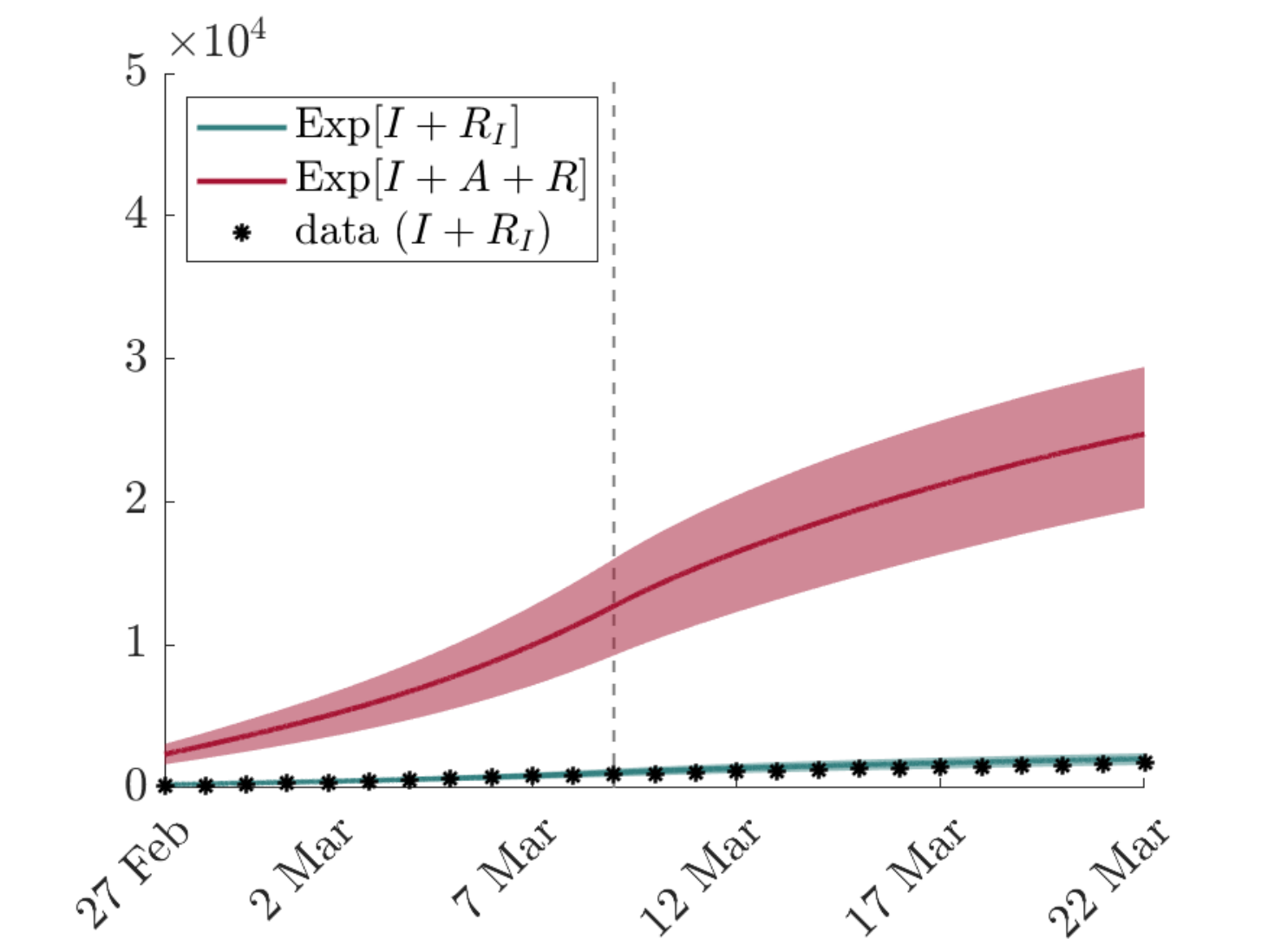}
\includegraphics[width=0.32\textwidth]{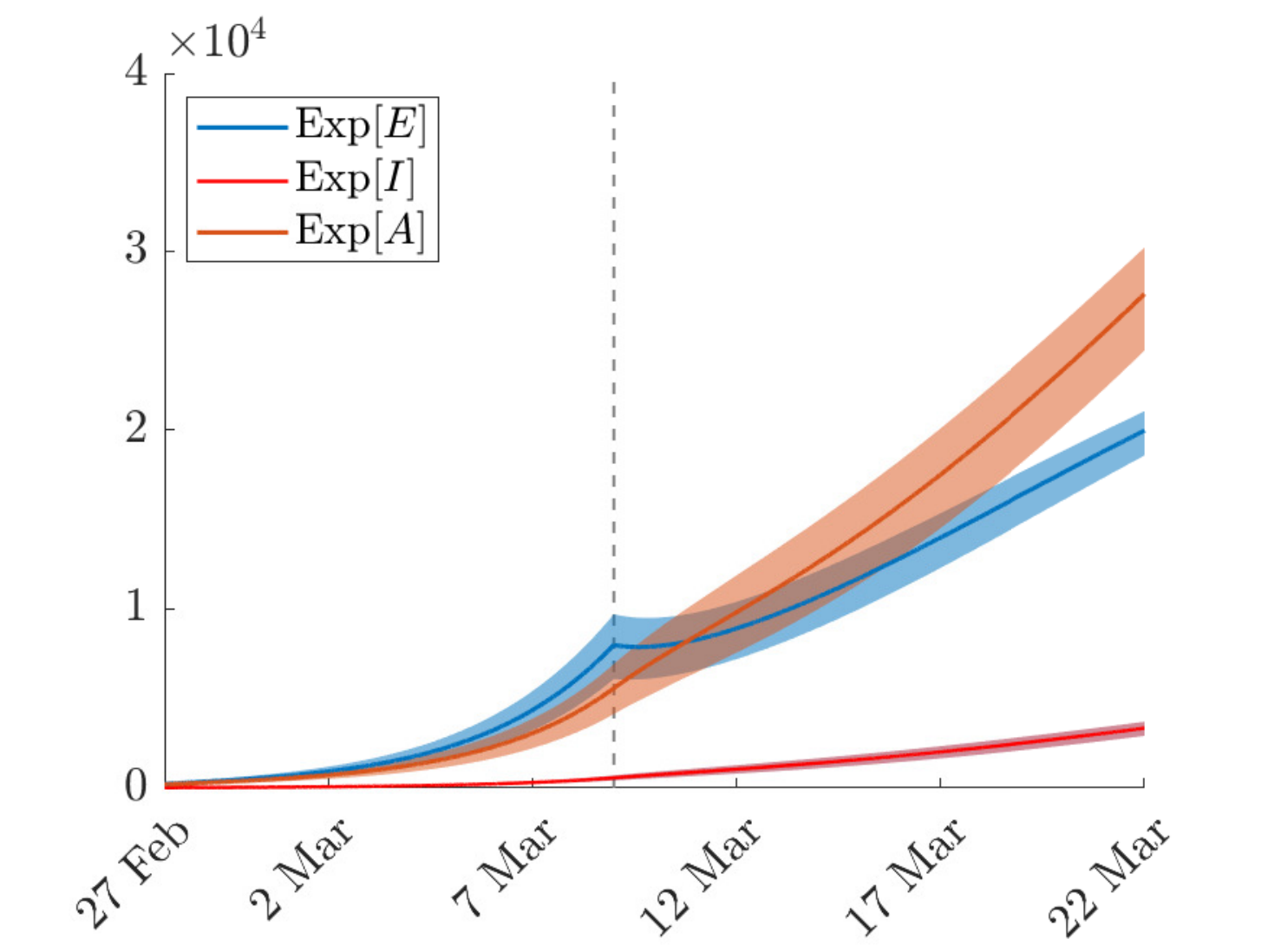}
\includegraphics[width=0.32\textwidth]{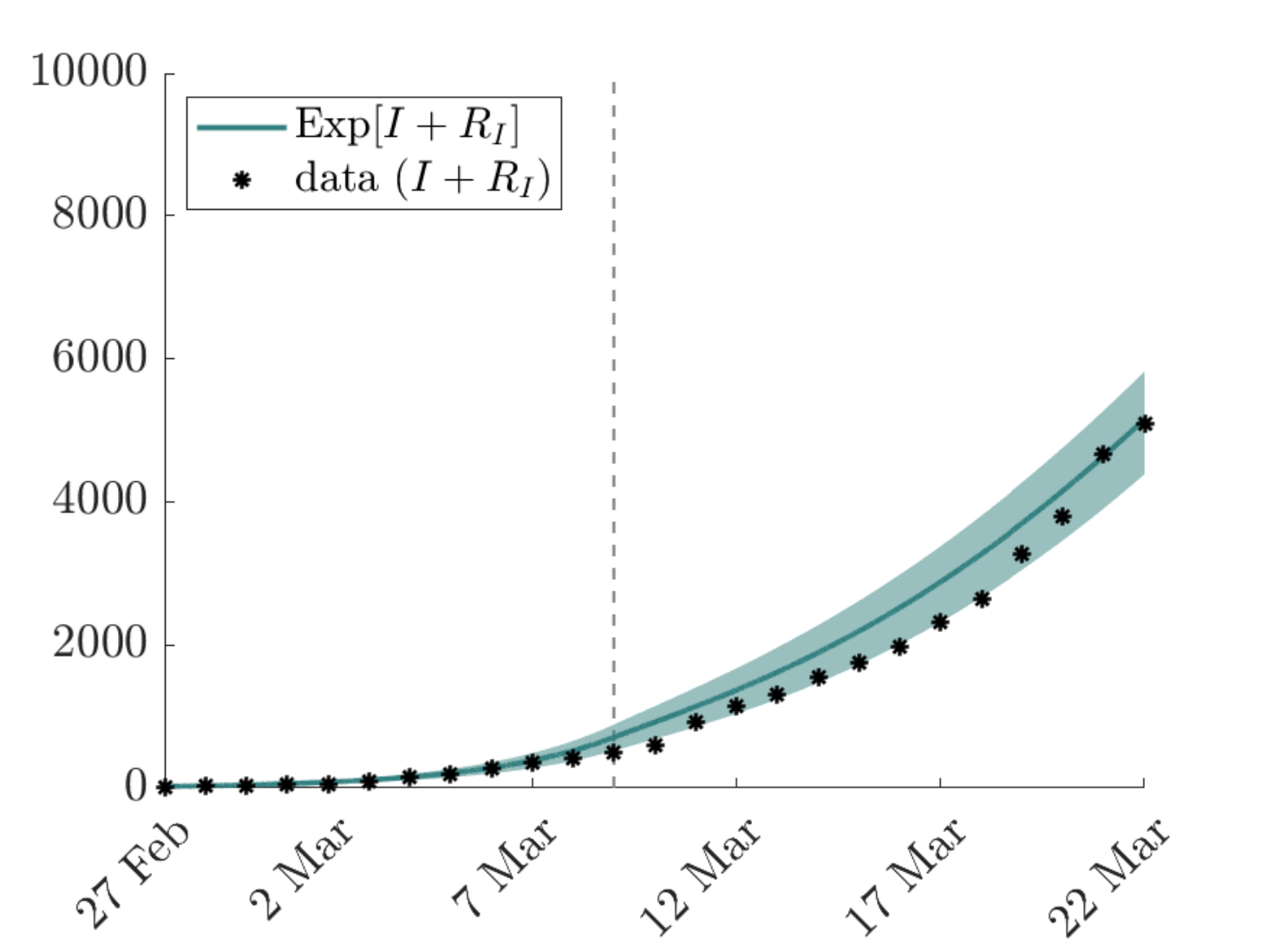}
\includegraphics[width=0.32\textwidth]{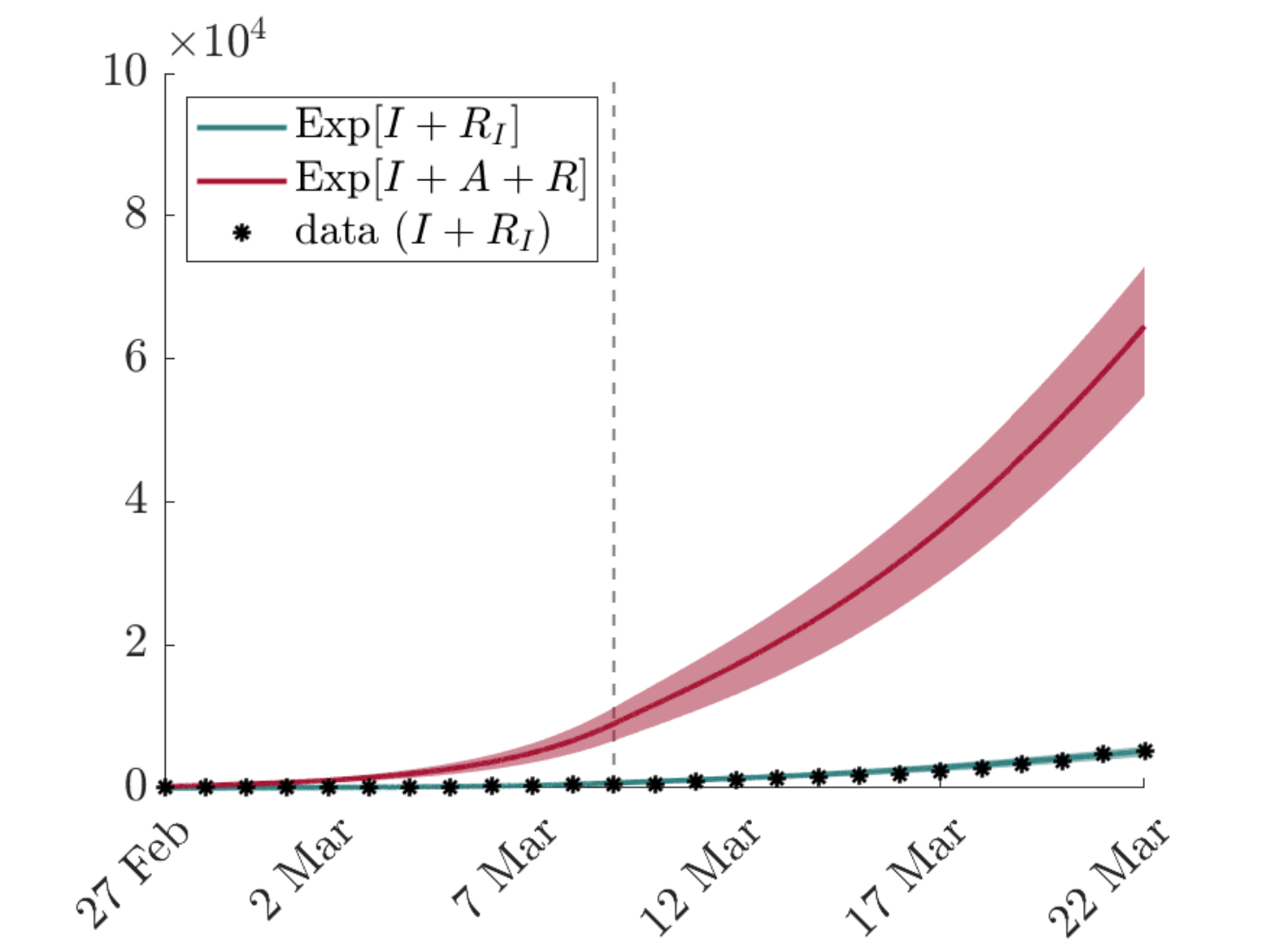}
\includegraphics[width=0.32\textwidth]{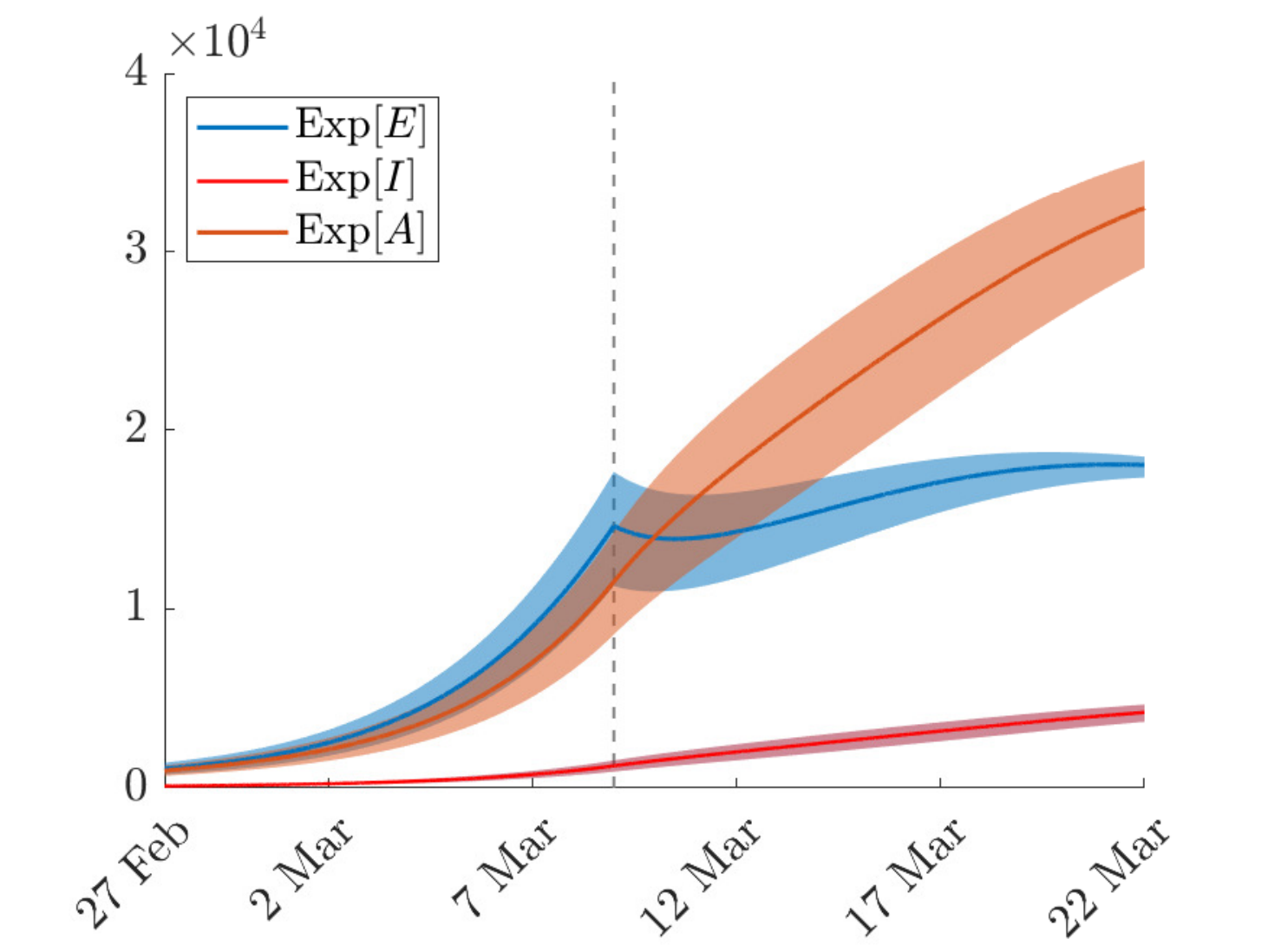}
\includegraphics[width=0.32\textwidth]{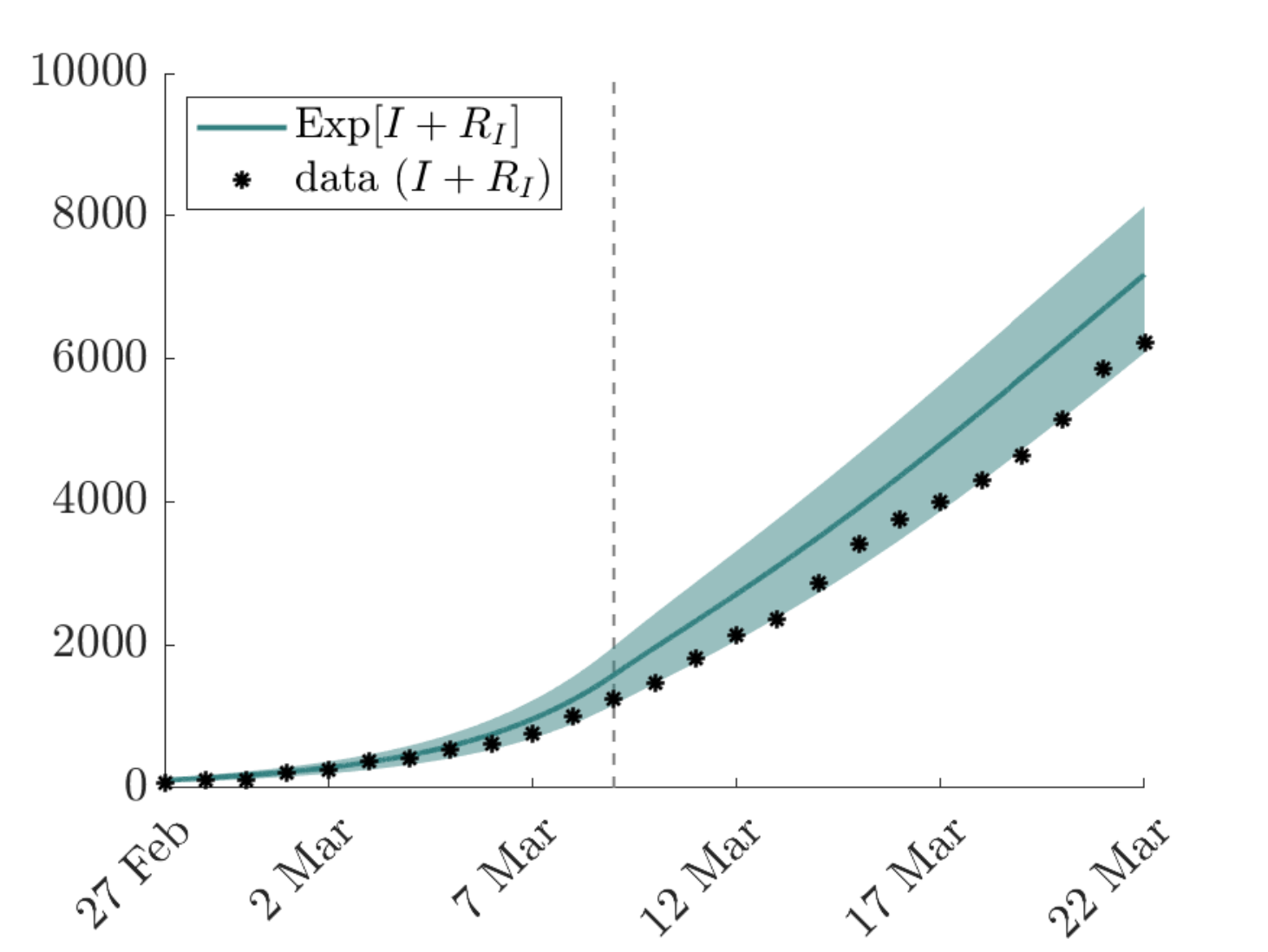}
\includegraphics[width=0.32\textwidth]{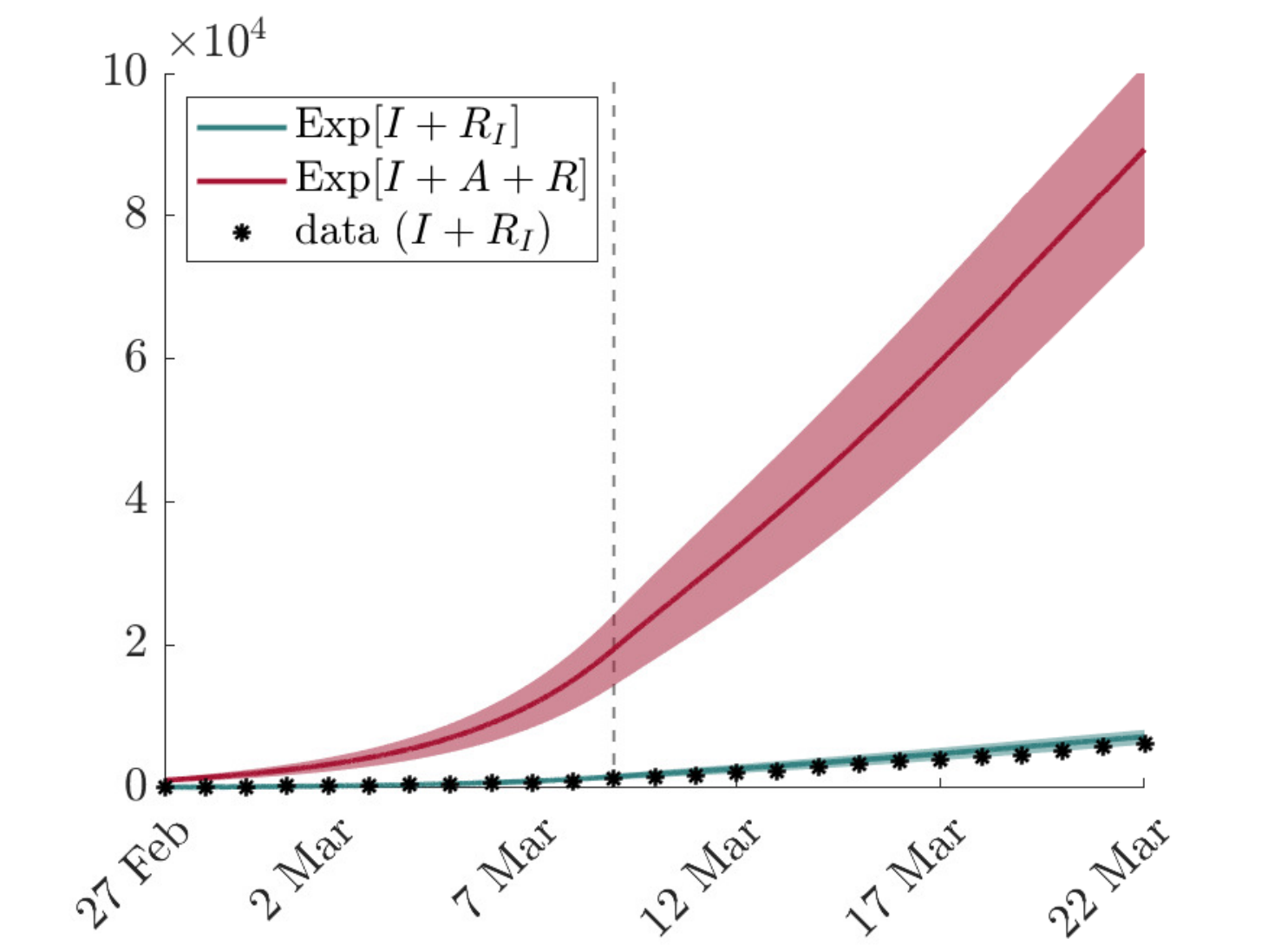}
\includegraphics[width=0.32\textwidth]{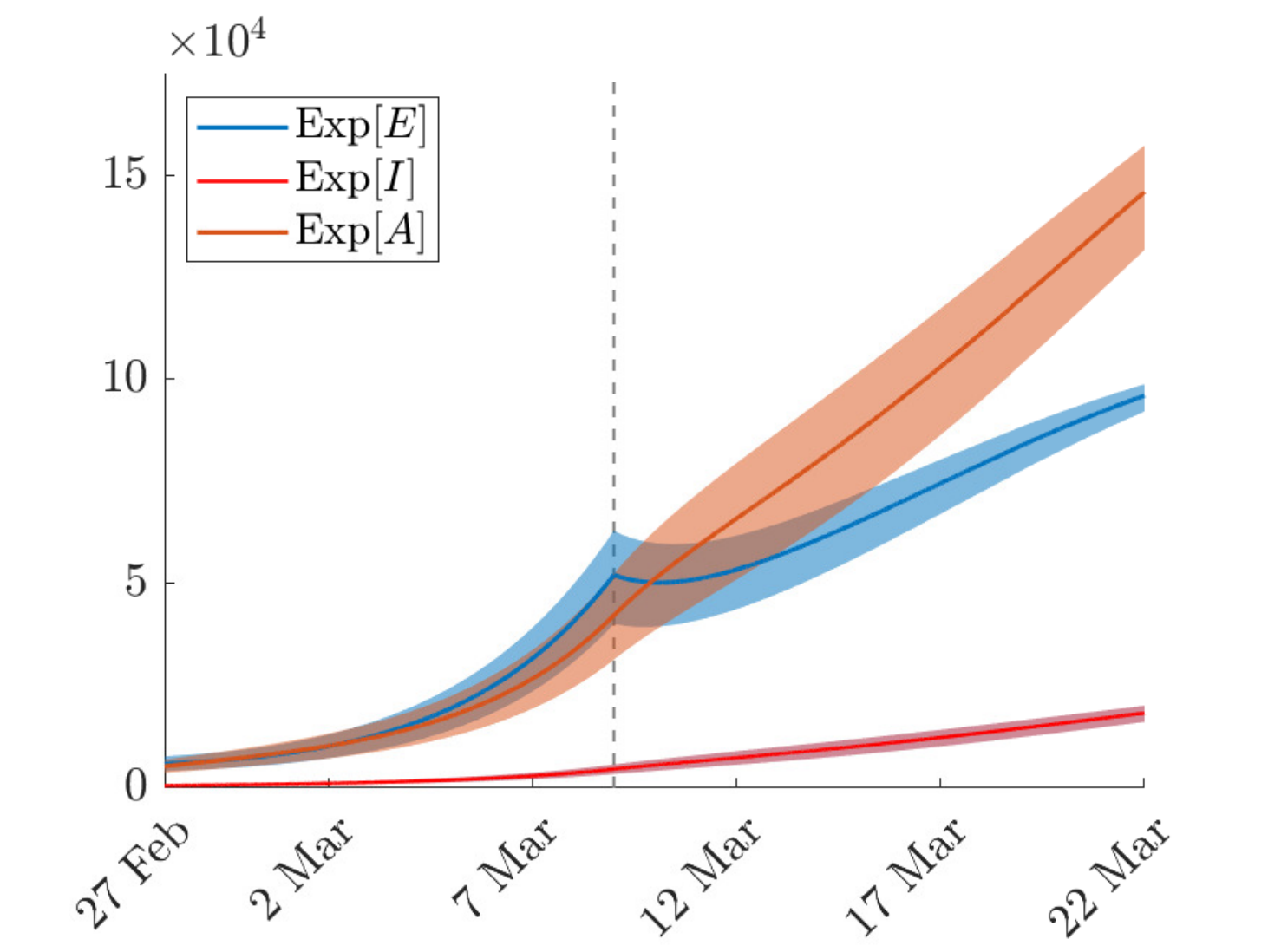}
\includegraphics[width=0.32\textwidth]{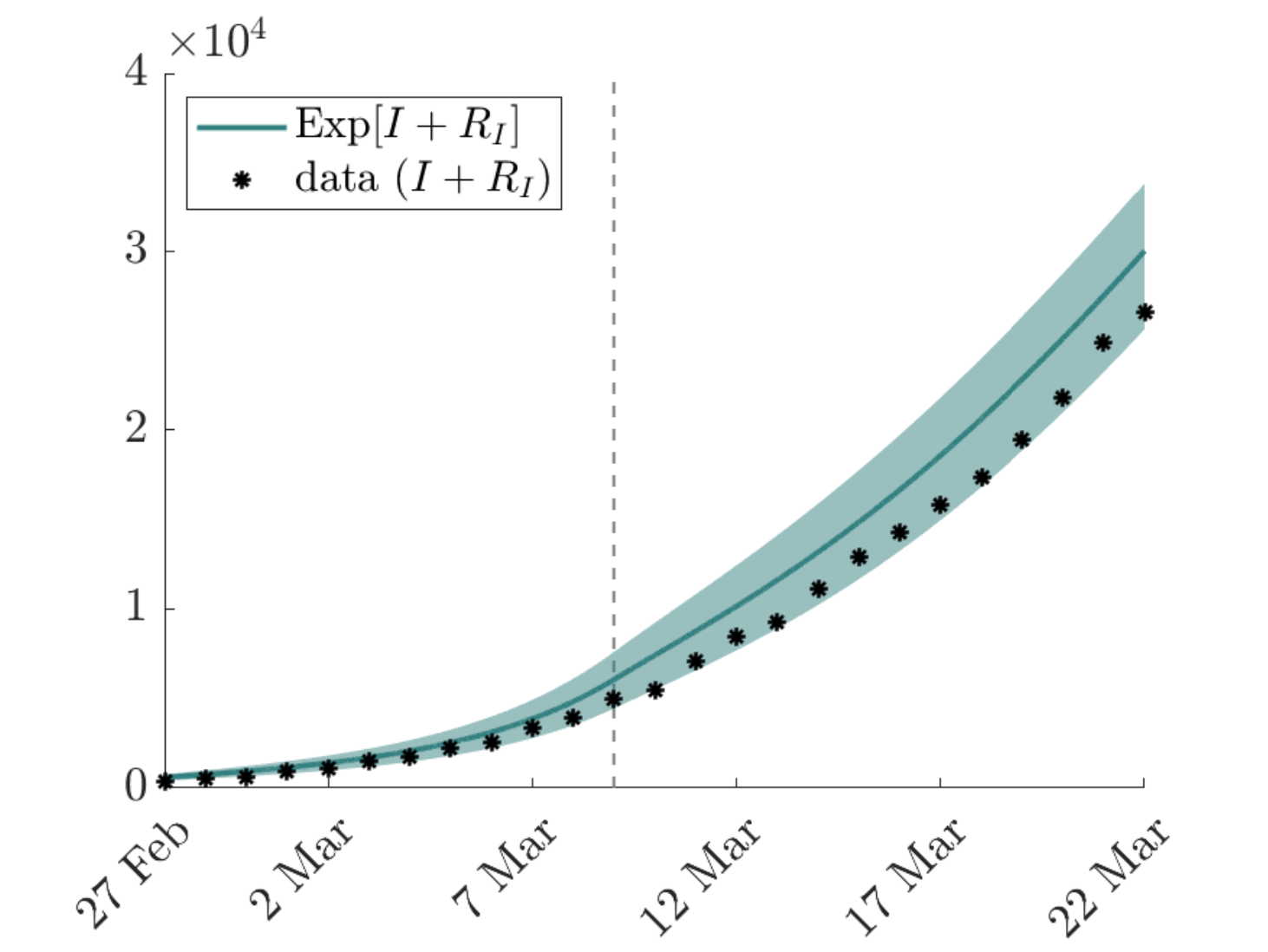}
\includegraphics[width=0.32\textwidth]{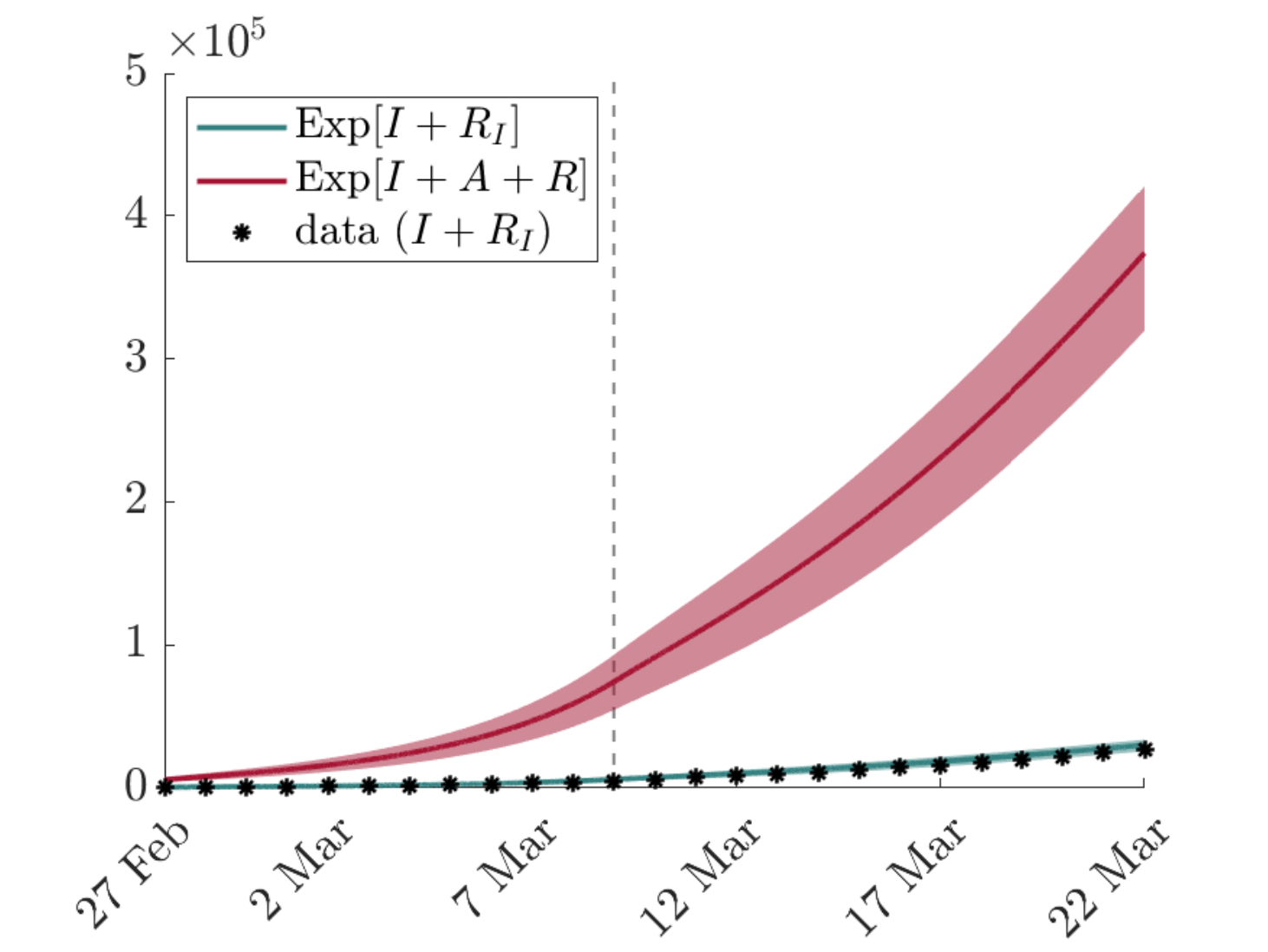}
\caption{Expected evolution Exp[$\cdot$] in time, with 95\% confidence intervals, for chosen representative cities, Lodi (first row), Milan (second row), Bergamo (third row), and the whole Lombardy network (fourth row), of: compartments $E$, $A$, $I$ (left); severe infectious ($I+R_I$) compared with data of cumulative infectious taken from the COVID-19 repository of the Civil Protection Department of Italy  (middle); severe infectious ($I+R_I$) with respect to the effective cumulative amount of total infectious people, including asymptomatic and mildly symptomatic individuals ($I+A+R$) (right).
Vertical dashed lines identify the onset of governmental lockdown restrictions.}
\label{fig.results_Lombardy}
\end{figure}
\begin{figure}[t!]
\centering
\includegraphics[width=0.45\textwidth]{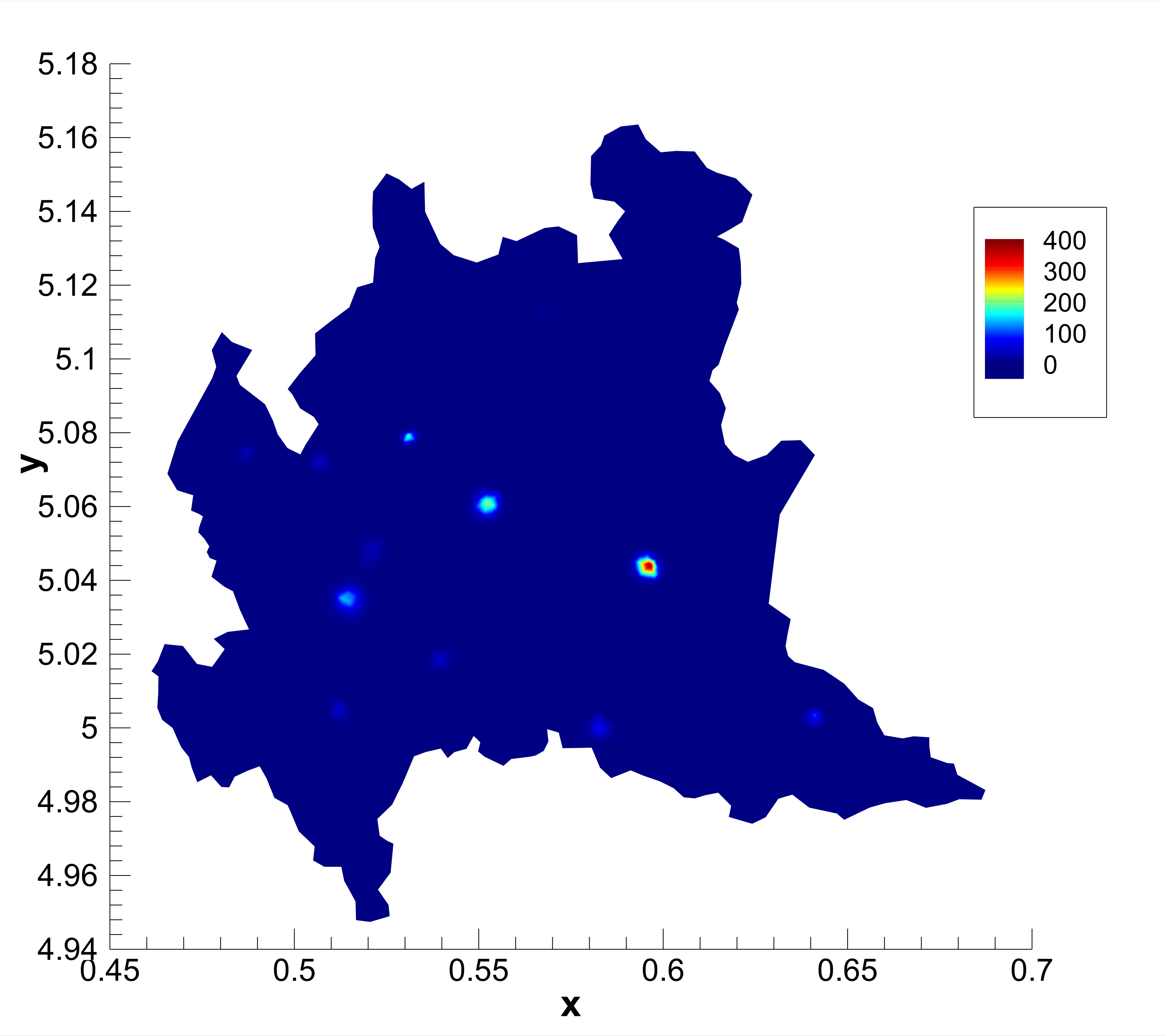}
\includegraphics[width=0.45\textwidth]{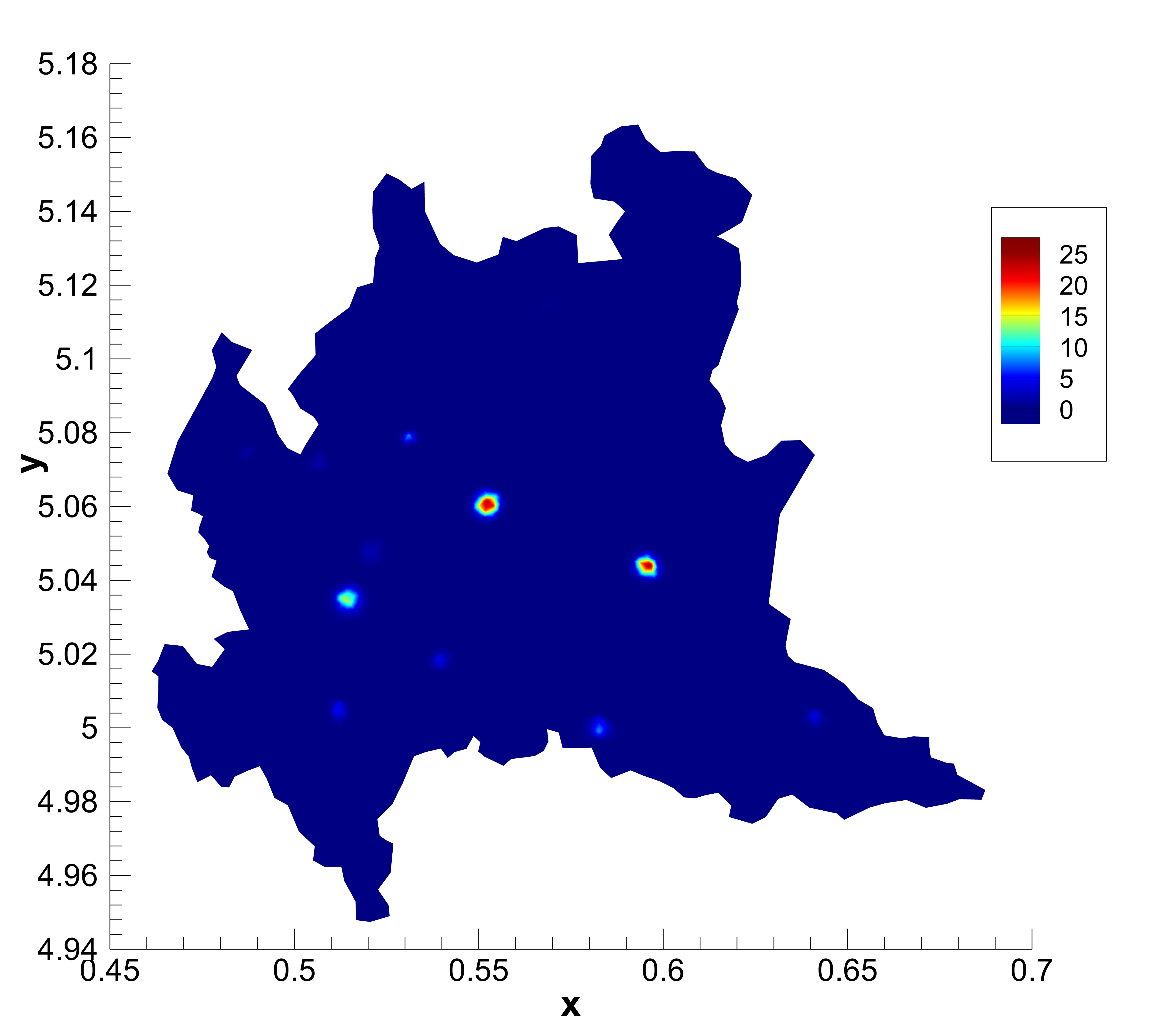}
\caption{Expectation (left) and variance (right) of the cumulative amount of infected people $E_T+A_T+I_T$ at the end of the simulation (March 22, 2020) in the Lombardy Region.}
\label{fig.results_2D}
\end{figure}

In Fig.~\ref{fig.results_Lombardy} (first column), the expected evolution in time of the infected individuals, together with 95\% confidence intervals, is shown for exposed $E$, highly symptomatic subjects $I$ and asymptomatic or weakly symptomatic people $A$, for three representative cities, namely Lodi, Milan and Bergamo, and the whole Lombardy Region. Here it is already appreciable the heterogeneity of the diffusion of the virus. Indeed, from the different y-axis scales adopted for the plot of the provinces, it can be noticed that Milan and Bergamo present a consistently higher contagion with respect to the one shown in Lodi.
From the same Fig. \ref{fig.results_Lombardy} (second column) it can be observed that the lower bound of the confidence interval of the cumulative amount in time of highly symptomatic individuals is in line with data reported by the Civil Protection Department of Italy. As expected, due to the uncertainty taken into account, the mean value of the numerical result in each city is higher than the registered one.
Also the comparison between the expected evolution in time of the cumulative amount of severe infectious with respect to the effective cumulative amount of total infectious people, including asymptomatic and mildly symptomatic individuals, is shown in Fig. \ref{fig.results_Lombardy} (third column). From this figure it is clear that the number of infections recorded during the first outbreak of COVID-19 in Lombardy represents a clear underestimation of the actual trend of infection suffered by the Region and by Italy as a whole, and how the presence of asymptomatic subjects, not detected, has affected the pandemic evolution. 
Results concerning the rest of the cities of the Region can be found in^^>\cite{bertaglia2021b}.

In Fig. \ref{fig.results_2D}, final expectation and variance of the cumulative amount of infected people, namely $E_T+A_T+I_T$, are reported in the 2D framework of Lombardy. If comparing Fig. \ref{fig.results_2D} (top left) with \ref{fig.computational_domain} (bottom left), it can be noticed that, at the end of March, the virus is no longer mostly affecting the province of Lodi and Cremona, but has been spread, arriving to hit most of all Brescia, Milan and Bergamo.

\section{Concluding remarks and research perspectives}
In this review paper, we presented a series of recent results obtained in the field of kinetic modeling applied to epidemiology. In particular, we focused on three main aspects: the influence of social features such as the number of contacts, wealth and age of individuals, the design of effective control techniques even in the presence of uncertain data, and finally, the impact on the pandemic of the movements of individuals both on urban and extra-urban scales. All these aspects proved essential in order to present realistic scenarios on the spread of the epidemic and in agreement with the observed data.  

The modeling approach presented here, although in some cases developed for the sake of simplicity on compartmental models with a very basic structure, can be easily extended, as analyzed in the last part of the survey, also to more realistic models for the spread of the COVID-19 epidemic. In particular, given the generality of the social structure modelling proposed in this survey, this opens interesting perspectives in future directions by going to evaluate the impact of additional features that can influence the evolution of the pandemic, such as the viral load^^>\cite{DellaMLoy,loy2021} or the spread of fake-news^^>\cite{Jona}. The former in particular plays a decisive role in analyzing the influence of the so-called super-spreaders^^>\cite{NATSS,Nielsen}, while the latter we have seen play a key role regarding the vaccination campaign^^>\cite{fake1,fake2}. 

\subsection{Data sources}
\label{sec:ds}
With respect to the numerical results presented in the simulations, specifically in Sections 2.15, 3.2.2, 4.1.6 and 4.2.4, the following data repositories were used. The GitHub repository of the Italian Civil Protection Department\footnote{\tt https://github.com/pcm-dpc/COVID-19}; the John Hopkins University GitHub repository^^>\cite{DDG}; Regione Lombardia, Italy, Commuters Data\footnote{\tt https://www.dati.lombardia.it/Mobilit-e-trasporti/Matrice-OD2020-Passeggeri/ hyqr-mpe2}; the Italian National Institute of Statistics, ISTAT\footnote{\tt https://demo.istat.it/}; Geographical Data from ISTAT\footnote{\tt https://www4.istat.it/it/archivio/209722}; Regione Emilia-Romagna, Italy, Commuters Data\footnote{\tt https://sasweb.regione.emilia-romagna.it/statistica/SceltaAnno.do?analisi= matPend2011\_2015}.

\begin{acknowledgement}
This work has been written within the
activities of GNFM and GNCS groups of INdAM (National Institute of High Mathematics). G.A., G.B., W.B., G.D. and  L. P. acknowledge the support of MIUR-PRIN Project 2017, No. 2017KKJP4X \emph{Innovative numerical methods for evolutionary partial differential equations and applications}. GA also acknowledges partial support from the Program Ricerca di Base 2019 of the University of Verona entitled ``Geometric Evolution of Multi-Agent Systems''. GB holds a Research Fellowship from INdAM. G.T. and M. Z. were partially supported by the MIUR Program (2018-2022), \emph{Dipartimenti di Eccellenza}, Department of Mathematics, University of Pavia. 
\end{acknowledgement}

\end{document}